%% file: 00_main.tex
\documentclass[12pt,a4paper]{report}
\usepackage[utf8]{inputenc}
\usepackage{fvextra}
\usepackage{csquotes}
\usepackage{float}
\usepackage{siunitx}

\input{00_thesis_setup.tex}

\begin{document}

\pagenumbering{roman}
\include{01_thesis_title_page/title_page}
\include{02_frontmatter/abstract}
\include{02_frontmatter/acknowledgments}

\tableofcontents
\listoffigures
\listoftables

\cleardoublepage
\pagenumbering{arabic}
\include{03_chapters/chapter1}
\include{03_chapters/chapter2}

\include{03_chapters/chapter3}

\include{03_chapters/chapter4}

\include{04_appendices/appendices}

\cleardoublepage
\printbibliography[heading=bibintoc, title={References}]

\end{document}

%% file: 00_thesis_setup.tex
\usepackage[T1]{fontenc}
\usepackage[english]{babel}
\usepackage{fvextra}
\usepackage{csquotes}
\usepackage{longtable}

\usepackage{tgtermes}
\usepackage[scaled=0.94]{helvet}
\usepackage{microtype}
\usepackage{fix-cm}

\usepackage[margin=3cm, marginparwidth=2cm]{geometry}
\usepackage{setspace}

\usepackage{amsmath,amssymb,amsthm,mathtools,bm}

\DeclareMathOperator*{\argmin}{arg\,min}

\usepackage{graphicx}
\usepackage{epstopdf}
\usepackage{subcaption}
\usepackage{csvsimple}
\usepackage{tabularx}
\usepackage{booktabs}
\usepackage{siunitx}
\usepackage{array}
\usepackage{multirow}
\usepackage{threeparttable}
\usepackage{pdflscape}
\usepackage{tikz}
\usepackage{pgfplots}
\pgfplotsset{compat=1.17}

\usepackage{xcolor}
\usepackage{minted}

\usepackage[hidelinks]{hyperref}
\usepackage{cleveref}

\usepackage[
  style=apa,
  backend=biber,
  giveninits=true,
  maxbibnames=10,
  minbibnames=1,
  maxcitenames=2,
  mincitenames=1,
  sorting=nyt,
  hyperref=true,
  backref=true  
  ]{biblatex}
\DeclareLanguageMapping{english}{english-apa}
\usepackage{titlesec}

\titleformat{\chapter}[display]
  {\Large\bfseries}
  {\chaptername\ \thechapter}
  {19pt}{}
\titlespacing*{\chapter}{0pt}{-1em}{2em}

\titleformat{\section}
  {\large\bfseries}
  {\thesection}
  {1em}{}
\titlespacing*{\section}{0pt}{1em}{0.5em}

\titleformat{\subsection}
  {\normalsize\bfseries\itshape}
  {\thesubsection}
  {1em}{}
\titlespacing*{\subsection}{1.5em}{0.75em}{0.5em}

\titleformat{\subsubsection}[hang]
  {\normalsize\itshape}
  {}
  {0pt}{}
\titlespacing*{\subsubsection}{3em}{0.5em}{0.2em}

\usepackage{longtable} %
\usepackage{lscape} %
\usepackage[table]{xcolor} %

\newcommand{\leading}{Feature - Leading Indicator}
\newcommand{\coincident}{Feature - Coincident Indicator}
\newcommand{\exogenous}{Feature - Exogenous Variable}
\newcommand{\otherVar}{Feature - Other Variable}
\newcommand{\laggedVar}{Lagged Variable - Feature}
\newcommand{\dummyVar}{Dummy Variable - Feature}

%% file: 01_thesis_title_page/title_page.tex
\begin{titlepage}

  \sffamily
  \bfseries
  \fontsize{11}{13.2}\selectfont

  \begin{center}
    \includegraphics[width=1.5736in,height=1.5736in]{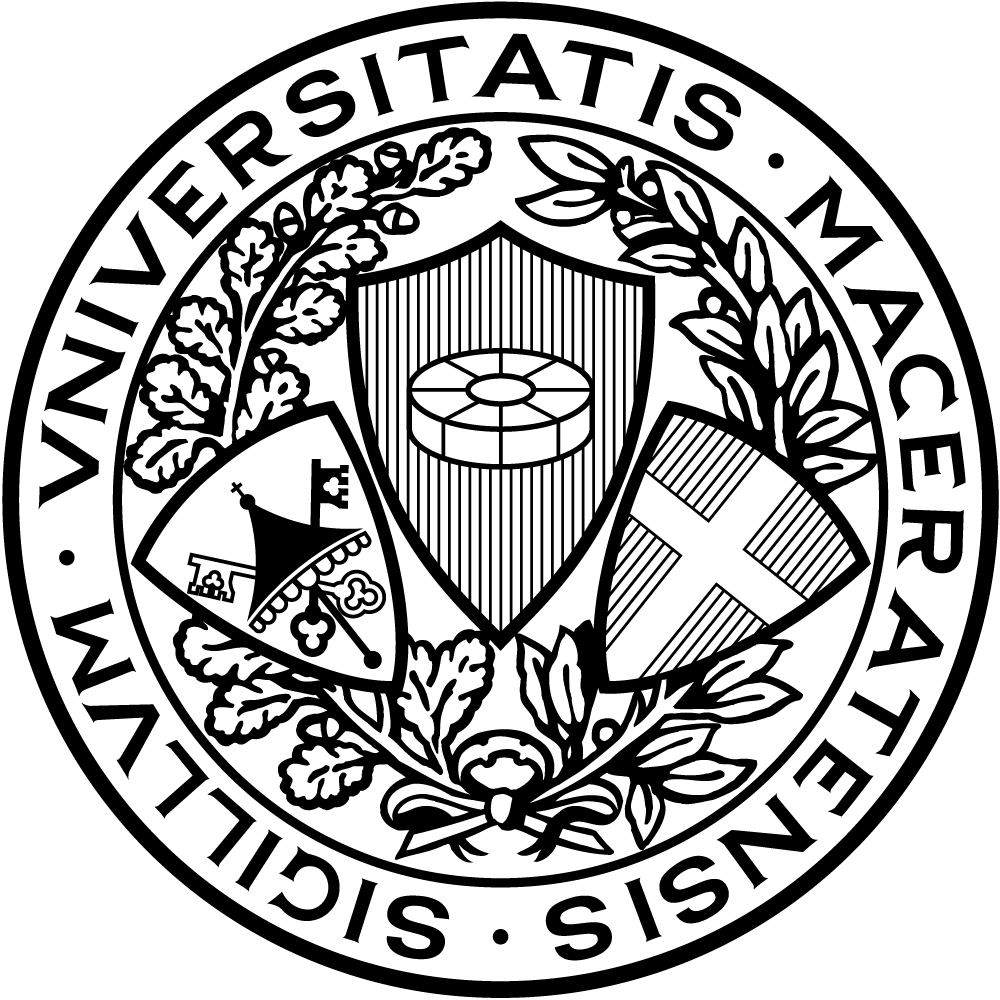}
  \end{center}

  \vspace{0.25cm}

  \begin{center}
    {\fontsize{14}{16.8}\selectfont UNIVERSITÀ DEGLI STUDI DI MACERATA}
  \end{center}

  \vspace{1.5cm}

  \begin{center}
    CORSO DI DOTTORATO DI RICERCA IN \\
    QUANTITATIVE METHODS FOR POLICY EVALUATION
  \end{center}

  \vspace{0.5cm}

  \begin{center}
    CICLO XXXVII
  \end{center}

  \vspace{1.25cm}

  \begin{center}
    OPENING THE BLACK BOX: \\
    NOWCASTING SINGAPORE'S GDP GROWTH \\
    AND ITS EXPLAINABILITY
  \end{center}

  \vspace{2.0cm}

  \begin{flushleft}
    \begin{tabular}{@{}l@{\quad}l@{}}
      SUPERVISORI DI TESI & \hspace{2.5cm} DOTTORANDO\\[5pt]
      Chiar.ma Prof.ssa Rosaria Romano & \hspace{2.5cm} Dott. Luca Attolico\\
      Chiar.mo Prof. Jamus Jerome Lim & \\
    \end{tabular}
  \end{flushleft}

  \vspace{1.25cm}

  \begin{flushleft}
    COORDINATORE\\
    Chiar.ma Prof.ssa Margherita Scoppola
  \end{flushleft}

  \vspace{1.7cm}

  \begin{center}
    ANNO 2025
  \end{center}

  \begin{center}
    \includegraphics[width=1.028in,height=0.7917in]{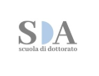}
  \end{center}

\end{titlepage}

%% file: 02_frontmatter/abstract.tex

\chapter*{Abstract}


Effective economic policymaking requires timely assessments of current economic conditions rather than relying solely on past-quarter data, especially in a dynamic and open economic system like Singapore. Promptly identifying turning points becomes critical, considering how international fluctuations can swiftly influence domestic conditions. This real-time assessment, often called "nowcasting," is particularly challenging in rapidly changing economic conditions.

The econometric approach for GDP growth nowcasting has dominated the forecasting landscape for years. Dynamic Factor Models (DFMs) applications in macroeconomic studies have guided substantial research on reducing high-dimensional datasets into a few latent factors that evolve over time. Nonetheless, these factors may introduce noise when distilling key signals from a vast pool of predictors, and they may not consistently guarantee the most substantial forecasting capabilities—ultimately affecting model interpretability. 

Traditional forecasting frameworks often encounter notable difficulties in handling structural breaks and crisis periods, emphasizing the need for flexible methods suited to volatile economic settings. Machine Learning (ML) models offer a highly effective path toward more accurate GDP growth nowcasting. Because these methods can approximate intricate nonlinear relationships, they frequently surpass older econometric techniques. In addition, recent advancements in explainable artificial intelligence (xAI) grant policymakers and researchers a clearer perspective on the internal mechanics of these models, enabling more targeted policy decisions. By integrating sub-period analyses, researchers can capture the heterogeneity of economic shocks, enhance real-time surveillance, and refine policy responses for smaller, highly exposed markets.

This research investigates how ML algorithms can improve the accuracy of nowcasting Singapore's GDP growth while clarifying the reasoning behind forecasts. Specifically, we introduce a structured pipeline designed to mitigate look-ahead bias, estimate model reliability via block bootstrap intervals, and highlight key features driving predictions. We also incorporate model combination techniques—such as simple averaging and weighted approaches—to boost forecast stability and monitor the dynamic weights assigned to individual models, thereby offering an interpretive lens for understanding which learners dominate at different points in time. These strategies are complemented by sub-period breakdowns that evaluate performance under varying market phases, revealing a broader perspective on robustness.

We analyze a set of approximately seventy variables, encompassing economic fundamentals (such as prices, production, trade, and investment), demographic factors (like population growth and age distribution), and societal indicators (e.g. employment rates). Data from the Singapore Department of Statistics and additional sources, including The Bank of Italy for external reference indices, span from 1990 Q1 to 2023 Q2. We implement penalized linear models (Lasso, Ridge, Elastic Net), dimensionality reduction approaches (Principal Component Regression, Partial Least Squares), ensemble learning methods (Random Forest, XGBoost), neural architectures (Multilayer Perceptron, Gated Recurrent Unit). Benchmarks such as a naïve Random Walk, an AR(3), and a Dynamic Factor Model are also included for comparison. A Bayesian optimization procedure is employed to select optimal hyperparameters within an expanding-window walk-forward validation framework, accommodating the temporal dependencies of macroeconomic time series.

Our findings provide reassurance about the stability and reliability of ML-based models in economic forecasting. These models consistently outperform benchmark forecasts, including official projections by major financial institutions, and demonstrate resilience across crisis episodes. On average, penalized linear models, dimensionality reduction approaches, and neural networks consistently achieve substantial reductions in nowcast errors—typically ranging from 40\% to 60\% compared to established benchmarks such as Random Walk, AR(3), and Dynamic Factor Models. When these strong learners are combined, forecasts achieve further improvements in nowcast errors, as shown by performance metrics.

In addition to GDP growth nowcasting, we employ explainability approaches appropriate to each family model—such as coefficient evolution, Variable Importance in the Projection, Gini-based and permutation-based measures, and Integrated Gradients—to enhance interpretability. By combining these techniques with block bootstrap procedures, we pinpoint which features exert the most significant impact on GDP variations and gauge the inherent uncertainty surrounding them—thereby illustrating, for instance, how changes in foreign trade components or prices can trigger significant shifts in growth projections. Although xAI techniques are gaining traction in fields like image recognition or consumer analytics, they remain relatively unexplored in macroeconomic contexts. This approach enables policymakers to have a deeper understanding of how shocks traverse the economic landscape.

Our study contributes to the broader discourse on ML-based macroeconomic forecasting by showing how sub-period analysis, uncertainty quantification, combined models, and xAI can operate in tandem. Through this integration, we achieve more substantial predictive accuracy and foster greater confidence and clarity in the outputs—factors of particular value for small-scale, internationally interconnected economies. Future work could further refine our methodology by encompassing high-frequency data or trialing innovative ML models that reinforce our framework's versatility and reliability.

\thispagestyle{empty} 
\cleardoublepage 

%% file: 02_frontmatter/acknowledgments.tex

\chapter*{Acknowledgments}

This thesis was made possible thanks to the institutional and personal support I received over the past years.

I am deeply grateful to the Università degli Studi di Macerata for funding and believing in this research project, and to ESSEC Business School, Asia-Pacific, for hosting me as a Visiting PhD Fellow and providing an intellectually stimulating environment in Singapore.

I wish to thank Dominique Lepore (Università Mercatorum) for immediately believing in my initial idea and for her invaluable advice in transforming it into a coherent research project. I am also grateful to the professors at Università Politecnica delle Marche---Maria Cristina Recchioni, Marco Gallegati (my master’s thesis supervisor), Emanuele Frontoni, Donato Iacobucci, and Mauro Gallegati---for their support and for the references that made my doctoral application possible.

A special word of thanks goes to my year in Hong Kong in 2019: Andrea Croci and the late Eric Szeto were true mentors and role models, and they profoundly shaped my relationship with the Asia-Pacific region.

My heartfelt thanks go to my doctoral supervisors, Rosaria Romano (Università degli Studi di Napoli Federico II) and Jamus Jerome Lim (ESSEC Business School Asia-Pacific), for their guidance, patience, and constant support throughout this work. I am equally indebted to Pierre Alquier (ESSEC Business School Asia-Pacific) and Andrea Bucci (Università degli Studi di Macerata) for their mentorship and expert guidance on econometrics and machine learning.

This thesis was conceived and developed as an autonomous research endeavor, with a strong orientation toward real-world applicability: a robust approach to nowcasting, a focus on uncertainty quantification, and interpretable methods that can be transferred beyond academia. I am grateful to all those, in Italy and in Singapore, who have supported me over the years, even simply through moral encouragement; your help has been invaluable and will not be forgotten.

\clearpage

%% file: 03_chapters/chapter1.tex
\chapter{Introduction and Literature Review}
\label{ch:introLitReview}

\section{\textbf{Bridging the Information Gap in Macroeconomics: The Importance of Nowcasting}}
\label{sec:1.1}

Timely and accurate macroeconomic indicators are fundamental for effective policymaking and private-sector investment decisions. However, official economic statistics---including key measures such as Gross Domestic Product (GDP), inflation rates, and employment levels---typically exhibit considerable publication delays. These delays introduce an "information gap", imposing significant challenges to policymakers, central banks, and financial institutions that must operate under heightened uncertainty. Crucial economic information frequently becomes publicly accessible only weeks or even months after the respective reference period, thereby limiting the effectiveness, precision, and responsiveness of economic interventions \parencite{GiannoneReichlinSmall2008,BokEtAl2018}.

Economic researchers have progressively embraced the methodological paradigm known as "nowcasting" to address this gap. Originating from meteorological nowcasting and subsequently adapted to macroeconomic applications, nowcasting aims explicitly to generate contemporaneous assessments of economic conditions. Unlike traditional nowcasting---which predominantly targets future periods---nowcasting leverages timely and readily available high-frequency data to estimate current economic activity, effectively bridging the information lag before official releases \parencite{BaumeisterEtAl2021,BanburaRunstler2011}. Consequently, nowcasting substantially reduces uncertainty, enhancing the accuracy and timeliness of economic decisions in both public and private sectors.

The global COVID-19 pandemic underscored the practical necessity and utility of robust nowcasting frameworks. During this period, traditional macroeconomic reporting frequencies proved insufficient, unable to capture rapid shifts in economic trajectories. As a result, prominent institutions such as the U.S. Federal Reserve heavily utilized advanced nowcasting methodologies, notably the GDPNow model from the Atlanta Fed \parencite{Higgins2014} and the New York Fed Staff Nowcast \parencite{BokEtAl2018}. These innovative frameworks successfully integrated unconventional real-time indicators—such as weekly unemployment insurance claims, mobility data, and financial market indices—to facilitate prompt, informed policy responses and clear communication with market stakeholders \parencite{CajnerEtAl2020,ForoniMarcellinoStevanovic2020}.

The benefits of nowcasting extend beyond public policy domains, providing substantial advantages for private investors, financial analysts, and corporate decision-makers. The ability to promptly assess economic conditions enables market participants to proactively manage risks, adjust investment portfolios, and refine strategic operations, thereby reducing informational asymmetries and improving overall market efficiency.

Nevertheless, despite their broad adoption and demonstrable benefits, nowcasting approaches face critical methodological limitations. Notably, these frameworks often lack sufficient interpretability and robust measures of predictive uncertainty. Policymakers and analysts increasingly emphasize the necessity for transparent and interpretable nowcasts, demanding clear insights into the underlying economic drivers of predictive outcomes. Therefore, further methodological advancements that explicitly incorporate model interpretability and rigorous uncertainty quantification are urgently required to enhance nowcasting's practical utility \parencite{PetropoulosMakridakis2020}.

In response to these methodological challenges, our research investigates advanced machine learning methodologies tailored to Singapore. Given Singapore's economic openness, the exceptional quality of its macroeconomic data, and heightened sensitivity to global economic fluctuations, it represents a uniquely suitable empirical setting. This research aims to improve nowcasting accuracy, interpretability, and uncertainty management, providing robust tools for policymakers and financial analysts in contexts characterized by structural uncertainties and rapid economic developments. The rationale and potential of these machine learning-based techniques are detailed in the following sections.

\section{\textbf{Machine Learning as a Catalyst for Nowcasting}}
\label{sec:1.2}

Machine learning (ML) techniques have rapidly gained prominence in macroeconomic nowcasting, driven by their distinctive ability to handle complex, high-dimensional data and effectively adapt to structural changes in economic dynamics.
Classical econometric frameworks such as autoregressive (AR), vector autoregressive (VAR), and dynamic factor models (DFM) commonly encounter challenges related to multicollinearity, dimensionality, and non-linear dynamics, particularly under structural breaks and volatile economic conditions \parencite{MedeirosEtAl2021,CoulombeEtAl2022}. In contrast, ML approaches effectively address these limitations by leveraging advanced computational frameworks and data-driven flexibility, significantly enhancing nowcasting accuracy and robustness in high-dimensional environments \parencite{KimSwanson2018}.

A key methodological advantage of ML lies in its ability to efficiently manage high-dimensional, heterogeneous datasets through built-in regularization, adaptive model complexity, and non-linear approximations. Specifically, penalized regressions (LASSO, Ridge, Elastic Net), dimensionality reduction methods (PCR, PLSR), ensemble algorithms (Random Forest, eXtreme Gradient Boosting), and neural networks (MLPs, GRUs) inherently adapt to complex patterns, accommodating data sparsity and non-linear interactions typically overlooked by linear models \parencite{ZouHastie2005,Breiman2001,ChenGuestrin2016,HewamalageBergmeirBandara2021}.

Recent empirical evidence underscores ML's superior performance relative to traditional econometric benchmarks, particularly during periods of economic instability. For example, \textcite{MedeirosEtAl2021} demonstrated enhanced nowcasting accuracy of ML techniques for inflation in volatile environments, while \textcite{CoulombeEtAl2022} confirmed ML's advantages in adapting dynamically to structural economic shifts. The COVID-19 pandemic notably exemplified ML's responsiveness to rapid regime changes, effectively integrating unconventional, high-frequency data sources such as mobility metrics, financial indicators, and digital transactions to achieve timely and accurate nowcasts \parencite{ForoniMarcellinoStevanovic2020,BarbagliaEtAl2023}.

The methodological robustness of ML approaches principally originates from three core attributes:
\begin{itemize}
    \item \textbf{Regularization and Variable Selection:} Penalized methods (e.g., LASSO, Elastic Net) automatically select relevant predictors, preventing overfitting and ensuring parsimonious, stable nowcasting outcomes \parencite{ZouHastie2005}.
    \item \textbf{Ensemble Learning:} Techniques like Random Forest and eXtreme Gradient Boosting aggregate multiple predictions, effectively reducing variance and capturing complex interactions inherent in macroeconomic data \parencite{Breiman2001,ChenGuestrin2016}.
    \item \textbf{Temporal Dynamics Modeling:} Neural network architectures (MLP, GRU) explicitly model sequential dependencies, enhancing adaptability to structural breaks and evolving economic regimes \parencite{ChoEtAl2014,HewamalageBergmeirBandara2021}.
\end{itemize}

Nevertheless, despite their empirical successes, ML models frequently suffer from limited interpretability and challenges related to uncertainty quantification. Policymakers and financial analysts increasingly prioritize transparency and robust uncertainty estimates, where traditional econometric methods still hold clear advantages. Consequently, integrating explainable artificial intelligence (XAI) techniques, such as Integrated Gradients or permutation importance measures, becomes critical for enhancing the transparency and credibility of ML-based nowcasts \parencite{LundbergEtAl2020,SundararajanTalyYan2017}. Indeed, recent scholarship emphasizes interpretability as a fundamental prerequisite for deploying ML models effectively in high-stakes economic and policy environments \parencite{Rudin2019}. Furthermore, the adoption of non-parametric uncertainty quantification methods, notably bootstrap procedures (e.g., stationary and block bootstrap), is essential to provide rigorous prediction intervals and quantify prediction uncertainty reliably \parencite{PolitisRomano1994,Li2021}.

Addressing these interpretability and uncertainty challenges represents a key advancement in ML-based nowcasting research. The methodological framework proposed in this study explicitly tackles these critical dimensions, aiming to strengthen the robustness, transparency, and practical relevance of nowcasts in highly dynamic macroeconomic environments.

\section{\textbf{Why Singapore? An Ideal Testbed}}
\label{sec:1.3}

Given its economic openness, sensitivity to global shocks, and exceptional data quality, Singapore provides an ideal empirical setting for methodological research on machine learning-based macroeconomic nowcasting.

Given its pronounced openness to trade and financial flows, Singapore rapidly transmits global shocks into domestic indicators like GDP and financial markets, highlighting the urgency for timely nowcasting methodologies.

Singapore's rigorous statistical framework ensures highly accurate and timely publication of key macroeconomic indicators \parencite{SingStatPortal}, significantly facilitating validation of advanced econometric and ML methodologies.

Despite these notable advantages, Singapore remains underrepresented in the nowcasting literature, especially regarding sophisticated ML frameworks. While advanced economies such as the United States or Eurozone have been extensively analyzed within empirical ML-based nowcasting studies, Singapore's case has scarcely been investigated, a gap implicitly confirmed by recent comprehensive reviews of nowcasting methodologies \parencite[e.g.,][]{ForoniMarcellinoStevanovic2020}. By explicitly addressing this gap, our research provides original insights into ML performance within an economic context uniquely sensitive to global disruptions.

Existing nowcasting practices within Singapore, i.e., the GDP Nowcasting model regularly published by DBS Bank \parencite{DBSNowcast2025}, predominantly rely on simpler year-over-year (YoY) methods and econometric frameworks lacking in sophistication. These models largely neglect advanced, interpretable ML approaches and do not adequately exploit the granular, quarterly-frequency data available. Introducing comprehensive ML methodologies---including penalized regressions, ensemble algorithms, and neural networks---to nowcast Singapore's quarterly GDP growth directly addresses this methodological shortcoming, promising substantial improvements in nowcasting accuracy, responsiveness, and transparency. Our research uniquely combines Singapore's economic sensitivity and exceptional data quality with advanced, interpretable ML methodologies---explicitly addressing critical gaps unaddressed by current practices and studies.

From a methodological perspective, Singapore's economic structure, combined with its high-quality data, offers an advantageous setting for rigorously evaluating ML models' predictive robustness and adaptability under varied economic conditions. The occurrence of recent economic shocks---such as the COVID-19 pandemic and geopolitical tensions affecting global supply chains---provides an ideal context to empirically assess how ML methodologies manage abrupt regime shifts and structural breaks. Such empirical evaluations yield vital methodological insights, extending applicability beyond the Singaporean context.

Its global financial prominence further enhances our research's practical and international relevance, benefiting policymakers, investors, and multinational firms through improved decision-making and risk management strategies. Indeed, addressing interpretability and uncertainty quantification is increasingly recognized as essential for the practical deployment of nowcasting models in policy-making contexts \parencite{PetropoulosMakridakis2020}.

By explicitly focusing on Singapore---an economically strategic yet understudied setting---this research bridges a crucial gap in existing nowcasting literature. It advances robust methodological frameworks with significant implications for global economic nowcasting practices.

\section{\textbf{Existing Approaches and Related Literature}}
\label{sec:1.4}

The literature on macroeconomic nowcasting features a diverse range of methodologies, broadly categorized into traditional econometric models and emerging machine learning (ML) techniques. We critically synthesize these approaches, highlighting their strengths and limitations, and position our research clearly within this established literature.

Dynamic Factor Models (DFM), extensively developed by \textcite{StockWatson2002,StockWatson2011,BaiNg2008}, represent one of the primary econometric tools for nowcasting. DFMs reduce data dimensionality by extracting common latent factors from large macroeconomic datasets. This approach enhances interpretability and nowcasting efficiency by summarizing numerous correlated variables into a few representative factors. Despite their widespread adoption, DFMs exhibit substantial methodological limitations. In particular, the selection of latent factors is inherently arbitrary and relies heavily on subjective econometric criteria, often leading to model misspecification or omitted variable bias. \textcite{BanburaEtAl2010} further underlines the complexities of factor selection, especially when dealing with large-scale datasets. Additionally, DFMs predominantly assume linear relationships among variables, failing to capture non-linear dynamics particularly pronounced during episodes of structural instability, such as the global financial crisis or the COVID-19 pandemic.

Other econometric approaches---autoregressive (AR) and vector autoregressive (VAR) models---are similarly prevalent in the nowcasting literature, particularly due to their straightforward implementation and clear interpretability. However, these methodologies suffer considerably from the "curse of dimensionality\footnote{The "curse of dimensionality" refers to the statistical and computational challenges arising from high-dimensional datasets, where the inclusion of numerous variables can deteriorate model performance and nowcasting accuracy due to sparsity, multicollinearity, and parameter instability \parencite{Bellman1961}.}", as the incorporation of numerous variables significantly diminishes model performance, resulting in parameter instability and nowcasting inefficiencies. Consequently, traditional econometric models often lack the robustness and adaptability to reflect rapid economic shifts adequately.

In response to these methodological limitations, recent literature increasingly explores ML-based nowcasting techniques. ML methodologies possess inherent advantages in managing complex, high-dimensional datasets characterized by intricate non-linear interactions and structural instabilities \parencite{Varian2014,MedeirosEtAl2021,CoulombeEtAl2022}. This methodological shift has been systematically documented by \textcite{KimSwanson2018}, who underscore ML's comparative advantages in predictive robustness. The principal ML methods in current nowcasting applications can be classified as follows:

\begin{itemize}
    \item \textbf{Penalized Regression Models (e.g., LASSO, Ridge, Elastic Net)}: These methods, originating from statistical learning, incorporate regularization to address multicollinearity and dimensionality issues, providing systematic variable selection and robust predictive performance. Nonetheless, their linear structure limits their ability to capture complex non-linear relationships fully.
    \item \textbf{Dimensionality Reduction Techniques (PCR, PLSR)}: Principal Component Regression and Partial Least Squares Regression effectively compress large datasets into fewer components, mitigating dimensionality and multicollinearity. However, similar to DFMs, their interpretability remains limited, and their linear assumptions restrict performance during pronounced structural breaks.
    \item \textbf{Ensemble Algorithms (Random Forest, eXtreme Gradient Boosting)}: Popularized by Breiman (2001) and Chen and Guestrin (2016), ensemble methods employ multiple decision trees to reduce nowcasting variance and capture non-linear relationships. Although effective, their "black-box" nature constrains interpretability, presenting notable challenges for policy-relevant applications.
    \item \textbf{Neural Networks (MLP, GRU)}: Deep learning models like Multi-layer Perceptrons (MLP) and Gated Recurrent Units (GRU) explicitly accommodate sequential data, offering superior adaptability to structural changes and non-linear dynamics \parencite{ChoEtAl2014,HewamalageBergmeirBandara2021}. Despite these advantages, their complexity introduces substantial interpretability concerns and increased risk of overfitting.
\end{itemize}

Recent methodological discourse increasingly highlights the importance of interpretability and uncertainty quantification in nowcasting applications. Traditional econometric models provide straightforward inferential frameworks (e.g., hypothesis tests, confidence intervals), enabling clear interpretability and robust uncertainty assessment. In contrast, ML methodologies often lack explicit statistical inference, necessitating alternative procedures---primarily non-parametric resampling methods such as stationary or block bootstrap---to deliver rigorous predictive uncertainty measures \parencite{PolitisRomano1994,Li2021}.

Moreover, recent literature emphasizes integrating explainable artificial intelligence (XAI) techniques into ML-based nowcasting frameworks. Popular methodologies such as Integrated Gradients, SHAP values, or permutation feature importance have been applied extensively to elucidate ML predictions, thereby increasing transparency and credibility for policymakers and financial analysts \parencite{LundbergEtAl2020,SundararajanTalyYan2017,Rudin2019}.

Our research explicitly acknowledges and addresses these methodological gaps, proposing a robust ML-based nowcasting framework tailored to the Singaporean economic context. The study systematically implements a comprehensive range of ML methodologies---penalized regressions, dimensionality reduction techniques, ensemble learning algorithms, and neural network models---rigorously evaluated through transparent interpretability methods and robust bootstrap-based uncertainty quantification. This approach substantially improves existing Singaporean practices, currently dominated by traditional econometric and simplified nowcasting techniques, thereby contributing original empirical insights and methodological advancements to the broader nowcasting literature.

\section{\textbf{Limitations of Traditional and Emerging Approaches}}
\label{sec:1.5}

Despite significant methodological advancements, traditional econometric and emerging machine learning (ML) approaches present noteworthy limitations in macroeconomic nowcasting. Critically examining these constraints provides valuable insights into their practical applicability and robustness, particularly under turbulent economic conditions.

Traditional econometric methods—dynamic factor models, autoregressive (AR), and vector autoregressive (VAR)—typically assume linear relationships and stationarity within data-generating processes. Such assumptions frequently become restrictive during pronounced non-linearities and abrupt structural shifts, limiting predictive accuracy and flexibility \parencite{StockWatson2011,BanburaEtAl2010,NgWright2013}. The global financial crisis, the COVID-19 pandemic, and geopolitical disruptions exemplify periods when traditional models often experienced pronounced nowcast deterioration due to their limited ability to dynamically adapt to extreme circumstances \parencite{StockWatson2011, BanburaEtAl2010}. Additionally, the subjectivity in selecting latent factors and variables, particularly within dynamic factor models, introduces risks of omitted-variable bias and model misspecification, thus adversely impacting nowcast stability \parencite{BaiNg2008}.

While conceptually superior in managing high-dimensional data and non-linear dynamics, ML methodologies exhibit their distinct limitations. Although theoretically adept at capturing complex economic interactions, empirical evidence on ML models' robustness remains mixed, especially during rapid regime shifts. Techniques such as Random Forest, eXtreme Gradient Boosting, and neural networks have demonstrated substantial predictive instability under abrupt economic changes \parencite{MedeirosEtAl2021, CoulombeEtAl2022}, raising concerns about their consistency across diverse economic scenarios \parencite{MedeirosEtAl2021,CoulombeEtAl2022,GiannoneLenzaPrimiceri2021}.

Interpretability also remains a prominent challenge for ML approaches. Methods such as ensemble learners and deep neural networks function predominantly as "black boxes," limiting transparency regarding internal decision processes. This opacity significantly diminishes their direct applicability in policy contexts, where the clarity of nowcasts and accountability are paramount. Recent advancements in explainable artificial intelligence (XAI), including Integrated Gradients, permutation importance, and SHAP values, partially mitigate these interpretability issues; however, achieving comprehensive transparency in ML-driven nowcasting remains an ongoing methodological challenge \parencite{Rudin2019, LundbergEtAl2020}.

Robust quantification of nowcast uncertainty presents another substantial obstacle within ML frameworks. Unlike traditional econometric methods, which offer well-established inferential techniques (confidence intervals, hypothesis testing), ML models primarily rely on algorithmic optimization, lacking direct statistical inference procedures. Thus, rigorous uncertainty estimation in ML necessitates alternative methods, such as non-parametric stationary or block bootstrap procedures. These bootstrap techniques, essential for valid uncertainty measures, must explicitly accommodate temporal dependencies and structural instabilities, complicating their methodological application and computational feasibility \parencite{PolitisRomano1994,Li2021}.

Moreover, predictive reliability in ML methods significantly depends on meticulous hyperparameter tuning, extensive computational resources, and rigorous cross-validation. ML models are prone to overfitting, highlighting the necessity of systematic validation procedures and careful preprocessing of heterogeneous high-frequency data to prevent misleading inferences and spurious relationships \parencite{HastieTibshiraniWainwright2015}.

While existing literature underscores ML's potential benefits, empirical evidence consistently indicates substantial heterogeneity in predictive robustness across economic contexts and methodological implementations. Systematic empirical analyses under diverse macroeconomic conditions, especially during structural breaks, remain indispensable for thoroughly understanding the comparative advantages and limitations of ML and traditional econometric methodologies.

Recognizing these gaps, our research explicitly incorporates advanced bootstrap methods for robust uncertainty quantification and sophisticated XAI techniques for enhanced interpretability. This methodological integration directly addresses the limitations identified in traditional econometric and emerging ML approaches, thereby significantly advancing the robustness, transparency, and practical utility of macroeconomic nowcasts under complex economic dynamics.

\section{\textbf{Objectives and Contributions of the Research}}
\label{sec:1.6}

This study advances the methodological frontier in macroeconomic nowcasting by addressing the following targeted research question: "\textit{How can machine learning (ML) and advanced econometric methods significantly enhance the quality, interpretability, and robustness of quarterly GDP growth nowcasts for Singapore, particularly during periods of structural instability?}" Motivated by clearly identified gaps within both traditional econometric and emerging ML frameworks, we propose an original, rigorous multi-model nowcasting framework tailored explicitly to Singapore's highly sensitive and globally interconnected economic context.

Specifically, our contributions extend the existing nowcasting literature along five principal dimensions:

\begin{itemize}
    \item \textit{Comprehensive Multi-Model Evaluation}. We rigorously deploy and comparatively evaluate multiple ML methodologies---including penalized regressions (LASSO, Ridge, Elastic Net), dimensionality-reduction techniques (PCR, PLSR), ensemble algorithms (Random Forest, eXtreme Gradient Boosting), and neural networks (MLP, GRU)---to explicitly mitigate model uncertainty and enhance predictive robustness during structural shifts (e.g., the COVID-19 pandemic).
    \item \textit{Rigorous Uncertainty Quantification via Block Bootstrap}. Addressing ML's limited inferential capabilities, we introduce block bootstrap techniques explicitly designed to handle temporal dependencies and significant structural breaks, including unprecedented economic shocks such as the COVID-19 pandemic. This methodological innovation significantly advances robust quantification of prediction intervals and feature-importance estimates, which are absent in existing ML-based nowcasting literature.
    \item \textit{Enhanced Interpretability through Model-Specific and XAI Techniques}. Given the policy-critical need for transparent nowcasts, we systematically implement both model-specific interpretability methods---such as coefficient-based feature importance for penalized regressions, gain importance for XGBoost, and mean decrease in impurity (MDI) for Random Forest---as well as Integrated Gradients for our neural models, a XAI technique rarely utilized in macroeconomic nowcasting. Combining these complementary approaches significantly enhances transparency and the practical applicability of complex ML forecasts in policy and economic contexts.
    \item \textit{Robust Model Selection via Model Confidence Set (MCS) and Nowcast Combinations}. We apply the Model Confidence Set methodology to systematically address model selection uncertainty, rigorously selecting statistically superior models. Furthermore, we implement advanced nowcast aggregation techniques (Weighted Average, Exponentially Weighted Average), substantially reducing individual model biases and enhancing overall nowcasting stability.
    \item \textit{Novel Empirical Insights from the Singaporean Context}. Our research offers unique empirical evidence on ML performance and adaptability in small, highly open economies by applying advanced ML methodologies specifically within Singapore's understudied yet strategically key context. This geographic-methodological contribution fills an explicit gap identified in recent comprehensive reviews on macroeconomic nowcasting, simultaneously providing a robust methodological benchmark potentially transferable to similar small, open economies frequently exposed to global structural disruptions.
\end{itemize}

These original methodological and empirical advancements significantly extend existing knowledge, providing robust, transparent, and operationally relevant nowcasting solutions to policymakers, central banks, and financial stakeholders in conditions of pronounced economic volatility and structural uncertainty.

\cleardoublepage %

%% file: 03_chapters/chapter2.tex

\chapter{Methodology and Pipeline}
\label{ch:methodologyPipeline}

\section{\textbf{Tools and Pipeline Organization}}
\label{sec:2.1}

\subsection{\textbf{\textit{Tools}}}
\label{subsec:tools}
We relied on a dual‐language and multi‐script workflow to execute our research, combining Python (version 3.x) and R (version 4.x) to exploit each environment's distinctive strengths. All scripts are managed within Spyder and RStudio integrated development environments. This section outlines the libraries employed, the families of nowcasting models under scrutiny, and the pipeline structure devised to accommodate predictive performance, prediction interval estimation, feature‐importance analysis, and full‐sample and sub‐period evaluations.

\vspace{0.35cm}

\phantomsection
\subsubsection*{\textbf{Families of Models and Evaluations}}
We analyzed four major families of nowcasting algorithms:

\begin{itemize}
    \item \textit{Penalized Linear Models:} LASSO, Ridge, and Elastic Net (EN);
    \item \textit{Dimension‐Reduction Models:} Principal Component Regression (PCR) and Partial Least Squares Regression (PLSR);
    \item \textit{Ensemble Learning Models:} Random Forest (RF) and eXtreme Gradient Boosting (XGB);
    \item \textit{Neural Models:} Multilayer Perceptron (MLP) and Gated Recurrent Unit (GRU) neural networks;
    \item \textit{Aggregation/Combination Models:} Simple Average (SA), Weighted Average (WA), and Exponentially Weighted Average (EWA), each generating aggregated forecasts from a subset of best‐performing models.
\end{itemize}

\vspace{0.35cm}

\phantomsection
\subsubsection*{\textbf{Programming Tools and Libraries}}

\vspace{0.2cm}
\noindent{}\textit{Common and General Libraries for Python-based Tools and Scripts}
\begin{itemize}
    \item \texttt{numpy} and \texttt{pandas}: supply the core data containers used in every pipeline stage.\\
        \textit{(i)} \texttt{numpy} arrays underpin all numerical routines (vectorized algebra, rolling operations, linear algebra);\\
        \textit{(ii)} \texttt{pandas} data-frames store the raw macro series, the recursively generated forecasts, and every intermediate diagnostic table; they also offer built‐in indexing and data manipulation;\\
    \item \texttt{matplotlib} and \texttt{seaborn}: provide visualizations for\\
        \textit{(i)} residual‐diagnostic panels;\\
        \textit{(ii)} forecast--vs.--actual plots;\\
        \textit{(iii)} stacked area charts tracking the time‐varying weights in forecast combinations.
    \item \texttt{statsmodels} and \texttt{scipy.stats}: supply econometric tests and distributional tools, such as\\
        \textit{(i)} ADF and Ljung--Box for unit‐root and serial‐correlation checks;\\
        \textit{(ii)} Shapiro--Wilk for normality assessment;\\
        \textit{(iii)} OLS wrappers, further augmented by a custom Lumley--Heagerty covariance for Giacomini--White predictive‐ability regressions.
    \item \texttt{scikit‐learn} (\texttt{sklearn}): offers general machine‐learning utilities. \texttt{StandardScaler} ensures homogeneous scaling of features before stationarity checks, whereas \linebreak \texttt{IsotonicRegression} imposes monotonic constraints in the WEAVE (Lumley--Heagerty) covariance approach;
    \item \texttt{logging}, \texttt{tracemalloc}, \texttt{random}, and \texttt{os} handle reproducibility and runtime monitoring:\\
        \textit{(i)} a single project‐wide random seed ensures replicable bootstrapping steps;\\ \textit{(ii)} \texttt{logging} maintains timestamped records of execution phases;\\ \textit{(iii)} \texttt{tracemalloc} assists in memory usage tracking;\\
        \textit{(iv)} \texttt{os} facilitates path management across different systems.
    \item \textit{openpyxl} or \texttt{xlsxwriter}: enable standardized workbook exports. Each file may hold dedicated sheets for:
    \begin{itemize}
        \item \textit{(i)} point forecasts;
        \item \textit{(ii)} RMSFE and coverage measures per sub‐period;
        \item \textit{(iii)} diagnostic test results, thus allowing external replication without rerunning the entire pipeline;
    \item \textit{(iv)} intermediate model outputs, such as MCS rankings, feature importances, and bootstrap resamples.
\end{itemize}
\end{itemize}

\vspace{0.2cm}
\noindent{}\textit{Specific Python Libraries for Scripted Model Families and Diagnostic Tests}
\begin{itemize}
    \item \textit{Penalized Linear Models (Lasso, Ridge, Elastic Net):}\\
    \texttt{sklearn.linear\_model} implements coordinate‐descent estimators, while \linebreak \texttt{statsmodels.api.OLS} may be used post‐selection for inference. Hyper‐parameters are explored through \texttt{optuna}, with parallelism via \texttt{joblib}. Confidence intervals for coefficients or forecasts are computed through a manual block‐bootstrap routine summarized in \texttt{scipy.stats};
    \item \textit{Dimension‐Reduction Models (Principal Component Regression, Partial Least Squares Regression):}\\
    \texttt{scikit-learn} libraries \texttt{sklearn.cross\_decomposition.PLSRegression} and \texttt{sklearn.decomposition.PCA} extract latent factors, while \texttt{scipy.linalg.svd} underpins the numerical routines. Bayesian hyper‐parameter search uses \texttt{optuna}. Predictive intervals and coefficient bands follow from manual block‐bootstrap sampling, summarized with \texttt{scipy.stats} for quantiles and confidence limits;
    \item \textit{Ensemble Learning Models (Random Forest, XGB):}\\
    \texttt{sklearn.ensemble.RandomForestRegressor} and \texttt{xgboost.XGBRegressor} \linebreak train the ensembles. Hyper‐parameters are optimized via \texttt{optuna}. Feature importance is derived from built‐in "feature importances" or permutations; predictive intervals and importance confidence ranges rely on in‐house block‐bootstrap logic coupled with \texttt{scipy.stats} percentile transforms;
    \item \textit{Neural Networks (MLP, GRU):}\\
    \texttt{torch} is employed for network layers, training loops, and \texttt{torch.optim.Adam} for parameter updates. Hyper‐parameter tuning uses \texttt{optuna}. Feature attribution is done via \texttt{captum}'s Integrated Gradients. Block‐bootstrap procedures for intervals and attribution bounds are coded internally, with \texttt{scipy.stats} providing quantiles;
    \item \textit{Confidence intervals and feature importance confidence ranges}:\\
    They rely on resampling techniques, including pair block-bootstrap implemented through \linebreak \texttt{arch.bootstrap.MovingBlockBootstrap}, with summary statistics generated \linebreak using \texttt{scipy.stats} for quantiles and confidence limits;
    \item \textit{Forecast Aggregation Schemes (Simple Average, Weighted Average, Exponentially Weighted Average):}\\
    Weights are computed in a fully manual way (\texttt{numpy}-based) and logged over time. No hyper‐parameter tuning is required, though each script saves the dynamic weight evolution and dominant‐model frequencies using \texttt{matplotlib};
    \item \textit{Diagnostic and Stability Tests:}\\
    ADF, Shapiro–Wilk, and Ljung–Box are from \texttt{statsmodels.tsa.stattools} and \texttt{statsmodels.stats.diagnostic}. Giacomini–White regressions use the custom \texttt{lumley\_heagerty.py} module, overriding default OLS covariances with the WEAVE correction for autocorrelated residuals.
\end{itemize}

\vspace{0.2cm}
\noindent{}\textit{Common and General Libraries for R-based Tools and Scripts}
\begin{itemize}
    \item \texttt{readxl} and \texttt{openxlsx}: simplify the process of importing and exporting data, ensuring the generation of loss matrices or summary tables can be read in R, and vice versa;
    \item \texttt{parallel}: enables multi‐core support in the Model Confidence Set bootstraps, distributing the sampling tasks across available cores;
    \item \texttt{stats}: offers core time‐series and sampling functions (e.g., \texttt{ar}, \texttt{filter}) plus descriptive statistics that are invoked throughout the R scripts;
    \item \texttt{base} usage: the Model Confidence Set (MCS) procedure is encapsulated in a custom S4 workflow\footnote{In R, S4 refers to a formal system for object-oriented programming. We employ it here to define the "Superior Set of Models" (SSM) class, which stores MCS outputs (i.e., rankings, p-values, bootstrap details) in a structured manner.}, formalized in an auxiliary source file and instantiated through the class "Superior Set of Models" (SSM), executing the Hansen–Lunde–Nason algorithm\footnote{Hansen, Lunde, and Nason \parencite{HansenLundeNason2011} developed an iterative algorithm which discards models found systematically inferior at a given significance level, thus retaining only those with statistically indistinguishable (superior) performance.}.
\end{itemize}

\vspace{0.2cm}
\medskip \noindent{}\textit{Specialized R Routines Implemented in the Project}
\begin{itemize}
    \item \textit{Model Confidence Set (MCS) Procedure}: a tailored S4 approach conducts the iterative removal of inferior models via block bootstrap replications of the test statistic. Specifically, it\\
        \textit{(i)} reads the matrix of losses (squared errors, in our case) for all candidate models,\\
        \textit{(ii)} generates B=10,000 contiguous‐block resamples to approximate the distribution of the test, and\\
        \textit{(iii)} prunes models until a final superior set emerges at a chosen confidence threshold (e.g., 10\%).
     The results, including p-values, ranking, and MCS membership, are stored in the SSM object and exported to a structured data file for documentation.
\end{itemize}

All scripts implement partial logging, memory cleanup, and a seed fix (e.g., \linebreak \texttt{np.random.seed(42)}, \texttt{os.environ["PYTHONHASHSEED"]="42"}, or \texttt{set.seed(42)})\footnote{The value "42" is often chosen as a "random seed" for reproducibility reasons as an Easter egg: in the humorous science fiction novel "The Hitchhiker's Guide to the Galaxy," British writer Douglas Adams identifies "42" as "the answer to life, the universe and everything" \parencite{Adams1979}. Since then, the number has become a recurring cultural reference in programming languages and technical documentation.} \linebreak to guarantee consistency throughout the entire pipeline.

\vspace{0.35cm}

\subsection{\textbf{\textit{Research workflow}}}
\label{subsec:workflow}
Our entire research pipeline comprises the following steps:

\begin{enumerate}
    \item \textbf{Data ingestion and stationarity checks.} We import the quarterly dataset, apply Augmented Dickey-Fuller tests to filter out non‐stationary features or transformations and incorporate shock dummy variables if needed;
    \item \textbf{Individual model training.} Each script related to the model families (Penalized Linear Models, Dimension‐Reduction Models, Ensemble Learning Models, Neural Networks) is executed, yielding point forecasts, sub‐period diagnostics (Shapiro–Wilk, Ljung–Box), and block‐bootstrap confidence intervals for forecast distributions or coefficient estimates;
    \item \textbf{Model Confidence Set.} We merge all loss columns into a single matrix and invoke the MCS procedure in R to remove systematically inferior models at a 10\% significance level, returning a final set of top‐performing models;
    \item \textbf{Combination of selected models.} The choice of best-performing individual models for each aggregated method is based on the MCS output, ensuring that only the top-performing models (those selected by the MCS procedure) are included in the ensemble combinations. This procedure iteratively discards inferior models based on bootstrap resampling, retaining only the top-performing models for the final aggregation. The three combined models we used are:
        \begin{itemize}
            \item \textit{Simple Average:} uniform weighting across MCS members;
            \item \textit{Weighted Average:} weighting inversely to partial RMSE up to each time step;
            \item \textit{Exponentially Weighted Average:} a two‐level scheme that firstly tries multiple $\eta$ parameters and then applies a second EWA across these "$\eta$‐experts."
        \end{itemize}
    \item \textbf{Benchmarking.} We compare each final model and aggregator to three benchmarks---Random Walk (RW), AR(3), and a Dynamic Factor Model (DFM)---and collect RMSFE ratios over the overall sample (Overall) and sub‐periods (Pre‐COVID, COVID, Post‐COVID, and Excluding‐COVID)\footnote{The "Overall" period encompasses the entire forecasting horizon analyzed, spanning from 2017 Q1 to 2023 Q2, and thus integrates all specified sub-periods, namely "Pre-COVID," "COVID," and "Post-COVID." This comprehensive timeframe captures the complete range of macroeconomic conditions, including stable economic phases and significant disruptions induced by the COVID-19 pandemic and subsequent recovery period.
    
    The "Excluding COVID" period covers the entire forecasting horizon from 2017 Q1 to 2023 Q2, analogously to the "Overall" period, but explicitly omits the quarters defined as the "COVID" sub-period. This exclusion isolates structural relationships between predictors and economic outcomes in periods free of the extreme economic disruptions associated with the COVID-19 pandemic, providing a reference for model stability and feature importance under comparatively standard economic conditions.};
    \item \textbf{Predictive‐ability tests.} Finally, Giacomini–White regressions are run in Python to assess whether each model or aggregator systematically outperforms the benchmark. We rely on a tailored Python procedure to incorporate a monotonic correlation structure in the residuals, culminating in tabulated Wald tests and p‐values stored in formatted data files.
\end{enumerate}

\begin{figure}[htbp]
\centering
\includegraphics[width=1.0\textwidth]{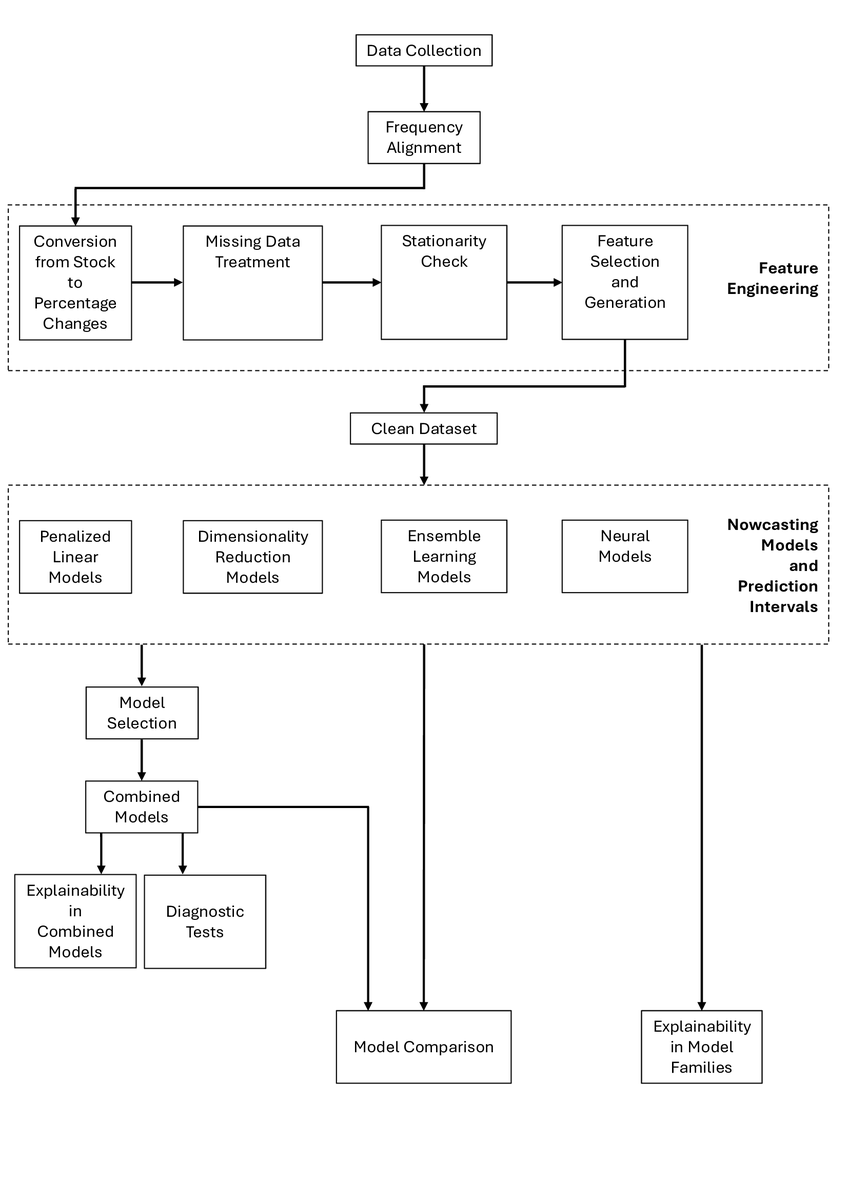}
\caption{\normalsize Research pipeline workflow}
\label{fig:pipeline}
\end{figure}

The overall workflow, summarized in Figure~\ref{fig:pipeline}.

\vspace{0.1cm}
This sequential approach ensures that each subtask---from forecast generation and intervals, through the MCS selection and final aggregation to the formal out‐of‐sample testing---remains consistent and reproducible across Python and R environments. The subsequent sections detail the core methodological elements, including data transformations, hyper‐parameter optimization, sub‐period evaluations of predictions and their reliability, feature‐importance analyses, and interpretability of the aggregated results.

\vspace{0.5cm}

\section{\textbf{Data and Preprocessing}}
\label{sec:2.2}
In this section, we detail the phases related to the construction of the dataset and the preprocessing techniques adopted to ensure consistency, uniformity, and reliability of the variables in the subsequent one-step-ahead nowcasting process. The central objective is twofold: on the one hand, to have a database adequately refined and free from distortions deriving from non-stationarity phenomena or non-uniform frequencies; on the other, to ensure that the transformation and standardization procedures strictly consider the temporal sequentiality, to avoid look-ahead bias (ex-post information contamination).
At the end of the initial acquisition operations, the original dataset had 95 candidate variables (features), with coverage from the first quarter of 1990 (1990 Q1) to the second quarter of 2023 (2023 Q2), for a total of 134 quarterly observations. Following the procedures of filling, cleaning, feature engineering, removal of features excessively incomplete or not plausibly linked to economic activity, and exclusion of series that proved to be non-stationary according to the augmented Dickey–Fuller (ADF) test, a core of 58 features has been reached. Five dummy variables were then added to these ones---three for seasonality ("Seasonality Quarter 1", "Seasonality Quarter 2", "Seasonality Quarter 3") and two to identify positive or negative economic shocks ("Past Negative Shocks of Singapore's GDP Quarter over Quarter Growth", "Past Positive Shocks of Singapore's GDP Quarter over Quarter Growth")---bringing the total number of explanatory variables to 63. Together with the target variable (i.e., the deflated GDP growth rate), the final dataset has 64 columns, continuously covering over 134 quarters. This configuration, the result of the balance between maximizing information and minimizing methodological uncertainty, constitutes the solid basis for the nowcasting analyses in the following chapters.

\vspace{0.35cm}

\phantomsection
\subsection{\textbf{\textit{Data Sources: Collection, Integration, and Indexing}}}
\label{sec:2.2.1}
To build the dataset, we collected data mainly from statistics produced by the Singapore Department of Statistics \parencite{SingStatPortal}. In particular, we used quarterly and monthly historical series relating to the main indicators of the Singapore economy: industrial production, foreign trade, price dynamics, real estate sector, household consumption, transport and airport movements, labor cost indices, balance of payments, exchange rates, business confidence indicators, as well as demographic variables. Each series required a preliminary acquisition, cleaning, and integration process into a single dataset, temporally indexed from the first quarter of 1990 to the most recent period covered by the official archives (second quarter of 2023, in our case).
In addition to the SingStat sources, we had to integrate two specific exchange rate series (Euro/SGD and Renminbi/SGD) from the Bank of Italy \parencite{BancaItaliaExchange} since the corresponding observations in SingStat were incomplete for the period of interest. In particular, for the Euro/SGD, we used a series that also covered the years before the formal introduction of the euro, using data referring to the ECU (European Currency Unit) and the one-to-one correspondence between ECU and euro starting from 1 January 1999. Similarly, in the case of the Renminbi, the SingStat source had gaps up to June 1993, so the series provided by the Bank of Italy offered a more extensive and homogeneous coverage, allowing us to preserve the historical coherence necessary for our analysis horizon.
Before proceeding with the transformations, we executed a preliminary check regarding the denomination, the frequency of release, and the correct association with the reference quarter of all the variables to avoid duplication or time misalignment errors, especially in the case of monthly series subsequently converted to quarterly frequency.

\vspace{0.35cm}

\phantomsection
\subsection{\textbf{\textit{Transformations and Feature Engineering}}}
\label{sec:2.2.2}
Once the variables were consolidated, we performed transformation and processing operations to make the series comparable and manage the different frequencies.

\vspace{0.35cm}

\phantomsection
\subsubsection{\textbf{\textit{Target Variable}}}
Singapore's GDP growth at constant prices is our forecasting target in the nowcasting analysis. We obtained the variable from the quarterly GDP data at current prices and the corresponding GDP deflator (base 2015 = 100).

First, we deflated the nominal values of GDP to obtain a series expressed in real terms and adjusted for variations in price levels. Subsequently, the deflated series was transformed into quarterly growth rates (quarter-over-quarter), calculating the percentage difference between the current and previous quarters.

\vspace{0.35cm}

\phantomsection
\subsubsection{\textbf{\textit{Frequency Alignment}}}
The collected series had heterogeneous frequencies: some were available with monthly periodicity (e.g., price indicators, exchange rates, meteorological variables), while others were already processed in quarterly format (e.g., composite confidence indices or GDP data). Since the target variable is on a quarterly basis, we decided to standardize all monthly series to the same frequency to ensure consistency in the subsequent modeling process. The wide availability of data and the relative ease of converting them made it unnecessary to use specific methodologies for managing mixed frequencies (such as MIDAS models\footnote{MIDAS (Mixed Data Sampling) models allow the combination of data observed at different sampling frequencies (e.g., monthly and quarterly) in one regression framework, without the need to convert all variables to the same frequency. For a detailed discussion, see \parencite{GhyselsEtAl2006}}) without compromising the robustness of the analysis. In detail, the choice of the aggregation method (sum, average, or end-of-period value) depends not only on the nature of the variable (e.g., stock vs. flow) but also on the original format of the SingStat publications. In the case of series already expressed as monthly cumulative (e.g., "Contracts Awarded" or "Duty-Paid Releases Of Tobacco"), the sum of the three corresponding months provides the quarterly value. The arithmetic mean over the quarter was calculated for variables released as a monthly average (such as exchange rates or specific price indices). For data cataloged at the end of each period (end-of-period), including "Currency In Circulation," the value recorded in the last month of each quarter was assumed. Although simplified, this approach adequately reflects the sources' structure and ensures a fully quarterly dataset, free from misalignments and gaps.

\vspace{0.35cm}

\phantomsection
\subsubsection{\textbf{\textit{Conversion from Stock to Percentage Changes}}}
Most of the initial variables reflect stocks or indicators on a nominal basis. We decided to convert these stocks into quarterly percentage changes to analyze the rates of change and the characteristics of macroeconomic nowcasting approaches. This practice not only grants comparability between variables with different reference scales but also allows for mitigating possible non-stationarity phenomena. The only exception concerns the availability of some series that already expressed quarterly differences in absolute terms (such as "Changes in Employment"). Since the original levels to calculate a percentage change were unavailable, we constructed an "Acceleration of Net Employment Flow" measure to track the speed with which net job creation varies. Although it is a second difference, we may interpret this indicator as the intensity of employment expansion or contraction.

\vspace{0.35cm}

\phantomsection
\subsubsection{\textbf{\textit{Missing Data and Variables Not Related to Economic Activity}}}
Some variables began on dates after 1990 Q1. To limit the introduction of noise and preserve the length of the time series, we eliminated those features with ten or more missing observations in the initial part of the sample (e.g., "Landed Properties Supply Change" or "Foreign Trade Index Change").
We opted for the forward-filling technique for variables with a few missing initial values (less than ten), i.e., translating the last known value to the subsequent quarters in which it was missing. We may justify this imputation by the hypothesis that, in the initial phase of the analysis period, there were no international or internal shocks to the Singapore economy that could cause structural changes that would make this technique inadequate.
Finally, we removed some variables that were poorly correlated with the performance of Singapore's real economy, such as elementary demographic data (change in live births, change in deaths) and meteorological variables (changes in air temperature and sunshine, relative humidity and rainfall) whose influence was neither plausible nor theoretically justifiable in a GDP growth forecasting context. This filtering allowed us to focus only on indicators intrinsically linked to macroeconomic developments.

\vspace{0.35cm}

\phantomsection
\subsubsection{\textbf{\textit{Stationarity}}}
Although the conversion to growth rates helps reduce the presence of trends or seasonality, we further verified the stationarity of each variable using the Augmented Dickey-Fuller (ADF) test, adopting a conventional significance level (5\%). We eliminated those series that remained non-stationary even after the transformation (including changes in personal saving rate and electricity generation). This approach reflects the need to build reliable predictive models, avoiding the persistence of trends or unit-roots generating misleading results in the estimation phase.

\vspace{0.35cm}

\phantomsection
\subsubsection{\textbf{\textit{Tracking Seasonality and Positive/Negative Economic Shocks}}}
To present the non-artificial nature of the features and the target variable and allow the model to recognize seasonal patterns through the associated coefficients, we chose to handle seasonal effects by creating dummy variables (i.e., "Seasonality Quarter 1", "Seasonality Quarter 2", "Seasonality Quarter 3") rather than proceeding with an ex-ante seasonal adjustment.

Regarding macroeconomic shocks, we defined two additional dummies (i.e., "Past Negative Shocks of  Singapore's GDP Quarter over Quarter Growth", "Past Positive Shocks of  Singapore's GDP Quarter over Quarter Growth") to mark quarters characterized by strongly negative GDP changes or particularly pronounced rebounds. To this end, we adopted a threshold-based criterion (–2.5\% to identify significant declines and +5\% to recognize robust recoveries), in line with the literature that uses percentage thresholds to distinguish episodes of contraction or exceptional growth \parencite{ClaessensKoseTerrones2011,HausmannPritchettRodrik2005,KannanScottTerrones2009}. 

Integrating these values with the history of significant economic and financial events for Singapore (e.g., the Asian financial crisis of the late 1990s, the Great Recession of 2007–2009, the COVID-19 pandemic) helps avoid ambiguous labeling between normal oscillations and shocks of significant magnitude, thereby more clearly distinguishing growth peaks from a context of statistical normality.\footnote{For an in-depth analysis of threshold-based approaches to identifying recessions and expansions, see also \parencite{BarroUrsua2008} and \parencite{BryBoschan1971}.}

\vspace{0.35cm}

\phantomsection
\subsubsection{\textbf{\textit{Iterative Standardization of Non-Dummy Variables}}}
\vspace{0.2cm}
The last phase of preprocessing is represented by the iterative standardization of continuous variables so that each forecast window reflects the actual information availability of the moment and avoid look-ahead bias (i.e., it does not benefit from future temporal knowledge). Specifically, at each nowcasting step, we separated the portion of data considered "training" up to that point and calculated the mean and standard deviation for each variable. These parameters were then applied to the validation and test subset to normalize the data consistently with the conditions present during training. The dummy variables were not involved in the standardization process, as they did not need to be scaled.
We repeated this mechanism iteratively as the training time horizon expanded, ensuring a correct simulation of the actual practice of nowcasting (in which models are regularly updated as new observations become available) and preventing look-ahead bias as a source of overestimation of the predictive performance.

\par{} 
The phases described in this section ensure that our input dataset's economic variables are consistent, stationary, and comparable while preserving historical sequentiality. The choices made regarding exclusion, imputation, and transformation of the variables constitute a balanced compromise between maximizing the amount of information available and safeguarding the statistical robustness of the nowcasting experiments illustrated in the following chapters.

\vspace{0.5cm}

\section{\textbf{Nowcasting: Methodologies, Baseline Models, and Hyperparameter Optimization}}
\label{sec:2.3}

\phantomsection
\subsection{\textbf{\textit{Expanding+Rolling Window Methodology}}}
\label{sec:2.3.1}
The integrated implementation of the Expanding Window and Rolling Window methodologies forms the basis of our Singapore GDP quarterly nowcasting framework \parencite{Tashman2000,InoueRossi2012,HyndmanAthanasopoulos2021}. This solution allows us to draw on the entire history of the data, maintaining a dynamic validation criterion that, at each iteration, provides updated feedback on the model's performance. At the same time, it is consistent with the need for a limited forecast horizon and accurate calibration of the hyperparameters, enhancing the robustness of the entire econometric analysis framework.

\begin{figure}[htbp]
\centering
\includegraphics[width=1.0\textwidth]{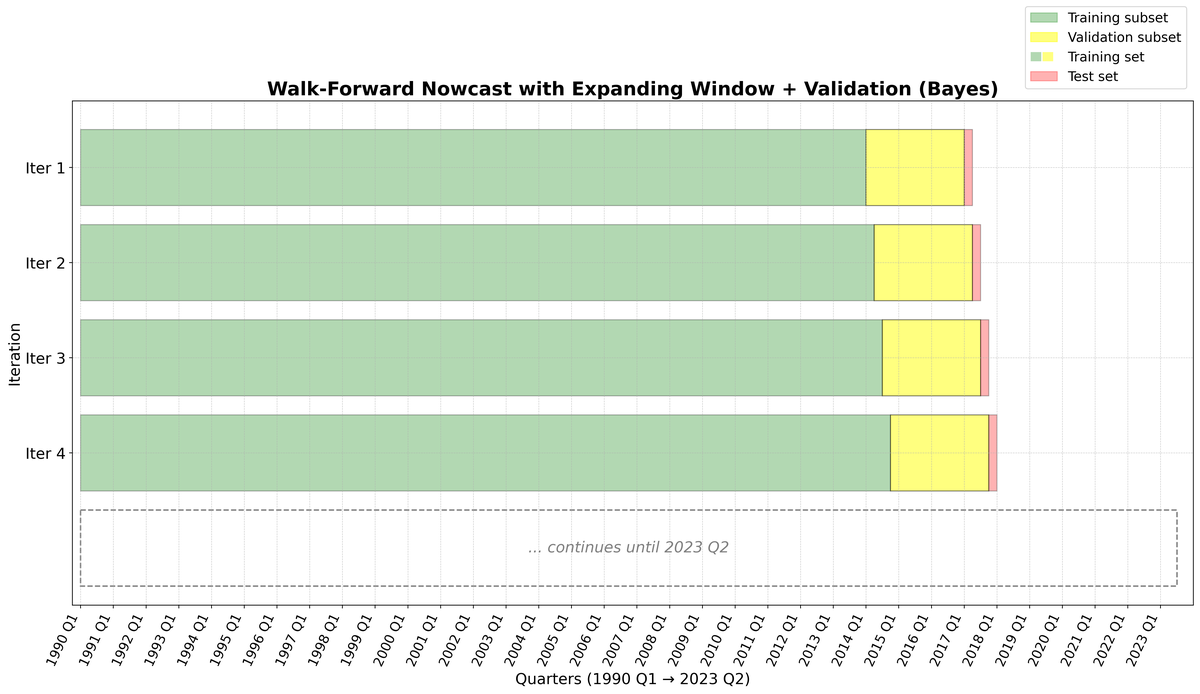}
\caption{\normalsize Walk-forward nowcast with expanding+rolling window procedure}
\label{fig:expanding_rolling_schematic}
\end{figure}

We adopt a walk-forward evaluation scheme with an expanding training window and a 12-quarter rolling validation window, followed by a one-step-ahead test quarter. This design preserves temporal ordering and avoids look-ahead bias while continuously assessing performance on the most recent data. The basic idea is to build a data window at each step that includes the training period and an updated validation set and then perform the estimate for a future quarter. This scheme preserves the flexibility needed to apply to different families of algorithms \parencite{Tashman2000}. A schematic of the Expanding+Rolling Window procedure is shown in \Cref{fig:expanding_rolling_schematic}.

The choice of a progressively expanding training window (Expanding Window) is based on the principle of valorizing the entire information heritage from the first available quarter to the last, without discarding remote data that might contain relevant structural signals. In a macroeconomic context, this approach is sometimes preferable to a pure Rolling Window since it facilitates the identification of long-term patterns in the presence of external shocks or cyclical changes that characterize open economies \parencite{StockWatson1996,ClementsHendry1998,PesaranTimmermann2007}.
In our implementation, we initially allocate approximately 80\% of the available quarters to the training stage (ending at Q4 2016). 
\footnote{An 80--20 division of the dataset is a common heuristic in time-series forecasting \parencite{HyndmanAthanasopoulos2021}, balancing the need for a sufficiently large training set with the requirement to retain enough observations for robust validation and testing.}
Subsequent iterations expand the training boundary by one quarter each time, preserving the entire economic history of Singapore \parencite{GiannoneReichlinSmall2008}.

In parallel, we applied the Rolling Window concept to define a validation subset that "slides" forward by one quarter at each iteration. In this way, the model is fine-tuned on an expanded training set but always validated on the last 12 quarters (3 years) immediately preceding the forecast \parencite{PesaranTimmermann2007,BanburaRunstler2011}. 
Choosing a 12-quarter window reflects a balance between reactivity to recent changes and stability of the validation error, a common practice in short-horizon macroeconomic forecasting \parencite{InoueRossi2012}. 
\footnote{Shorter validation windows (e.g., 4--8 quarters) often yield higher variance in the validation metric, while significantly longer windows may dilute the impact of recent structural shifts. Three years is frequently adopted by institutions and empirical studies, as it provides enough recent data without overwhelming the validation with older observations \parencite{ClarkMcCracken2009}.}

The integration between the Expanding Window and the Moving Window thus represents a trade-off between maximizing historical information and evaluating the model on data consistent with the recent economic situation. Extending the training window allows us to capture long-term patterns, while the moving validation portion highlights any instability or regime changes \parencite{ClarkMcCracken2009}.

Once hyperparameters are selected based on the last 12 quarters of validation, the model is then re-fitted on the entire training-plus-validation window to incorporate all available data. This final fit forms the basis for generating the one-step-ahead forecast on the immediately following quarter, thus ensuring that the model parameters reflect the most up-to-date information collected prior to the test horizon.

At the end of each iteration, the forecast is made for the immediately following quarter (the Test Set Window). The result is a one-step-ahead forecasting mechanism that maximizes the use of historical information and isolates a subset of data to fine-tune the hyperparameters. Furthermore, this nowcasting procedure aligns with the operational practice of economic research centers and the mainstream literature on quarterly forecasts since it systematically updates the estimates for each newly available quarter \parencite{GiannoneReichlinSmall2008,BanburaRunstler2011}.

\vspace{0.35cm}

\phantomsection
\subsubsection{\textbf{\textit{Expanding Training Set Window}}}
\vspace{0.2cm}
In the first iteration, we set 2016 Q4 as the final date of the initial raining Set Window (green+yellow area), covering about 80\% of the initial observations. Each subsequent iteration extends the training subset boundary (green area) by an additional quarter, including progressively newer observations.
This incremental approach leverages the entire economic history available for Singapore while ensuring that the model receives fresh data at each step \parencite{GiannoneReichlinSmall2008}.

\vspace{0.35cm}

\phantomsection
\subsubsection{\textbf{\textit{Rolling Validation Subset Window}}}
\vspace{0.2cm}
We used the last 12 quarters (three years) of the Training Set Window to fine-tune the hyperparameters during each iteration. This Validation Subset Window (yellow area) shifts one quarter forward after each step, ensuring continuous performance examination of recent data \parencite{PesaranTimmermann2007,InoueRossi2012}. 
Twelve quarters are frequently cited in the macro-forecasting literature as a window length that balances the need to detect short-run regime shifts without discarding essential cyclical information \parencite{ClarkMcCracken2009,BanburaRunstler2011}.

\vspace{0.35cm}

\phantomsection
\subsubsection{\textbf{\textit{One-step-ahead Nowcasting and Test Set Window}}}
\vspace{0.2cm}
Once the training-validation block is built, we use the immediately following quarter as the Test Set Window (pink area) for the out-of-sample evaluation. Adopting a one-step-ahead horizon is standard in nowcasting exercises: it reduces error propagation across multiple steps and focuses on the next immediate quarter, which is typically of greatest policy and practical interest \parencite{Tashman2000,MarcellinoStockWatson2006}.

\paragraph{} 
The combination of Expanding Window in training and Rolling Window in validation meets the requirements of a quarterly nowcasting system, where the progressive availability of new information consolidates the model's robustness \parencite{GiannoneReichlinSmall2008}. The 12-quarter Validation Window offers a balance between continuity of estimates and attention to the most recent signals, while the single-step forecast horizon reflects the immediate macroeconomic outlook \parencite{BanburaRunstler2011}. It is worth recognizing that exceptional shocks or sudden regime shifts could attenuate the framework's ability to capture new conditions promptly \parencite{PesaranTimmermann2007}. In such circumstances, the progressive accumulation of data may be insufficient to compensate for structural changes \parencite{ClarkMcCracken2009}. Nonetheless, the iterative mechanism illustrated here retains a high level of adaptability, as it incorporates the most recent observations at each round and, if necessary, allows for the reconsideration of variables or model configurations \parencite{GiraitisKapetaniosPrice2013}.

\vspace{0.35cm}

\phantomsection
\subsection{\textbf{\textit{Hyperparameter Tuning in the Nowcasting Process}}}
\label{sec:2.3.2}
Validation and selection of hyperparameters are crucial steps in our nowcasting framework, as the optimal configuration of the prediction model's hyperparameters significantly affects forecast accuracy.

In our setup, we integrate hyperparameter optimization into the iterative Expanding+Rolling data scheme described in the previous subsection. This procedure unfolds in several steps, ranging from data loading and validation-split creation to the final choice of the hyperparameter set that minimizes the mean squared error (MSE) on the validation set.

\vspace{0.35cm}

\phantomsection
\subsubsection{\textbf{\textit{Hyperparameter Optimization: Overview and Rationale for Bayesian Methods}}}
\vspace{0.2cm}
Hyperparameter fine-tuning in a predictive model (e.g., the hidden dimension of a neural network, the regularization strength, or the learning rate) includes various strategies, for example:

\begin{itemize}
  \item \textit{Grid Search}: It exhaustively evaluates a grid of parameter combinations. Although conceptually straightforward, it becomes infeasible for large or continuous parameter spaces as the number of combinations rapidly grows.
  \item \textit{Random Search}: This approach randomly samples parameter sets. It typically outperforms grid search in high-dimensional spaces but does not systematically leverage prior search outcomes.
  \item \textit{Genetic Algorithms}: They "evolve" a population of candidate solutions (i.e., hyperparameter vectors) through mutation and crossover operators inspired by natural selection \parencite{Holland1975,Goldberg1989,Mitchell1998}. While they can explore large search spaces, they often require numerous generations and careful tuning of evolutionary parameters (e.g., population size or mutation rate).
\end{itemize}

While each method has its merits, they can become computationally intensive in higher-dimensional scenarios. Grid Search suffers from combinatorial explosion, Random Search fails to incorporate past trial outcomes systematically, and Genetic Algorithms demand multiple generations and specialized parameter tuning. By contrast, Bayesian Optimization updates its search distribution after every trial, directing the search toward the most promising hyperparameter regions and reducing the computational overhead in our nowcasting setting \parencite{SnoekEtAl2012,GustafssonEtAl2023}.

To thoroughly explore potentially optimal model configurations, we adopt broad hyperparameter ranges for all methods tested\footnote{Recall that these methods are: LASSO, Ridge, Elastic Net, Principal Component Regression, Partial Least Square Regression, Random Forest, eXtreme Gradient Boosting, Multilayer Perceptron, Gated Recurrent Unit.}.
Theoretically, wide intervals help prevent overlooking unusual but beneficial hyperparameter values in macroeconomic contexts \parencite{BergstraEtAl2011,ShahriariEtAl2016}. Narrow ranges can bias or restrict the model to local optima, especially when relationships are nonlinear or the predictor set is large \parencite{SnoekEtAl2012}. By contrast, allowing hyperparameters to span orders of magnitude (e.g., for regularization or hidden dimensions) gives the algorithm the flexibility to discover diverse solutions.
Practically, macroeconomic nowcasting often involves heterogeneous data, from short, low-frequency series to many high-dimensional indicators \parencite{BanburaRunstler2011,Varian2014}. Broader ranges increase robustness, letting TPE and pruning mechanisms find "extreme" values that might be useful under regime shifts or strong nonlinearities \parencite{CerqueiraEtAl2020}. Bayesian optimization also prunes less promising configurations early \parencite{SnoekEtAl2012,AkibaEtAl2019}, focusing the search on promising regions without high computational costs. That is especially relevant for rolling or expanding-window procedures, where limiting hyperparameters could exacerbate misspecification risks \parencite{Tashman2000,ProbstWrightBoulesteix2019}.

In the specific case of quarterly nowcasting for Singapore's deflated GDP, with a dataset of 63 explanatory variables and 134 total observations, we need a method that efficiently uses each evaluation and quickly prioritizes promising hyperparameter configurations, thus preventing excessive computational burden. Recent evidence from large Bayesian VARs highlights how careful hyperparameter selection (e.g., shrinkage priors) can markedly improve macroeconomic forecast accuracy\footnote{See also the foundational work on BVARs in \parencite{Litterman1986}.} \parencite{BanburaEtAl2010,CarrieroClarkMarcellino2015a,GiannoneEtAl2015,CimadomoEtAl2022}.

\textit{Bayesian Optimization}, specifically the \textit{Tree-structured Parzen Estimator} (TPE) method from the Python library Optuna, offers an attractive compromise since it updates a surrogate model at each new trial (where a trial, within each iteration, corresponds to a training and validation attempt using a given set of hyperparameters, here represented by the vector \(\boldsymbol{\theta}\)). This surrogate model steadily focuses on the most promising regions of the hyperparameter space.

\vspace{0.35cm}

\phantomsection
\subsubsection{\textbf{\textit{Core Principles of Bayesian Optimization}}}
\vspace{0.2cm}
Bayesian optimization aims to progressively locate those regions in the hyperparameter space that yield superior results without resorting to an exhaustive or purely random search. In general terms, consider defining an objective function:

\begin{equation}
\mathcal{L}(\boldsymbol{\theta}) 
= \mathrm{MSE}_{\mathrm{val}}(\boldsymbol{\theta})
= \frac{1}{n_{\mathrm{val}}} \sum_{i=1}^{n_{\mathrm{val}}}
\Bigl(y_{i}^{(\mathrm{val})} - \hat{y}_{i}^{(\mathrm{val})}(\boldsymbol{\theta})\Bigr)^{2}
\end{equation}

where \(\boldsymbol{\theta} \in \mathcal{H}\) is a hyperparameter configuration (e.g., the hidden dimension of a neural network, a regularization coefficient, or the learning rate), and \(\mathcal{H}\) is the space of all admissible hyperparameter combinations. In \(\hat{y}_{i}^{(\mathrm{val})}(\boldsymbol{\theta})\), the model configured by \(\boldsymbol{\theta}\) yields the prediction for the \(i\)-th observation in the validation set. In contrast, \(n_{\mathrm{val}}\) is the number of observations in that validation subset. Bayesian optimization iteratively constructs a surrogate model of \(\mathcal{L}(\boldsymbol{\theta})\), updating it whenever we obtain a new MSE value for a particular set of hyperparameters.

We employ the Tree-structured Parzen Estimator algorithm in Optuna\footnote{The TPE algorithm was originated in the Hyperopt library and subsequently integrated into other frameworks, including Optuna.}. TPE splits trials into two classes: those with a "good" error (below a threshold \(\gamma\)) and those with a "less good" error (above or equal to \(\gamma\)). The threshold \(\gamma\) is a dynamic threshold used by TPE to classify trials (i.e., the different hyperparameter sets tested), distinguishing those with a validation error under \(\gamma\) from those with worse performance. Unlike a static approach, \(\gamma\) is often determined by selecting some fraction (e.g., 10\%) of the lowest errors so far, and it updates as new results are generated. TPE then constructs two conditional probability distributions:

\begin{equation}
\ell(\boldsymbol{\theta}) = p\!\bigl(\boldsymbol{\theta} \mid (\mathrm{MSE}<\gamma)\bigr)
\end{equation}

\begin{equation}
g(\boldsymbol{\theta}) = p\!\bigl(\boldsymbol{\theta} \mid (\mathrm{MSE}\ge \gamma)\bigr)
\end{equation}

and selects the next hyperparameter configuration \(\boldsymbol{\theta}\) by maximizing the ratio \(\ell(\boldsymbol{\theta}) / g(\boldsymbol{\theta})\). Consequently, the algorithm focuses on upcoming trials in regions of the hyperparameter space that, based on prior observations, appear most promising. This approach is especially effective with a limited dataset (134 quarters, in our case) since each evaluation provides valuable information immediately incorporated into the surrogate model. In practice, the algorithm works in this way:

\begin{itemize}

    \item \textit{Number of Trials \((n_{\mathrm{trials}})\):} a maximum number of evaluations (for instance, 60) is set for each time-window iteration. If \(n_{\mathrm{trials}}=60\), then as many as 60 training+validation cycles---each using a different set of hyperparameters---can occur within that single iteration. The algorithm finally identifies the \(\boldsymbol{\theta}^{*}\) combination that minimizes the average MSE across the entire validation set.
    \item \textit{Validation Error Computation:} in each trial, the model is trained on the whole training set \((X_{\mathrm{train}}, y_{\mathrm{train}})\) up to a maximum number of iterations---or until the solver's convergence criterion is met---according to the model's specified configuration. It then generates 12 forecasts (one per quarter in the validation set) and computes the mean squared error (MSE) across these 12 points. Based on that MSE, TPE determines whether the trial is "good" (\(\mathrm{MSE} < \gamma\)) or "less good."
\end{itemize}

\subsubsection*{\textit{Pruning Mechanism (Median Pruner) and Reducing Computational Overhead}}
The algorithm applies a pruning mechanism, i.e., early termination of trials whose partial performance is significantly worse than the median of completed trials at the same training epoch. Pruning avoids expending excessive computational resources on clearly suboptimal configurations. Specifically, the "Median Pruner" in Optuna compares the cumulative validation MSE in the early epochs of each trial to the median MSE recorded among completed trials. If the trial in progress lags well behind that median, Median Pruner truncates the trial prematurely. This strategy proves highly beneficial in our context, where the limited sample (134 quarters) and the large number of features (63) make it necessary to conserve computational resources. Additionally, a predefined number of "warm-up" trials run to completion, gathering baseline statistics for TPE and balancing exploration against potential exploitation of already-discovered good solutions.

Hence, there can be two levels of early termination in each trial:

\begin{itemize}
  \item \textit{Internal early stopping}, if the validation MSE fails to improve for 30 epochs,
  \item \textit{Pruning}, if the trial's partial results are significantly above the median of previous trials.
\end{itemize}

\subsubsection*{\textit{Iterative Procedure under the Expanding+Rolling Framework}}
As explained in Section~2.3.1, at each time-window, iteration the data are partitioned as follows:

\begin{itemize}
  \item \textit{Training set} \((X_{\mathrm{train}}, y_{\mathrm{train}})\): covers the observations from the initial period up to a certain limit, expanded at each step;
  \item \text{Validation set} \((X_{\mathrm{val}}, y_{\mathrm{val}})\): includes the last 12 quarters of the training window, used for hyperparameter evaluation;
  \item \textit{Test set} \((X_{\mathrm{test}}, y_{\mathrm{test}})\): the immediately following quarter, excluded from both hyperparameter choice and validation, thereby yielding an out-of-sample forecast.
\end{itemize}

In each iteration, TPE explores up to \(n_{\mathrm{trials}}\) hyperparameter configurations internal to that same time window. For each trial:

\begin{enumerate}
  \item The model is trained on \((X_{\mathrm{train}}, y_{\mathrm{train}})\) with the proposed hyperparameters \(\boldsymbol{\theta}\).
  \item The validation error is computed on \((X_{\mathrm{val}}, y_{\mathrm{val}})\).
  \item The pruning procedure can halt those trials with notably poor performance.
  \item After exhausting (or pruning) all \(n_{\mathrm{trials}}\), TPE selects the \(\boldsymbol{\theta}^{*}\) that minimizes \(\mathrm{MSE}_{\mathrm{val}}(\boldsymbol{\theta})\).
\end{enumerate}

Afterward, the model is retrained on \(\bigl(X_{\mathrm{train}} \cup X_{\mathrm{val}}, y_{\mathrm{train}} \cup y_{\mathrm{val}}\bigr)\) using \(\boldsymbol{\theta}^{*}\), generally with the same training parameters (up to 200 epochs, early stopping at 30). This final model then produces an out-of-sample forecast on \((X_{\mathrm{test}}, y_{\mathrm{test}})\). Shifting the time window by one quarter completes one iteration; repeating this procedure yields a sequence of out-of-sample forecasts that aligns with real-world nowcasting practices.

\par{}  
Adopting TPE alongside a pruning mechanism (Median Pruner) effectively handles the search for hyperparameter configurations over a broad space, even with relatively few quarterly observations \parencite{GustafssonEtAl2023}. TPE rapidly directs trials toward promising regions, reducing the full training required compared to a fully random or exhaustive search; pruning halts configurations with high error, saving computational resources. Overall, this synergy iteratively improves the out-of-sample forecasts while maintaining a pipeline sufficiently agile to be updated every quarter.

\vspace{0.35cm}

\phantomsection
\subsection{\textbf{\textit{Handling Shock Dummies to Prevent Look-Ahead Bias}}}
\label{sec:2.3.3}
Modeling macroeconomic shocks in a nowcasting context plays a crucial role in correctly interpreting economic evolution. In such a context, using dummy variables that signal disruptive events of a positive or negative nature---such as sudden financial crises, pandemics, or unexpected production recoveries---allows us to emphasize exceptional circumstances that usual cyclical trends cannot recognize \parencite{Salkever1976,ClementsHendry1996,CarrieroEtAl2022}. However, such variables must be rigorously managed to ensure that the validation and forecasting process does not incorporate future information that is not yet available during the analysis (look-ahead bias) \parencite{GiannoneLenzaPrimiceri2021,LenzaPrimiceri2022}.

During the Expanding+Rolling iterative procedure and in order to prevent any temporal bias, we adopted the following strategy:

\begin{enumerate}
    \item \textbf{Validation subset (12 quarters).} For the first eleven quarters of validation, the shock dummies maintain their effective value, reflecting the availability of complete historical data in retrospect; in the last validation quarter, the dummies are forced to zero, simulating the scenario in which forecasters did not know impending extraordinary events when they generate the validation estimates.
    \item \textbf{Test set (1 quarter).} Likewise, during the out-of-sample prediction quarter, we neutralize the shock dummies. That ensures that the model's forecast does not inadvertently exploit knowledge of any future disturbance that has not yet appeared in the observable data.
\end{enumerate}

\par{}  
By neutralizing shock dummies only in the last validation quarter and in the test quarter, the procedure reflects a genuine forecasting scenario in which new shocks are unknown until they manifest in real data. At the same time, the model can still learn from historically observed shocks, thereby retaining any insights gained from past anomalous events. This combination of selective neutralization and rolling estimation promotes robust and bias-free short-term projections, thus enhancing the credibility of the overall nowcasting framework.

\vspace{0.35cm}

\phantomsection
\subsection{\textbf{\textit{Model Families and Benchmarks}}}
\label{sec:2.3.4}
The absence of an official nowcasting model for Singapore's deflated quarterly GDP growth motivates the adoption of three essential benchmarks, alongside which we apply several families of statistical-mathematical models. On the one hand, we establish a comparison with elementary or standard references (Random Walk, AR(3) selected from various orders, and a specially configured Dynamic Factor Model); on the other hand, we evaluate several types of models that, due to their structure and theoretical assumptions, present characteristics potentially suitable for a small, dynamic and highly integrated economy.

In this section, we describe the fundamental mathematical-statistical principles of benchmarks and nowcasting models, grouped according to their distinct estimation approaches and theoretical premises to provide a solid methodological basis essential to evaluate the effectiveness of each approach:

\begin{itemize}
    \item \textbf{the benchmarks}, which act as fundamental baselines for evaluating the performance of our nowcasting models;
    \item \textbf{the families of nowcasting models}, grouped based on their respective estimation criteria and structural assumptions:
    \begin{itemize}
        \item \textit{Penalized Linear Models}, which exploit regularization (e.g., Lasso, Ridge, Elastic Net) as a way to handle high-dimensional data and mitigate overfitting;
        \item \textit{Dimensionality Reduction Models}, which employs techniques (e.g., Principal Component Regression, Partial Least Squares Regression) to project the original dataset onto a lower-dimensional subspace, preserving the core informational content;
        \item \textit{Ensemble Learning Models}, based on the combination of multiple elementary predictors (such as Random Forest or XGBoost) exploiting their internal diversity in order to improve stability and accuracy of predictions;
        \item \textit{Neural Models}, which adopt flexible hyperparametric architectures (Multilayer Perceptron, Gated Recurrent Unit) capable of approximating non-linear relationships and capturing temporal dependencies.
    \end{itemize}
\end{itemize}

\vspace{0.5cm}

\phantomsection
\subsubsection{\textbf{\textit{Benchmarks}}}

\vspace{0.35cm}

\underline{{Random Walk}}
\vspace{0.2cm}\\
The Random Walk (RW) or "na\"ive forecast" offers a basic yet essential strategy for predicting GDP growth without imposing structural constraints or including exogenous regressors \parencite{Muth1960, MakridakisEtAl1998, Tashman2000}. Its core assumption is that the optimal predictor for period \(t+1\) is simply the observed value at time \(t\), thereby forgoing any additional hypothesis about underlying dynamics.

Formally, the model assumes a unit-root process:

\[
y_{t+1} = y_{t} + \varepsilon_{t+1},
\]

\noindent{}where:

\begin{itemize}
\item \(y_{t}\) represents the target variable (e.g., quarterly GDP growth),
\item \(\varepsilon_{t+1}\) is a stochastic error term with zero mean and constant variance.
\end{itemize}

Consequently, the one-step-ahead forecast is given by:

\[
\hat{y}_{t+1 \mid t} = y_{t}.
\]

This specification implies that any permanent shock is not reabsorbed and, in the absence of further information, the method cannot distinguish a long-term economic trend from a temporary fluctuation \parencite{NelsonPlosser1982, StockWatson1996}.

The forecast procedure can be summarized in three steps:

\begin{enumerate}
    \item Define an initial historical window containing \(y_{1}, y_{2}, \dots, y_{t}\).
    \item At time \(t\), assign \(\hat{y}_{t+1 \mid t} = y_{t}\), positing that no supplementary information alters the most recent observed change.
    \item Once the actual data \(y_{t+1}\) becomes available, compute the forecast error 
    \(\hat{y}_{t+1 \mid t} - y_{t+1}\) to assess predictive performance.
\end{enumerate}

\par{}  
The effectiveness of the Random Walk heavily depends on the data-generating process. If the time series predominantly follows a stochastic pattern, the na\"ive forecast can be highly competitive vis-\`a-vis more complex techniques \parencite{MakridakisEtAl1998, Tashman2000}. However, in the presence of pronounced deterministic trends or clearly defined cyclical patterns, this approach may underestimate or overestimate structural changes, as it merely projects the latest observation forward without modeling any trend component. Because it requires no parameter or hyperparameter estimation, the Random Walk serves as a fundamental benchmark in forecasting exercises \parencite{MakridakisSpiliotisAssimakopoulos2022, MakridakisSpiliotisAssimakopoulos2022}. Its minimalism highlights the gains in predictive accuracy offered by more sophisticated methods, making improvements more credible once compared with a technique that, by definition, integrates no additional information beyond the last observed data point.

\vspace{0.35cm}

\underline{{Autoregressive Models}}
\vspace{0.2cm}\\
Autoregressive (AR) models rank among the most widespread approaches for analyzing economic time series, as they assume that the current value of a variable (e.g., quarterly deflated GDP growth)
depends linearly on its past observations. The general form of an AR($p$) model is:

\[
y_{t} = \alpha + \phi_{1} \, y_{t-1} + \phi_{2} \, y_{t-2} + \cdots + \phi_{p} \, y_{t-p} + \varepsilon_{t},
\]

\noindent{}where:

\begin{itemize}
    \item $y_{t}$ denotes the endogenous variable, in this case, the deflated quarterly GDP growth at time $t$;
    \item $\alpha$ is the intercept that captures the systematic mean level of the series;
    \item $\phi_{1}, \phi_{2}, \dots, \phi_{p}$ are the autoregressive coefficients, each measuring the influence exerted by the corresponding lag;
    \item $\varepsilon_{t}$ is a random error term with zero mean and constant variance, representing shocks not accounted for by the model.
\end{itemize}

Such a structure enables the model to capture persistence over time in an economic process,
especially when prior shocks continue to affect subsequent observations.

An AR(3) model expresses the current value of a time series (here, Singapore's deflated quarterly GDP growth) as a linear combination of its three most recent past values. Formally, this can be written as:

\[
y_{t} = \alpha + \phi_{1}\,y_{t-1} + \phi_{2}\,y_{t-2} + \phi_{3}\,y_{t-3} + \varepsilon_{t},
\]

\noindent{}where:

\begin{itemize}
    \item \(y_{t}\) is the deflated GDP growth at time \(t\);
    \item \(\alpha\) denotes the long-run average component (intercept);
    \item \(\phi_{1}, \phi_{2}, \phi_{3}\) capture the influence of the first, second, and third lag, respectively;
    \item \(\varepsilon_{t}\) represents residual shocks or noise unaccounted for by the three lags \parencite{MarcellinoStockWatson2006, Hamilton1994}.
\end{itemize}

This specification goes beyond the simplicity of, for instance, an \textit{AR(1)} or a Random Walk \parencite{NelsonPlosser1982}, as it can capture short-run dynamics over multiple quarters yet remains more parsimonious than higher-order autoregressive models.

The estimation and forecast procedure for an AR(3) typically follows these steps:

\begin{enumerate}
    \item \textbf{Defining the historical window.} An initial dataset \(\{y_{1}, y_{2}, \dots, y_{T}\}\) is isolated. If a rolling estimation approach is applied, the earliest points may be omitted to allow for progressive model updating \parencite{BanerjeeMarcellino2006}.
    \item \textbf{Model fitting.} At each iteration, one estimates the intercept \(\alpha\) and the coefficients \(\phi_{1}, \phi_{2}, \phi_{3}\) via maximum likelihood or least squares, under standard assumptions of white noise errors \parencite{Hamilton1994}.
    \item \textbf{One-step-ahead forecasting.} Substituting \(\hat{\alpha}, \hat{\phi}_{1}, \hat{\phi}_{2}, \hat{\phi}_{3}\) and the last three observed values \(y_{t}, y_{t-1}, y_{t-2}\) into the AR(3) equation gives the forecast \(\hat{y}_{t+1}\).
    \item \textbf{Forecast error evaluation.} Once the actual \(y_{t+1}\) is released, \(\hat{y}_{t+1} - y_{t+1}\) quantifies the prediction error for that quarter \parencite{MarcellinoStockWatson2006}\footnote{In our analysis of Singapore's GDP nowcasting, we tested multiple AR orders, from AR(1) to AR(4), and compared them by the Root Mean Squared Forecast Error (RMSFE). Overall, AR(3) exhibits the lowest RMSFE, although in certain sub-periods (Pre-COVID, COVID, Post-COVID, Excluding-COVID) other orders may marginally outperform it. Hence, AR(3) is a balanced choice for the entire sample, as it captures up to three-lag dynamics without the additional parameters of higher-order models. Nonetheless, large structural shifts---like those witnessed during the pandemic---can alter which AR order best aligns with specific phases of the economy \parencite{LenzaPrimiceri2022, GiannoneLenzaPrimiceri2021, MarcellinoSchumacher2010}.}.
\end{enumerate}

\par{}  
In our context, AR(3) offers a middle-ground strategy that often outperforms simpler models, such as AR(1) or the Random Walk \parencite{NelsonPlosser1982}, while remaining more tractable than AR processes of higher order. Its practical effectiveness, however, ultimately depends on the stability of the underlying economic environment and on how frequently large, unanticipated shocks occur.

\vspace{0.35cm}

\underline{{Dynamic Factor Models}}
\vspace{0.2cm}\\
Dynamic Factor Models (DFM) are a widely used tool in macroeconomic analysis. They allow the synthesizing of a large set of variables that describe the dynamic structure of an economic system in a small number of latent factors.

Starting from the studies of \parencite{BaiNg2002}, which introduce optimal selection criteria for the number of factors, and of \parencite{StockWatson2002}, which show how a limited set of principal components can explain most of the macroeconomic variability, DFMs have evolved by integrating different solutions for the management of incomplete data and heterogeneous frequencies. In particular, \parencite{DozGiannoneReichlin2011,DozGiannoneReichlin2012} employ a state-space approach \footnote{In the state-space approach, we represent a system by a transition equation for the dynamics of latent variables (states) and an observation equation that links the states themselves to the empirical data. A Kalman filter continuously updates the states, even with missing data or non-homogeneous release timing.} and an Expectation--Maximization (EM) algorithm \footnote{EM procedures alternate an Expectation phase, in which the distribution of unobserved variables is estimated based on the current parameters, with a Maximization phase, in which the parameters are updated to maximize the conditional likelihood. This iteration continues until convergence.}, combined with the Kalman filter, to address problems of ragged edge\footnote{With ragged edge, we indicate the situation in which some variables end before others or have different publication delays, generating an irregular structure in the last points of the sample.} \parencite{BanburaRunstler2011} or time series with mixed frequency.

This type of approach provides maximum flexibility. However, in contexts where the data are aligned on a single frequency and are complete, these tools may exceed the actual operational needs and be computationally expensive without frequency discrepancies or significant missing data.

From a theoretical point of view, a DFM assumes that the vector

\(
x_{t} \in \mathbb{R}^{N}
\),

the vector of the variables observed at time \(t\), is composed of common factors and idiosyncratic components according to the equation:

\[
x_{t} = \Lambda f_{t} + \varepsilon_{t},
\]

\noindent{}where:

\begin{itemize}
\item \(f_{t} \in \mathbb{R}^{r}\) represents the latent factors (with \(r \ll N\)),
\item \(\Lambda\) indicates the factor loading matrix and
\item \(\varepsilon_{t}\) contains the portion of variability specific to each variable (not explained by the factors)\footnote{In many \parencite{StockWatson2011} implementations, \(\varepsilon_{t}\) is often treated as simple white noise or dynamically modeled.}.
\end{itemize}

In this work, we adopt a "light" DFM model with a more essential structure than the state-space one. In particular, we extracted the factors \(f_{t}\) through Principal Component Analysis. This choice reflects the perspective of \parencite{StockWatson2002}, according to which a few principal factors can explain most of the macroeconomic behavior. To allow for dynamic modulation of the unrelated errors, we consider a structure

\[
\varepsilon_{t} = \Phi(L)\,\varepsilon_{t-1} + e_{t},
\]

\noindent{}where:

\begin{itemize}
\item \(\Phi(L)\) is a polynomial in the lag\footnote{For example, with order \(p\), we have \(\Phi(L) = 1 - \phi_{1}L - \phi_{2}L^{2} - \cdots - \phi_{p}L^{p}\). Applying \(\Phi(L)\) to \(\varepsilon_{t}\) allows us to introduce dependencies on past observations, making the idiosyncratic an autoregressive process.} of empirically selected order
\item \(e_{t}\) is a white noise error term.
\end{itemize}

Therefore, we preferred an AR for the idiosyncratic term\footnote{In this autoregressive model, the \(\varepsilon_{t}\) variables can retain persistences not captured by the factors. Since we have completed and without frequency mismatches in the quarterly series, we considered the adoption of iterative estimates of the latent states unnecessary.}, avoiding using a completely state-space model since the data were already aligned and did not present missing values that needed to be estimated iteratively.
To automatically search for the optimal number of factors \(r\) and the order of \(\Phi(L)\) for each quarter of nowcasting, we used the Bayesian-like iterative optimization process on a sliding validation window and pruning described in section~\ref{sec:2.3.2}.
We omitted using Kalman filters and EM procedures since our series were already quarterly and free of missing data. This condition made unnecessary the data reconciliation or imputation mechanisms proposed in \parencite{DozGiannoneReichlin2011,DozGiannoneReichlin2012}.
During the analysis, we then directly integrated the dummy variables of seasonality and economic shocks into the regression component that links the factors to the target variable without adopting more sophisticated state-space smoothing techniques.

\par{}  
This approach is consistent with the reference literature. However, it simplifies the management of any temporal misalignments and missing data, exploiting a dataset in which the quarterly frequency is already uniform.

\vspace{0.5cm}

\phantomsection
\subsubsection{\textbf{\textit{Penalized Linear Models}}}
\vspace{0.2cm}
Penalized linear models offer a structured approach to controlling overfitting in contexts where numerous regressors may exhibit strong collinearity or marginal relevance. In the one-step-ahead quarterly nowcasting setting for a small open economy, introducing specific shrinkage terms in the objective function helps reconcile flexibility and parsimony. This mechanism prevents extraneous variables from overshadowing essential economic signals, enhancing predictive accuracy \parencite{Tibshirani1996,HoerlKennard1970,ZouHastie2005}.

At the core of this family are techniques that impose different regularization strategies. Lasso, for instance, incorporates an L1 penalty, which drives some coefficients to zero and facilitates automatic variable selection. Conversely, Ridge employs an L2 penalty, shrinking all coefficients continuously while retaining each predictor. Elastic Net combines L1 and L2 components, balancing feature selection with robust shrinkage. These frameworks adapt to evolving economic relationships and integrate new indicators without excessively complicating the model's structure. Their adaptability is particularly relevant when policy shifts or global disturbances arise, highlighting their utility in dynamic environments.

Several design principles distinguish penalized linear models from unpenalized regressions:

\begin{itemize}
    \item \textit{Regularization Control}: adjusting the penalty strength governs the trade-off between bias and variance, thus mitigating overfitting.
    \item \textit{Sparse or Dense Solutions}: Lasso-induced sparsity excludes uninformative regressors, whereas Ridge retains them at reduced magnitudes.
\end{itemize}

Penalized linear models thus represent robust tools for macroeconomic forecasting, especially when data availability expands incrementally. They can handle variations in sample size and shifting data patterns without resorting to intricate architectures. While these methods typically employ linear specifications and do not inherently capture highly nonlinear dynamics, they often balance predictive performance and simplicity. With careful and incremental hyperparameter tuning, they yield short-horizon forecasts that absorb shocks yet remain attuned to underlying trends. This robustness underscores their efficacy in capturing the dynamic nature of economic systems, frequently outperforming unregularized alternatives in volatile and high-dimensional contexts \parencite{SmeekesWijler2018,UematsuTanaka2019}.

\vspace{0.35cm}

\underline{{Least Absolute Shrinkage and Selection Operator}}
\vspace{0.2cm}\\
Least Absolute Shrinkage and Selection Operator (LASSO) is a regression technique for forecasting models when dealing with several explanatory variables \parencite{Tibshirani1996, ZouHastie2005}. It effectively manages many explanatory variables, including some that may not substantially influence the prediction. In this scenario, LASSO is theoretically helpful for selecting the most significant variables and avoiding overfitting\footnote{The term overfitting refers to excessive adaptation of the model to the training data, which compromises its generalization ability \parencite{SmeekesWijler2018}.}
the model to the data, thus optimizing real-time forecasts.

The use of L1 regularization distinguishes LASSO \parencite{HastieTibshiraniWainwright2015}. In this context, the parameter $\lambda$ imposes a penalty equal to the sum of the absolute values of the coefficients $\bigl(\sum_{j=1}^{p}\lvert \beta_{j}\rvert\bigr)$. This mechanism, often referred to as "shrinkage penalty," progressively reduces the coefficients of some variables until they are zero, thereby excluding the less relevant ones. By increasing $\lambda$, the extent of this reduction increases, producing a more accentuated "sparseness\footnote{"Sparsity" is the property of L1 penalty to drive some coefficients exactly to zero, thus eliminating unimportant features. This principle was originally introduced by \parencite{Tibshirani1996}, and later extended by \parencite{ZouHastie2005}, demonstrating its practical advantages in high-dimensional regression scenarios.}" effect.

In the case of models with multiple regressors, where there is a high risk of overfitting, the LASSO allows maintaining a good balance between the estimates' precision and the specification's simplicity \parencite{Zou2006, UematsuTanaka2019}. LASSO differs from traditional linear regressions by automatically selecting variables based on coefficient values, thereby highlighting the most predictive variables \parencite{MeiShi2024}.

From a theoretical point of view, LASSO\footnote{In this discussion, we follow the convention used by the \texttt{scikit-learn} library \parencite{scikit-learn, sklearnLasso}, which inserts the term $\tfrac{1}{2n}$ to normalize the sum of squares and estimates an intercept parameter $\beta_{0}$ separately.} is based on minimizing a cost function that combines the estimation error with an L1 penalty, proportional to the sum of the absolute values of the coefficients. In summary form, the function to minimize is:

\[
\hat{\boldsymbol{\beta}}
= \argmin_{\boldsymbol{\beta}}
\Bigl\{
\,\underbrace{\frac{1}{2n} \sum_{i=1}^{n} \bigl(y_{i} - \beta_{0} - \sum_{j=1}^{p} \beta_{j}\,x_{ji}\bigr)^{2}}_{\text{error term (reduced MSE)}}
+\underbrace{\lambda \sum_{j=1}^{p} \lvert \beta_{j} \rvert}_{\text{L1 regularization}}
\Bigr\},
\]

\noindent{}where:

\begin{itemize}
    \item $\tfrac{1}{2n}\sum_{i=1}^{n}\bigl(y_{i}-\beta_{0}-\sum_{j=1}^{p}\beta_{j}\,x_{ji}\bigr)^{2}$ represents the measure of the estimation error (a "reduced form of MSE\footnote{The expression "reduced form of MSE" denotes that, instead of using the canonical formula $\tfrac{1}{n}\sum_{i=1}^{n}(\dots)^2$, the coefficient $\tfrac{1}{2n}$ is introduced in front of the squared residuals. This change neither alters the location of the minimum nor the interpretation of the mean squared error, but it simplifies the derivatives and numerical handling during optimization.}");
    \item $\lambda \sum_{j=1}^{p} \lvert \beta_{j}\rvert$ is the L1 penalty (regularization term), which controls how much the coefficients are shrunk \parencite{HastieTibshiraniWainwright2015}.
\end{itemize}

\noindent{}More specifically,

\begin{itemize}
    \item $y_{i}$ denotes the observed dependent variable;
    \item $x_{i}$ represents the vector of independent variables (features) for observation $i$;
    \item $\beta_{j}$ indicates the coefficients to be estimated;
    \item $\lambda$ is the regularization parameter regulating the intensity of the L1 penalty (the higher $\lambda$, the more coefficients are reduced to zero).
\end{itemize}

Our approach to modeling with LASSO follows these fundamental steps, emphasizing that the penalty parameter $\lambda$ is determined through an iterative process:
\begin{enumerate}
    \item \textbf{Selection of the parameter $\lambda$ by optimization and pruning.} The determination of $\lambda$, which controls the intensity of the L1 regularization, takes place at each iteration through a Bayesian optimization process based on the TPE and pruning techniques, implemented in Optuna \parencite{SnoekEtAl2012, GustafssonEtAl2023}. In each temporal block of data (train+validation), the algorithm searches the space of possible values of $\lambda$ and selects the one minimizing the mean square error on the validation subset. This procedure is repeated quarter by quarter, thus adapting progressively to the latest economic conditions \parencite{SmeekesWijler2018};
    \item \textbf{Model training and forecast generation.} Once $\lambda$ is selected in the single optimization step, the LASSO model is trained on the combined training and validation data, then used to generate the forecast for the test quarter (or period). After each forecast, the training dataset is extended by including the most recent observations, and the process of hyperparameter optimization and training is repeated to account for the continuous inflow of available information. This sequential strategy — where $\lambda$ is updated with each sample expansion — allows the model to keep track of evolving economic dynamics and reduces the risk of over-fitting in the long run.
\end{enumerate}

\par{}  
Adopting LASSO in this context allows us to choose $\lambda$ in a robust manner, minimizing the validation error and improving the model's predictive ability in future data \parencite{HastieTibshiraniWainwright2015, UematsuTanaka2019, MeiShi2024}.

\vspace{0.35cm}

\underline{{Ridge}}
\vspace{0.2cm}\\
Ridge Regression is a forecasting method used in regression analysis that involves several explanatory variables \parencite{HoerlKennard1970,ExterkateEtAl2016,KimSwanson2014,SmeekesWijler2018}. It efficiently handles cases where a large number of variables affect the outcome, even when none of them have a particularly strong effect on their own. In this context, using the L2 penalty allows for keeping all the regressors, even the less relevant ones, while ensuring rigorous control over overfitting by reducing the amplitude of the coefficients \parencite{HastieTibshiraniFriedman2009}.

The distinctive element of Ridge lies in the penalty that equals the square of the coefficients \(\bigl(\sum_{j=1}^{p} \beta_{j}^{2}\bigr)\). This approach, known as the "L2 type shrinkage penalty," incrementally lowers the coefficients' values without eliminating them completely. By raising \(\alpha\), the degree of contraction experienced by the coefficients increases, thus helping to reduce the model's variance in situations that may be prone to multicollinearity.

In the presence of multiple regressors, the Ridge maintains a balance between the estimates' precision and the model's structural simplicity since each variable remains included but suffers a penalty proportional to its size. As a result, the influence of less significant predictors is reduced without implying their complete removal.

From a theoretical point of view, the Ridge\footnote{As with other penalties, in \texttt{scikit-learn} we follow the convention of introducing the factor \(\tfrac{1}{2n}\) in front of the sum of the squared residuals and of estimating the intercept \(\beta_{0}\) separately \parencite{scikit-learn,sklearnRidge}.}
minimizes a cost function that combines the estimation error with an L2 penalty proportional to the sum of the squared coefficients. In summary form, the function to minimize is:

\[
\hat{\beta}
= \arg\min_\beta
\Bigl\{
\,\underbrace{\tfrac{1}{2n}\sum_{i=1}^{n}
\bigl(y_{i} - \beta_{0} - \sum_{j=1}^{p} \beta_{j}\,x_{ji}\bigr)^{2}}_{\text{estimation error (reduced form of MSE)}}
\;+\;
\underbrace{\alpha \sum_{j=1}^{p}\beta_{j}^{2}}_{\text{L2 regularization}}
\Bigr\},
\]

\noindent{}where:

\begin{itemize}
    \item \(\tfrac{1}{2n}\sum_{i=1}^{n}\bigl(y_{i}-\beta_{0}-\sum_{j=1}^{p}\beta_{j}\,x_{ji}\bigr)^{2}\) represents the measure of the estimation error (i.e., a "reduced" form of MSE);
    \item \(\alpha \sum_{j=1}^{p}\beta_{j}^{2}\) is the L2 penalty, which establishes the extent of the coefficients' contraction.
\end{itemize}

More specifically,

\begin{itemize}
    \item \(y_{i}\) indicates the observed dependent variable;
    \item \(x_{i}\) represents the vector of independent variables for observation \(i\);
    \item \(\beta_{j}\) denotes the coefficients to be estimated;
    \item \(\alpha\) is the regularization parameter, which regulates the intensity of the L2 penalty (the greater \(\alpha\), the stronger the contractions).
\end{itemize}

Our Ridge implementation follows these key steps, emphasizing that the parameter \(\alpha\) is determined through an iterative process:

\begin{enumerate}
    \item \textbf{Selection of the \(\alpha\) parameter through optimization and pruning.} The determination of \(\alpha\), which governs the intensity of the L2 penalty, was conducted at each iteration through a Bayesian optimization process based on the TPE approach and pruning techniques, implemented in Optuna \parencite{SnoekEtAl2012}. In each time block of data (train+validation), the algorithm explored the space of possible values of \(\alpha\), choosing the one that minimized the mean squared error on the validation subset. This procedure was repeated each quarter, gradually adapting to the most recent economic conditions \parencite{MedeirosEtAl2021};
    \item \textbf{Model training and forecast generation.} Once the value of \(\alpha\) was determined for each optimization phase, the Ridge model was trained on the combined training and validation data, then used to generate the forecast for the test quarter (or period). After each forecast, the training dataset was extended to include the most recent observations, and the calibration and training process was repeated to handle the constant influx of new information. This sequential strategy — in which \(\alpha\) is redefined at each sample expansion — makes the model responsive to evolving economic dynamics and counteracts any over-fitting phenomena in the long run.
\end{enumerate}

\par{} 
The calibration and validation mechanism in our model, which uses Ridge regression as a "prediction engine," allows the configuration of the hyperparameter \(\alpha\) while preserving the model's robustness for future data, thanks to effective shrinkage in reducing multicollinearity among the regressors \parencite{ExterkateEtAl2016,KimSwanson2014}.

\vspace{0.35cm}

\underline{{Elastic Net}}
\vspace{0.2cm}\\
Elastic Net (EN) is a regression methodology that integrates the L1 type penalty and the L2 type penalty in a single framework \parencite{ZouHastie2005,HastieTibshiraniWainwright2015}. This provides a flexible structure that combines the advantages of LASSO regression (partial selection of variables by zeroing some coefficients) with those of Ridge regression (variance containment thanks to the contraction of parameters). This double penalty is particularly effective in models with numerous regressors since it mitigates multicollinearity problems \parencite{SmeekesWijler2018,UematsuTanaka2019} and gradually eliminates the less significant predictors.

From a theoretical point of view, Elastic Net\footnote{The \texttt{scikit-learn} library \parencite{scikit-learn,sklearnElasticNet} applies by default a \(\tfrac{1}{2n}\) factor to the squared residuals and estimates the \(\beta_{0}\) intercept separately.} minimizes the following cost function:

\[
\hat{\boldsymbol{\beta}}
\,=\,
\argmin_{\boldsymbol{\beta}}
\Bigl\{
\underbrace{\frac{1}{2n}\sum_{i=1}^{n}
\Bigl(y_{i} - \beta_{0} - \sum_{j=1}^{p}\beta_{j}\,x_{ji}\Bigr)^{2}
}_{\text{error term (reduced MSE)}}
\;+\;
\underbrace{\alpha\Bigl[
 (1-\gamma)\sum_{j=1}^{p}\beta_{j}^{2}
 \;+\;
 \gamma \sum_{j=1}^{p}\lvert\beta_{j}\rvert
 \Bigr]}_{\text{mixed penalty L2 and L1}}
\Bigr\},
\]

\noindent{}where:

\begin{itemize}
 \item \(\frac{1}{2n}\sum_{i=1}^{n}
\Bigl(y_{i} - \beta_{0} - \sum_{j=1}^{p}\beta_{j}\,x_{ji}\Bigr)^{2}\) represents the estimate error component (reduced form of the \(\mathrm{MSE}\));
 \item \(\alpha\Bigl[(1-\gamma)\sum_{j=1}^{p}\beta_{j}^{2} \;+\; \gamma\sum_{j=1}^{p}\lvert\beta_{j}\rvert\Bigr]\) is the penalty term that mixes L2 and L1.
\end{itemize}

More specifically:

\begin{itemize}
      \item \(\alpha>0\) governs the overall intensity of the penalty (i.e., the amount of regularization imposed on the coefficients);
      \item \(0 \le \gamma \le 1\) determines the relative weight of L1 versus L2. If \(\gamma=1\), the algorithm becomes equivalent to pure LASSO; if \(\gamma=0\), it coincides with Ridge regression;
      \item since \(\beta_{0}\) is excluded from the penalty, both the intercept and the coefficients are handled consistently within this mixed penalization framework.
\end{itemize}

Our Elastic Net implementation follows these key steps, emphasizing that the parameters \(\alpha\) and \(\gamma\) are determined through an iterative process:

\begin{enumerate}
    \item \textbf{Searching for the \(\alpha\) and \(\gamma\) hyperparameters with optimization and pruning.} We calibrate the Elastic Net over multiple time steps using an iterative procedure for calibration and validation, regularly updating the \(\alpha\) and \(\gamma\) hyperparameters in light of the most recent data. In each data block (union of a training subset and validation window), the model searches for the values of \(\alpha\) and \(\gamma\) that minimize the mean squared error on the validation set, adopting a Bayesian optimization approach based on TPE (Tree-structured Parzen Estimator) \parencite{BergstraEtAl2011,AkibaEtAl2019} and possible pruning techniques. This procedure ensures an accurate selection of the penalty levels \parencite{KockMedeirosVasconcelos2020} and a constant adaptability of the model to systematic changes in the regressors;
    \item \textbf{Final training and forecast generation.} Once the optimal parameters are identified in a single block, the model is recalibrated on the entire training sample (\textit{training+validation}) and used to generate the forecast. Subsequently, the \textit{training} set is extended with the most recent observations, and the iteration is repeated, progressively aligning the Elastic Net regression to the dynamic characteristics of the analyzed phenomenon \parencite{PesaranTimmermann2007}.
\end{enumerate}

\par{} 
The sequential update strategy allows the Elastic Net to adjust its penalty parameters dynamically, progressively combining the L1 and L2 components. This approach improves the forecasting capacity of future data since it mitigates the risks of overfitting and maintains high performance in contexts characterized by continuous variations in the regressors.

\vspace{0.5cm}

\phantomsection
\subsubsection{\textbf{\textit{Dimensionality Reduction Models}}}
\vspace{0.2cm}
Dimensionality reduction techniques prioritize parsimony and robustness by mapping a potentially large set of interconnected variables onto fewer latent factors \parencite{JolliffeCadima2016,StockWatson2002,BoivinNg2006}. In quarterly nowcasting for a small open economy in Southeast Asia, such as Singapore, these methods reduce excessive noise and mitigate multicollinearity, thereby improving forecast stability. By transforming high-dimensional datasets into fewer orthogonal or covariance-driven components, dimensionality reduction models limit the risk of overfitting and accommodate unforeseen policy shifts or external shocks.

Principal Component Regression (PCR) and Partial Least Squares Regression (PLSR) exemplify this approach. PCR first identifies a subset of principal components that capture the bulk of variance in the predictor space \parencite{JolliffeCadima2016}, then fits a regression on these uncorrelated axes. This procedure attenuates issues arising from large predictor pools, helping the model focus on essential signals. Meanwhile, PLSR targets components that maximize the covariance between predictors and the target variable \parencite{Wold1985,MehmoodEtAl2012}, making it particularly appealing when the data contain many near-redundant series \parencite{BoivinNg2006}. By directly emphasizing predictive relevance, PLSR can outperform variance-only methods in scenarios where traditional factor-based models might include uninformative directions.

Several design elements characterize these methods:
\begin{itemize}
  \item \textit{Orthogonality}: PCR relies on principal components that are mutually uncorrelated, enhancing coefficient stability.
  \item \textit{Target-driven factors}: PLSR identifies latent directions aligned with the response variable, reinforcing predictive accuracy.
  \item \textit{Shrinkage control}: in PCR, modern implementations often include penalties (e.g., Ridge penalization) to limit excessive parameter growth.
  \item \textit{Adaptive calibration}: periodic re-estimation in an expanding window setup allows the system to integrate evolving patterns and structural changes.
\end{itemize}

Dimensionality reduction provides a more parsimonious representation, as superfluous features are filtered out. This streamlined perspective proves advantageous for nowcasting tasks in dynamic environments, where robust forecasts are often required. Moreover, these models do not impose the strict assumptions of purely linear frameworks, enabling flexible adaptations to evolving economic conditions.

\vspace{0.35cm}

\underline{{Principal Component Regression}}
\vspace{0.2cm}\\
The Principal Component Regression (PCR) constitutes a hybrid methodological framework that combines the dimensionality reduction of a set of predictors with the estimation of a linear regression model. In particular, the PCR extracts a subset of principal components starting from a potentially large and high-dimensional, multicollinearity-prone (i.e., the presence of strong linear correlations among two or more predictors) set of regressors. These components are constructed as orthogonal linear combinations of the original predictors---meaning they are uncorrelated weighted sums of the input variables, each capturing a specific direction of maximal variance in the data. By concentrating most of the original information into a smaller set of uncorrelated components, PCA effectively reduces the problem's dimensionality and mitigates issues such as multicollinearity \parencite{JolliffeCadima2016,sklearnPCR}.

Once the original regressors are projected onto the new orthogonal axes defined by these components, linear regression is performed using the resulting transformed variables, known as principal components (or scores). The classical formulation of PCR uses Ordinary Least Squares (OLS)\footnote{Ordinary regression (OLS) does not include any penalty term; it minimizes only the sum of the squared residuals.}---a method first proposed by \parencite{Massy1965}---which, however, may suffer from overfitting or high variance in the presence of noisy components. In traditional PCR, the components \(\mathbf{Z}\) are used within a simple OLS regression on the dependent variable \(\mathbf{y}\); such approaches have been successfully applied in macroeconomic forecasting \parencite{StockWatson2002}. In our implementation, we replace the OLS estimator with a Ridge regression \parencite{HoerlKennard1970,sklearnRidge}, thereby introducing an L2 penalty term that regularizes the magnitude of the coefficients and extends the standard PCR framework to enhance robustness and generalization. Recent studies have shown that Ridge regression can be a viable alternative to traditional PCR in high-dimensional settings \parencite{DeMolGiannoneReichlin2008, GiannoneLenzaPrimiceri2021}.

The Principal Component Analysis (PCA) itself relies on the spectral decomposition of the regressor matrix \(\mathbf{X} \in \mathbb{R}^{n \times p}\) after appropriate centering or standardization. This process yields eigenvectors and eigenvalues\footnote{Eigenvectors are directions that remain unchanged when a linear transformation is applied. Eigenvalues are scalars that quantify the amount of stretching, shrinking, or mirroring along those directions.} that characterize the data structure. Formally, one begins by computing the sample covariance (or correlation) matrix:

\[
\mathbf{S} = \frac{1}{n-1} \mathbf{X}^\top \mathbf{X},
\]

\noindent{}and solving the eigenvalue problem:

\[
\mathbf{S}\,\mathbf{v}_r = \lambda_r\,\mathbf{v}_r,
\]

\noindent{}which yields a set of orthogonal eigenvectors \(\mathbf{v}_r \in \mathbb{R}^{p}\), each associated with an eigenvalue \(\lambda_r\). The eigenvectors define the directions of maximal variance in the data, while the eigenvalues indicate how much variance is explained in each direction.

We retain the first \(k\) components that capture most of the total variance by ordering the eigenvectors according to their associated eigenvalues in decreasing order. Letting \(\mathbf{V}_k \in \mathbb{R}^{p \times k}\) denote the matrix whose columns are the top \(k\) eigenvectors, the matrix of scores---i.e., the transformed data in the reduced component space---is obtained as:

\[
\mathbf{Z} = \mathbf{X}\,\mathbf{V}_k.
\]

Each row of \(\mathbf{Z}\) corresponds to an observation expressed in terms of the new orthogonal axes. The columns of \(\mathbf{Z}\) are uncorrelated by construction, facilitating subsequent modeling steps and alleviating problems arising from multicollinearity in the original data.

In "traditional" PCR, the principal components \(\mathbf{Z}\) are used within a simple OLS regression on the dependent variable \(\mathbf{y}\). However, OLS may yield estimates with high variance when the data are noisy. Moreover, even if all variables carry some marginal signal, purely OLS-based estimates can become unstable if each predictor contributes only slightly but is still relevant; in such cases, leaving minor coefficients unshrunk can inflate variance.

Instead, our second stage adopts a Ridge regression in the principal component domain. Specifically, let \(\boldsymbol{\gamma} = (\gamma_1, \dots, \gamma_k)\) be the vector of Ridge coefficients on the principal components \(\mathbf{Z}\). Formally:

\[
\hat{\boldsymbol{\gamma}}_{\mathrm{Ridge}}
\;=\;
\argmin_{\boldsymbol{\gamma}}
\Bigl\{
\underbrace{\tfrac{1}{2n}\,\|\mathbf{y} - \mathbf{Z}\,\boldsymbol{\gamma}\|^2}_{\text{estimation error (MSE form)}}
\;+\;
\underbrace{\lambda\sum_{j=1}^{k}\gamma_{j}^{2}}_{\text{L2 penalty}}
\Bigr\},
\]

\noindent{}where \(\lambda\) controls the intensity of the regularization and \(\mathbf{Z}\) is the matrix of the first \(k\) principal components of \(\mathbf{X}\). Notably, the penalty \(\lambda \sum_{j=1}^{k} \gamma_j^2\) acts directly on these coefficients in the (orthogonal) component space, rather than on the original predictors. The intercept \(\gamma_0\) is estimated separately (and typically excluded from the penalty) if mean-centering is performed. Once \(\hat{\boldsymbol{\gamma}}_{\mathrm{Ridge}}\) is obtained, the corresponding coefficients in the original feature space arise via the loadings \(\mathbf{V}_k\), namely:
\[
\hat{\boldsymbol{\beta}}_{\mathrm{PCR}}
\;=\;
\mathbf{V}_k\,\hat{\boldsymbol{\gamma}}_{\mathrm{Ridge}},
\]

\noindent{}which effectively re-expresses the penalized coefficients back to the domain of the initial predictors.

This variant of PCR (with Ridge regression in the second stage) can limit the estimates' variance while maintaining the benefits of dimensionality reduction. By imposing \(\lambda \sum_{j=1}^{k} \gamma_{j}^{2}\) on the coefficients \(\boldsymbol{\gamma}\), the Ridge regression shrinks large parameter estimates toward zero, thereby mitigating the risk of overfitting that could arise even after dimensionality reduction. Indeed, while the projection onto a reduced space alleviates multicollinearity by extracting orthogonal components, some of those components may still capture variance that is not strictly relevant for prediction. The L2 penalty thus serves as an additional safeguard by discouraging overly large weights and containing the impact of noise. This mechanism is especially beneficial in high-dimensional contexts, where the number of potential regressors---and hence principal components---may be considerable and the sample size is limited. Consequently, combining PCA with a Ridge penalty reduces the dimensional complexity and provides a second layer of shrinkage, promoting more stable and robust forecasts.

In our nowcasting context, PCR is adopted with an iterative approach on time blocks, where at each sample expansion, both the number \(k\) of principal components and the regularization parameter \(\lambda\) of the L2 regression are recalibrated. In each window (train+validation), the procedure follows two fundamental phases:

\begin{enumerate}
\item \textbf{Search for hyperparameters with Bayesian optimization and pruning.} The model explores the space of values for \(k\) (number of components) and \(\lambda\) employing a TPE (Tree-structured Parzen Estimator) approach \parencite{BergstraEtAl2011}. At each iteration, the algorithm finds the combination \((k,\,\lambda)\) that minimizes the mean squared error on the validation set. In the case of unpromising trials, a pruning mechanism halts the evaluation prematurely, thereby reducing computational overhead.
\item \textbf{Final training and prediction.} Once the optimal values of \(k\) and \(\lambda\) are determined, the PCA is performed on the entire train+validation subset, in order to produce the principal components \(\mathbf{Z}\). A Ridge regression model is then applied to these components, and predictions are made for the test period.
\end{enumerate}

At the end of each prediction, the training window is expanded to include the last tested observation, and the entire hyperparameter calibration and training cycle is repeated. This "expanding window" strategy allows constant updating of the principal component decomposition and the degree of L2 penalization to reflect any structural changes in the underlying economic relationships \parencite{BanburaRunstler2011}.

\par{} 
PCR emerges as a flexible forecasting engine that leverages dimensionality reduction and linear regression. The adoption of Ridge regression in our context (instead of the traditional OLS estimation) enhances the model's robustness, mitigates the risk of overfitting while maintaining a high predictive capacity, and is in line with recent findings in high-dimensional macroeconomic forecasting \parencite{GiannoneLenzaPrimiceri2021}. Furthermore, the iterative hyperparameter calibration process, which employs Bayesian techniques, enables the model to adapt to changes in data \parencite{ChinnMeunierStumpner2023}.

\vspace{0.35cm}

\underline{{Partial Least Squares Regression}}
\vspace{0.2cm}\\
Partial Least Squares Regression (PLSR) originates from the work of \parencite{Wold1985} and serves as a regression strategy that helps reduce the dimensionality of the covariate system while identifying the latent directions that maximize the covariance between the regressors and the dependent variable \parencite{MehmoodEtAl2012}. This procedure is advantageous in the presence of datasets with potential multicollinearity or a high number of variables, since it concentrates the relevant information in a small number of uncorrelated factors while preserving the relationship with the response.

From an applicative point of view, PLSR offers a compromise between the reduction of dimensionality (characteristic of component methods) and the focus on the relationships with the variable of interest. Compared to other linear techniques, PLSR does not only perform a decomposition of the regressor matrix \(\mathbf{X}\)\footnote{The notation \(\mathbf{X}\in\mathbb{R}^{n\times p}\) indicates \(n\) observations and \(p\) regressors; in PLSR, the dependent variable \(\mathbf{y}\) is treated in parallel, identifying the components with maximum covariance between \(\mathbf{X}\) and \(\mathbf{y}\).}, but also builds latent components that aim to explain both the variance of \(\mathbf{X}\) and, simultaneously, the variability of \(\mathbf{y}\).

This feature differentiates PLSR, for example, from Principal Component Regression. In PCR, the extraction of the components occurs by considering only the maximization of the variance of \(\mathbf{X}\), without explicitly taking into account \(\mathbf{y}\). If some aspects of \(\mathbf{X}\) explain a large portion of variance but are poorly correlated with the target variable, they can still be included among the principal components, slowing or hindering the predictive capacity \parencite{MehmoodEtAl2012}. On the contrary, PLSR integrates the variable of interest in the same decomposition procedure, selecting informative directions that correlate with \(\mathbf{y}\). This approach tends to generate latent factors more relevant from the viewpoint of prediction, making PLSR an effective option in situations where the mere explanation of the variance of the regressors does not suffice to guarantee adequate model performance.

Furthermore, PLSR proves particularly robust when regressors exhibit strong collinearity or when the number of variables is high compared to the available observations. In such cases, focusing on components that maximize \(\mathrm{cov}(\mathbf{X}, \mathbf{y})\) helps exclude directions with high variance but low predictive relevance, which could otherwise distort the estimate \parencite{KraemerSugiyama2011}. Consequently, the set of latent factors tends to be better suited for capturing meaningful relationships between the regressors and the output variable, mitigating the risk of overfitting.

Thanks to this mechanism, PLSR can provide a competitive advantage in regression tasks by extracting dimensions that both explain data variance and contribute significantly to estimating the dependent variable. In nowcasting or macroeconomic forecasting contexts, where numerous potentially informative series are often available, PLSR supports a more direct focus on the components that truly matter for the phenomenon of interest \parencite{KellyPruitt2015, GroenKapetanios2016, Stamer2024}, thus allowing the model to concentrate on latent structures that meaningfully enhance predictive ability.

PLSR\footnote{In the implementation based on \texttt{scikit-learn} \parencite{scikit-learn-pls}, the \texttt{PLSRegression} class jointly handles the regressors and the dependent variable, typically excluding the intercept from the penalization.} searches for a sequence of latent factors \(\mathbf{t}_{1}, \mathbf{t}_{2}, \dots, \mathbf{t}_{a}\) such that the projection of \(\mathbf{X}\) and \(\mathbf{y}\) onto these factors is maximally correlated \parencite{Wold1985}. Formally, let \(\mathbf{w}_{j}\) be the weight vectors that define each factor \(\mathbf{t}_{j} = \mathbf{X}\,\mathbf{w}_{j}\). Once \(a\) such factors are determined, the model can be written as:

\[
y
\,=\,
\beta_{0}
\;+\;
\sum_{j=1}^{a}
t_{j}\,c_{j}
\;+\;
\varepsilon,
\quad
\mathbb{E}\bigl[\varepsilon \mid \mathbf{X}\bigr]
\,=\,0
\]

\noindent{}where \(\mathbf{c}_{j}\) denotes the coefficient associated with the \(j\)-th factor. The number of components \(a\) is a critical hyperparameter, as it dictates the level of dimensional reduction and the balance between explanatory capability and model complexity. The estimation error is typically evaluated via the Mean Squared Error (MSE):

\[
\mathrm{MSE}
\;=\;
\frac{1}{n}
\sum_{i=1}^{n}
\bigl(y_{i}
- \hat{y}_{i}\bigr)^{2}.
\]

The search for latent factors proceeds iteratively, maximizing the covariance \(\mathrm{cov}(\mathbf{t}_{j}, \mathbf{u}_{j})\) between the projections \(\mathbf{t}_{j}\) (from \(\mathbf{X}\)) and \(\mathbf{u}_{j}\) (from \(\mathbf{y}\)) at each step \parencite{MehmoodEtAl2012}.

Consistent with our other approaches, we employed PLSR as a "prediction engine" to generate one-step-ahead estimates through an expanding window technique. In each time block, the dataset was split into a training+validation subset and a test observation. The selection of the number of components \(a\) was carried out via Bayesian optimization based on TPE \parencite{BergstraEtAl2011, AkibaEtAl2019, SnoekEtAl2012}, which:

\begin{enumerate}
\item \textbf{Explored the space of possible configurations} for \(a\) (and other parameters, if relevant), evaluating their mean squared error on the validation portion;
\item \textbf{Identified the combination of hyperparameters} that minimized the validation error, halting in advance the least promising attempts (pruning).
\end{enumerate}

After this procedure was completed, we recalibrated the model using the entire span (training+validation) with the optimal number of components, then generated the forecast for the test portion. Upon each sample expansion, the window was extended to include the newly available observation, and the calibration--validation cycle was repeated in full. These progressive hyperparameter updates yielded a flexible framework that adapted to the evolving nature of the data.

\par{} 
Adopting PLSR in this framework allows combining the selection of latent components and the optimization of their number in an iterative way. This methodology provides a model able to capture significant linear relationships even in the presence of potentially collinear regressors and updates the structure of the forecasting system with the arrival of new information.

\vspace{0.5cm}

\phantomsection
\subsubsection{\textbf{\textit{Ensemble Learning Models}}}
\vspace{0.2cm}
Ensemble learning methods combine multiple predictors into a single forecasting framework, leveraging the diversity among base learners to enhance predictive accuracy and stability \parencite{Breiman2001,ChenGuestrin2016}. These approaches offer notable benefits in the context of quarterly one-step-ahead nowcasting for a small open economy such as Singapore. By aggregating distinct trees or boosted sequences of estimators, ensemble models can more effectively accommodate sudden policy shifts, external shocks, and other structural changes than purely linear specifications \parencite{Yoon2021,ChuQureshi2023}. Their adaptability is further heightened by their capacity to handle many potentially correlated variables without imposing rigid functional forms.

A key advantage of ensemble techniques lies in their ability to reduce the variance of individual learners through either bagging or boosting. Random Forest, for instance, grows multiple trees independently on bootstrap samples and averages their outputs, producing stable point estimates with limited sensitivity to outliers \parencite{Breiman2001}. Conversely, XGBoost trains trees sequentially, with each new iteration refining previous residuals through gradient-based error correction \parencite{ChenGuestrin2016}. Although bagging and boosting differ in managing errors and variance, both methods can address intricate, nonlinear patterns frequently encountered in macroeconomic series.

The methodological structure of these models is characterized by:
\begin{itemize}
    \item \textit{Randomization}: Bagging (e.g., Random Forest) applies sampling with replacement and random feature selection, enabling robust inferences even under high multicollinearity.
    \item \textit{Sequential refinement}: Boosting (e.g., XGBoost) iteratively improves on residuals, reducing bias while incorporating regularization to avoid overfitting.
    \item \textit{Hyperparameter control}: Tuning tree depth, learning rates, and sub-sampling balances model complexity and generalization capacity.
    \item \textit{Nonlinearity}: Through tree-based splits, ensemble methods capture threshold effects and regime switches naturally.
\end{itemize}

Ensemble learning strategies adapt systematically to new observations when implemented in a rolling or expanding window setup. The model selection procedure involves identifying optimal hyperparameters by maximizing performance on a validation segment and then retraining on updated datasets. Such continuous recalibration proves particularly relevant for small and open economies, where shifts in global demand, exchange rates, or trade policies can alter traditional forecasting relationships. The combination of variance reduction, nonparametric flexibility, and straightforward interpretability of split structures makes ensemble learning a compelling paradigm for nowcasting tasks, complementing traditional factor models and more recent neural network approaches.

\vspace{0.35cm}

\underline{{Random Forest}}
\vspace{0.2cm}\\
The Random Forest (RF) is an ensemble learning technique based on bagging \parencite{Breiman2001,BiauScornet2016}, where multiple decision trees are trained independently on bootstrap subsamples of the original dataset. At each tree node, a subset of regressors is randomly selected \parencite{ProbstWrightBoulesteix2019}, thereby ensuring internal diversification among the various trees, reducing the variance of the single model, and mitigating the risks of overfitting. Furthermore, unlike some penalized regressions that require an explicit functional hypothesis, the Random Forest can capture potential nonlinear relationships between macroeconomic variables while maintaining robustness to correlations among predictors. Such methodological flexibility proves advantageous in nowcasting analyses, where multiple potentially correlated series and heterogeneous indicators often render the specification of a strictly linear model problematic.

From a theoretical standpoint, the Random Forest is composed of \(B\) decision trees. Each tree \(T_{b}\) \(\bigl(b=1,\dots,B\bigr)\) is built starting from a bootstrap sample of the training data \parencite{Breiman2001}, and a random subset of available features is considered at each node to find the optimal split. In terms of prediction, each of these \(B\) trees provides an estimate \(\hat{y}_{b}(\mathbf{x}_{i})\) for the observation \(\mathbf{x}_{i}\). Aggregation then occurs by averaging the predictions of the individual trees, i.e.:

\[
\hat{y}_{i}
\,=\,
\frac{1}{B}
\sum_{b=1}^{B}
T_{b}(\mathbf{x}_{i}).
\]

The training objective can be formalized as the maximization (or minimization with opposite sign) of a function related to the root mean square error (\(\mathrm{RMSE}\)). More precisely, we define:

\[
\mathrm{Obj}
=
\underbrace{-\,\sqrt{\frac{1}{n}\sum_{i=1}^{n}
\Bigl(
   \underbrace{y_{i}}_{\text{actual value}}
   \;-\;
   \underbrace{\frac{1}{B}\sum_{b=1}^{B}T_{b}\bigl(\mathbf{x}_{i}\bigr)}_{\text{ensemble-based forecast}}
\Bigr)^{2}
}}_{\text{negative RMSE (score function)}}.
\]

\noindent{}where:

\begin{itemize}
\item \(y_{i}\) is the actual value of target variable;
\item \(\tfrac{1}{B}\sum_{b=1}^{B} T_{b}\bigl(\mathbf{x}_{i}\bigr)\) represents the Random Forest's aggregate prediction.
\end{itemize}

By combining the outputs of all \(B\) trees, the Random Forest reduces variance compared with a single tree and, through randomization both in the observations (bootstrap) and in the features (\texttt{max\_features}), it preserves model stability even in scenarios characterized by high multicollinearity.\footnote{In high-dimensional macroeconomic contexts, this bagging framework and the random subspace selection ($m_{\mathrm{try}}$) can effectively mitigate collinearity and overfitting \parencite{BiauScornet2016,ProbstWrightBoulesteix2019}. Such diversity-promoting mechanisms bear a conceptual resemblance to the feature selection or shrinkage strategies adopted by penalized linear models.}

The main hyperparameters of the Random Forest regulate each tree's complexity and define the sampling strategies \parencite{ProbstWrightBoulesteix2019}. In particular, within our modeling framework\footnote{In brackets we have indicated the names of the hyperparameters used in our model in Python \parencite{sklearnRandomForestRegressor}.}:

\begin{itemize}
    \item $B$ (\texttt{n\_estimators}): higher values tend to reduce variance, albeit at the cost of increased computational demands;
    \item $d_{\max}$ (\texttt{max\_depth}): limiting tree depth controls overfitting by preventing excessively complex splits;
    \item $n_{\mathrm{split}}$ and $n_{\mathrm{leaf}}$ (\texttt{min\_samples\_split} and \texttt{min\_samples\_leaf}): set the minimum number of samples required to split a node or create a leaf, significantly influencing a single tree's variance;
    \item $m_{\mathrm{try}}$ (\texttt{max\_features}): determines how many regressors are selected at each split, fostering internal ensemble diversity and counteracting correlations among predictors;
    \item $\mathrm{boot}$ (\texttt{bootstrap}): determines whether each tree is trained on subsets of observations drawn with or without replacement, promoting internal diversity within the ensemble model;
    \item $\kappa$ (\texttt{criterion}): sets the splitting criterion (e.g., mean squared error or absolute error), thus affecting the model's sensitivity to outliers or unusual residual distributions.
\end{itemize}

Calibrating these parameters enables practitioners to balance bias and variance, flexibly adapt to the dataset's features, and ensure satisfactory predictive performance \parencite{Breiman2001,ProbstWrightBoulesteix2019}.

In our procedure, we implemented the Random Forest as a "prediction engine" within a sequential protocol. In each time block, we split the data into a train+validation segment (to calibrate the aforementioned hyperparameters) and a single test period corresponding to the subsequent observation. We relied on a Bayesian search strategy for hyperparameter optimization \parencite{SnoekEtAl2012,AkibaEtAl2019,ShahriariEtAl2016}, enhanced by early "pruning" of less promising configurations. Once the optimal values were identified, we retrained the model on the same train+validation subset and generated the forecast for the test. The dataset was then expanded to incorporate the newly predicted observation, and the entire process was repeated at every fresh period. This "expandable window" approach enabled the Random Forest to adapt progressively to shifts in macroeconomic conditions, reducing the likelihood of prolonged misspecification and supporting more reliable estimates over time \parencite{CerqueiraEtAl2020}.

\par{} 
Integrating the Random Forest with a Bayesian hyperparameter search permits exploring a vast configuration space, leveraging the trees' inherent capacity to detect nonlinear patterns. By iteratively updating these parameters at each dataset expansion, the model achieves greater stability in its estimates without relinquishing algorithmic flexibility, making the Random Forest a competitive option in nowcasting scenarios that demand dynamic and consistent forecasts.

Despite its advantages, our application uses an i.i.d. bootstrap for sampling, rather than a time-series block bootstrap \parencite{Kunsch1989}\footnote{In mathematical terms, if we indicate with $\{(x_t,y_t)\}_{t=1}^T$ our series, a block bootstrap could consist in building a succession of blocks $\{B_j\}$ of length $L$, where $B_j = \{(x_{(j-1)L+1}, y_{(j-1)L+1}), \dots, (x_{jL}, y_{jL})\}$. By sampling such blocks with replacement (possibly managing the remainder when $T$ is not a multiple of $L$), a new artificial series $\tilde{\mathcal{D}}$ of length $T$ is constructed, preserving part of the internal temporal correlation. Each Random Forest tree would then be trained on a subsample $\tilde{\mathcal{D}}_{b}$ generated via this block logic.}. The latter could better account for autocorrelation in quarterly macroeconomic data but would considerably increase computational times, already high due to ensemble methods and hyperparameter optimization, especially when timely nowcasts are needed on a quarter-by-quarter basis \parencite{BiauScornet2016,ProbstWrightBoulesteix2019}. Implementing a fully time-series-friendly Random Forest, incorporating explicit lag features and a block bootstrap for each tree, further inflates data dimensionality and necessitates more elaborate data manipulations. Given the one-step-ahead nature of our nowcasting exercise, we opt for the simpler i.i.d.\ bootstrap, balancing real-time feasibility with robust predictive performance \parencite{Varian2014,MedeirosEtAl2021}.

We acknowledge that this strategy does not fully preserve temporal dependence in the training process and may underreact to abrupt regime shifts \parencite{CarrieroClarkMarcellino2015b}. Nonetheless, adopting an expanding-window validation protocol and sequential hyperparameter tuning alleviates much of the risk of overfitting past regimes, as newly available data gradually refine the model \parencite{HyndmanAthanasopoulos2021}. Moreover, this streamlined approach has proven effective in multiple forecasting competitions, such as the M5 Competition, where computational overhead is a constraint and many participants successfully employed simpler i.i.d.\ or rolling holdout strategies \parencite{MakridakisSpiliotisAssimakopoulos2022}. Future research could explore whether the gains from a block bootstrap or additional time-series-oriented structures outweigh the added complexity in contexts characterized by significant autocorrelation or frequent structural breaks.

\vspace{0.35cm}

\underline{{Gradient Boosting with XGBoost (eXtreme Gradient Boosting)}}
\vspace{0.2cm}\\
The Gradient Boosting methodology is an ensemble learning technique based on a sequential boosting approach. This approach combines multiple "weak trees" to reduce the forecast error progressively. The key idea is to correct, at each iteration, the residual errors committed by the previous model, thus reducing the bias and strengthening the predictive capabilities.

In this context, eXtreme Gradient Boosting (XGBoost or XGB) \parencite{ChenGuestrin2016,XGBoostDocs2023} integrates powerful regularization mechanisms, controlled rate iterative learning, and parallel computing facilities while preserving the ability to capture non-linear relationships between regressors and macroeconomic output. This flexibility presents advantages in datasets with multiple potentially correlated variables, where a purely linear model could prove limiting.

Formally, XGB builds an ensemble of \(M\) regression trees, each denoted as \(f_{m}\). The set of such trees produces an aggregate prediction:

\[
\hat{y}_{i}
\,=\,
\sum_{m=1}^{M}
f_{m}\bigl(\mathbf{x}_{i}\bigr),
\]

\noindent{}where \(f_{m}\bigl(\mathbf{x}_{i}\bigr)\) represents the single prediction provided by the \(m\)-th tree for the observation \(\mathbf{x}_{i}\). 
The optimization focuses on minimizing an error metric, such as RMSE (Root Mean Squared Error), which can be transformed into an objective function to be maximized by placing a negative sign in front of the square root. Concretely:

\[
\mathrm{Obj}
\,=\,
\underbrace{-\,\sqrt{\frac{1}{n}\sum_{i=1}^{n}
\Bigl(
\underbrace{y_{i}}_{\text{actual value}}
\;-\;
\underbrace{\sum_{m=1}^{M}f_{m}\bigl(\mathbf{x}_{i}\bigr)}_{\text{sum of the predictions of the M trees}}
\Bigr)^{2}
}}_{\text{score function (RMSE with negative sign)}},
\]

\noindent{}where:

\begin{itemize}
\item \(y_{i}\) indicates the actual value of the dependent variable;
\item \(\sum_{m=1}^{M} f_{m}\bigl(\mathbf{x}_{i}\bigr)\) synthesizes the sequential combination of the predictions provided by the individual trees.
\end{itemize}

In addition to the minimization of the loss function, XGB also incorporates a penalty term on the complexity of each tree, aiming to reduce overfitting. Specifically, the regularization includes an L2 penalty governed by \(\lambda\) (\texttt{reg\_lambda}), which penalizes large magnitudes of the leaf weights:

\[
\Omega(f_{m}) 
\,=\,
\gamma\,T_{m}
\;+\;
\frac{\lambda}{2}\sum_{j=1}^{T_{m}}\bigl(w_{j}\bigr)^{2},
\]

\noindent{}where:

\begin{itemize}
    \item \(T_{m}\) is the number of leaves of the \(m\)-th tree;
    \item \(\gamma\) (\texttt{gamma}) is a parameter that penalizes the addition of new leaves;
    \item \(w_{j}\) denotes the weight (i.e.\ the constant prediction) associated with each leaf \(j\);
    \item \(\lambda\) (\texttt{reg\_lambda}) penalizes large leaf weights \(w_{j}\)\footnote{This L2 penalty acts as a form of "weight decay," analogous to that seen in neural networks \parencite{KroghHertz1992,HastieTibshiraniWainwright2015}, shrinking large coefficients to reduce overfitting.}.
\end{itemize}

By incorporating \(\Omega(f_{m})\) into the global objective, XGB discourages both an excessive number of leaves and excessively large leaf weights, thereby mitigating variance and helping the model generalize better.

Each tree \(f_{m}\) operates as a step function that partitions the space of variables into distinct regions, assigning a constant estimate \(w_{j}\) to each region (leaf). The structure of these trees—splits, depth, and leaf values—is learned iteratively, correcting, at each iteration, the residual errors of its predecessors. However, the L2 penalty shrinks large leaf coefficients, thus reducing the risk of overfitting to high-frequency noise. As previously discussed, this shrinkage also helps preserve smaller yet potentially informative effects by proportionally reducing leaf weights rather than nullifying them altogether.

In our study, we implement XGB as a "prediction engine" within a sequential protocol analogous to that adopted for the Random Forest \parencite{ChuQureshi2023,MasiniMedeirosMendes2023}. In each time block, the dataset is divided into a train+validation subset (used to calibrate the hyperparameters) and a single test observation corresponding to the subsequent period.

For hyperparameter tuning, we rely on a Bayesian search strategy based on TPE\parencite{BergstraEtAl2011,AkibaEtAl2019,OptunaDocs2023}, augmented by a pruning mechanism that halts evaluations showing limited promise. Specifically, we explore the following main hyperparameters (with the names used in the script in brackets):
\begin{itemize}
  \item \(M\) (\texttt{n\_estimators}): total number of boosted trees;
  \item \(d_{\max}\) (\texttt{max\_depth}): maximum depth of each tree;
  \item \(\eta\) (\texttt{learning\_rate}): rate at which new trees correct existing forecasts;
  \item \(\gamma\) (\texttt{gamma}): minimum gain threshold for splitting a node;
  \item \(w_{\min}\) (\texttt{min\_child\_weight}): minimum number of samples required for a split;
  \item \(\mathrm{col}_{\mathrm{bytree}}\) (\texttt{colsample\_bytree}): fraction of predictors to consider in each tree;
  \item \(\mathrm{sub\_s}\) (\texttt{subsample}): fraction of observations sampled before growing each tree;
  \item \(\lambda\) (\texttt{reg\_lambda}): coefficient for the L2 penalty term \(\Omega(f_{m})\).
\end{itemize}

Once the best configuration of these hyperparameters is identified, we retrain the model on the combined train+validation subset and generate the forecast for the test observation. The dataset is then expanded to include the newly predicted quarter (or period), and the entire process is repeated at each fresh time step. This expanding window approach enables XGB to adapt gradually to structural shifts in the macroeconomic environment and to reduce the likelihood of long-lasting misspecification, thereby enhancing its predictive reliability.

\par{} 
This XGB approach, therefore, merges the predictive capacity of sequential trees with an accurate iterative error correction mechanism while maintaining control over the complexity thanks to targeted regularization parameters. Furthermore, the continuous updating of hyperparameters, made possible by Bayesian optimization, ensures a constant adaptation of the model to macroeconomic series changes without compromising the overall estimates' robustness.

\vspace{0.5cm}

\phantomsection
\subsubsection{\textbf{\textit{Neural Models}}}
\vspace{0.2cm}
Neural networks \parencite{Hornik1989,GoodfellowBengioCourville2016} offer a flexible framework for capturing nonlinear dynamics, complex interactions among variables, and evolving economic conditions. In the specific context of one-step-ahead nowcasting for a small and open economy such as Singapore, their capacity to adapt to rapidly changing environments proves advantageous. This section focuses on two architectures: the Multilayer Perceptron (MLP) and the Gated Recurrent Unit (GRU) \parencite{ChoEtAl2014}, each equipped with regularization mechanisms to ensure robustness against overfitting.

Neural models differ from linear or factor-based approaches by incorporating activation functions, 
memory cells, or gating structures that introduce nonlinearity. Such features are crucial when 
macroeconomic indicators shift abruptly due to exogenous shocks, policy interventions, or other 
regime changes. Indeed, recent studies highlight the competitiveness of neural architectures in 
macroeconomic contexts characterized by volatility \parencite{AlmosovaAndresen2023,HewamalageBergmeirBandara2021}. 
Neural networks respond to these shifts by updating their internal parameters in a rolling or 
expanding manner, thus preserving the ability to learn new data patterns as they arise.

In designing neural networks for quarterly nowcasting, there are several guiding considerations:

\begin{itemize}
    \item \textit{Nonlinear Relationships.} Traditional regression models rely on linearity assumptions or static factor structures, which may not fully capture the diverse and potentially intricate linkages that govern aggregate economic outcomes. Neural layers accommodate nonlinearities by design, thus more readily uncovering hidden patterns or threshold effects.
    \item \textit{Regularization Strategies.} Given the limited sample size typical of quarterly data, shrinkage methods, and dropout help prevent the over-parametrization of the model. These techniques penalize large weights or stochastically exclude neurons, protecting the model from spurious correlations in historical data.
    \item \textit{Temporal Evolution.} While MLPs treat each observation independently, GRU models incorporate memory states that track historical context, providing a natural handle on time-lagged effects. This property is especially relevant in small open economies, where external shocks, trade flows, and policy decisions can induce persistent fluctuations.
    \item \textit{Hyperparameter Optimization.} Choosing the right network depth, hidden dimensionality, and learning rate is critical. We employ a Bayesian search strategy with pruning to explore these configurations efficiently. The model is then retrained on updated samples at each step, enhancing adaptability to unforeseen structural breaks.
\end{itemize}

Although neural approaches can be more opaque than penalized regressions, the iterative recalibration of weights and the inclusion of interpretability strategies (as already discussed in this study) facilitate a clearer understanding of which predictors drive the nowcast. Theoretically, the synergy between nonlinearity, sequence modeling, and automatic hyperparameter tuning renders neural models a competitive alternative in scenarios marked by high volatility or nonstationary data-generating processes.

\vspace{0.35cm}
\underline{Multilayer Perceptron}
\vspace{0.2cm}\\
The Multilayer Perceptron (MLP) is a feedforward neural network architecture\footnote{A feedforward neural network is a class of networks in which signals flow solely from inputs to outputs, with no feedback or internal loops.} \parencite{Hornik1989} designed to address regression problems that may exhibit nonlinear relationships\footnote{Nonlinear relationships are interactions between variables in which a slight change in a regressor can produce non-constant (or non-proportional) changes in the dependent variable.} between the regressors and the variable of interest. Unlike pure linear models, the MLP introduces nonlinear activation functions\footnote{Nonlinear activation functions (e.g., Rectified Linear Unit (ReLU) or sigmoid) transform the output of a neuron after the linear combination of the inputs, allowing to capture more complex relationships.} \parencite{GlorotBordesBengio2011} and multiple hidden layers\footnote{Hidden layers are the intermediate levels between the input and the output of the network, where internal representations of the data are elaborated.} (hidden layers), allowing the representation of even complex links between economic inputs.

Adopting an MLP for nowcasting is particularly appropriate when the number of variables is high, and it is suspected that non-trivial interactions may influence the target variable (e.g., GDP growth). Furthermore, including regularization mechanisms (such as dropout\footnote{This technique randomly "turns off" some units during the training process, helping to prevent the model from passively storing the noise in the historical data.} \parencite{SrivastavaEtAl2014}) helps mitigate overfitting, favoring a robust adaptation in the face of unstable macroeconomic dynamics.

In such settings, it is common that some predictors become irrelevant or partially redundant, thus introducing noise and complicating the learning process. Regularization mechanisms such as dropout are advantageous in addressing extraneous or spurious regressors that overshadow the genuinely significant ones. These mechanisms foster a more parsimonious representation of the data.

Formally, the MLP expresses the relationship

\[
y_{i}
\,=\,
f\!\bigl(\mathbf{x}_{i};\,\boldsymbol{\theta}\bigr)
\;+\;
\varepsilon_{i},
\quad
\mathbb{E}\bigl[\varepsilon_{i} \mid \mathbf{x}_{i}\bigr]
\,=\,0,
\]

\noindent{}where:

\begin{itemize}
  \item \(\displaystyle y_{i}\) is the target variable (e.g., quarterly GDP growth), expressed as a sum of a systematic component and a random noise term;
  \item \(\displaystyle f(\mathbf{x}_{i};\,\boldsymbol{\theta})\) is the function learned by the network, which maps regressors \(\mathbf{x}_{i}\) to a predicted value of \(y_i\);
  \item \(\displaystyle \varepsilon_{i}\) represents the unpredictable part (noise);
  \item \(\displaystyle \mathbb{E}[\varepsilon_{i}\mid \mathbf{x}_{i}] = 0\) assumes no systematic bias remains in the residuals once \(\mathbf{x}_{i}\) is known;
  \item \(\mathbf{x}_{i}\in\mathbb{R}^{p}\) is the vector of regressors;
  \item \(\boldsymbol{\theta}\) collects all the weights and biases of the network.
\end{itemize}

Each neuron\footnote{A neuron is the fundamental processing unit of a neural network, in which an activation function follows a linear combination of inputs.} linearly combines the inputs and applies an activation function (e.g., ReLU) to introduce nonlinearity. If a hidden layer has \(h\) neurons, each neuron \(j\) processes

\[
z_{j}
\,=\,
\sigma\!\Bigl(\sum_{k=1}^{d} w_{jk}\,x_{k} \;+\; b_{j}\Bigr),
\]

\noindent{}where:

\begin{itemize}
  \item \(\sigma(\cdot)\) is the activation function;
  \item \(w_{jk}\) and \(b_{j}\) are the parameters to optimize.
\end{itemize}

Connecting multiple layers in sequence creates a hierarchy of representations useful for capturing complex patterns.

In our empirical setup, we configure the MLP with a single hidden layer (or a limited number of layers). Although deeper architectures can approximate highly intricate relationships, in macroeconomic nowcasting with relatively short samples, additional hidden layers do not necessarily improve generalization and can exacerbate overfitting \parencite{Hornik1989,GlorotBordesBengio2011}. A shallower network is often sufficient to capture the bulk of nonlinearities while retaining computational efficiency.

The estimate of \(\boldsymbol{\theta}\) occurs by minimizing a cost function consisting of the mean squared error (MSE):

\[
\min_{\boldsymbol{\theta}}
\Bigl\{
    \frac{1}{n}\sum_{i=1}^{n}
    \bigl(y_{i} - f(\mathbf{x}_{i};\,\boldsymbol{\theta})\bigr)^{2}
\Bigr\}
\]

The backpropagation process\footnote{Backpropagation is the procedure by which the prediction error is propagated backward, allowing the partial derivatives of the error to be computed with respect to each weight and bias.} \parencite{RumelhartHintonWilliams1986} calculates the gradients\footnote{The gradient is the vector of the partial derivatives of a function (in this case, the error) with respect to the variables to be optimized (weights and biases). It indicates the direction of maximum growth of the function itself and, by reversing its direction, the model iteratively reduces the error.} of this MSE with respect to each weight and bias, allowing them to be updated iteratively through an optimizer (e.g., Adam) \parencite{KingmaBa2015}. 

In deep networks or datasets with a large number of variables, overfitting may emerge. Regularization strategies (i.e., dropout) are used to counteract overfitting, restricting the influence of unhelpful or purely noisy features. 
Hence, although the MLP is in principle flexible enough to incorporate multiple hidden layers, we typically keep the network shallow in nowcasting tasks to avoid excessive complexity and make training more stable in limited-sample scenarios.

Our empirical study implements the MLP as a "forecasting engine" according to a rolling protocol with an expandable window \parencite{Tashman2000}. During each iteration, we explore the following main hyperparameters (with the names used in the script in brackets):
\begin{itemize}
\item \(\alpha_{\mathrm{hd}}\) (\texttt{hidden\_dim}): the number of hidden units (neurons) per layer;
\item \(\alpha_{\mathrm{nl}}\) (\texttt{num\_hidden\_layers}): the total number of hidden layers in the MLP;
\item \(\alpha_{\mathrm{dr}}\) (\texttt{dropout\_rate}): the percentage of neurons "turned off" at each epoch to prevent overfitting;
\item \(\eta\) (\texttt{lr}): the learning rate of the optimizer, which determines the size of the weight update steps;
\item \(\beta\) (\texttt{batch\_size}): the size of the mini-batches used in training.
\end{itemize}

Also, in this case, the search for optimal parameters occurs during each iteration through the Bayesian optimization approach based on the TPE \parencite{BergstraEtAl2011,SnoekEtAl2012,AkibaEtAl2019}. The algorithm explores the space of configurations and, for each one, estimates the MSE on a validation subset. The search process is accelerated by "pruning" mechanisms that stop the least promising hyperparameter combinations early. 
At the end of the exploration, the combination of hyperparameters that minimizes the validation error is selected. To prevent excessive overfitting, we halt training when there is no improvement in validation error for a specified number of epochs.
Once the optimal parameters are determined, the MLP is recalibrated on the entire training+validation block, then the out-of-sample forecast for the test quarter is generated.
The last observation (just forecasted) is incorporated into the dataset, and the next block is moved on, repeating the entire process at the new time horizon.
In this way, the model adjusts its weights and hyperparameters at each sample expansion, remaining sensitive to possible shocks or regime changes and reducing the probability of misspecification over time.

\par{} 
MLP, integrated with Bayesian optimization and regularization strategies, combines the ability to capture nonlinear relationships and adapt quickly to macroeconomic framework changes. Despite the lower transparency of network parameters compared to penalized linear models (such as LASSO or Ridge), the iterative selection of hyperparameters and the adoption of mechanisms such as dropout ensure a balance between flexibility and robustness, mitigating the impact of potentially irrelevant regressors.
Frequent updating allows the system to respond promptly to structural changes, preserving good generalization to future data or unexpected economic scenarios.

\vspace{0.35cm}
\underline{Gated Recurrent Unit}
\vspace{0.2cm}\\
Gated Recurrent Units (GRUs) are an architecture belonging to the class of recurrent networks\footnote{Recurrent networks (RNNs) are distinguished from feedforward models by their feedback loops that allow the store of information about the previous state, providing a dynamic structure to the forecast.} \parencite{ChoEtAl2014,ChungGulcehreChoBengio2014}, designed to capture dynamic relationships between economic variables along the time axis. The introduction of the gating mechanism\footnote{In the GRU, the update gates (\(z_{t}\)) and the reset gates (\(r_{t}\)) regulate how the new information is combined with the previous state, mitigating phenomena such as the vanishing gradient typical of traditional RNNs.} allows to efficiently synthesize and transmit the memory between successive states, maintaining relatively low computational costs compared to other more complex recurrent networks \parencite{JozefowiczZarembaSutskever2015}.

A key motivation in using GRUs for regression is the need to contain the influence of any irrelevant variables, which could introduce noise in the estimation phases. In this context, it is appropriate to include forms of regularization, such as the L2 penalty (or weight decay) \parencite{KroghHertz1992}, to penalize the excessive growth of coefficients associated with non-informative predictors. In parallel, adopting dropout guarantees further control over overfitting, randomly switching off some neurons during the training phase \parencite{SrivastavaEtAl2014,GalGhahramani2016,ZarembaSutskeverVinyals2014}. Thanks to these measures, GRUs offer flexibility and simultaneously limit the risk of excessively including superfluous components, favoring a reduction in the complexity of the model \parencite{MerityKeskarSocher2018}.

Formally, the GRU model represents the relation

\[
y_{i}
\,=\,
f\!\bigl(\mathbf{x}_{i};\,\boldsymbol{\theta}\bigr)
\;+\;
\varepsilon_{i},
\quad
\mathbb{E}\bigl[\varepsilon_{i} \mid \mathbf{x}_{i}\bigr]
\,=\,0,
\]

\noindent{}where:

\begin{itemize}
    \item \(\displaystyle y_{i}\) is the target variable (here, quarterly GDP growth), modeled as the sum of a systematic component and a random noise;
    \item \(\displaystyle f(\mathbf{x}_{i};\,\boldsymbol{\theta})\) is the function learned by the network, mapping the regressors \(\mathbf{x}_{i}\) to a predicted value;
    \item \(\varepsilon_{i}\) is a noise term, capturing the unpredictable part of \(y_{i}\);
    \item \(\mathbb{E}[\varepsilon_{i}\mid \mathbf{x}_{i}] = 0\) states that, on average, the noise has zero mean once \(\mathbf{x}_{i}\) is given (no systematic bias remains in the residual);
    \item \(\mathbf{x}_{i}\in\mathbb{R}^{p}\) denotes the vector of regressors;
    \item \(\boldsymbol{\theta}\) collects the set of parameters (weights, biases, and gating constructs) to be optimized.
\end{itemize}

In GRUs, at each instant \(t\), the \textit{latent output} \(\mathbf{h}_{t}\) selectively updates the information coming from the \textit{previous state} \(\mathbf{h}_{t-1}\) and from the current input \(\mathbf{x}_{t}\).

\begin{itemize}
    \item \(\mathbf{h}_{t}\) (\textit{latent output}) is the hidden state not directly observed but internally updated by the gating mechanism;
    \item \(\mathbf{h}_{t-1}\) (\textit{previous state}) is the hidden state carried over from the preceding time step;
    \item \(\mathbf{z}_{t}\) (\textit{update gate}) controls how much of the old state is retained vs.\ replaced by new input information;
    \item The \textit{reset} part of \(\mathbf{h}_{t-1}\) determines how much past information is forgotten before computing \(\tilde{\mathbf{h}}_{t}\).
\end{itemize}

In particular, the update gate \(\mathbf{z}_{t}\) balances the stored and new components:

\[
\mathbf{h}_{t}
\,=\,
(1 - \mathbf{z}_{t}) \,\circ\, \mathbf{h}_{t-1}
\;+\;
\mathbf{z}_{t} \,\circ\, \tilde{\mathbf{h}}_{t},
\]

\noindent{}with \(\tilde{\mathbf{h}}_{t}\) calculated by a linear combination of \(\mathbf{x}_{t}\) and the "reset" part of the previous state. The operator \(\circ\) indicates the element-by-element product. The number of hidden units and the number of layers filled determine the representativeness of the model.

In this study, we are employing a single hidden layer for the GRU.\footnote{A 
similar design choice applies to the MLP in the previous section, wherein a 
single hidden layer suffices to capture nonlinear relationships while controlling 
overfitting in limited-sample environments \parencite{Hornik1989,GlorotBordesBengio2011}}. Although deeper recurrent architectures can be beneficial under extensive data availability, several empirical findings suggest that, for quarterly macroeconomic nowcasting, additional layers do not necessarily improve accuracy and may increase overfitting risk \parencite{HewamalageBergmeirBandara2021,AlmosovaAndresen2023}. A single-layer GRU often suffices to capture short-run dynamics in limited-sample environments, preserving computational efficiency and converging more reliably \parencite{ZarembaSutskeverVinyals2014}.

The set of parameters \(\boldsymbol{\theta}\) is estimated by minimizing a cost function consisting of the mean squared error (MSE) plus an L2 penalty

\[
\min_{\boldsymbol{\theta}}
\Bigl\{
    \frac{1}{n}\sum_{i=1}^{n}
    \bigl(y_{i} - f(\mathbf{x}_{i};\,\boldsymbol{\theta})\bigr)^{2}
    \;+\;
    \lambda_{\mathrm{L2}}\;\|\boldsymbol{\theta}\|^{2}
\Bigr\}
\footnote{Similarly to what we did in MLP's section, here we refer to the regularization parameter as $\lambda_{\mathrm{L2}}$ to highlight the neural-network weight decay perspective, but its function is analogous to $\lambda$ in a linear Ridge model.}
\]

\noindent{}using backpropagation through time \parencite{Werbos1990}. Backpropagation is the algorithm that computes the gradient of the error with respect to all trainable parameters, systematically propagating the errors backward from the output to earlier layers (and across time steps in a recurrent network).

Backpropagation through time gradually propagates the error through the past recursive iterations of the network and allows weights to be updated using optimizers (e.g., Adam) with an L2 penalty. In addition to the dropout, these regularization elements help to contain the possible proliferation of coefficients relevant only to some data segments, preventing an uncontrolled growth of the effective dimensionality.

In the empirical study adopted here, the estimation of the GRU model followed a rolling iterative forecasting protocol with an expandable window: at each sample expansion, the best hyperparameters were automatically identified thanks to TPE \parencite{BergstraEtAl2011,AkibaEtAl2019} that evaluated candidate configurations based on the in-sample validation results. The search process was further accelerated by "pruning" mechanisms, which interrupted less promising trials early and thus reduced the computational overhead.

The search variables included are:

\begin{itemize}
\item \(\alpha_{\mathrm{hd}}\) (\texttt{hidden\_dim}), i.e. the hidden dimensionality of each GRU layer;
\item \(\alpha_{\mathrm{nl}}\) (\texttt{num\_layers}), the number of overlapping recurrent layers;
\item \(\alpha_{\mathrm{do}}\) (\texttt{dropout\_rate}), the share of neurons stochastically excluded during training;
\item \(\lambda_{\mathrm{L2}}\) (\texttt{l2\_reg}): the weight decay coefficient controlling the intensity of L2 regularization;
\item \(\eta\) (\texttt{lr}), the learning rate of the optimizer;
\item \(\beta\) (\texttt{batch\_size}), the size of mini-batches in stochastic descent.
\end{itemize}

The search for hyperparameter configurations occurred at each estimation block, stopping training early (early stopping) if no tangible improvements in validation emerged. Once the optimal values were determined, the model was recalibrated on the entire training+validation set and produced the forecast for the out-of-sample horizon. Subsequently, it was advanced by one quarter, including the new historical information and the entire procedure was repeated. This iterative scheme guaranteed a constant updating of the network, allowing it to adapt to any structural changes and containing the risks of misspecification.

\par{} 
The GRU configured as a "predictive engine" is capable of exploiting temporal persistence, with a lighter internal structure than LSTMs, but equally suitable for modeling non-linear connections between macroeconomic inputs and the target variable. The dynamic optimization approach---through progressive sample expansion and Bayesian selection of hyperparameters---ensures its operational flexibility, maintaining a balance between learning ability and generalization robustness.

\vspace{0.5cm}

\section{\textbf{Prediction Intervals around nowcasts}}
\label{sec:2.4}
Constructing "prediction intervals" in macroeconomic nowcasting allows to quantify the uncertainty associated with each point estimate \parencite{Chatfield1993, Tashman2000}. In contexts where the residuals can be considered approximately linear and stationary\footnote{Residuals are defined as "linear" when they can be described by linear relationships with their lags or with the past values of explanatory variables (the so-called "state variables"). "Stationary" means those processes whose statistical properties (mean, variance, autocorrelation) remain constant over time.}, more traditional methods can be adopted, such as the residual-based Bootstrap (or, if appropriate, the sieve bootstrap in AR structures\footnote{Bootstrap is a method for estimating the distribution of a parameter (or estimate) by resampling the original data with replacement.

"Time series bootstrap" methods consider the autocorrelation present in time series (i.e., subsequent values may depend on previous ones). Therefore, resampling procedures (e.g., "block bootstrap") preserve the order of the data or group contiguous intervals to reproduce the temporal structure more realistically.

In the "residual-based bootstrap," the estimated residuals from a model are sampled and fed back into the equations to generate new simulated paths.

The "sieve bootstrap" is a variant typically used in AR contexts, where the residuals are obtained from an autoregressive process approximating the dynamics of the series.}). However, when the analysis is extended to multiple nonlinear models or structural breaks are suspected, these techniques risk not correctly capturing the data's complexity \parencite{Kunsch1989}. Thus, in the presence of potential nonlinearities and structural changes, the pair block bootstrap \parencite{Masarotto1990,PanPolitis2016} offers the possibility of generating numerous alternative training trajectories while maintaining the temporal dependence in the data and being agnostic to the shape of the residuals, thus more suitable for a multi-model scenario and subject to possible regime shifts. Furthermore, it allows monitoring the evolution of the estimated uncertainty at each forecasting iteration, thus providing a measure of the predictive robustness even in the presence of endogenous variations and potential structural shocks \parencite{Grigoletto1998}.

Consider any predictive model that, starting from regressors \(\mathbf{x}_t \in \mathbb{R}^p\), produces an estimate \(\widehat{y}_t\) of the macroeconomic variable \(y_t\). Formally, we assume that

\[
y_{t}
\,=\,
f\!\bigl(\mathbf{x}_{t};\,\boldsymbol{\theta}\bigr)
\;+\;
\varepsilon_{t},
\quad
\mathbb{E}\bigl[\varepsilon_{t}\mid \mathbf{x}_{t}\bigr]
\,=\,
0,
\]

\noindent{}where:

\begin{itemize}
\item \(\displaystyle y_t\) is the real value of the quantity of interest (in our case, the change in GDP);
\item \(\displaystyle f(\mathbf{x}_{t};\,\boldsymbol{\theta})\) represents the functional mapping estimated by the model, subject to various parameters \(\boldsymbol{\theta}\);
\item \(\varepsilon_t\) indicates the noise component, free of conditioned bias and not explicitly modeled;
\item \(\displaystyle \mathbb{E}[\varepsilon_t \mid \mathbf{x}_t] = 0\), a condition that guarantees the absence of bias (zero mean value of the error) once \(\mathbf{x}_t\) is known, i.e. \(\varepsilon_t\) reflects exclusively noise, without systematic tendencies to over- or underestimate the phenomenon.
\end{itemize}

The overall uncertainty around the forecast \(\widehat{y}_t\) depends not only on the variability of the errors \(\varepsilon_t\) but also on possible limitations of the model \((f)\), on the small size of the available sample and any breakpoints (e.g., crises, shocks, legislative changes) \parencite{ClarkRavazzolo2015, PesaranPettenuzzoTimmermann2006}.

In a preliminary analysis, we applied a Bai-Perron test\footnote{The Bai-Perron test is a procedure to identify multiple breakpoints in a regression model. Using information criteria (AIC, BIC) evaluates whether the regression function should be split into segments with different parameters and determines the optimal number of breaks.} using Akaike Information Criterion (AIC) and Bayesian Information Criterion (BIC) on the entire historical series of the quarterly GDP growth rate (from 1990 Q1 to 2023 Q2). According to these criteria, no significant structural breaks emerged since the model with zero breaks was preferred\footnote{Specifically: for \(m=0\) breaks we found RSS = 998.853 (AIC = 273.175, BIC = 278.971); for \(m=1\) break yielded RSS = 945.221 (AIC = 275.668, BIC = 281.412); increasing \(m\) to 2--5 further raised both AIC and BIC. Consequently, the no-break (\(m=0\)) model remained preferred under both criteria.}.

However, this outcome may reflect specific features of Singapore's economy---such as its relatively small size, high degree of openness, and history of rapid cyclical adjustments---which can limit the power of break tests in detecting short-lived or swiftly reversed shocks \parencite{GiraitisKapetaniosPrice2013}. Despite the Bai-Perron test preferring a model with zero breaks, many studies adopt ex-ante break dates \parencite{ClementsHendry1996} whenever major global crises are known to have potentially altered economic regimes. Such an approach is economically motivated rather than purely statistical, justified in contexts where some shocks, though real, prove too transient or compensated to manifest as strong, persistent breaks in the data.

Due to the well-known global economic shocks (Asian financial crisis of the late 1990s, the September 11 attacks in 2001, SARS, the 2008 crisis, the advent of COVID-19, and the Russian-Ukrainian conflict), we decided to impose certain structural breaks ex-ante corresponding to the quarters in which such events have likely affected global and Singaporean macroeconomic dynamics, regardless of the statistical test \parencite{ClementsHendry1996}. This choice is justified by the economic interest in assessing the possible impact of such shocks. It also reflects a model design approach intended to preserve relevant historical information, even if purely statistical criteria do not emphasize these breaks.

In our analysis and for each iteration, we divided the procedure into the following steps:

\begin{enumerate}
\item \textbf{Subdivision of the train+validation sample into sub-segments.}
For each iteration, we subdivided the train+validation time interval (up to the last quarter considered in-sample) into multiple sections/sub-segments, separating them at the historical-economic breaks considered crucial (including 1997 Q3, 2001 Q1, 2003 Q1, 2008 Q3, 2020 Q1, 2022 Q1). Although the Bai-Perron test did not confirm these breaks statistically, we opted to maintain them ex-ante to reflect real-world events (such as global crises or shocks) that could affect the Singapore economy. This approach preserves economic coherence, allowing the distinction of pre- and post-crisis phases \parencite{PesaranPettenuzzoTimmermann2006}.

\item \textbf{Creation of contiguous blocks for each subsegment.}
In each subsegment, the data \((\mathbf{x}_t, y_t)\) were grouped into contiguous blocks of fixed size (4 quarters) to safeguard any internal serial autocorrelation. Instead of randomly shuffling individual observations, we considered consecutive data groups, reflecting the temporal dynamics of an entire year of observations\footnote{Such a block size (one full year) preserves typical annual seasonality in macroeconomic series and reduces the risk of "breaking" within-year correlations, as recommended by \parencite{MakridakisEtAl1998}.} \parencite{Masarotto1990}.

Instead of randomly shuffling individual observations, we considered consecutive data groups, reflecting the temporal dynamics of an entire year of observations. We set the block size to 4 quarters (one year) to preserve potential seasonality within each sampled block and to maintain key short-run dependencies that often manifest at an annual frequency in macroeconomic series (see \parencite{Masarotto1990,PanPolitis2016}). This approach reduces the risk of breaking crucial within-year correlations that could otherwise bias the bootstrap re-sampling.

\item \textbf{Sampling with block replacement.}
Several blocks of the same length (4 quarters) were randomly extracted from each subsegment, with replacement, until a quantity of data similar to the original one of the subsegment was reached. In this way, a new sample \(\bigl(\mathbf{x}_\mathrm{boot}, y_\mathrm{boot}\bigr)\) was created that preserves the structure of intra-block dependencies but varies in terms of order or frequency of the blocks themselves. Unlike a simple row shuffling, this method does not assume that the errors are independent and identically distributed (i.i.d.) \parencite{ThombsSchucany1990}.

\item \textbf{Concatenation of the resampled subsegments.}
The "bootstrapped" subsegments (i.e., reassembled into blocks) were then concatenated in sequence, forming an "alternative" train+validation dataset compared to the initial one, but with the same length and with a similar temporal subdivision, at least in the macro-phases. As a result, we obtained a complete and ready-to-be-used training set \parencite{PanPolitis2016}.

\item \textbf{Retraining the model and nowcast on the test quarter.}
Once the dataset \(\bigl(\mathbf{x}_\mathrm{boot}, y_\mathrm{boot}\bigr)\) was defined, the model used as a "predictive engine" (e.g., Elastic Net, PLSR, GRU) was recalibrated on the resampled data. Once training was completed, the estimate \(\widehat{y}_\mathrm{B}\) for the test quarter was generated. The operation was repeated \(\mathrm{B}\) = 1000 times, producing a distribution of predictions \(\{\widehat{y}^{(b)}\}_{b=1}^B\). In this way, multiple "predictive scenarios" are collected under different reorganizations to reflect the potential variability of the train+validation data \parencite{HewamalageBergmeirBandara2021}.

\item \textbf{Construction of the confidence interval.}
The set of estimates \(\{\widehat{y}^{(b)}\}_{b=1}^B\) was then ordered, and the quantiles associated with the desired levels (e.g., 2.5\% and 97.5\%), defined as:

\begin{itemize}
\item \(\underline{q}_{\alpha}
\;=\;
\text{quantile }\!\bigl(\{\widehat{y}^{(b)}\},\, \alpha\bigr)\) 
  (lower quantile, i.e.\ 2.5\%)
\item \(\overline{q}_{\alpha}
\;=\; 
\text{quantile }\!\bigl(\{\widehat{y}^{(b)}\},\, 1-\alpha\bigr)\) 
  (upper quantile, i.e.\ 97.5\%)
\end{itemize}

\noindent{}which results in the Confidence Interval (Prediction Interval, PI) around the nowcast

\[
\mathrm{PI}_{1-2\alpha}(t)
\;=\;
\Bigl[\,
\underline{q}_{\alpha},\,\overline{q}_{\alpha}
\Bigr].
\]

This interval captures the sampling uncertainty generated by a potential undersampling of the starting data and the variability introduced by the possible presence (or imposition) of structural breaks \parencite{Chatfield1993}.

\end{enumerate}

These intervals are estimated within an Expanding+Rolling protocol with hyperparameter selection through TPE and pruning (and early stopping, in the case of neural networks). In particular:

\begin{itemize}
\item at each sample expansion, the best hyperparametric configuration is identified in the train+validation subset, training the model on updated historical data \parencite{ShahriariEtAl2016,BergstraEtAl2011,AkibaEtAl2019};
\item once the hyperparameters have been fixed, the nowcast for the following quarter (test quarter) is produced;
\item finally, to evaluate the sensitivity of the forecast (and the corresponding confidence interval) to potential breaks, the pair block bootstrap was repeated on the same train+validation window, obtaining the distribution of predictions and the related extreme quantiles.
\end{itemize}

The process continues iteratively, moving forward one quarter until covering the entire range of out-of-sample forecasts, ensuring consistency with the Expanding+Rolling strategy \parencite{BanburaEtAl2010}.

Additionally, to evaluate how predictive uncertainty evolves under different macroeconomic regimes, we compute the average width of the prediction intervals across several sub-periods (Pre-COVID, COVID, Post-COVID, Excluding-COVID) and the entire forecast horizon (Overall). This breakdown highlights how global disruptions (e.g., COVID-19) alter the scale of forecast uncertainty compared to more stable phases, offering a more granular assessment of each model's sensitivity to structural shocks.

\par{} 
The "prediction intervals around nowcasts" thus calculated preserve the series' internal correlation, integrate the effect of historical-economic breaks (even if not confirmed by the Bai-Perron statistical test), and provide empirical uncertainty measures valid on multiple models, reflecting the degree of risk associated with each nowcast. At the same time, these intervals serve as a crucial verification criterion for measuring forecast stability, contributing to a more robust and informed comparative analysis between different methodologies \parencite{MakridakisSpiliotisAssimakopoulos2022}.

\vspace{0.5cm}

\section{\textbf{Feature Importance and Confidence Intervals}}
\label{sec:2.5}

\phantomsection
\subsection{\textbf{\textit{Explainability Methods per Model Family}}}
\label{sec:2.5.1}
Constructing a coherent interpretability framework for macroeconomic nowcasting requires reconciling each model's internal logic with the temporal dependencies inherent in time-series data. Our primary objective is twofold. First, we aim to align the feature importance measure with the specific structure of every model, whether linear or nonlinear, so that we retain the model's characteristic effects (e.g., shrinkage, gradient-based splits, or latent-factor extractions). Second, we seek to determine whether distinct estimation procedures converge on similar rankings of predictors, thus providing a robust cross-check of our nowcasting results.

We choose not to adopt computationally intensive methods, such as time-series variants of SHapley Additive exPlanations (SHAP)\footnote{SHAP (Shapley Additive Explanations) is a "game-theoretic" approach to model interpretation in which the contribution of each explanatory variable is computed based on "Shapley values," i.e., the average marginal effect of including that variable across all possible coalitions, as derived from cooperative game theory. The so-called "time-series SHAP" methods extend this algorithm by introducing dynamic baselines---which replace a single static reference with time-varying or regime-specific points of comparison, thereby adapting the measure of each variable's impact to evolving macroeconomic conditions---and similar adaptations for capturing temporal dependencies and regime changes typical of macroeconomic forecasting. See Lundberg and Lee \parencite{LundbergLee2017} for the theoretical foundations of SHAP, and Janizek et al. \parencite{JanizekTsunodaLundberg2021} for applications to time-series predictions.}, for two main reasons:

\begin{itemize}
    \item hardware constraints, since model-agnostic interpretability tools may require extensive resampling or repeated forward passes to compute contributions for each feature and each time point;
    
    \item our preference for model-specific techniques directly leveraging the estimation principles already embedded within each forecast.
\end{itemize}

Instead, we relied on measures compatible with rolling and expanding time windows, preserving seasonality and potential structural shifts while minimizing computational overhead. That maintained the temporal structure of our data, which is necessary for capturing real-world macroeconomic regimes.

We divided our methodology into a triple analysis to gain an overall view of the variables' importance:

\begin{itemize}
    \item \textit{Global analysis}, where all the final values of feature importance, relative to the entire period, are accumulated, and their average (or other appropriate aggregate statistic) is calculated. That allows us to highlight which features are overall important beyond any structural variations in the economic system;

    \item \textit{Analysis by sub-periods}, breaking down the observed horizon into macro-phases identified based on specific events or situations. For each sub-period, we calculate the average importance scores for the forecasts falling within that time interval. In this way, any differences in the role of specific features are highlighted: for example, some of them could prove critical in the contraction phases, while others could emerge in the recovery phase;

    \item \textit{Quarterly analysis}, where we keep track of iteration by iteration, generating a real historical series of the feature importance (quarter-by-quarter evolution). That makes it possible to promptly identify gradual or sudden variations in the relevance of specific predictors, highlighting any moments of transition or macroeconomic shocks.
\end{itemize}

In each of these perspectives (global, sub-periods, and quarterly), we sorted the features by the absolute value of their importance scores. Finally, we identified the top 10 for immediate ranking and interpretability. This final step simplifies the comparison of relevant predictors, facilitating an at-a-glance overview of which variables dominate under various economic conditions or time frames.

This triple perspective extends well beyond a single global viewpoint, enabling in-depth exploration of how variables may fluctuate in importance when subjected to structural breaks or nonlinearities. By retaining the internal rationale of each model's fitting procedure, we capture a finer-grained picture of the drivers behind our forecasting. Consequently, our interpretability framework offers a scalable and consistent method across various model classes and provides valuable insights into the temporal progression of each feature's influence. This approach is especially pertinent in macroeconomic contexts, where economic indicators can shift rapidly due to policy changes, crises, or other exogenous shocks, calling for nuanced and computationally feasible interpretability solutions.

\vspace{0.35cm}

\phantomsection
\subsubsection{\textbf{\textit{Coefficient-Based Feature Importance in Penalized Linear Models}}}
\vspace{0.2cm}
In penalized linear models (e.g., LASSO \parencite{Tibshirani1996,sklearnLasso}, Ridge \parencite{HoerlKennard1970,sklearnRidge}, and Elastic Net \parencite{ZouHastie2005,HastieTibshiraniWainwright2015,sklearnElasticNet}), the estimated coefficients are direct indicators of the influence exerted by each explanatory variable on the forecasts. From this perspective, we can consider the Coefficient-Based Feature Importance (CBFI) as an interpretability measure for penalized linear models. It consists in interpreting the value (positive or negative) of each coefficient as a signal of the direction of the effect and its magnitude as a measure of the intensity of the impact on the target.

This approach hinges on the idea that, in a linear model 

\begin{equation}
\mathbf{y} = \mathbf{X}\boldsymbol{\beta} + \boldsymbol{\varepsilon},
\end{equation}

\noindent{}the coefficient $\beta_{j}$ associated with each predictor $\mathbf{X}_{j}$ reflects the sensitivity of the dependent variable to that predictor \parencite{Hamilton1994,HastieTibshiraniFriedman2009}. In the case of penalized models, the estimation process introduces a constraint (of type L1, L2, or hybrid) that tends to reduce---or, in some cases, cancel---the coefficients associated with the less relevant features. Formally, for instance, a LASSO model solves

\begin{equation}
\min_{\boldsymbol{\beta}} \biggl\{
\frac{1}{2n}\|\mathbf{y}-\mathbf{X}\boldsymbol{\beta}\|^2 
\;+\;
\lambda \sum_j \lvert \beta_{j}\rvert
\biggr\},
\end{equation}

\noindent{}where $\lambda$ controls the degree of shrinkage, pushing less important $\beta_{j}$ toward zero \parencite{Tibshirani1996,Zou2006}.

Consequently, those regressors characterized by coefficient values $\hat{\beta}_j$ significantly different from zero in terms of amplitude assume a central role in explaining the variability of the target quantity. In contrast, coefficients close to zero suggest a lower relevance. Furthermore, this approach maintains a model-specific connotation by relying entirely on the linear regression structure and the shrinkage effects of regularization \parencite{SmeekesWijler2018,UematsuTanaka2019}.

In our linear models, the procedure for calculating the importance of the variables via CBFI follows these steps:
\begin{enumerate}
    \item \textbf{Iterated estimation of the penalized model.} 
    After selecting the corresponding optimal hyperparameters, regression fitting (e.g., LASSO, Ridge, or Elastic Net) is performed during each iteration \parencite{BergstraEtAl2011,AkibaEtAl2019}.

    \item \textbf{Extraction and storage of coefficients.} 
    At the end of each iteration, the coefficients associated with only the non-dummy features are recorded. For each predictor, we note the numerical value of the coefficient, its sign (positive or negative), and the corresponding magnitude in absolute value. That allows us to trace its direction and intensity immediately \parencite{MullerGuido2016}.

    \item \textbf{Aggregation over time.} 
    The collected coefficients are subsequently managed according to three perspectives:
    \begin{itemize}
        \item \textit{Global}: all the iterations available on the entire time sample are considered, computing the average coefficient of each variable, e.g.\ 
        \begin{equation}
           \overline{\beta}_{j}^{(\mathrm{global})} 
           \;=\; \frac{1}{|\mathcal{T}_{\mathrm{all}}|}\,\sum_{t \,\in\,\mathcal{T}_{\mathrm{all}}}
           \hat{\beta}_{j}^{(t)},
        \end{equation}
        where $\mathcal{T}_{\mathrm{all}}$ indicates the set of all iterations in the sample \parencite{Salkever1976,MakridakisEtAl1998};
        
        \item \textit{By sub-periods}: the sample is divided into relevant economic phases (Pre-COVID, COVID, Post-COVID, Excluding-COVID), and the average of the coefficients is calculated only for the iterations whose test-quarters fall in each sub-period. Formally, if $\mathcal{T}_{p}$ is the subset of iterations referring to the $p$-th sub-period, then
        \begin{equation}
           \overline{\beta}_{j}^{(p)} 
           \;=\; \frac{1}{|\mathcal{T}_{p}|}\,\sum_{t\,\in\,\mathcal{T}_{p}}
           \hat{\beta}_{j}^{(t)}.
        \end{equation}
        \item \textit{Quarterly}: the entire sequence of iteration-based coefficients is retained, forming a time series for each feature. Formally, for feature \(j\) we define
        \begin{equation}
        \mathcal{Q}_j 
        \,=\, 
        \Bigl(\hat{\beta}_{j}^{(1)}, \,\hat{\beta}_{j}^{(2)}, \,\dots, 
        ,\hat{\beta}_{j}^{(n)}\Bigr),
        \end{equation}
        where \(\mathcal{Q}_j\) is the time sequence of the coefficients related to the variable estimated at each iteration and \(\hat{\beta}_{j}^{(t)}\) is the coefficient obtained at iteration \(t\). This quarter-by-quarter evolution facilitates the prompt detection of gradual or sudden changes in a variable's importance and potential shifts in the underlying macroeconomic regime \parencite{ClementsHendry1999,PesaranPettenuzzoTimmermann2006}.
    \end{itemize}

    \item \textbf{Definition of the relevance ranking.} 
    In each of the above perspectives (global, sub-periods, and quarterly), the features are ordered based on the absolute value of the average coefficient. Thus, the predictors that present values significantly different from zero in modulus (i.e., \ large $|\beta_j|$) emerge as the most influential \parencite{BelloniChernozhukovHansen2014,KapetaniosZikes2018}.
\end{enumerate}

\noindent
The choice of using a Coefficient-Based Feature Importance in linear contexts with the regularization logic adopted by penalized models proves advantageous and consistent because:
\begin{itemize}
    \item it offers a feature-simple selection criterion based on a single summary value (the coefficient or its average over time) \parencite{Tibshirani1996,HoerlKennard1970,HastieTibshiraniWainwright2015};
    \item it does not require post hoc interpretative techniques since the coefficients are easily interpretable in economic or statistical terms \parencite{ZouHastie2005,SmeekesWijler2018};
    \item it directly reflects the penalizing dynamics imposed by the model. The L1 and/or L2 regularizers push the less important coefficients towards zero (zeroing or significant reduction), suggesting a negligible contribution in a given temporal context and facilitating the identification of a limited number of relevant predictors \parencite{Zou2006,KimSwanson2018};
    \item it is largely justified in our analysis framework, both for the need for a global overview of the main economic determinants and for the interest in isolating significant differences between different phases of macroeconomic evolution \parencite{CarrieroClarkMarcellino2015a,KapetaniosZikes2018,AliajCiganovicTancioni2023}.
\end{itemize}

\vspace{0.35cm}

\phantomsection
\subsubsection{\textbf{\textit{Coefficient-Based Feature Importance in Principal Component Regression}}}
\vspace{0.2cm}
Also in PCR \parencite{Massy1965,Jolliffe1982,JolliffeCadima2016}, the estimated coefficients are direct indicators of each explanatory variable's influence on the forecasts, albeit through an intermediate transformation. From this perspective, we can also view the Coefficient-Based Feature Importance as an interpretability tool for PCR models. Specifically, each coefficient's sign signals the direction of the effect, and its magnitude measures the intensity of the impact on the target once those coefficients are re-expressed in the space of the original predictors.

This approach revolves around the idea that, in a standard linear relationship
\begin{equation}
\mathbf{y} = \mathbf{X}\boldsymbol{\beta} + \boldsymbol{\varepsilon},
\end{equation}
the coefficient $\beta_{j}$ reflects how sensitive the dependent variable is to changes in predictor $\mathbf{X}_{j}$ \parencite{Hamilton1994,HastieTibshiraniFriedman2009}. However, under PCR, we first reduce $\mathbf{X}$ to $\mathbf{Z}$---a set of orthogonal principal components \parencite{JolliffeCadima2016}---and then fit a regression (a penalized one using Ridge, in our case) of the form:
\begin{equation}
\min_{\boldsymbol{\gamma}}
\Bigl\{ 
\tfrac{1}{2n}\|\mathbf{y} - \mathbf{Z}\boldsymbol{\gamma}\|^2
\;+\;
\lambda \sum_{k} \gamma_{k}^2
\Bigr\},
\end{equation}

\noindent{}where:

\begin{itemize}
    \item $\boldsymbol{\gamma}$ are the coefficients in the principal component space;
    \item $\lambda$ controls the degree of shrinkage.
\end{itemize}

Only after projecting these $\hat{\gamma}_{k}$ back to the original domain (via the loadings from the PCA) \parencite{Jolliffe1982}, we obtain a set of $\hat{\beta}_{j}$ corresponding to each original predictor. As in penalized models \parencite{SmeekesWijler2018}, large absolute values of $\hat{\beta}_{j}$ reveal the most relevant features, whereas small (or near-zero) coefficients indicate minor relevance. In this manner, PCR retains a model-specific interpretation anchored in a linear structure, though partially mediated by the dimension-reduction step \parencite{Massy1965}.

In our PCR framework, the procedure for calculating the variables' importance via CBFI proceeds through the following steps:
\begin{enumerate}
    \item \textbf{Iterated estimation of the PCR model.} 
    We perform the PCR fitting at each iteration after selecting the optimal number of components and penalty hyperparameters (if used) \parencite{BergstraEtAl2011,AkibaEtAl2019}. That includes:
    \begin{itemize}
        \item applying Principal Component Analysis to the current training set of $\mathbf{X}$;
        \item estimating a penalized regression of $\mathbf{y}$ on the principal components $\mathbf{Z}$.
    \end{itemize}

    \item \textbf{Projection and storage of coefficients.} 
    At the end of each iteration, we map the estimated coefficients $\hat{\gamma}_{k}$ in the Principal Component space back to the domain of the original predictors using the loadings matrix. Hence, for each non-dummy predictor, we obtain a numerical value $\hat{\beta}_{j}$, its sign, and its magnitude in absolute value, thereby identifying direction and intensity \parencite{MullerGuido2016}.

    \item \textbf{Aggregation over time.} 
    The resulting $\hat{\beta}_{j}$ values are then processed according to three distinct perspectives (similarly to what we did in penalized linear models):
    \begin{itemize}
        \item \textit{Global}: we compute the average coefficient for each variable over all iterations. Formally,
        \begin{equation}
           \overline{\beta}_{j}^{(\mathrm{global})} 
           \;=\; \frac{1}{|\mathcal{T}_{\mathrm{all}}|}\sum_{t \,\in\, \mathcal{T}_{\mathrm{all}}} \hat{\beta}_{j}^{(t)},
        \end{equation}
        where $\mathcal{T}_{\mathrm{all}}$ indicates the set of available iterations \parencite{Salkever1976,MakridakisEtAl1998}.
        \item \textit{By sub-periods}: the sample is split into meaningful macroeconomic phases (Pre-COVID, COVID, Post-COVID, Excluding-COVID). Within each sub-period $p$, we average the coefficients obtained during the iterations that fall into that sub-period:
        \begin{equation}
           \overline{\beta}_{j}^{(p)} 
           \;=\; \frac{1}{|\mathcal{T}_{p}|}\sum_{t\,\in\,\mathcal{T}_{p}} \hat{\beta}_{j}^{(t)},
        \end{equation}
        enabling a comparison of variable importance across different economic regimes \parencite{StockWatson2007,PesaranTimmermann2007}.
        \item \textit{Quarterly}: for each feature, we consider the full-time series of iteration-based coefficients:
        \begin{equation}
        \mathcal{Q}_j 
        \;=\; 
        \Bigl(\hat{\beta}_{j}^{(1)},\;\hat{\beta}_{j}^{(2)},\;\dots,\;\hat{\beta}_{j}^{(n)}\Bigr),
        \end{equation}
        which allows the detection of gradual shifts or abrupt changes in a variable's importance over the rolling sequence \parencite{ClementsHendry1999}.
    \end{itemize}

    \item \textbf{Definition of the relevance ranking.} 
    The analysis ranks the features according to the absolute values of their average coefficients across different perspectives (global, sub-periods, quarterly). Predictors with large $\lvert \beta_j \rvert$ stand out as key drivers of the target, whereas those with coefficients near zero are deemed less influential \parencite{BelloniChernozhukovHansen2014,KapetaniosZikes2018}.
\end{enumerate}

Implementing a Coefficient-Based Feature Importance in the context of Principal Component Regression offers a natural extension of the logic employed in penalized linear models with an added dimension-reduction step. In particular:
\begin{itemize}
    \item it allows the adoption of a succinct selection criterion based on a single summary value (the back-projected coefficient or its temporal average) \parencite{JolliffeCadima2016};
    \item it preserves interpretability by restoring each Principal Component coefficient to the space of the original explanatory variables, obviating the need for additional post hoc explanations \parencite{ZouHastie2005,SmeekesWijler2018};
    \item it aligns with the regularization framework of Ridge regression in the Principal Component domain, guiding coefficient shrinkage and focusing attention on the most relevant components and, ultimately, on the most relevant predictors \parencite{HoerlKennard1970,KimSwanson2018};
    \item it is justified in scenarios requiring both a global perspective on dominant factors and a deeper scrutiny of distinct macroeconomic segments, enabling the identification of shifts in the role of explanatory variables over time \parencite{StockWatson2007,PesaranTimmermann2007,CarrieroClarkMarcellino2015a}.
\end{itemize}

This additional projection step does not alter the fundamental meaning of $\hat{\beta}_{j}$ regarding its direction and magnitude; instead, it embeds the penalty-induced shrinkage within the dimensionality-reduced space \parencite{Jolliffe1982}. Therefore, PCR's Coefficient-Based Feature Importance approach remains consistent with its penalized linear counterparts while accommodating the PCA transformation.

\vspace{0.35cm}

\phantomsection
\subsubsection{\textbf{\textit{Contribution of Components via Variable Importance in Projection in Partial Least Squares Regression}}}
\vspace{0.2cm}
In Partial Least Squares Regression (PLSR) \parencite{Wold1985,MehmoodEtAl2012,Abdi2010}, the extracted latent components are direct indicators of the influence exerted by each explanatory variable on the forecasts, albeit through an intermediate step tied to dimensionality reduction. From this perspective, the Variable Importance in Projection (VIP) \parencite{AkarachantachoteChadchamSaithanu2014,ChongJun2005} is an interpretability measure for PLSR. It consists of interpreting the VIP value of each feature as a signal of the intensity (magnitude) of its impact on the target. By definition, VIP values are always positive. Hence, they quantify only the magnitude of a predictor's contribution to the latent factors and do not convey any information about the sign of the relationship with the target.

This approach hinges on the idea that, in a PLS framework
\begin{equation}
\mathbf{X} = \mathbf{T}\mathbf{P}^\top + \boldsymbol{E}
\quad\text{and}\quad
\mathbf{y} = \mathbf{T}\mathbf{q} + \boldsymbol{f},
\end{equation}

\noindent{}the latent factors $\mathbf{T}$ reflect the directions in $\mathbf{X}$ that maximally covary with $\mathbf{y}$ \parencite{KraemerSugiyama2011}. Consequently, those regressors characterized by large VIP scores assume a central role in explaining the variability of the target quantity. In contrast, VIPs close to or below unity suggest a lower relevance \parencite{AkarachantachoteChadchamSaithanu2014,ChongJun2005}. Furthermore, this approach maintains a model-specific connotation by relying entirely on the PLS decomposition, which aligns $\mathbf{X}$ and $\mathbf{y}$ in a shared latent space.

As described in our PLSR forecasting section, this estimation routinely adopts an expanding or rolling window. Thus, we recalibrate the model with newly available observations in each iteration before measuring VIP \parencite{FuentesPoncelaRodriguez2015}. That consistency ensures that the interpretative measure aligns with our one-step-ahead forecast procedure.

In our PLSR setting, the procedure for calculating the importance of the variables via VIP follows these steps:

\begin{enumerate}
    \item \textbf{Iterated estimation of the PLSR model.} 
    After selecting the corresponding optimal hyperparameters (e.g., the number of PLS components), we perform regression fitting during each iteration. In this phase---consistently with the rolling or expanding window described earlier---the model extracts latent factors from $\mathbf{X}$ that are most predictive of $\mathbf{y}$.

    \item \textbf{Extraction and storage of VIP scores.} 
    At the end of each iteration, the VIP scores of only the non-dummy features are recorded. For each predictor, we note the numerical value of the VIP, which captures its relative contribution to the latent factors underlying the forecasts \parencite{MehmoodEtAl2012}.

    \item \textbf{Aggregation over time.} 
    The collected VIP scores are subsequently managed according to three perspectives:
    
    \begin{itemize}
        \item \textit{Global}: we average each variable's VIP over all the iterations on the entire time sample, e.g.,
        \begin{equation}
           \overline{\mathrm{VIP}}_{j}^{(\mathrm{global})} 
           \;=\; \frac{1}{|\mathcal{T}_{\mathrm{all}}|}\,\sum_{t \,\in\,\mathcal{T}_{\mathrm{all}}}
           \mathrm{VIP}_{j}^{(t)},
           \label{eq:globalAvgVIP}
        \end{equation}
        where $\mathcal{T}_{\mathrm{all}}$ indicates the set of all iterations \parencite{Salkever1976,MakridakisEtAl1998}. This global average highlights the variable's overall relevance across the entire sample.

        \item \textit{By sub-periods}: the sample is divided into relevant macroeconomic phases (Pre-COVID, COVID, Post-COVID, Excluding-COVID), and the average VIP is computed only for the iterations corresponding to each sub-period. Formally, if $\mathcal{T}_{p}$ is the subset of iterations referring to the $p$-th sub-period, then
        \begin{equation}
           \overline{\mathrm{VIP}}_{j}^{(p)} 
           \;=\; \frac{1}{|\mathcal{T}_{p}|}\,\sum_{t\,\in\,\mathcal{T}_{p}}
           \mathrm{VIP}_{j}^{(t)}.
        \end{equation}
        That allows us to detect how a feature's influence changes across distinct economic contexts.

        \item \textit{Quarterly}: the entire sequence of iteration-based VIP scores is retained, forming a time series for each feature. Concretely, for feature \(j\) we define
        \begin{equation}
        \mathcal{V}_j
        \;=\;
        \Bigl(\mathrm{VIP}_{j}^{(1)},\,
              \mathrm{VIP}_{j}^{(2)}, \,
              \dots,\,
              \mathrm{VIP}_{j}^{(t)}\Bigr),
        \label{eq:quarterlyEvolutionVIP}
        \end{equation}
        where \(\mathcal{V}_j\) represents the temporal progression of VIP estimates at each iteration. This quarter-by-quarter evolution facilitates the prompt recognition of gradual or sudden shifts in a variable's importance.
    \end{itemize}

    \item \textbf{Definition of the relevance ranking.} 
    The features are ordered based on their average VIP in the above perspectives (global, sub-periods, and quarterly). Predictors that present high \(\mathrm{VIP}_{j}\) values indicate a notable influence on the latent factors driving the forecasts, whereas variables with lower VIPs appear less relevant. This ranking helps isolate the few most influential regressors.
\end{enumerate}

The choice of using VIP-based feature importance in a PLSR context proves advantageous and consistent because:
\begin{itemize}
    \item it offers a straightforward selection criterion based on a single summary value (the VIP or its average) \parencite{Wold1985,MehmoodEtAl2012};
    \item it does not require post hoc interpretative techniques since the VIP is intrinsically connected to the latent factors discovered by PLS \parencite{KraemerSugiyama2011};
    \item it directly reflects the covariance-driven dynamics of PLS, prioritizing features that best align with the explanatory structure of $\mathbf{y}$ \parencite{KimSwanson2018};
    \item it is largely justified in our forecasting framework, both for obtaining a global overview of dominant economic determinants and for isolating significant differences across diverse macroeconomic regimes \parencite{FuentesPoncelaRodriguez2015};
    \item it preserves a model-specific connotation by relying on PLS decomposition, ensuring that each predictor's importance is measured consistently with the structure used for predictive modeling.
\end{itemize}

The emphasis on maximizing covariance rather than solely capturing variance does not alter the fundamental role of each predictor with respect to the target; instead, it leverages the latent-factor structure of PLSR to highlight the most relevant explanatory directions. Therefore, the VIP-based assessment of feature importance remains fully coherent with the logic adopted in penalized linear models and PCR while incorporating an explicit alignment of \(\mathbf{X}\) and \(\mathbf{y}\) that enhances interpretability and predictive accuracy.

\vspace{0.35cm}

\phantomsection
\subsubsection{\textbf{\textit{Impurity-Based Feature Importance in Random Forest}}}
\vspace{0.2cm}
In Random Forest models \parencite{Breiman2001,BreimanFriedmanOlshenStone1984}, the Impurity-Based Feature Importance (also known as Mean Decrease in Impurity, MDI) directly indicates the influence exerted by each explanatory variable on the forecasts. From this perspective, the Impurity-Based Feature Importance is a model-specific interpretability measure for Random Forests. It consists of interpreting the average reduction in node-level impurity\footnote{In a regression tree, "impurity" measures how heterogeneous the target values are within a node. Typically, for node $\ell$ containing $t$ observations $\{y_1,\dots,y_t\}$, impurity can be defined via the variance of the $y_i$ (equivalent to the MSE). An alternative measure is the mean absolute deviation from the median (MAE), which may be more robust to outliers. In both cases, reducing impurity leads to more homogeneous nodes and enhanced predictive performance. For classification trees, criteria such as Gini or entropy are instead employed \parencite{BreimanFriedmanOlshenStone1984}.} (for instance, mean squared error for regression trees) associated with each predictor, thus capturing the magnitude of its impact on the target \parencite{MullainathanSpiess2017}.

This approach hinges on the idea that, in a Random Forest:

\begin{equation}
\mathrm{RF} 
\;=\; 
\Bigl\{\mathrm{Tree}_1,\;\mathrm{Tree}_2,\;\dots,\;\mathrm{Tree}_B\Bigr\},
\end{equation}

\noindent{}each \(\mathrm{Tree}_b\) is independently built on a bootstrap subsample, and the splitting at each node is chosen to maximize the decrease in impurity. In a regression setting, we can define a succinct formalism for the MDI of the variable \(j\) as:

\begin{equation}
\mathrm{MDI}_{j} \;=\; 
\frac{1}{B} 
\sum_{b=1}^{B}
\left[
\sum_{\ell \,\in\, \mathcal{L}_b(j)} 
\Delta(\mathrm{Impurity})_{\ell}
\right],
\end{equation}

\noindent{}where:

\begin{itemize}
    \item \(B\) is the total number of trees;
    \item \(\mathcal{L}_b(j)\) denotes the set of nodes in the \(b\)-th tree that use feature \(j\) for splitting;
    \item \(\Delta(\mathrm{Impurity})_{\ell}\) quantifies the local reduction in MSE at node \(\ell\).
\end{itemize}

In regression contexts, Random Forests may adopt either "squared error" or "absolute error" as the node-level impurity measure. In our study, the choice between these two criteria is determined dynamically via the TPE-based hyperparameter optimization \parencite{ShahriariEtAl2016,BorupChristensenMuhlbachNielsen2023}. During each iteration, our model tries both options and selects the one minimizing the validation mean squared error:

\begin{itemize}
    \item \textit{Squared Error Criterion:} if the algorithm chooses "squared error," then the node-level impurity corresponds to the mean of squared residuals with respect to the sample mean. Specifically, if a node \(\ell\) contains \(n\) observations \(\{y_1, \dots, y_n\}\) with average \(\bar{y}\), its impurity is
    
    \begin{equation}
    \mathrm{Impurity}(\ell)
    \;=\;
    \frac{1}{n}
    \sum_{i=1}^{n}
    (y_i - \bar{y})^2.
    \end{equation}
    When splitting \(\ell\) on feature \(X_j\) into left and right child nodes, we define
    \begin{equation}
    \Delta(\mathrm{Impurity})_{\ell}
    \;=\;
    \mathrm{Impurity}(\ell)
    \;-\;
    \Bigl[
        \tfrac{n_L}{n}\,\mathrm{Impurity}(\ell_L)
        \;+\;
        \tfrac{n_R}{n}\,\mathrm{Impurity}(\ell_R)
    \Bigr],
    \end{equation}
    
    where \(n_L\) and \(n_R\) are the cardinalities of the two child nodes of the parent node \(n\) (\(n = n_L + n_R\)). The total importance for \(X_j\) accumulates these \(\Delta(\mathrm{Impurity})_{\ell}\) over all relevant splits in all trees of the ensemble.

    \item \textit{Absolute Error Criterion:} if the algorithm chooses "absolute error," the impurity is based instead on the mean of absolute deviations from the median \(\tilde{y}\). In that case,
    
    \begin{equation}
    \mathrm{Impurity}(\ell)
    \;=\;
    \frac{1}{n}
    \sum_{i=1}^{n}
    \bigl|\,y_i - \tilde{y}\bigr|,
    \end{equation}
    with the node-level decrease in impurity \(\Delta(\mathrm{Impurity})_{\ell}\) computed analogously as the difference between parent and children nodes' weighted absolute-error impurity.
\end{itemize}

Thus, Mean Decrease in Impurity (MDI) is conceptually similar in both cases: each split contributes its local decrease in impurity \(\Delta(\mathrm{Impurity})\) to the variable causing that partition, and we average these contributions across all trees to obtain

\[
\mathrm{MDI}_{j} 
\;=\;
\frac{1}{B}
\sum_{b=1}^{B}
\sum_{\ell \,\in\,\mathcal{L}_b(j)}
\Delta(\mathrm{Impurity})_{\ell}.
\]

Here, \(B\) is the total number of trees, and \(\mathcal{L}_b(j)\) denotes the set of splits using \(X_j\) in the \(b\)-th tree. The only distinction is whether impurity is measured as variance-like (squared error) or deviation-like (absolute error), and the hyperparameter-tuning routine selects whichever yields a superior validation performance.

Regressors with high MDI values are central determinants of the overall predictive performance. On the other hand, regressors with near-zero MDI scores indicate low relevance. This method is closely tied to the model's specific features because it depends solely on how decisions are made through the tree-splitting process in the Random Forest \parencite{Breiman2001}.

In line with the principles illustrated in the previous sections (Coefficient-Based Feature Importance in penalized linear models, PCR, and PLSR), the procedure adopted to estimate the importance of the variables via Impurity-Based Feature Importance consists of four phases:

\begin{enumerate}
    \item \textbf{Iterated estimation of the Random Forest model.}
    After selecting the optimal hyperparameters (for example, number of trees and maximum depth), we estimate the Random Forest at each iteration, associating each iteration with a specific time block (rolling or expanding). In each ensemble of trees, predictors that achieve splits with more significant error reduction at the nodes acquire higher importance \parencite{BorupChristensenMuhlbachNielsen2023}.

    \item \textbf{Extraction and storage of impurity-based importances.}
    At the end of each iteration, we obtain an MDI vector \(\hat{I}_{j}\) for all the non-dummy features. Each value corresponds to the average reduction in impurity attributable to a given variable \(\mathbf{X}_j\) in the set of trees in the forest. We record these values to monitor each predictor's contribution and trace its evolution. However, it is worth noting that highly correlated predictors can inflate the MDI \parencite{StroblEtAl2007,NembriniKonigWright2018}, similar to issues discussed with other methodologies in cases of multicollinearity. Alternative approaches, such as permutation-based importances, can mitigate correlation biases albeit at a higher computational cost.

    \item \textbf{Aggregation over time.}
    Analogous to the approach illustrated for coefficients (CBFI) or VIP in PLSR, the MDI values collected at each iteration are then handled according to three perspectives:
    \begin{itemize}
        \item \textit{Global}: we consider all the iterations over the entire sample, computing the average MDI for each variable, for instance
        
        \begin{equation}
           \overline{I}_{j}^{(\mathrm{global})}
           \;=\;
           \frac{1}{|\mathcal{T}_{\mathrm{all}}|}
           \sum_{t\,\in\,\mathcal{T}_{\mathrm{all}}}
           \hat{I}_{j}^{(t)},
        \end{equation}
        
        where \(\mathcal{T}_{\mathrm{all}}\) is the set of all available iterations \parencite{MullainathanSpiess2017}. That offers a comprehensive view of the importance of \(\mathbf{X}_j\) over the entire period.

    \item \textit{By sub-periods}: if the sample is split into macroeconomic phases (Pre-COVID, COVID, Post-COVID, Excluding-COVID), we compute the average \(\hat{I}_{j}\) only for the iterations falling into each \(p\)-th sub-period, formally
    
        \begin{equation}
           \overline{I}_{j}^{(p)}
           \;=\;
           \frac{1}{|\mathcal{T}_{p}|}
           \sum_{t\,\in\,\mathcal{T}_{p}}
           \hat{I}_{j}^{(t)},
        \end{equation}
        
        thus identifying whether the contribution of each predictor is higher or lower in certain economic contexts.

        \item \textit{Quarterly}: for each variable \(\mathbf{X}_j\), we retain the entire sequence of iteration-based MDI values, defining
        
        \begin{equation}
        \mathcal{M}_j
        \;=\;
        \Bigl(
        \hat{I}_{j}^{(1)},
        \,\hat{I}_{j}^{(2)},
        \,\dots,
        \,\hat{I}_{j}^{(t)}
        \Bigr),
        \end{equation}
        
        thereby highlighting gradual or abrupt shifts in \(\mathbf{X}_j\)'s importance on a quarterly basis \parencite{BorupChristensenMuhlbachNielsen2023}.
    \end{itemize}

    \item \textbf{Definition of the relevance ranking.}
    The variables are ordered based on the mean MDI value in all three perspectives (global, sub-periods, quarterly). Predictors with substantially high \(\overline{I}_{j}\) stand out as central determinants, whereas those with near-zero values indicate a marginal influence.
\end{enumerate}

The choice of using Impurity-Based Feature Importance (Mean Decrease in Impurity) in a Random Forest context proves advantageous and consistent because:

\begin{itemize}
    \item it provides a single-value criterion derived directly from the ensemble's internal splitting mechanism;
    \item it does not require post hoc interpretative tools since the MDI emerges from the calculation of the impurities on the nodes \parencite{BreimanFriedmanOlshenStone1984};
    \item it accounts for nonlinear relationships and possible interactions, thanks to the hierarchical structure of regression trees \parencite{MedeirosEtAl2021};
    \item it is justified in our iterative framework, both for obtaining a global overview of primary economic determinants and for revealing significant differences across macroeconomic sub-periods \parencite{BorupChristensenMuhlbachNielsen2023};
    \item it remains a model-specific measure, fully reflecting how each predictor reduces the error in node partitions and thus integrating seamlessly into the rolling or expanding estimation procedure.
\end{itemize}

\vspace{0.35cm}

\phantomsection
\subsubsection{\textbf{\textit{Gain-Based Feature Importance in XGBoost}}}
\vspace{0.2cm}
In XGBoost models, the Gain-Based Feature Importance (sometimes referred to simply as "Gain Importance") \parencite{ChenGuestrin2016, XGBoostDocs2023} directly indicates the influence exerted by each explanatory variable on the forecasts. From this perspective, the Gain-Based Feature Importance is a model-specific interpretability measure for boosting methods \parencite{LundbergEtAl2020,ZhouHooker2021}. It consists of interpreting the average improvement in the objective function (such as a regularized mean squared error) associated with each predictor at split nodes, thus capturing the magnitude of its impact on the target. In line with the interpretative rationale observed in linear and factor-based approaches \parencite{HastieTibshiraniFriedman2009,Jolliffe1982,Wold1985}, this measure offers a more adaptable perspective on variable significance, mainly when the data-generating process includes nonlinearities or intricate interactions. However, like other model-specific importance, it only pinpoints which features most effectively reduce the loss function without disclosing whether their relationship with the outcome is positive or negative \parencite{StroblEtAl2007,NembriniKonigWright2018,ZhouHooker2021}. Consequently, it complements methods such as Coefficient-Based Feature Importance, Principal Component Regression, and Partial Least Squares Regression, especially under conditions of strong nonlinearity.

This approach centers on the concept that, in an XGBoost model:

\begin{equation}
\mathrm{XGB}
\;=\;
\sum_{b=1}^{B} 
\mathrm{Tree}_b,
\end{equation}

each \(\mathrm{Tree}_b\) is trained sequentially to reduce a differentiable loss function \(\mathcal{L}\) (often with \(L_2\) regularization), and the splitting at each node is chosen to maximize the gain, which measures the drop in the overall loss once a node is partitioned into two child nodes. Formally, for a node \(\ell\) split on feature \(j\), the gain can be expressed as

\begin{equation}
\mathrm{Gain}_{\ell} 
\;=\;
\mathrm{Score}(\ell)
\;-\;
\Bigl[
\mathrm{Score}(\ell_L)
\;+\;
\mathrm{Score}(\ell_R)
\Bigr]
\;-\;
\Gamma,
\end{equation}

where:

\begin{itemize}
    \item \(\mathrm{Score}(\ell)\) is the node-specific value derived from first- and second-order gradient statistics\footnote{In a typical regression setting with squared-error loss (and $L_2$ regularization), the node-level score can be approximated by
    
    \[
    \mathrm{Score}(\ell) \;=\; -\, \frac{\bigl(\sum_{i \in \ell} g_i\bigr)^2}{\sum_{i \in \ell} h_i + \lambda},
    \]
    
    where $g_i$ and $h_i$ denote the first- and second-order gradients of the loss function evaluated on each observation $i$ in node $\ell$, and $\lambda$ is the $L_2$ regularization coefficient. For classification tasks (e.g., logistic loss), XGBoost employs the corresponding gradient statistics of the logistic objective, preserving the same notion that a higher (more negative) score indicates a better local fit \parencite{ChenGuestrin2016}.};
    \item \(\ell_L\) and \(\ell_R\) denote the left and right child nodes, respectively;
    \item \(\Gamma\) is a regularization penalty that discourages overly complex splits\footnote{In some references, $\Gamma$ is denoted by $\gamma$ and represents the minimum loss reduction required to partition a leaf node further, thereby controlling overfitting. A larger $\Gamma$ (or $\gamma$) penalizes additional splits that offer only marginal gains.}.
\end{itemize}

Thus, the Gain-Based Feature Importance (GBI) of the variable \(j\) in regression settings with XGBoost can be summarized as:

\begin{equation}
\mathrm{GBI}_j
\;=\;
\frac{1}{B}
\sum_{b=1}^{B}
\left[
\sum_{\ell \,\in\, \mathcal{L}_b(j)} 
\mathrm{Gain}_{\ell}
\right],
\end{equation}

\noindent{}where:

\begin{itemize}
    \item \(B\) is the total number of boosting rounds (or trees);
    \item \(\mathcal{L}_b(j)\) denotes the set of nodes in the \(b\)-th tree where feature \(j\) is used to split;
    \item \(\mathrm{Gain}_{\ell}\) quantifies the local improvement in the objective function at node \(\ell\).
\end{itemize}

In regression contexts, XGBoost commonly employs a squared-error objective with \(L_2\) regularization \parencite{ChenGuestrin2016,XGBoostDocs2023}, but it may also adopt alternative loss formulations. For example:

\begin{itemize}
    \item \textit{Squared Error Criterion:} if the algorithm chooses a squared-error loss, the node-level \(\mathrm{Score}(\ell)\) stems from the sum of squared residuals and their second-order derivatives, with an additional \(L_2\) regularization term applied. The gain \(\mathrm{Gain}_{\ell}\) thus measures how much splitting on \(X_j\) at node \(\ell\) reduces the regularized MSE-based objective.

    \item \textit{Absolute Error Criterion:} alternatively, an absolute-error-oriented loss (or a pseudo-Huber variant) can be employed. In that scenario, \(\mathrm{Score}(\ell)\) relies on gradient statistics originating from an \(L_1\)-like loss, and the gain quantifies the improvement in minimizing the absolute deviations at the node level.
\end{itemize}

Hence, different objective functions alter the exact \(\mathrm{Score}(\ell)\) calculation but preserve the overarching concept that each split's \(\mathrm{Gain}\) indicates how much that split (and hence the feature used) advances the model's predictive accuracy. The hyperparameter-tuning routine (e.g., TPE-based) can systematically compare these objectives \parencite{BergstraEtAl2011,AkibaEtAl2019,ShahriariEtAl2016} and retain the one yielding the best validation performance.

In our study, we typically adopt a squared-error loss with \(L_2\) regularization, but alternative objectives (including absolute-error-based functions) may also be employed if indicated by the validation results. The choice of hyperparameters (e.g., number of boosted trees, maximum depth of each tree, learning rate, regularization terms) is similarly handled via TPE-based optimization \parencite{BergstraEtAl2011,AkibaEtAl2019}, mirroring the procedure described for Random Forest. The model determines the best combination of these hyperparameters during each iteration and re-estimates the XGBoost ensemble, yielding updated gain-based importance scores.

High GBI values indicate central determinants for the predictive performance. Conversely, regressors with near-zero GBI scores have a marginal influence. Like other model-specific importances, the Gain-Based measure depends solely on the gradient-based splitting decisions within the ensemble and does not reveal the sign of the underlying relationship.

Following the same principles illustrated in previous sections (Coefficient-Based Feature Importance in penalized linear models, PCR, and PLSR), the procedure adopted to estimate the importance of the variables via Gain-Based Feature Importance consists of four phases:

\begin{enumerate}
    \item \textbf{Iterated estimation of the XGBoost model.}
    After selecting the optimal hyperparameters (e.g., number of boosting rounds, maximum depth, and learning rate), we fit the XGBoost model at each iteration, associating each iteration with a specific time block (rolling or expanding). Within each ensemble, predictors that yield splits with more significant loss reduction gain higher importance.

    \item \textbf{Extraction and storage of gain-based importances.}
    At the end of each iteration, we obtain a GBI vector \(\hat{I}_{j}\) for all non-dummy features. Each value corresponds to the average gain attributable to a given variable \(\mathbf{X}_j\). We record these values to monitor each predictor's contribution. As in Random Forest, highly correlated predictors may inflate GBI, and permutation-based methods or additional regularization adjustments might address such correlation biases at higher computational cost \parencite{StroblEtAl2007,NembriniKonigWright2018}.

    \item \textbf{Aggregation over time.}
    Analogous to the approach used for Coefficient-Based Feature Importance or VIP in PLSR, the GBI values from each iteration are examined under three perspectives:
    \begin{itemize}
        \item \textit{Global}: we compute the average GBI for each variable over all iterations,
        \begin{equation}
           \overline{I}_{j}^{(\mathrm{global})}
           \;=\;
           \frac{1}{|\mathcal{T}_{\mathrm{all}}|}
           \sum_{t\,\in\,\mathcal{T}_{\mathrm{all}}}
           \hat{I}_{j}^{(t)},
        \end{equation}
        offering a comprehensive assessment of how relevant \(\mathbf{X}_j\) is throughout the entire sample.

        \item \textit{By sub-periods}: if the sample is divided into macroeconomic phases (Pre-COVID, COVID, Post-COVID, Excluding-COVID), we calculate 
        \begin{equation}
           \overline{I}_{j}^{(p)}
           \;=\;
           \frac{1}{|\mathcal{T}_{p}|}
           \sum_{t\,\in\,\mathcal{T}_{p}}
           \hat{I}_{j}^{(t)},
        \end{equation}
        thereby highlighting whether each predictor's importance increases or decreases in specific economic contexts.

        \item \textit{Quarterly}: for each variable, we retain the entire sequence 
        \begin{equation}
        \mathcal{M}_j
        \;=\;    \Bigl(\widehat{I}_{j}^{(1)},\,\widehat{I}_{j}^{(2)},\,\dots\,\widehat{I}_{j}^{(t)}\Bigr),
        \end{equation}
        of iteration-based GBI scores, revealing gradual or abrupt shifts in \(\mathbf{X}_j\)'s influence.
    \end{itemize}

    \item \textbf{Definition of the relevance ranking.}
    The variables are ordered according to their mean GBI in all three perspectives (global, sub-periods, quarterly). Predictors with substantially high \(\overline{I}_{j}\) consistently appear as key drivers in the XGBoost ensemble, whereas those with near-zero scores have negligible effects on the model.
\end{enumerate}

Adopting a Gain-Based Feature Importance approach in XGBoost offers several advantages:

\begin{itemize}
    \item it provides a single-value criterion that arises naturally from the gradient-boosted splitting mechanism;
    \item it encapsulates nonlinear relationships and potential interactions due to the sequential refinement of residuals;
    \item it aligns well with rolling or expanding frameworks, offering both an overall view of key economic determinants and insights on how these determinants evolve across different macroeconomic regimes;
    \item it remains strictly model-specific, reflecting how each predictor contributes to the boosted ensemble's objective function at the node level.
\end{itemize}

\par{}
Impurity-Based Feature Importance (MDI) in Random Forest and Gain-Based Feature Importance (GBI) in XGBoost represent model-specific approaches to interpretability. They quantify how much each variable contributes to improving the model's performance by reducing errors (measured by chosen loss measure---an impurity measure or a gradient-based objective) at different decision points (split nodes). However, these methods do not indicate whether a variable has a positive or negative effect on the target outcome. While they do not show direct cause-and-effect relationships, they can effectively capture complex and nonlinear connections, especially in macroeconomic situations where simpler linear models might be inadequate.

Therefore, MDI and GBI offer an alternative and complementary insight to CBFI in penalized and Principal Component regressions, and VIP in Partial Least Squares Regression into which predictors matter the most under conditions of high complexity or time-varying structures while at the same time requiring careful handling of potential correlation biases \parencite{StroblEtAl2007,NembriniKonigWright2018}.

\vspace{0.35cm}

\phantomsection
\subsubsection{\textbf{\textit{Integrated-Gradients-Based Feature Importance in MLP and GRU}}}
\vspace{0.2cm}
In Multilayer Perceptron (MLP) and Gated Recurrent Unit (GRU) models, the Integrated-Gradients-Based Feature Importance (IG) quantifies the contribution of each explanatory variable to the neural network's forecast. From this perspective, the Integrated Gradients (IG) technique is a model-specific interpretability measure for differentiable models. It relies on computing the integral of the gradients of the network output concerning the input features, thus capturing how each predictor influences the final prediction \parencite{SundararajanTalyYan2017}.

This approach hinges on the concept that, in a neural network \( F \):

\[
F(\mathbf{x}) \;=\; \widehat{y},
\]

\noindent{}the partial derivatives \(\tfrac{\partial F}{\partial x_j}\) indicate how small perturbations in the input feature \(x_j\) affect the output \(\widehat{y}\). The Integrated Gradients formula extends this idea by evaluating and averaging those partial derivatives along a straight-line path from a reference baseline \(\mathbf{b}\) to the actual input \(\mathbf{x}\).

In practice, we define a linear trajectory

\[
\mathbf{x}(\alpha) \;=\; \mathbf{b} \;+\; \alpha \,\bigl(\mathbf{x} - \mathbf{b}\bigr),
\quad \alpha \,\in\, [0,1],
\]

where \(\mathbf{b}\) is often a zero vector or another neutral reference. The Integrated Gradient \(\mathrm{IG}_j(\mathbf{x})\) of feature \(j\) is then given by

\[
\mathrm{IG}_j(\mathbf{x})
\;=\;
\bigl(x_j - b_j\bigr)
\;\times\;
\int_{0}^{1}
\frac{\partial F\bigl(\mathbf{x}(\alpha)\bigr)}{\partial x_j}
\,d\alpha.
\]

Hence, the total importance of predictor \(j\) is derived by multiplying the difference \(\bigl(x_j - b_j\bigr)\) by the average gradient along that path. A numerical approximation typically divides \(\alpha\) into discrete steps and sums the corresponding gradients.

In both MLP and GRU models, this technique quantifies the magnitude of each input's impact on the final output \parencite{FreeboroughVanZyl2022,PaudelEtAl2023}, without indicating its sign (positive or negative effect). Consequently, it complements methods such as Coefficient-Based Feature Importance, Principal Component Regression, or Partial Least Squares by addressing highly nonlinear or interaction-heavy data-generating processes.

A key practical decision involves choosing the baseline \(\mathbf{b}\). In many applications, \(\mathbf{b}\) is set to the zero vector, but alternative references include the average of the training set or any meaningful neutral point in the feature space. Similarly, the discretization of \(\alpha\) (i.e., the step size or number of steps) influences both the computational burden and the numerical precision of the integrated gradients.

In analogy to the Random Forest (MDI) and XGBoost (GBI) frameworks, the neural network model (MLP or GRU) is fitted at each iteration of a rolling or expanding window. Formally, we define:
\begin{equation}
\mathrm{NN} 
\;=\; 
\Bigl\{\mathrm{NN}^{(1)},\;\mathrm{NN}^{(2)},\;\dots,\;\mathrm{NN}^{(T)}\Bigr\},
\end{equation}
where each \(\mathrm{NN}^{(t)}\) is either an MLP or a GRU trained on a specific time block. We assume the hyperparameters (e.g., number of layers, hidden dimension, learning rate, regularization) are selected via TPE-based optimization or a similar approach. Once the network parameters are fixed for iteration \(t\), the calculation of IG-based importances proceeds through four main steps:

\begin{enumerate}
    \item \textbf{Train the model}: estimate \(\mathrm{NN}^{(t)}\) on the designated training (and possibly validation) set within the rolling or expanding window.
    \item \textbf{Compute Integrated Gradients}: for each non-dummy predictor \(j\), evaluate \(\mathrm{IG}_j^{(t)}(\mathbf{x})\) over an appropriate subset of observations, using the chosen baseline \(\mathbf{b}\) and step discretization in \(\alpha\). 
    \item \textbf{Aggregate into a single importance score}: optionally average the individual-observation attributions, obtaining a scalar \(\widehat{I}_{j}^{(t)}\) for each variable \(j\). 
    \item \textbf{Record and store}: collect the resulting IG-based importances into a vector \(\bigl[\widehat{I}_{1}^{(t)},\dots,\widehat{I}_{p}^{(t)}\bigr]\) for subsequent analysis.
\end{enumerate}

At the conclusion of iteration \(t\), the vector \(\widehat{I}_{j}^{(t)}\) is available for all non-dummy features. These values measure how much each variable \(\mathbf{X}_j\) contributes, on average, to the neural network's output relative to \(\mathbf{b}\). We preserve these IG values to:

\begin{itemize}
    \item monitor the role of each predictor over time,
    \item detect potential instability of the importance measure in the presence of collinearity,
    \item compare integrated gradients across distinct macroeconomic contexts.
\end{itemize}

In neural networks, correlated inputs may lead to partial redundancies in IG scores, analogous to how collinearity affects MDI and GBI, albeit via a different allocation mechanism governed by backpropagation.

Consistent with the principles adopted for Random Forest and XGBoost, we aggregate the IG values across all iterations according to three perspectives:

\begin{itemize}
    \item \textit{Global}: we compute the mean of \(\widehat{I}_{j}^{(t)}\) over every iteration, resulting in
    
    \begin{equation}
    \overline{I}_{j}^{(\mathrm{global})}
    \;=\;
    \frac{1}{|\mathcal{T}_{\mathrm{all}}|}
    \sum_{t\,\in\,\mathcal{T}_{\mathrm{all}}}
    \widehat{I}_{j}^{(t)},
    \end{equation}
    
    which captures a holistic view of how \(\mathbf{X}_j\) influences the forecast over the full sample.

    \item \textit{By sub-periods}: if the sample is split into distinct macroeconomic phases (e.g., Pre-COVID, COVID, Post-COVID, Excluding-COVID), we similarly define
    
    \begin{equation}
    \overline{I}_{j}^{(p)}
    \;=\;
    \frac{1}{|\mathcal{T}_{p}|}
    \sum_{t\,\in\,\mathcal{T}_{p}}
    \widehat{I}_{j}^{(t)},
    \end{equation}
    
    facilitating a comparison of each predictor's relevance in different economic environments.

    \item \textit{Quarterly}: we store the entire sequence
    
    \begin{equation}
    \mathcal{M}_j
    \;=\;    \Bigl(\widehat{I}_{j}^{(1)},\,\widehat{I}_{j}^{(2)},\,\dots\,\widehat{I}_{j}^{(t)}\Bigr),
    \end{equation}
    
    thus enabling an examination of gradual or abrupt shifts in \(\mathbf{X}_j\)'s importance on a quarter-by-quarter basis.
\end{itemize}

Following the same logic adopted for MDI or GBI, we establish an importance ranking by ordering the features according to their average IG score in the perspectives described above (global, sub-periods, quarterly). Predictors with high \(\overline{I}_{j}\) values consistently appear as principal contributors, whereas variables whose IG scores remain close to zero signify a limited effect on the neural network's output. Because IG measures how the output changes relative to the baseline, these values are strictly model-specific and do not provide information on whether increments in \(\mathbf{X}_j\) associate with a positive or negative effect on the target.

Adopting an integrated-gradients-based feature importance in neural networks offers several benefits:

\begin{itemize}
    \item it provides a single-value criterion derived from backpropagation gradients, thus cohering with the model's core optimization process;
    \item it accommodates nonlinear relationships and interactions, reflecting the flexible structure of MLPs and GRUs;
    \item it naturally aligns with rolling or expanding frameworks, yielding both a comprehensive perspective of dominant predictors and a window-by-window analysis of importance variation;
    \item it clarifies how each predictor alters the fitted function from a baseline state to the actual input, forming a conceptual parallel to "impurity reduction," yet grounded in gradients rather than splits.
\end{itemize}

Similarly to other model-specific metrics (MDI, GBI), IG-based importances do not inherently convey the sign of the underlying relationship, nor are they entirely immune to correlation-related distortions. Nevertheless, this methodology remains a powerful tool for illuminating predictors' intricate and evolving significance in deep or recurrent neural architectures \parencite{Rudin2019}.

\vspace{0.35cm}

\phantomsection
\subsection{\textbf{Confidence Intervals for Feature Importance via Pair Block Bootstrap}}
\label{sec:2.5.2}
In this study, we calculated confidence intervals for feature importance across all models, including linear ones, instead of relying on p-values. This approach provides a consistent measure of uncertainty for feature importance, regardless of the type of model used. Specifically, p-values are not readily computable for models such as XGBoost or GRU, making it challenging to assess the stability of feature importance in those cases \parencite{WassersteinLazar2016}. By employing confidence intervals for all models, we ensure a common foundation for comparison, allowing us to evaluate the stability of feature importance across a diverse range of model types. This methodological choice strengthens the comparability of our results, offering a unified framework for interpreting feature importance across different predictive models \parencite{Shmueli2010,Molnar2022}.

The construction of "confidence intervals for feature importance" represents an essential step in quantifying the uncertainty surrounding the contribution of each predictor in macroeconomic models. This process provides a more robust interpretation of the role of variables in the model's predictive capacity \parencite{BreimanTwoCultures2001}.

In contexts where residuals exhibit temporal dependencies or structural breaks, advanced methods such as the block bootstrap \parencite{Kunsch1989,PolitisRomano1994,Lahiri2003,PanPolitis2016,Masarotto1990} estimate these confidence intervals, offering a more reliable measure of feature importance. Specifically, we adopt a pair block bootstrap, meaning that each resampled block contains both the explanatory variables $\mathbf{x}_t$ and the target variable $y_t$, thus preserving the simultaneous temporal structure of input-output data. As previously described in Section~\ref{sec:2.4}, where a pair block bootstrap was used for prediction intervals, this technique ensures that the temporal structure of the series is respected, thereby providing a resampling procedure agnostic to the distribution of the residuals.

The block bootstrap procedure can be broken down into the following steps:

\begin{enumerate}
    \item \textbf{Subdivision of the dataset into sub-segments.} In each iteration, the training and validation dataset (up to the last in-sample quarter) is divided into sub-segments corresponding to key historical economic breaks, such as global crises or other major economic shocks. This segmentation is based on ex-ante economic knowledge of the periods during which significant regime shifts occurred rather than solely relying on statistical tests (such as the Bai-Perron test). By adopting this approach, we ensure that periods affected by global disruptions (e.g., the 2008 financial crisis or the COVID-19 pandemic) are adequately considered in calculating feature importance, even if these breaks were not statistically detected. This step is crucial for capturing potential structural changes in the data that might influence the importance of the features.

    \item \textbf{Creation of contiguous blocks for each subsegment.} Within each sub-segment, the data pairs $(\mathbf{x}_t, y_t)$ are grouped into contiguous blocks of fixed size (e.g., four quarters). The choice of a block size corresponding to one year (four quarters) is intended to preserve the seasonality and internal serial autocorrelation typical of macroeconomic time series. This block size ensures that intra-annual dependencies are maintained, which is especially important when modeling macroeconomic data, where seasonal patterns play a significant role \parencite{MakridakisEtAl1998}.

    \item \textbf{Sampling with block replacement.} Several blocks, each of the same length, are randomly extracted from each subsegment, with replacement, until the size of the bootstrap sample matches that of the original subsegment. This sampling method guarantees that temporal dependencies within the blocks are preserved while allowing the order and frequency of the blocks to vary between iterations. The key feature of this step is that it ensures the temporal dependence across blocks is maintained, which is essential for accurately assessing feature importance over time, as temporal relationships are critical in macroeconomic forecasting \parencite{PolitisRomano1994,Lahiri2003}.

    \item \textbf{Concatenation of the resampled subsegments.} The resampled blocks are concatenated to form an alternative training and validation dataset. Although this new dataset preserves the overall length and temporal structure of the original dataset, it is rearranged in such a way as to reflect different configurations of the data. This step ensures that the variability in feature importance is captured through multiple resampled datasets, reflecting potential shifts in the underlying data-generating process \parencite{PanPolitis2016}.

    \item \textbf{Retraining the model and importance calculation.} Once the bootstrap sample $(\mathbf{x}_\mathrm{boot}, y_\mathrm{boot})$ has been generated, the model is retrained on this new dataset. The feature importance values are recalculated for each bootstrap iteration. This process is repeated $B$ times (typically $B = 1000$) to generate a distribution of feature importance scores. Each iteration yields a new estimate of feature importance, which accounts for potential changes in the data structure and provides insight into the uncertainty associated with each feature's contribution to the model \parencite{WagerHastieEfron2014,MentchHooker2016,IshwaranLu2019,WilliamsonEtAl2023,AltmannEtAl2010}.

    \item \textbf{Construction of the confidence interval for feature importance.} After obtaining the feature importance values from each bootstrap iteration, the values are ordered, and the quantiles corresponding to the desired confidence level (e.g., 2.5\% and 97.5\%) are calculated. The lower and upper quantiles denoted as $\underline{q}_{\alpha}$ and $\overline{q}_{\alpha}$, respectively, define the confidence interval for each feature's importance:
    
    \[
    \mathrm{CI}_{1-2\alpha}(\text{Feature})
    = [ \underline{q}_{\alpha}, \overline{q}_{\alpha} ]
    \]
    
    These confidence intervals reflect the uncertainty in the estimated feature importance, considering both the inherent variability due to sampling and the potential impact of structural shifts or regime changes. The construction of these intervals allows for a more nuanced evaluation of feature stability over time, capturing the effect of external shocks such as global economic crises or pandemics \parencite{BergmeirHyndmanBenitez2016,Li2021}.
\end{enumerate}

This procedure is also applied iteratively within an Expanding+Rolling framework. The best hyperparameter configuration is determined in each iteration, and the model is retrained on the updated historical data. After the hyperparameters are fixed, feature importance for the next test set is computed, and the block bootstrap is again employed to evaluate the sensitivity of the feature importance estimates to possible breaks or variations in the training data. This iterative process ensures that feature importance measures are continuously refined and accurately reflect changes in the data structure over time.

Furthermore, the average width of the confidence intervals for feature importance is calculated across multiple sub-periods (Pre-COVID, COVID, Post-COVID, Excluding-COVID). That enables an assessment of how structural changes, such as economic crises or pandemics, influence the uncertainty associated with the feature importance measures. The analysis allows us to evaluate the robustness of feature importance estimates under varying macroeconomic conditions and to understand how global disruptions impact the model's sensitivity to predictors. By comparing these intervals across different phases, we gain valuable insights into the dynamic nature of feature importance over time.

\par{} 
The "confidence intervals for feature importance" calculated in this manner provide reliable, valid uncertainty measures across various models. These intervals offer a deeper understanding of the variability in feature importance, serving as an essential tool for assessing the stability of feature importance across different model configurations and historical periods. This process contributes to a more informed and robust analysis of the role of predictors in macroeconomic forecasting.

\vspace{0.5cm}

\section{\textbf{Model Selection and Construction of Combined Models}}
\label{sec:2.6}

\phantomsection
\subsection{\textbf{\textit{Identifying Superior Models for Combination via the Model Confidence Set}}}
\label{sec:2.6.1}
The selection of a coherent subset of forecasting models constitutes a critical prerequisite for effective ensemble methods. In particular, macroeconomic forecasting often encompasses multiple candidate specifications (linear and nonlinear, parametric and nonparametric), each potentially suited to different phases of the economic cycle. Consequently, identifying a "Superior Set of Models" (SSM)\footnote{The term "Superior Set of Models" (SSM) refers to the subset of candidate models that, according to the "Model Confidence Set" (MCS), cannot be statistically rejected as inferior at a chosen significance level. In our application, we set this level to $\alpha$ = 0.10. \parencite{White2000,Hansen2005,HansenLundeNason2011}} in a statistically rigorous manner is essential for mitigating forecast uncertainty and avoiding the inclusion of systematically underperforming models. In this section, we adopt the "Model Confidence Set" (MCS) framework \parencite{HansenLundeNason2011} to fulfill this task. We first outline the theoretical underpinnings of the MCS approach and then detail its step-by-step implementation. Our exposition aims to clarify how the MCS can be applied effectively in a time-series context, ensuring that temporal dependence is duly respected.

A Model Confidence Set is a statistical procedure identifying the best subset of models based on predictive accuracy. It tests, in an iterative and statistically controlled manner, whether any model in a pool is inferior\footnote{"Inferior" here indicates that a model's predictive loss is significantly higher than that of at least one competing model. Specifically, we say that model $M_j$ is "inferior" if the p-value computed from the MCS bootstrap procedure falls below our significance threshold $\alpha$ = 0.10. In other words, "significantly higher" means that the estimated difference in losses (relative to at least one other model) is unlikely to have arisen by chance under the joint null hypothesis of equal predictive ability. \parencite{HansenLundeNason2011}} to the others by a statistically significant margin.\footnote{A "statistically significant margin" is any observed difference in mean losses between two models (or between one model and the group) that, after block bootstrap resampling, yields a test statistic with an associated p-value < $\alpha$. In our empirical implementation, $\alpha$ = 0.10, so significance implies that we reject the null hypothesis of equal performance at the 10\% level.} In the presence of $K$ candidate models $\{M_1,\dots,M_K\}$, the procedure proceeds as follows:

\begin{enumerate}
    \item \textbf{Computes pairwise loss differentials.} Let $\ell_t(M_j)$ denote the loss function for model $M_j$ at time $t$, where $t=1,\dots,N$. We focus on squared errors in our application (though the approach accommodates other loss measures). For each pair $(M_i,M_j)$, define the pairwise difference
    
    \[
      d_t(M_i, M_j) \;=\; \ell_t(M_i) \;-\; \ell_t(M_j).
    \]
    
    Intuitively, $d_t(M_i, M_j)$ indicates whether model $M_i$ is performing better (lower loss) or worse (higher loss) than model $M_j$ at each time $t$. Aggregating these differences over $t=1,\dots,N$ provides a basis for measuring whether $M_i$ systematically outperforms $M_j$.

    \item \textbf{Performs a joint hypothesis test.} Using a block bootstrap scheme to account for potential autocorrelation in the series $\{d_t(M_i,M_j)\}_{t=1}^N$, the MCS procedure tests if at least one model exhibits significantly larger losses than all others. This approach addresses the multiple-testing issue (i.e., controlling the family-wise error rate) and prevents inflating Type I error across many comparisons \parencite{White2000,Hansen2005}.

    \item \textbf{Eliminates inferior models.} A user-chosen significance level $\alpha$ (e.g., $0.10$) governs the threshold for deciding whether a model is significantly worse. If a model is found consistently inferior to at least one competitor across the bootstrap replications, it is removed from the set. The test is then iterated on the remaining models until no further rejections occur.
\end{enumerate}

In adopting this iterative screening, the MCS obviates the need for ex-ante assumptions about model validity; instead, it relies on evidence from the block bootstrap to identify those models that do not show persistent inferiority relative to others \parencite{AiolfiTimmermann2006}. 

Below, we summarize the main conceptual steps followed in our methodological application:

\begin{enumerate}
    \item \textbf{Definition of the loss matrix.}
    We gather and organize the out-of-sample forecasting errors (in our case, squared errors) into an $N\times K$ loss matrix, where each column corresponds to one of the $K$ candidate models, and each $N$ row represents a specific time point (e.g., a quarter).

    \item \textbf{Block bootstrap for test statistics.}
    To check if a model $M_j$ is distinctly inferior to others, we employ a block bootstrap with a chosen block size $k$. In our empirical study, we set $k=4$ quarters to preserve any annual pattern or seasonal correlation. We replicate the bootstrap $B$ times (with $B=10{,}000$ in our final specification) to ensure a stable estimate of the distribution of the pairwise differentials. Statistics such as "Tmax" or "TR"\footnote{Both Tmax and TR are test statistics frequently used in MCS frameworks. Tmax refers to the maximum absolute t-statistic of the loss differentials across all model pairs, thereby identifying whether any single model stands out as worse relative to the rest. TR instead focuses on the range (or span) of these t-statistics, effectively capturing whether the overall spread of model performance is large enough to conclude that one model is outperformed by others. In each bootstrap replication, these test statistics are recalculated, yielding an empirical distribution against which the observed statistics are tested.} are computed in each replication to ascertain whether any model significantly exceeds a certain loss threshold.

    \item \textbf{Iterative elimination and significance.}
    If a model repeatedly fails the test (i.e., it shows statistically higher average loss at level $\alpha=0.10$), it is expunged from the current set of models. The algorithm continues until all remaining models cannot be rejected as inferior, producing the final SSM.

    \item \textbf{Rankings and p-values.}
    For interpretability, the MCS returns a ranking of models based on their performance metrics (e.g., the average differential or a t-based ordering) and a global p-value indicating how strongly the data support the joint hypothesis that no surviving model is worse than the others. In practical terms, a model included in the SSM is considered not distinctly inferior to the top performers, thereby qualifying for subsequent combination in the nowcasting ensemble \parencite{SamuelsSekkel2017,AmendolaEtAl2020}.
\end{enumerate}

By structuring the procedure in these steps, we ensure that each model is subjected to a consistent, statistically rigorous evaluation. The block bootstrap is especially vital in macroeconomic contexts, where serial correlation in forecast errors is the rule rather than the exception.

The MCS focuses on comparative performance across multiple forecasts: it does not proclaim a single, absolute "best model," but rather identifies those models for which there is insufficient evidence of inferiority\footnote{"Insufficient evidence of inferiority" means that, within the bootstrap-based hypothesis testing framework, the loss differentials do not exhibit a consistent or statistically robust indication that a given model is outperformed by any other model in the set at $\alpha=0.10$. \parencite{HansenLundeNason2011}} relative to the rest of the group. This balanced perspective is aligned with common practices in forecast evaluation, especially in real-time macroeconomic settings characterized by model uncertainty.

Including only the surviving models from the final SSM in subsequent ensemble methods has several advantages:
\begin{itemize}
    \item \textit{Statistical reliability:} because the procedure controls for multiple comparisons in a bootstrap-based manner, the set of retained models is less likely to contain consistently poor performers \parencite{White2000,HansenLundeNason2011}.
    \item \textit{Efficiency and clarity:} dropping inferior models curtails the risk of "averaging down" when forecasts are combined, as a large pool of candidates can dilute improvements \parencite{BatesGranger1969}.
    \item \textit{Flexibility for real-time updating:} as new data become available, the MCS can be re-estimated, and the SSM can change accordingly if certain models degrade or others improve in predictive accuracy.
\end{itemize}

While the MCS does not claim that any included model is definitively superior in all circumstances, it ensures that none of the retained models shows statistically validated inferiority at the chosen level (i.e., $\alpha=0.10$). In other words, the methodology offers a robust filter, grounded in time-series bootstrap inference, before forming aggregated nowcasts. This systematic screening aligns with our broader principle of reducing model risk by leveraging iterative hypothesis testing and bootstrap-based diagnostics, and it is computationally more demanding than standard pairwise tests but furnishes a more reliable subset of forecast candidates. 

In the subsequent sections, we illustrate how these retained models (i.e., the final Superior Set of Models) are combined through various aggregation techniques, including unweighted and weighted averaging. Such a two-stage strategy---MCS selection followed by forecast aggregation---helps balance the competing goals of robust performance (through screening) and informational breadth (by blending multiple specifications).

\vspace{0.35cm}

\phantomsection
\subsection{\textbf{\textit{Combining Superior Models}}}
\label{sec:2.6.2}
The search for robust predictive accuracy often transcends the reliance on a single specification, even when the model under scrutiny emerges from rigorous selection procedures such as the Model Confidence Set (MCS). Indeed, the practice of combining forecasts traces back to seminal studies arguing that a suitable aggregation of predictions from multiple parsimonious models can outperform (in terms of bias or variance reduction) any single best contender\footnote{In the literature, this approach is frequently referred to as "forecast combination methods," "aggregation models," or "ensemble models." In the present work, we adopt the term "combination models" to avoid confusion with standard machine-learning "ensemble methods," such as random forests or gradient boosting, which we already employ as base learners in certain specifications.}\parencite{BatesGranger1969}. Recent refinements reinforce this intuition, suggesting that pooled forecasts mitigate specification risk in dynamic and potentially unstable contexts \parencite{Timmermann2006}.

In macroeconomic scenarios, structural shifts, regime changes, and incomplete information can undermine the stability of individual estimators. Aggregating the predictions of multiple "superior" models---each capturing distinct data features or leveraging different identification strategies---can help offset model-specific fragilities. Furthermore, exogenous shocks or evolutions in data collection processes may cause an abrupt deterioration in the performance of even well-tuned estimators. A pooled strategy, by design, spreads the risk of relying excessively on a single specification. Consequently, these combination models provide a pragmatic safeguard, especially where the sample size is modest, or the pattern of economic indicators is subject to frequent fluctuations \parencite{AiolfiTimmermann2006}.

In the present work, we focus on three principal techniques for combining the forecasts of the MCS-retained models:

\begin{itemize}
    \item \textit{Simple Average (SA)}: it assigns uniform weights to each model, thus offering a baseline benchmark. Although it overlooks differential predictive accuracy across models, it remains widely adopted for its transparency and resistance to abrupt overfitting.
    \item \textit{Weighted Average (WA)}: it constructs a set of weights that scale inversely with the historical error of each model. Hence, those estimators that consistently exhibit lower losses receive higher weights when generating new out-of-sample predictions.
    \item \textit{Exponential Weighted Average (EWA)}: it extends this idea by applying a dynamic exponential decay to the cumulative errors, thereby emphasizing more recent performances. In certain variants, the procedure itself is nested into a "meta-layer" (Meta-EWA), where several possible decay parameters (\(\eta\)) compete through an additional weighting mechanism. 
\end{itemize}

While differing in the explicit formulation of the weighting scheme, these three methods share a common architecture: they revisit and aggregate the predictions at each time step, recording residuals and partial metrics (e.g., MSFE, RMSFE, MAFE) in a rolling or iterative fashion. Such a parallel structure enables a coherent comparison of strengths, limitations, and overall forecast accuracy across the aggregation approaches.

The next subsections elaborate on each of the aforementioned combination strategies. For each, we proceed as follows:

\begin{enumerate}
    \item \textbf{Theoretical Rationale and General Functioning.} We illustrate the conceptual foundation of each "combination model," focusing on its objectives and the high-level mechanisms by which it aims to consolidate multiple base forecasts;
    \item \textbf{Combination and Weighting Scheme.} We describe the specific method by which the individual model outputs are aggregated, emphasizing how weights are assigned or updated over time in each approach;
    \item \textbf{Implementation in Our Empirical Setting.} We summarize how the combination procedure is integrated into our overall forecasting framework, indicating if it is executed in a purely offline fashion\footnote{We refer to a procedure as "offline" (batch) if it loads the entire dataset ex-post, simulating the forecasting process over all observations already available.
    
    In an "online" (or streaming) approach, the model continuously updates its parameters or weights as soon as new data points arrive without reprocessing all past observations.
    
    In this study, each combination method operates in a batch mode: the data for the entire sample (e.g., 2017 Q1--2023 Q2) is loaded at once, and the step-by-step evolution of weights is replayed retroactively. If new observations become available, the entire procedure must be rerun. Our choice aligns with typical macroeconomic data releases, which appear at discrete intervals rather than in a high-frequency streaming format.} or through a rolling re-estimation protocol. We also outline any relevant hyperparameters, where applicable.
\end{enumerate}

\par{} 
These approaches offer a nuanced perspective on balancing multiple models' relative merits and minimizing misspecification's impact on final nowcasts. By systematically updating (or recalibrating) the weighting structure, we aim to accommodate potential breaks in economic relationships while preserving the benefits of diversification. In the subsequent subsections, we detail each technique's design and present the respective empirical findings in a standardized format, thus ensuring a direct comparability of results. We also compare WA forecasts with both simpler and more advanced approaches in order to determine whether partial RMSE-based weighting yields measurable improvements in predictive stability under diverse economic conditions.

\vspace{0.35cm}

\phantomsection
\subsubsection{\textbf{\textit{Simple Average}}}
\vspace{0.2cm}
A Simple Average (SA) is the most straightforward strategy for aggregating multiple model forecasts. It assigns each base learner (model) the same weight, thereby avoiding discrimination based on historical performance or model complexity. The underlying rationale is to minimize the risk of over-reliance on a single specification by evenly distributing the contribution of each model in the final forecast \parencite{BatesGranger1969, Clemen1989, SmithWallis2009}. Consequently, the SA approach offers a transparent and stable benchmark, particularly valuable when the pool of candidate models exhibits comparable performance or when no clear ex-ante indication suggests favoring a specific learner. Additionally, in contexts characterized by abrupt regime shifts or data outliers, uniform weighting can provide a robust baseline that reduces the influence of sudden performance degradations from any one component model \parencite{MakridakisEtAl2020_M4}.

From a statistical standpoint, the Simple Average can reduce variance by blending forecasts from distinct estimation methods, data transformations, or structural assumptions \parencite{Armstrong2001, StockWatson2004, Timmermann2006}. This property is especially desirable in situations marked by parameter instabilities. Moreover, the mechanism does not require frequent re-estimation of weights. Once the set of base models has been determined, each forecast is given an identical portion of the overall "voting power." Although the method disregards potential differences in predictive accuracy, it often yields robust outcomes in empirical settings where economic or financial indicators are subject to substantial noise. In principle, it would be possible to introduce a dynamic aspect to the weights, but this would deviate from the uniform philosophy at the core of SA. In our script implementation, partial error metrics (such as MSFE or RMSFE) are tracked at each time step purely for diagnostic or comparative purposes, without feeding back into the weights. As such, any deterioration in the performance of a specific model does not prompt a reallocation of weights \parencite{HyndmanAthanasopoulos2021, ChenHood2020}.

Let the MCS procedure select a subset of $M$ forecasting models, indexed by $m \in \{1,2,\dots,M\}$. At each forecasting period $t$, each model $m$ produces a predicted value $f_{m,t}$. Under the Simple Average aggregator, the final combination $\widehat{y}_{t}$ is defined as a uniform mean of all base forecasts:
\[
\widehat{y}_{t} \;=\; \frac{1}{M}\sum_{m=1}^{M} f_{m,t}.
\]
Hence, all models are assigned the weight $1/M$, regardless of their historical performance or estimation strategy. When the size of the model set $M$ remains fixed over the evaluation window, the aggregation proceeds identically for all time steps \parencite{GenreEtAl2010, Zhemkov2021}.

Partial metrics, such as MSFE or RMSFE, are often computed at each time step to assess accuracy patterns or to compare with more complex weighting schemes. However, these statistics do not feed back into the weights for SA, since the method preserves the same assignment ($1/M$) throughout the sample. Thus, no dynamic mechanism adjusts any coefficient in response to recent forecast errors. This design choice allows a direct comparison with methods such as Weighted Average or Exponentially Weighted Average, which do update their weights as new data arrive \parencite{MakridakisHibon2000, Hansen2008, White2000}.

This weighting scheme offers two principal advantages:

\begin{itemize}
    \item it is computationally simple: once model outputs are obtained, the final prediction requires no further parameters or recursive updates;

    \item it provides a strong baseline against which more refined methods (e.g., Weighted Average or Exponentially Weighted Average) may be tested. Supposing a given model begins to deteriorate due to structural breaks or unforeseen economic shocks, its impact on the combined forecast remains limited, as its influence is capped at $1/M$ of the total prediction. However, an important caveat of SA is its insensitivity to differences in individual accuracy: should one model exhibit systematically higher predictive power, the Simple Average offers no mechanism to grant that model a proportionally larger weight.
\end{itemize}

In this study, the SA method is applied offline (batch mode) to the models retained by the preceding MCS-based selection. Specifically, once the MCS step determines which learners are statistically superior under the chosen loss function, each of these $M$ models contributes equally to generate the final aggregated nowcast. The procedure operates as follows:

\begin{enumerate}
    \item \textbf{Initialization.} Identify the subset of base models deemed superior by the MCS. No additional hyperparameters or performance tracking are required for the Simple Average method.
    \item \textbf{Uniform Aggregation.} At each time $t$, compute each model's forecast $f_{m,t}$; sum these predictions and divide by $M$ to obtain the SA forecast, $\widehat{y}_{t}$.
    \item \textbf{Batch Application.} Because the weighting remains fixed, there is no re-estimation of weights once the set of models is chosen. If new observations become available and the set of superior models is unchanged, the SA forecast is updated simply by averaging the newly generated model predictions.\footnote{Should the sample period be extended, all data points are reprocessed ex-post in a batch fashion. This reprocessing does not alter the uniform nature of the combination. Additional logging and error metrics (e.g., RMSFE, MAFE) are stored for robustness but do not modify any weighting.}
    \item \textbf{Final Nowcast.} Record $\widehat{y}_{t}$ for all $t$ in the evaluation sample to compare against realized outcomes. Although the average is constant in weighting, it can still incorporate underlying structural differences indirectly if the base models capture distinct aspects of the data. During this step, partial error metrics are computed at each $t$ to facilitate subsequent comparison with other aggregation methods. Period-specific analyses (including sub-period evaluations) are also carried out, mirroring the approach used for Weighted Average or Exponentially Weighted Average, without any effect on the fixed uniform weights.
\end{enumerate}

The Simple Average provides a robust and parsimonious benchmark by obviating the need for a dynamic adjustment mechanism. Its forecasts may, in certain instances, rival or surpass those derived from more elaborate procedures, especially in moderate samples or under conditions where no single approach maintains a lasting predictive edge. Later sections compare this uniform aggregator with adaptive schemes, investigating whether more sophisticated weighting improves accuracy and stability across diverse economic environments.

\vspace{0.35cm}

\phantomsection
\subsubsection{\textbf{\textit{Weighted Average}}}
\vspace{0.2cm}
The Weighted Average (WA) combination strategy adjusts each model's influence according to its observed performance over a historical window \parencite{Clemen1989,Timmermann2006}. In contrast to a simple average, WA grants proportionally significant weight to models displaying lower cumulative forecasting errors \parencite{BatesGranger1969,StockWatson2004}. It relies on the idea that models with consistently minor deviations from actual outcomes are more likely to capture the prevailing economic dynamics.

The core principle of WA is that the accuracy of each model, when measured and tracked over multiple periods, furnishes a credible basis for guiding future weights \parencite{Armstrong2001,Atiya2020}. In practice, one first defines a performance metric, such as the partial root mean square error (RMSE), which reflects how effectively a model has predicted the target variable from the initial period up to the current time. Models with lower partial RMSE are considered more reliable and receive higher weights.

Let there be \(M\) distinct models and \(d_{m,t}\) the in-sample forecast error (in our case, the squared difference between observed and predicted) for model \(m\) at time \(t\). A partial RMSE, spanning the interval \([1, \dots, t]\), can be inverted to yield a factor \(\alpha_{m,t}\) that increases with model quality\footnote{In this context, "increases with model quality" means that lower partial RMSE values correspond to higher \(\alpha_{m,t}\). Hence, a model exhibiting a smaller RMSE up to time \(t\) is deemed to have performed better historically, thus warranting a proportionally greater weight in the combination.}:

\[
\alpha_{m,t}
=
\frac{1}{\mathrm{RMSE}_{m,t}}
\quad
\text{where}
\quad
\mathrm{RMSE}_{m,t}
=
\sqrt{\frac{1}{t}
\sum_{\tau=1}^{t}
d_{m,\tau}}.
\]

If \(\mathrm{RMSE}_{m,t}\) is especially small, \(\alpha_{m,t}\) becomes correspondingly large, signifying robust past performance. After computing \(\alpha_{m,t}\) for each model, one applies a normalization step to ensure the sum of the weights to unity:

\[
w_{m,t}
=
\frac{\alpha_{m,t}}{\sum_{j=1}^{M}\alpha_{j,t}}.
\]

The coefficient \(w_{m,t}\) represents the proportion of the total ensemble weight allocated to model \(m\) at time \(t\).

Once these weights are established, the combined WA forecast for time \(t\) is formed as a convex linear blend of the base models:

\[
\widehat{y}_{t}
=
\sum_{m=1}^{M}
\,
w_{m,t}
\,
f_{m,t},
\]

\noindent{}where \(f_{m,t}\) is the prediction generated by model \(m\). Whenever new data becomes available, each model's partial RMSE is updated to include the latest forecast error, and a fresh set of weights is derived. This routine encourages models with deteriorating accuracy to lose weight, while those maintaining a strong alignment with the observed values see their weight increase.

The main advantage of WA lies in its direct, performance-driven rationale: the better a model has performed recently, the more heavily it is relied upon for the ensemble forecast. Should a model's forecasts begin to diverge from actual realizations, its partial RMSE grows, automatically reducing its relative weight. Conversely, a model whose predictions remain close to observed values achieves a smaller partial RMSE, thereby gaining a larger share of the overall combination \parencite{GenreEtAl2013}.

In this study, we apply WA to the base models retained by the Model Confidence Set (MCS) \parencite{HansenLundeNason2011}. The key steps are:

\begin{enumerate}
    \item \textbf{Performance Tracking}. At each quarterly time \(t\), we calculate the partial RMSE for each model, comparing its forecasts with realized outcomes from the initial period up to \(t\).
    \item \textbf{Inverse RMSE}. For each model, we invert its partial RMSE so that lower RMSE values translate into higher \(\alpha_{m,t}\).
    \item \textbf{Normalization}. We divide each model's \(\alpha_{m,t}\) by the sum of all \(\alpha_{j,t}\) values, ensuring the weights sum to one.
    \item \textbf{Aggregation}. The final WA forecast at time \(t\) is the weighted sum of the base models' predictions, with weights determined simultaneously.
\end{enumerate}

We perform these steps in an offline (batch) fashion \parencite{Tashman2000}, retrospectively simulating how the weights evolve each quarter. Each time new macroeconomic data become available, the entire sample is reprocessed, allowing the method to reflect changes in model performance over differing economic regimes.

The principal strength of WA lies in its simplicity and interpretability: the weights have a tangible meaning, namely that models performing better in the recent past receive proportionally higher emphasis. Moreover, the weights adjust automatically when a model's accuracy degrades, discouraging continued reliance on outdated performance.

Nonetheless, WA might fail to capture abrupt regime shifts if partial RMSE takes too long to incorporate recent, more severe errors. In highly volatile environments, a swifter reweighting mechanism may be necessary: supposing that multiple models exhibit nearly indistinguishable performance, their weights can converge to a near-uniform distribution, effectively approximating a simple average and thereby limiting the added value that a more selective weighting scheme would otherwise provide.

The Weighted Average (WA) combination adopted in this study adheres to these guidelines:

\begin{itemize}
    \item \textit{Initialization:} each model's partial RMSE is set to zero, and uniform weights are assigned for the first nowcast.
    \item \textit{RMSE Update:} at each quarterly time \(t\), compute the new forecast error for each model and incorporate it into the partial RMSE, thereby accumulating performance information.
    \item \textit{Weighting and Normalization:} invert the partial RMSE to obtain a measure of each model's predictive accuracy, then normalize across all models so that the weights sum to unity.
    \item \textit{Final Aggregation:} produce the combined nowcast as a weighted sum of the MCS-selected base models, with weights reflecting their respective performance records.
\end{itemize}

This procedure allows a quarter-by-quarter reconstruction of how partial RMSE and weights would have evolved in real-time. Whenever new macroeconomic data become available, the entire procedure is repeated so that the WA weighting mechanism remains responsive to recent patterns of model accuracy.

\vspace{0.35cm}

\phantomsection
\subsubsection{\textbf{\textit{Exponentially Weighted Average}}}
\vspace{0.2cm}
The Exponentially Weighted Average (EWA) constitutes a flexible mechanism for combining multiple forecasting models, each distinguished by its own pattern of predictive errors \parencite{LittlestoneWarmuth1994,DevaineEtAl2013}. At a conceptual level, EWA assigns greater weight to those models that have demonstrated superior accuracy in the most recent past. Formally, at every time step, the method incorporates an exponential decay factor into each model's cumulative loss, progressively attenuating the impact of older forecast errors and allowing the ensemble to adjust promptly when new information arrives.

Macroeconomic series are prone to regime changes and structural breaks that can weaken the predictive capabilities of a single forecast combination with static weights \parencite{ClarkMcCracken2010,TianAnderson2014}. The EWA algorithm addresses this limitation by weighting models according to their latest performance, penalizing recent inaccuracies more heavily than older ones \parencite{Timmermann2006}. It is grounded in the notion that, under evolving economic dynamics, an estimator's most reliable gauge of current accuracy is its error profile in the latest observations \parencite{Rossi2021}. 

From a formal perspective, let \(m\in\{1,\dots,M\}\) index the base models selected via the Model Confidence Set (MCS). At time \(t\), each model \(m\) accumulates a loss \(\mathcal{L}_{m,t}\), which could be measured as a sum of squared forecast errors up to period \(t\). An exponential weighting function then transforms these cumulative losses into a set of normalized coefficients:

\[
w_{m,t}
\,=\,
\frac{\exp\!\bigl[-\,\eta\,\mathcal{L}_{m,t}\bigr]}{\sum_{j=1}^{M} \exp\!\bigl[-\,\eta\,\mathcal{L}_{j,t}\bigr]},
\]

\noindent{}where \(\eta>0\) denotes a \textit{learning rate} (or decay parameter). It determines the speed at which the EWA scheme adapts to new forecast errors: larger \(\eta\) values amplify the penalty for recent mistakes, enabling quick revisions of weights; smaller values yield a more gradual adjustment. The coefficient \(w_{m,t}\) represents the fraction of the total weight allocated to model \(m\) at time \(t\). In all cases, these weights remain nonnegative (\(0 \,\le w_{m,t}\le 1\)) and sum to unity (\(\sum_{m=1}^M w_{m,t} = 1\)), thus yielding a valid convex combination. Because \(w_{m,t}\) depends inversely on \(\mathcal{L}_{m,t}\) through the exponential term, models that exhibit smaller cumulative losses (i.e., better historical performance) receive higher weights and those with larger losses see their weights diminished.

To generate a forecast at time \(t\), EWA calculates:

\[
\widehat{y}_{t}
\,=\,
\sum_{m=1}^{M}
\,w_{m,t}\,f_{m,t},
\]

\noindent{}where \(f_{m,t}\) is the one-step-ahead forecast from model \(m\). Once the actual \(y_t\) is observed, each model's loss (e.g., squared error, \((y_t - f_{m,t})^2\)) is updated accordingly, and the new cumulative losses feedback into the weighting scheme. EWA proves particularly valuable in contexts where distinct models exhibit comparative advantages in different macroeconomic regimes. An approach that excels during expansions may temporarily lose favor if it underperforms during downturns, allowing more robust models in recessive phases to gain increased weight in the ensemble \parencite{RafteryEtAl2010,deRooijEtAl2014}. Hence, EWA systematically adjusts the relative influence of each model as the economic environment evolves without relying on prespecified thresholds or manual interventions.

In our analysis, we employ EWA to merge the forecasts of those models (also referred to as "experts" \parencite{CesaBianchiLugosi2006} retained by the MCS procedure. At each quarterly step, we:

\begin{enumerate}
    \item \textit{Compute the per-model loss} based on squared deviations from the observed target;
    \item \textit{Accumulate the losses} to form \(\mathcal{L}_{m,t}\) for each model \(m\);
    \item \textit{Derive exponential weights} by applying \(\exp\bigl[-\eta\,\mathcal{L}_{m,t}\bigr]\) to each cumulative loss;
    \item \textit{Normalize} these weights so that \(\sum_{m=1}^{M} w_{m,t}=1\);
    \item \textit{Construct the nowcast} as the weighted sum of the base forecasts.
\end{enumerate}

While EWA can naturally operate in an online context---updating weights at each incoming observation---we implement it offline (batch mode) for consistency with our quarterly data releases \parencite{Tashman2000}. Specifically, in each experiment, we retroactively "simulate" the time evolution of weights by sequentially introducing historical observations and recomputing the ensemble's forecast at each step.

We further enhance EWA by including a "meta-layer\footnote{We refer to it as "Meta-EWA" because, after computing separate EWA forecasts for multiple values of \(\eta\), we treat these parallel EWA aggregators themselves as "meta-experts" in a second-tier exponential scheme. This meta-layer automatically learns which decay parameter performs better at each stage, attenuating the risk of using a single \(\eta\) poorly suited to the entire sample.}" that accounts for multiple values of \(\eta\). Each distinct learning rate \(\eta_j\) is treated as defining a separate EWA aggregator, wherein the base experts' weights are updated according to \(\eta_j\). Concretely, for each \(\eta_j\in\{\eta_1,\eta_2,\dots,\eta_K\}\):

\begin{itemize}
    \item We maintain a running cumulative loss for every base expert (as above) but apply \(\eta_j\) to obtain a specialized weighted forecast at each \(t\).
    \item This procedure yields \(K\) parallel EWA-based predictions, each reflecting a different speed of adaptation.
\end{itemize}

Subsequently, we interpret these \(K\) aggregated forecasts as "meta-experts," each with its own incurred error (i.e., the difference between its aggregated forecast and the realized outcome). A second-tier exponential weighting framework then assigns mass across these \(\eta\)-based aggregators, using an analogous rule:

\[
\omega_{j,t}
\,=\,
\frac{\exp\!\bigl[-\,\lambda\,\widetilde{\mathcal{L}}_{j,t}\bigr]}{\sum_{h=1}^{K}\exp\!\bigl[-\,\lambda\,\widetilde{\mathcal{L}}_{h,t}\bigr]},
\]

\noindent{}where \(\widetilde{\mathcal{L}}_{j,t}\) denotes the cumulative forecast loss for aggregator \(j\) up to time \(t\), and \(\lambda\) is an additional learning rate controlling how aggressively we switch among different \(\eta_j\)-forecasters. By combining these \(\eta\)-aggregators via a second-level EWA, the "Meta-EWA" mechanism mitigates the risk of misspecification that arises from fixing a single decay parameter \(\eta\). This layered scheme automatically identifies the \(\eta_j\) that best adapts to the prevailing data pattern, assigning a proportionally higher weight to its associated aggregator \parencite{Wolpert1992,Breiman1996,MonteroMansoEtAl2020}.

Regarding uniform weighting, EWA provides a more dynamic mechanism that promptly favors models with lower recent errors. Consequently, EWA's adaptability may result in superior out-of-sample performance in volatile or rapidly shifting environments \parencite{BallingsEtAl2019}. Because of these qualities, EWA remains widely adopted whenever the real-time responsiveness of weights is deemed essential. Its recursive structure ensures the ensemble accommodates new information without discarding the historical record entirely. As a result, it aligns well with macroeconomic nowcasting contexts, where the timely integration of evolving economic indicators is of primary concern.

The EWA combination in this study adheres to the following protocol:

\begin{itemize}
    \item \textit{Initialization}: set all models' losses to zero and assign uniform weights.
    \item \textit{Cumulative update}: at time \(t\), compute each model's new squared error and add it to the respective loss term.
    \item \textit{Weighting and normalization}: convert the updated losses into exponential weights, then normalize.
    \item \textit{Final aggregation}: generate the integrated forecast as a weighted sum of the MCS-selected base predictions.
    \item \textit{Meta-EWA extension}: in some iterations, allow multiple \(\eta\)-based EWA forecasts to compete and apply an additional exponential weighting to choose among them adaptively.
\end{itemize}

\par{} 
These steps are repeated over the entire sample in an offline simulation, enabling a quarter-by-quarter reconstruction of how weights and forecasts would have evolved in real-time. Whenever new macroeconomic data become available, we rerun this offline procedure so that the EWA weights remain anchored to recent error patterns. In subsequent sections, we compare EWA outcomes with those from simpler weighting schemes, thereby assessing whether exponential adaptation enhances predictive accuracy under various macroeconomic conditions.

\vspace{0.35cm}

\phantomsection
\subsection{\textbf{\textit{Dynamic Weighting Mechanisms and Explainability in Nowcast Combination Models}}}
\label{sec:2.6.3}
Weighted Average (WA) and Exponential Weighted Average (EWA) models offer robust frameworks for combining forecasts, capturing the dynamic contributions of individual models over time. These aggregation techniques are especially beneficial in time-series forecasting, where understanding each model's evolving relative importance enhances the accuracy and interpretability of predictions. By focusing on their shared architecture, we can highlight the conceptual transparency of these methods, which allows for a clear explanation of how each nowcasting model's relative importance changes as economic conditions evolve.

Both WA and EWA follow a similar process for weight assignment, which can be broadly outlined in the following steps \parencite{AiolfiTimmermann2006, GenreEtAl2013, Hansen2008}:

\begin{enumerate}
    \item \textit{Update:} at each time step $t$, the procedure updates the forecast errors of the individual base models, quantifying how closely each model approximates the observed outcome. This step is the primary input to the subsequent weight adjustment \parencite{DieboldMariano1995}.
    \item \textit{Normalize:} the performance indicators (such as root mean square errors, cumulative losses, or exponential losses) are then mapped to a common scale, ensuring that the resultant weights fall within the $[0,1]$ interval and sum to unity. This normalization ensures that the weights are comparable across models, regardless of their scale or magnitude \parencite{RadchenkoEtAl2023}.
    \item \textit{Reweight:} the normalized weights are then used to combine the forecasts of all base models. This reweighting reflects how each model's recent performance influences its share in the final combined prediction, forming a dynamically updated forecast \parencite{SamuelsSekkel2017}.
\end{enumerate}

In this shared framework, Bates and Granger \parencite{BatesGranger1969} originally proposed calculating each model's relative importance by inverting its partial RMSE (Weighted Average, WA). A later work introduced exponential decay factors to reduce the influence of older forecast errors (Exponentially Weighted Average, EWA) \parencite{DevaineEtAl2013, LittlestoneWarmuth1994}. Despite these mathematical differences, the interpretability of both methods is rooted in the same principle: if a model performs poorly over a specific period, its weight diminishes, and vice versa.

To explicitly explain how the weights are calculated and evolve at each time step \( t \), we break down the process as follows:

1. \textit{Forecast Errors Calculation (Update):} the forecast errors for each model \( m \) are computed based on the difference between the model's prediction \( f_{m,t} \) and the observed target \( y_t \). The forecast error is given by:

\[
d_{m,t} = |f_{m,t} - y_t|
\]

   For WA, the error is simply the root mean square error (RMSE) calculated over a rolling window of observations. For EWA, the errors are accumulated into a cumulative loss function, which is then exponentially decayed as follows:

\[
\mathcal{L}_{m,t} = \sum_{\tau=1}^{t} \exp\left(-\eta(t-\tau)\right) \cdot d_{m,\tau}
\]

2. \textit{Normalization of Performance Indicators (Normalize):} the forecast errors are then normalized to ensure the final weights are on a comparable scale. This step is critical for WA and EWA, ensuring that all weights are positive and sum to 1. For WA, this is done by computing the inverse of the RMSE for each model:

\[
w_{m,t}^{WA} = \frac{1 / \text{RMSE}_{m,t}}{\sum_{j=1}^{M} 1 / \text{RMSE}_{j,t}}
\]

   For EWA, the cumulative losses are normalized using the exponential loss formula:

\[
w_{m,t}^{EWA} = \frac{\exp\left(-\eta \mathcal{L}_{m,t}\right)}{\sum_{j=1}^{M} \exp\left(-\eta \mathcal{L}_{j,t}\right)}
\]

3. \textit{Reweighting Forecasts (Reweight):} once the normalized weights are calculated, they are applied to the individual forecasts of each model. This reweighting reflects the relative performance of each model, with better-performing models being assigned a higher weight. The combined forecast for the time step \( t \) is then given by:

\[
\widehat{y}_{t} = \sum_{m=1}^{M} w_{m,t} f_{m,t}
\]

This methodology ensures that the weights are updated in real-time (or on a rolling basis in a batch setting), reflecting the dynamic contribution of each model's forecast based on its recent accuracy.

Our analysis used "stacked area plots" to visually represent how model weights evolve and WA and EWA's interpretability. These plots allow a clear visual interpretation of the evolution of the weights assigned to each model at every time step as new data becomes available:

\begin{itemize}
    \item each area in the plot represents the weight evolution of a single model in the total nowcast;
    \item as models gain or lose influence based on their performance, the size of their respective areas adjusts accordingly.
\end{itemize}

Using this visualization technique, we can track how each model adapts to changes in the economic environment. For example, a sudden increase in the weight of a specific model may suggest that it is better suited to the current economic regime. In contrast, a decrease in its weight may indicate underperformance. These visual tools enhance the interpretability of both WA and EWA models, allowing users to gain deeper insights into the underlying decision-making process.

Although both WA and EWA share the core logic of \textit{Update $\rightarrow$ Normalize $\rightarrow$ Reweight}, they differ in how they penalize recent errors. WA relies on an inversion of errors, where the model's weight is inversely proportional to its RMSE \parencite{GenreEtAl2013}. In contrast, EWA applies an exponential decay factor to cumulative forecast errors, gradually reducing the influence of older data \parencite{DevaineEtAl2013}. This difference makes EWA more sensitive to abrupt changes in model accuracy, allowing it to respond more quickly to deteriorations in predictive performance. On the other hand, WA can appear more stable, requiring a more extended period of consistent underperformance before a significant weight adjustment occurs \parencite{Atiya2020}.

In contrast to WA and EWA, the Simple Average (SA) aggregator assigns an equal weight to each model and does not modify these weights over time. As a result, SA lacks the dynamic dimension that enables the interpretability of WA and EWA. While SA is transparent in its equal allocation of weights across models, it does not provide insight into which models perform better at any given stage since the weights remain fixed \parencite{Clemen1989, StockWatson2004}. Therefore, while SA promotes diversity by giving equal importance to all models, it does not allow for the interpretation of how individual model performance affects the aggregated forecast.

In our framework, WA and EWA models provide a comprehensible method for aggregation model interpretability by revealing a time-dependent set of weights that indicate which models are deemed most reliable at each forecasting period \parencite{HansenLundeNason2011, SamuelsSekkel2017}. These methods enable an understanding of how the combined forecast is formed and offer a quantitative measure of the evolving trust placed in each model. These features foster transparency that is particularly valuable in economic and financial contexts, where understanding the sources of predictive power is as important as achieving high levels of forecast accuracy \parencite{MakridakisHibon2000, StockWatson2011, WangEtAl2023}.

\vspace{0.5cm}

\section{\textbf{Diagnostic Tests and Predictive Ability Tests}}
\label{sec:2.7}

\phantomsection
\subsection{\textbf{\textit{Diagnostic Tests for Combined Models}}}
\label{sec:2.7.1}
The robustness of macroeconomic forecasting models, particularly those based on combined approaches such as Simple Average (SA), Weighted Average (WA), and Exponentially Weighted Average (EWA)\parencite{Clemen1989,GenreEtAl2013,WangEtAl2023}, necessitates thorough diagnostic checks. These tests ensure that the models' residuals do not exhibit systematic patterns that would undermine the validity of the forecasts. In this section, we focus on two widely employed diagnostic tests: the Shapiro-Wilk test for normality\parencite{ShapiroWilk1965} and the Ljung-Box test for autocorrelation in the residuals\parencite{LjungBox1978}. Both tests were implemented in the context of the model combination strategies discussed earlier, offering a systematic evaluation of their predictive performance.

\phantomsection
\subsubsection{\textbf{\textit{The Shapiro-Wilk Test for Normality}}}
The Shapiro-Wilk test\parencite{ShapiroWilk1965,BarrowKourentzes2016} is a statistical method used to assess whether a sample of data comes from a normally distributed population. This test is particularly relevant in time-series forecasting, where the assumption of normally distributed residuals often underpins various diagnostic checks and the construction of confidence intervals. In the context of combined models, the Shapiro-Wilk test examines the normality of residuals derived from the aggregate forecasts of the SA, WA, and EWA methods.

The Shapiro-Wilk test evaluates the null hypothesis that the model's residuals are normally distributed. It is based on the calculation of a test statistic \( W \), which compares the observed distribution of the residuals with a theoretical normal distribution. A value of \( W \) close to 1 indicates that the residuals do not significantly deviate from normality, while a value considerably less than 1 suggests non-normality.

In our implementation, the residuals from each combination model—SA, WA, and EWA—are extracted after the predictions are generated. The Shapiro-Wilk test is then applied to these residuals, assessing whether they follow a normal distribution. This check is crucial because many forecasting models, particularly those relying on statistical techniques such as ordinary least squares or maximum likelihood estimation, assume normality in the residuals for valid inference.

\[
W = \frac{\left( \sum_{i=1}^n a_i x_i \right)^2}{\sum_{i=1}^n (x_i - \bar{x})^2}
\]

\noindent{}where:

\begin{itemize}
    \item \( x_i \) are the ordered sample values of the residuals;
    \item \(\bar{x} \) is the sample mean;
    \item \( a_i \) are constants derived from the expected values of the order statistics of a normal distribution.
\end{itemize}

In time-series forecasting, especially in our context of nowcasting (from 2017 Q1 to 2023 Q2), the normality of residuals is often a desirable but not always attainable property. The presence of economic shocks (e.g., the COVID-19 pandemic)\parencite{CarrieroEtAl2022} can lead to distributions with heavy tails or skewness, which deviates from normality. Despite these challenges, the Shapiro-Wilk test remains useful for preliminary diagnostics. Although it assumes independence of residuals (an assumption that is often violated in time-series data, where residuals can exhibit autocorrelation), its application can still highlight significant deviations from normality.

In this study, we justify using the Shapiro-Wilk test as a first diagnostic tool for evaluating residuals, even though it does not account for serial correlation. However, the test's application is more reliable, considering that the data used in this study were iteratively standardized to avoid look-ahead bias and that stationarity issues were fully addressed. Despite the economic shocks during the period, this preparation ensures the test is a valid preliminary check for normality in the residuals. It provides a simple and effective way to detect non-normality, which could suggest underlying issues such as model misspecification or inadequate handling of non-linearity in the data.

\subsubsection{\textbf{\textit{The Ljung-Box Test for Autocorrelation}}}
The Ljung-Box test\parencite{LjungBox1978} is a diagnostic tool used to detect autocorrelation in the residuals of a time series model. Autocorrelation in residuals suggests that the model has failed to capture some of the time series structure, which could lead to biased or inconsistent forecasts. This test is crucial when dealing with time series data, where residual autocorrelation may indicate model misspecification or an inadequate model structure.

The Ljung-Box test evaluates the null hypothesis that no autocorrelation exists at any lagged values up to a specified maximum lag \( q \). The test statistic \( Q \) is computed as:

\[
Q = n(n + 2) \sum_{k=1}^q \frac{\hat{\rho}_k^2}{n-k}
\]

\noindent{} where:

\begin{itemize}
    \item \( \hat{\rho}_k \) is the sample autocorrelation at lag \( k \);
    \item \( n \) is the number of residuals.
\end{itemize}

Under the null hypothesis of no autocorrelation, \( Q \) follows a chi-squared distribution with \( q \) degrees of freedom. A significant value of \( Q \) (i.e., larger than the critical value from the chi-squared distribution) indicates the presence of autocorrelation in the residuals, suggesting that the model may not have fully accounted for the time-series structure.

We applied the Ljung-Box test to the residuals of the combined forecasts produced by the SA, WA, and EWA models. The presence of autocorrelation in these residuals would indicate that the aggregation process might not be optimal and that further refinements in model selection or combination may be necessary.

For each combination model—SA, WA, and EWA—the residuals are first calculated as the difference between the actual observations and the model forecasts. These residuals are then subjected to the Shapiro-Wilk test for normality and the Ljung-Box test for autocorrelation. The implementation steps are as follows:

\begin{enumerate}
    \item \textit{Residual Calculation:} for each combination model, compute the residuals by subtracting the model's forecast from the actual observed values;
    \item \textit{Normality Test (Shapiro-Wilk):} apply the Shapiro-Wilk test to the residuals to check for normality. A non-significant result suggests that the residuals are normally distributed;
    \item \textit{Autocorrelation Test (Ljung-Box):} perform the Ljung-Box test on the residuals to detect any autocorrelation at different lags. A significant result indicates that the residuals exhibit autocorrelation and that model improvements are required.
\end{enumerate}

The results of these diagnostic tests provide valuable insights into the reliability of the combined models' forecasts. In theory, a failure to meet the assumptions of normality or autocorrelation in the residuals would suggest potential issues in the model structure, necessitating further refinement or reconsidering of the combination approach\parencite{Tashman2000}.

\par{}
Applying the Shapiro-Wilk and Ljung-Box tests to the residuals of the combined models (SA, WA, EWA) ensures that the residuals are approximately normal and uncorrelated—conditions essential for the validity of many econometric techniques used in forecasting. The outcomes of these tests allow for the identification of potential issues within the aggregated models, guiding subsequent refinements, and enhancing the reliability and robustness of the forecasts. This process is crucial for providing accurate and dependable insights and informed macroeconomic decision-making.

\vspace{0.35cm}

\phantomsection
\subsection{\textbf{\textit{Predictive Ability Tests for Model's Comparison}}}
\label{sec:2.7.2}
The evaluation of predictive performance is essential in macroeconomic forecasting, where consistently assessing models' relative accuracy and ability to outperform benchmark models is critical. This section discusses the application of the "Giacomini-White test"\parencite{GiacominiWhite2006} to compare the predictive ability of the best individual models selected through the Model Confidence Set, and the combined models in comparison with our three reference models (Random Walk, AR(3), and Dynamic Factor Model). We aim to evaluate whether our individual and combined models (SA, WA, EWA) significantly improve predictive accuracy over these selected benchmarks.

The Giacomini-White test is a statistical method used to compare the predictive performance of a candidate model against a benchmark, testing whether there are significant differences in forecasting accuracy. Building on earlier work on forecast accuracy tests such as the Diebold-Mariano test\parencite{DieboldMariano1995} and the Reality Check\parencite{White2000}, it focuses on conditional predictive ability rather than unconditional comparisons\parencite{InoueRossi2012}. 

The null hypothesis of the Giacomini-White test posits that there is no systematic difference in predictive ability between the candidate and reference models. The alternative hypothesis suggests that at least one model outperforms the other statistically. To implement the test, we compute the difference in forecast errors between the candidate and benchmark models and then regress these differences using ordinary least squares (OLS)\parencite{Hansen2005}. 

A key component of this test is using a robust covariance estimator to account for heteroskedasticity and autocorrelation in the residuals. Specifically, we employ the "Lumley-Heagerty covariance matrix" (also called WEAVE, Weighted Empirical Adaptive Variance Estimation)\footnote{The Lumley-Heagerty covariance matrix is a robust "sandwich estimator" designed to adjust the standard errors from an OLS regression for both heteroskedasticity and autocorrelation. It achieves that by applying adaptive weights---often derived through isotonic regression to enforce a monotonic decay---to the autocovariance of the residuals. These weights are then used to construct a more reliable covariance matrix estimator, thereby improving inference accuracy when model assumptions regarding constant variance and uncorrelated errors are violated. For the sake of clarity, a "sandwich estimator" is a succinct term for a robust variance estimator that typically appears in the product form bread $\times$ meat $\times$ bread, where "bread" represents the inverse Hessian of the regression model, and "meat" captures the empirical covariance of the residuals, thus accounting for possible misspecifications and error correlations.}\parencite{LumleyHeagerty1999}, which adjusts for autocorrelation and heteroskedasticity, ensuring reliable statistical inference.

The Giacomini-White test performed in this study follows these steps:

\begin{enumerate}
    \item \textbf{Loss Differential Calculation:} for each best individual and combined model, we first calculate the loss differential, which is the difference between the squared errors of the model and the reference model:
    \[
    d_t = \text{Error of model} - \text{Error of benchmark}
    \]
    where the errors are typically measured as squared forecast errors (e.g., $\epsilon^2_t = (y_t - \hat{y}_t)^2$).
    
    \item \textbf{Regress Loss Differentials:} the loss differential is then regressed using OLS, where the dependent variable is the loss differential $d_t$, and the independent variable is typically an intercept (although optional instruments can be used for additional refinements).
    
    \item \textbf{Application of Lumley-Heagerty Covariance:} the covariance of the OLS residuals is adjusted using the Lumley-Heagerty covariance matrix. This adjustment accounts for potential autocorrelation and heteroskedasticity in the forecast errors, providing a more accurate covariance estimate.
    
    \item \textbf{Wald Test:} the Wald test is conducted on the regression coefficients (including the intercept). A rejection of the null hypothesis (that all coefficients are zero) suggests a significant difference in predictive performance between the model and the benchmark.
    
    \item \textbf{Significance Evaluation:} the results of the Wald test are used to determine whether the difference in predictive accuracy between the model and the reference model is statistically significant. A low p-value (typically below 0.05) indicates that the model significantly outperforms the reference model.
\end{enumerate}

The results of the Giacomini-White test for each best individual and aggregated model are summarized by the "Wald statistic" and the associated p-value\footnote{The R-squared values are not reported, as the primary focus of the Giacomini-White test is on the Wald test and the significance of the intercept term. Since only the intercept is used in the regression model to capture the mean difference in forecast errors, the R-squared is expected to be low and is not critical for interpreting the test results.}. These results provide valuable insights into the relative performance of the models:

\begin{itemize}
    \item \textit{Wald Statistic:} a large Wald statistic indicates a strong rejection of the null hypothesis, suggesting a significant difference in predictive performance between the model and the benchmark.
    \item \textit{p-Value:} the p-value associated with the Wald test indicates the statistical significance of the result. A p-value below 0.05 suggests that the model provides a significantly better prediction than the reference model.
    \item \textit{Intercept Estimate:} the intercept estimate in the regression represents the average difference in forecast errors between the model and the reference model. A negative intercept suggests the model consistently provides lower forecast errors than the benchmark, while a positive intercept suggests the opposite.
\end{itemize}

The outcomes of these tests are used to determine whether the best individual models and the combination models significantly outperform the reference models in terms of average predictive accuracy. A failure to reject the null might also indicate forecast breakdowns or model instability if combined with further analysis\parencite{GiacominiRossi2009}. However, negative, significant intercepts in the Wald test results here indicate that the models generally outperform the benchmarks in forecasting accuracy, offering robust evidence of their predictive superiority.

In summary, the Giacomini-White test for model comparison, applied in conjunction with the Lumley-Heagerty covariance estimator, provides a robust framework for evaluating the predictive ability of different forecasting models. By comparing the selected best individual models and the combined models against benchmark models, we can confidently assess whether these models offer statistically significant improvements in forecasting accuracy. Using the Wald test and the Lumley-Heagerty covariance ensures that our results are robust to heteroskedasticity and autocorrelation, which are common in time-series data.

\cleardoublepage 

%% file: 03_chapters/chapter3.tex
\chapter{Empirical Results and Discussion}
\label{ch:empiricalResultsDiscussion}
[opening remarks]

\section{\textbf{Empirical Results}}
\label{sec:3.1}

\subsection{\textbf{\textit{Predictive Accuracy of Individual Models}}}
\label{subsec:accuracymodels}
In this subsection, we present the results related to the nowcast accuracy of individual models, evaluating both the Mean Square Forecast Error (MSFE), Root Mean Square Forecast Error (RMSFE), and Mean Absolute Forecast Error (MAFE) metrics in absolute value (Table~\ref{tab:absolutePerformance}, first part) and the values of RMSFE normalized to the three benchmarks considered (Random Walk, AR(3), Dynamic Factor Model), as illustrated in the second table (Table~\ref{tab:relativePerformance}) in correspondence with different macroeconomic regimes. 

The main objective is to highlight how each approach manages to contain the nowcast error (MSFE, RMSFE, MAFE) in different macroeconomic contexts and quantify the relative benefits (relative gain) compared to the benchmarks. Below, we provide a summary, isolating the most relevant trends in the overall nowcast period (Overall) and the sub-periods (Pre-COVID, COVID, Post-COVID, Excluding COVID).

The mechanisms underlying these differences and the methodological and operational implications will be explored in section \ref{sec:3.2}.

\vspace{0.35cm}

\phantomsection
\subsubsection{\textbf{\textit{Absolute results}}}
\vspace{0.2cm}
The first table (\textit{Forecast accuracy metrics across models and sub-periods}) reports the main error measures (MSFE, RMSFE, and MAFE) for each family of models. Although multiple metrics are provided, we focus on the RMSFE as it heavily penalizes large deviations and proves particularly relevant in highly volatile phases, typically observed during macroeconomic crises or turbulent periods. Below, we summarize each model family's behavior across the entire prediction horizon (Overall) and the four sub-periods (Pre-COVID, COVID, Post-COVID, and Excluding COVID), maintaining a qualitative perspective consistent with Table~\ref{tab:absolutePerformance}.

{
\fontsize{9pt}{10pt}\selectfont
\begin{table}[htbp]
    \centering
    \caption{Forecasting performance metrics (MSFE, RMSFE, and MAFE) of models across sub-periods}
    \label{tab:absolutePerformance}

    \resizebox{0.67\textwidth}{!}{%
    \begin{tabular}{lrrrrr}
    
    \multicolumn{6}{l}{\textbf{Penalized linear models}} \\
    \multicolumn{6}{c}{\textbf{LASSO}} \\
    \midrule
    \textbf{Metric} & \textbf{Overall} & \textbf{Pre-COVID} & \textbf{COVID} & \textbf{Post-COVID} & \textbf{Excluding COVID} \\
    \cmidrule(lr){2-6}
    MSFE  & 4.973 & 0.395 & 26.038 & 2.042 & 1.144 \\
    RMSFE & 2.230 & 0.628 & 5.103 & 1.492 & 1.069 \\
    MAFE  & 1.164 & 0.538 & 3.194 & 1.104 & 0.795 \\
    \addlinespace

    \multicolumn{6}{c}{\textbf{Ridge}} \\
    \midrule
    \textbf{Metric} & \textbf{Overall} & \textbf{Pre-COVID} & \textbf{COVID} & \textbf{Post-COVID} & \textbf{Excluding COVID} \\
    \cmidrule(lr){2-6}
    MSFE  & 2.091 & 0.654 & 3.518 & 3.245 & 1.831 \\
    RMSFE & 1.446 & 0.808 & 1.876 & 1.801 & 1.353 \\
    MAFE  & 0.978 & 0.635 & 1.184 & 1.306 & 0.940 \\
    \addlinespace

    \multicolumn{6}{c}{\textbf{Elastic net}} \\
    \midrule
    \textbf{Metric} & \textbf{Overall} & \textbf{Pre-COVID} & \textbf{COVID} & \textbf{Post-COVID} & \textbf{Excluding COVID} \\
    \cmidrule(lr){2-6}
    MSFE  & 1.669 & 0.599 & 3.582 & 2.189 & 1.322 \\
    RMSFE & 1.292 & 0.774 & 1.893 & 1.480 & 1.150 \\
    MAFE  & 0.936 & 0.598 & 1.276 & 1.206 & 0.875 \\
    \addlinespace
    \addlinespace[1em]  %
    
    \multicolumn{6}{l}{\textbf{Dimensionality reduction-based models}} \\
    \multicolumn{6}{c}{\textbf{Principal component regression}} \\
    \midrule
    \textbf{Metric} & \textbf{Overall} & \textbf{Pre-COVID} & \textbf{COVID} & \textbf{Post-COVID} & \textbf{Excluding COVID} \\
    \cmidrule(lr){2-6}
    MSFE  & 2.146 & 0.823 & 1.002 & 4.192 & 2.355 \\
    RMSFE & 1.465 & 0.907 & 1.001 & 2.047 & 1.534 \\
    MAFE  & 1.164 & 0.825 & 0.902 & 1.675 & 1.211 \\
    \addlinespace

    \multicolumn{6}{c}{\textbf{Partial least squares regression}} \\
    \midrule
    \textbf{Metric} & \textbf{Overall} & \textbf{Pre-COVID} & \textbf{COVID} & \textbf{Post-COVID} & \textbf{Excluding COVID} \\
    \cmidrule(lr){2-6}
    MSFE  & 2.152 & 0.579 & 5.274 & 2.791 & 1.584 \\
    RMSFE & 1.467 & 0.761 & 2.297 & 1.671 & 1.259 \\
    MAFE  & 1.038 & 0.653 & 1.433 & 1.342 & 0.966 \\
    \addlinespace

    \addlinespace[1em]  %

    \multicolumn{6}{l}{\textbf{Ensemble learning}} \\
    \multicolumn{6}{c}{\textbf{Random forest}} \\
    \midrule
    \textbf{Metric} & \textbf{Overall} & \textbf{Pre-COVID} & \textbf{COVID} & \textbf{Post-COVID} & \textbf{Excluding COVID} \\
    \cmidrule(lr){2-6}
    MSFE  & 7.048 & 0.954 & 33.962 & 3.596 & 2.155 \\
    RMSFE & 2.655 & 0.977 & 5.828 & 1.896 & 1.468 \\
    MAFE  & 1.750 & 0.841 & 4.991 & 1.545 & 1.161 \\
    \addlinespace

    \multicolumn{6}{c}{\textbf{eXtreme Gradient Boosting}} \\
    \midrule
    \textbf{Metric} & \textbf{Overall} & \textbf{Pre-COVID} & \textbf{COVID} & \textbf{Post-COVID} & \textbf{Excluding COVID} \\
    \cmidrule(lr){2-6}
    MSFE  & 6.655 & 0.739 & 35.912 & 2.051 & 1.335 \\
    RMSFE & 2.580 & 0.860 & 5.993 & 1.432 & 1.156 \\
    MAFE  & 1.673 & 0.787 & 5.294 & 1.287 & 1.015 \\
    \addlinespace

    \addlinespace[1em]  %

    \multicolumn{6}{l}{\textbf{Neural networks}} \\
    \multicolumn{6}{c}{\textbf{Multilayer perceptron}} \\
    \midrule
    \textbf{Metric} & \textbf{Overall} & \textbf{Pre-COVID} & \textbf{COVID} & \textbf{Post-COVID} & \textbf{Excluding COVID} \\
    \cmidrule(lr){2-6}
    MSFE  & 7.967  & 0.777 & 41.666 & 3.114 & 1.840 \\
    RMSFE & 2.823  & 0.882 & 6.455  & 1.765 & 1.356 \\
    MAFE  & 1.537  & 0.691 & 5.059  & 1.144 & 0.897 \\
    \addlinespace

    \multicolumn{6}{c}{\textbf{Gated recurrent unit}} \\
    \midrule
    \textbf{Metric} & \textbf{Overall} & \textbf{Pre-COVID} & \textbf{COVID} & \textbf{Post-COVID} & \textbf{Excluding COVID} \\
    \cmidrule(lr){2-6}
    MSFE  & 5.410  & 0.405 & 29.705 & 1.698 & 0.992 \\
    RMSFE & 2.326  & 0.636 & 5.450  & 1.303 & 0.996 \\
    MAFE  & 1.305  & 0.557 & 4.159  & 1.061 & 0.786 \\
    \end{tabular}%
    } %
\end{table}
} %

\begin{enumerate}
    \item \textbf{Penalized linear models.}
        \begin{itemize}
        \item \textit{Average robustness and performance.} Ridge and Elastic Net exhibit moderate and stable error levels across the Overall prediction period and perform particularly well during low-volatility sub-periods (Excluding COVID). Conversely, LASSO achieves competitive performance exclusively during stable phases but its prediction errors during the COVID period markedly degrade its overall accuracy.
        \item \textit{Resistance to shocks.} During the COVID phase, Ridge and Elastic Net demonstrate relative robustness, experiencing only moderate error increases compared to the more stable periods. In sharp contrast, LASSO shows pronounced vulnerability to this turbulent regime, dramatically escalating prediction errors.
        \end{itemize}
    
    \item \textbf{Dimensionality reduction-based models.}
        \begin{itemize}
        \item \textit{General trend.} Both models, based on factorial projections, exhibit error metrics comparable to or slightly higher than penalized models in stable sub-periods (Pre-COVID and Excluding COVID). However, PCR and PLSR display diverging behaviors across different phases, particularly when shocks occur.
        \item \textit{Crisis segment.} During the COVID period, PCR demonstrates remarkable robustness, effectively containing prediction errors, while PLSR experiences a substantial deterioration. Conversely, in the Post-COVID phase, PCR's performance worsens significantly, whereas PLSR exhibits a clear recovery, reducing its nowcast error notably.
        \end{itemize}
        
    \item \textbf{Ensemble learning models.}
        \begin{itemize}
        \item \textit{Overall performance.} Both models achieve relatively good accuracy during stable periods (Pre-COVID, Excluding COVID), though performing less favorably than penalized and Dimensionality reduction methods. However, substantial prediction errors during the COVID crisis significantly compromise their overall performance, making them among the worst-performing methods across the entire evaluation period.
        \item \textit{COVID and Post-COVID.} In the presence of crises, both models record significant error increases. In particular, Random Forest and XGB show a partial recovery in the time window following the COVID sub-period.
        \end{itemize}
        
    \item \textbf{Neural models.}
        \begin{itemize}
        \item \textit{Normal vs. crisis.} Neural networks perform well in stable periods but deteriorate substantially during the COVID phase. Notably, MLP registers one of the highest errors among all considered models. Conversely, despite a substantial increase in prediction errors during COVID, GRU shows greater robustness and faster error reduction in the subsequent Post-COVID period.
        \item \textit{Average Overall Results.} Neural networks' overall performances remain generally superior to Random Forest but remain behind penalized linear models and Dimensionality reduction methods. GRU consistently outperforms MLP, showing notably better overall stability and accuracy.
        \end{itemize}
\end{enumerate}

The absolute results, therefore, highlight that Ridge and Elastic Net consistently obtain the lowest and most stable error levels in most circumstances, particularly in phases characterized by volatility and shocks (COVID). PCR exhibits remarkable robustness, specifically during the COVID crisis, but deteriorates notably in the subsequent phase, while PLSR behaves oppositely, suffering significantly during COVID but recovering thereafter. Conversely, LASSO, ensemble methods (RF and XGB), and neural networks show significant nowcasting deterioration during turbulent periods. Among neural networks, GRU demonstrates relatively greater resilience compared to MLP.

\vspace{0.35cm}

\phantomsection
\subsubsection{\textbf{\textit{Relative values to benchmarks}}}
The second table (Prediction accuracy relative to benchmarks across sub-periods (RMSFE ratios)), Table~\ref{tab:relativePerformance}, provides a comparison of the ratios between the RMSFE of each model and that of the three reference benchmarks (Random Walk, AR(3), Dynamic Factor Model). A ratio < 1 indicates an improvement to the benchmark, while a value > 1 corresponds to a relative deterioration.

	{
		\fontsize{9pt}{10pt}\selectfont
		\begin{table}[htbp]
			\centering
			\caption{Forecast performance (RMSFE) of models relative to benchmarks (Random Walk, AR(3), Dynamic Factor Model) to benchmarks across sub-periods (RMSFE ratios)}
			\label{tab:relativePerformance}
			
			\resizebox{0.98\textwidth}{!}{%
				\begin{tabular}{l|rrrrr}
					
					\multicolumn{1}{l|}{\textbf{RMSFE, relative to Random walk}} & \multicolumn{5}{l}{} \\
					& \textbf{Overall} & \textbf{Pre-COVID} & \textbf{COVID} & \textbf{Post-COVID} & \textbf{Excluding COVID} \\
					\cline{1-6}
					\textbf{Penalized linear models} &       &       &       &       &  \\
					LASSO & 0.427 & 0.215 & 0.446 & 0.493 & 0.367 \\
					Ridge & 0.277 & 0.276 & 0.164 & 0.621 & 0.464 \\
					Elastic net & 0.247 & 0.264 & 0.166 & 0.510 & 0.394 \\
					\cline{1-6}
					\textbf{Dimensionality reduction-based models} &       &       &       &       &  \\
					Principal component regression & 0.280 & 0.310 & 0.088 & 0.706 & 0.526 \\
					Partial least squares regression & 0.281 & 0.260 & 0.201 & 0.576 & 0.432 \\
					\cline{1-6}
					\textbf{Ensemble learning models} &       &       &       &       &  \\
					Random forest & 0.508 & 0.334 & 0.510 & 0.654 & 0.504 \\
					eXtreme Gradient Boosting & 0.494 & 0.294 & 0.524 & 0.494 & 0.396 \\
					\cline{1-6}
					\textbf{Neural networks models} &       &       &       &       &  \\
					Multi-layer perceptron & 0.540 & 0.301 & 0.565 & 0.609 & 0.465 \\
					Gated recurrent units & 0.445 & 0.217 & 0.477 & 0.449 & 0.342 \\
					\cline{1-6}
					
					\multicolumn{6}{l}{} \\
					
					\multicolumn{1}{l|}{\textbf{RMSFE, relative to AR(3)}} & \multicolumn{5}{l}{} \\
					& \textbf{Overall} & \textbf{Pre-COVID} & \textbf{COVID} & \textbf{Post-COVID} & \textbf{Excluding COVID} \\
					\cline{1-6}
					\textbf{Penalized linear models} &       &       &       &       &  \\
					LASSO & 0.636 & 0.334 & 0.660 & 0.728 & 0.557 \\
					Ridge & 0.412 & 0.430 & 0.243 & 0.918 & 0.705 \\
					Elastic net & 0.368 & 0.411 & 0.245 & 0.754 & 0.599 \\
					\cline{1-6}
					\textbf{Dimensionality reduction-based models} &       &       &       &       &  \\
					Principal component regression & 0.418 & 0.482 & 0.129 & 1.044 & 0.800 \\
					Partial least squares regression & 0.418 & 0.404 & 0.297 & 0.851 & 0.656 \\
					\cline{1-6}
					\textbf{Ensemble learning models} &       &       &       &       &  \\
					Random forest & 0.757 & 0.519 & 0.754 & 0.967 & 0.765 \\
					eXtreme Gradient Boosting & 0.735 & 0.457 & 0.775 & 0.730 & 0.602 \\
					\cline{1-6}
					\textbf{Neural networks models} &       &       &       &       &  \\
					Multi-layer perceptron & 0.804 & 0.469 & 0.835 & 0.899 & 0.707 \\
					Gated recurrent units & 0.663 & 0.338 & 0.705 & 0.664 & 0.519 \\
					\cline{1-6}
					
					\multicolumn{6}{l}{} \\
					
					\multicolumn{1}{l|}{\textbf{RMSFE, relative to Dynamic factor model}} & \multicolumn{5}{l}{} \\
					& \textbf{Overall} & \textbf{Pre-COVID} & \textbf{COVID} & \textbf{Post-COVID} & \textbf{Excluding COVID} \\
					\cline{1-6}
					\textbf{Penalized linear models} &       &       &       &       &  \\
					LASSO & 0.555 & 0.507 & 0.561 & 0.535 & 0.530 \\
					Ridge & 0.360 & 0.653 & 0.206 & 0.674 & 0.670 \\
					Elastic net & 0.321 & 0.625 & 0.208 & 0.554 & 0.569 \\
					\cline{1-6}
					\textbf{Dimensionality reduction-based models} &       &       &       &       &  \\
					Principal component regression & 0.364 & 0.733 & 0.110 & 0.767 & 0.760 \\
					Partial least squares regression & 0.365 & 0.614 & 0.253 & 0.626 & 0.623 \\
					\cline{1-6}
					\textbf{Ensemble learning models} &       &       &       &       &  \\
					Random forest & 0.660 & 0.789 & 0.641 & 0.710 & 0.727 \\
					eXtreme Gradient Boosting & 0.642 & 0.694 & 0.659 & 0.536 & 0.572 \\
					\cline{1-6}
					\textbf{Neural networks models} &       &       &       &       &  \\
					Multi-layer perceptron & 0.702 & 0.712 & 0.710 & 0.661 & 0.672 \\
					Gated recurrent units & 0.579 & 0.514 & 0.600 & 0.488 & 0.493 \\
					\cline{1-6}
					
				\end{tabular}%
			} %
		\end{table}
	}

\begin{enumerate}
    \item \textbf{Compared to Random Walk.}
        \begin{itemize}
        \item \textit{Overall.} Penalized linear (Ridge, Elastic Net) and Dimensionality reduction-based models (PCR, PLSR) show, overall, ratios significantly lower than 1. Other approaches (ensemble learning and neural networks) remain below the unit threshold, although with slightly less pronounced margins.
        \item \textit{Pre-COVID.} In this stable window, most models---especially penalized ones and dimensionality reducers---realize a very marked advantage over random walk. Neural networks and ensembles confirm competitive performances, even if the gap from RW is, in some cases, less wide than that of penalized ones.
        \item \textit{COVID.} During the crisis, all methods worsened in absolute terms. However, PCR achieves the lowest relative error (ratio below 0.1), followed closely by Ridge and Elastic Net, highlighting exceptional robustness even in this turbulent phase. The other solutions (e.g., LASSO, XGB, GRU) suffer more significant increases in error but generally remain superior to the benchmark.
        \item \textit{Post-COVID.} In the periods following the most acute phase, some models (e.g., LASSO, XGB, and GRU) obtain more evident relative improvements, always offering better performance than RW. Ridge and particularly PCR remain below the unit threshold but demonstrate significant deterioration in their relative ratios, especially PCR.
        \item \textit{Excluding COVID.} If we eliminate the pandemic window, all models maintain a clear advantage over RW. The penalized ones remain particularly stable, while ensembles and neural networks can show slightly larger oscillations but remain on ratios < 1.
        \end{itemize}

        \begin{figure}[H]
        \centering
        \begin{subfigure}[t]{0.49\textwidth}
            \includegraphics[width=\textwidth]{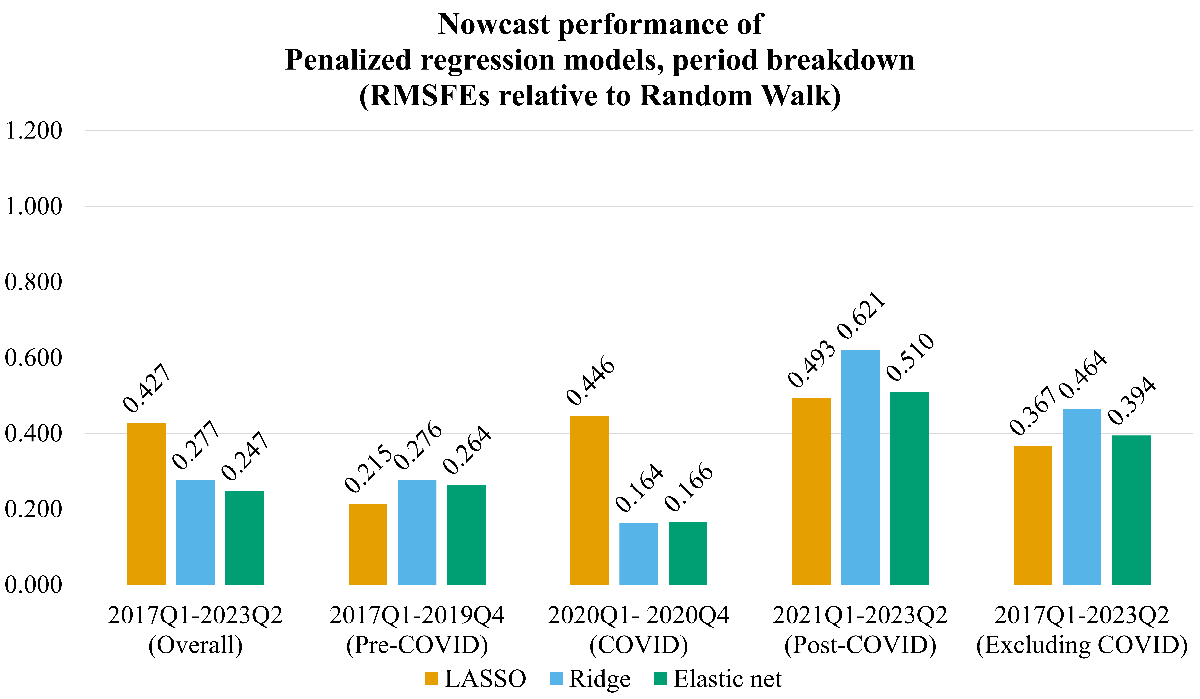}
            \caption{Linear models}
        \end{subfigure}
        \begin{subfigure}[t]{0.49\textwidth}
            \includegraphics[width=\textwidth]{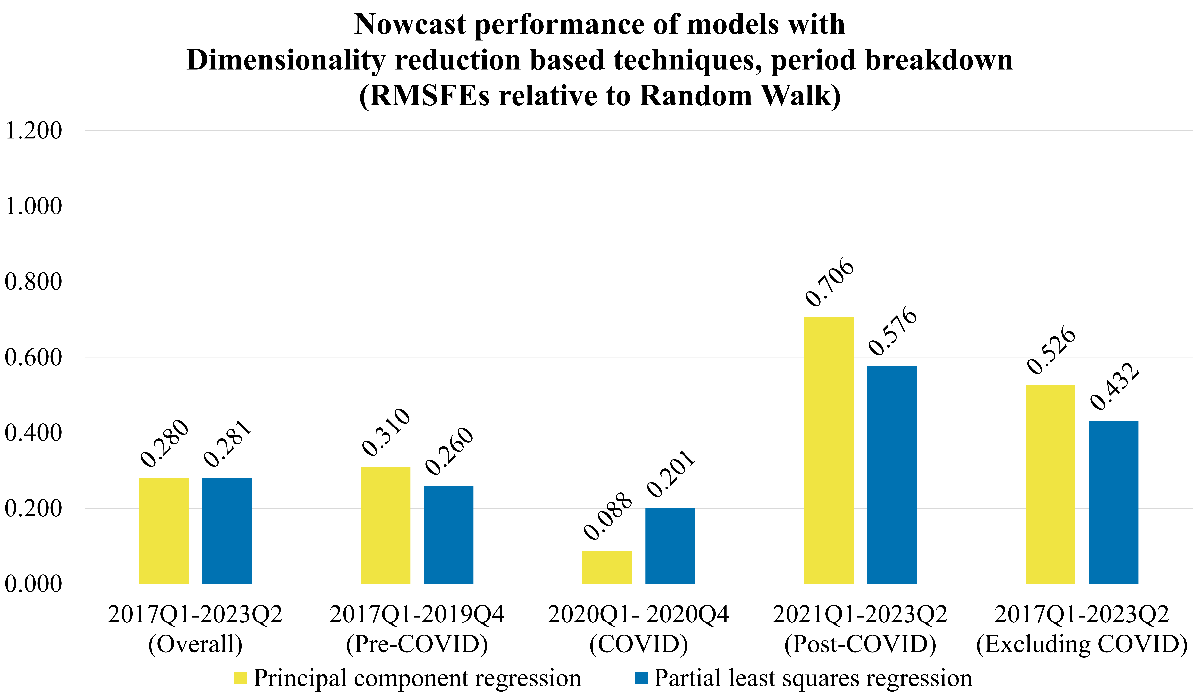}
            \caption{Dimensionality reduction models}
        \end{subfigure}
        
        \vspace{0.2cm}
        
         \begin{subfigure}[t]{0.49\textwidth}
            \includegraphics[width=\textwidth]{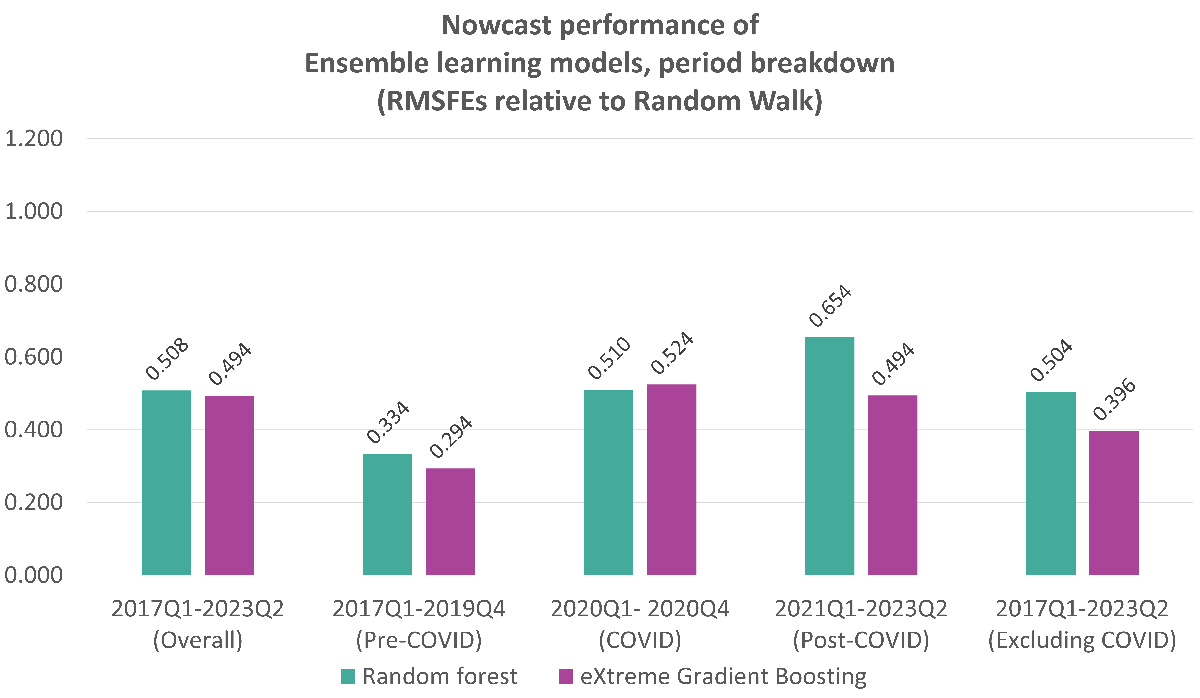}
            \caption{Ensemble learning models}
        \end{subfigure}
        \begin{subfigure}[t]{0.49\textwidth}
            \includegraphics[width=\textwidth]{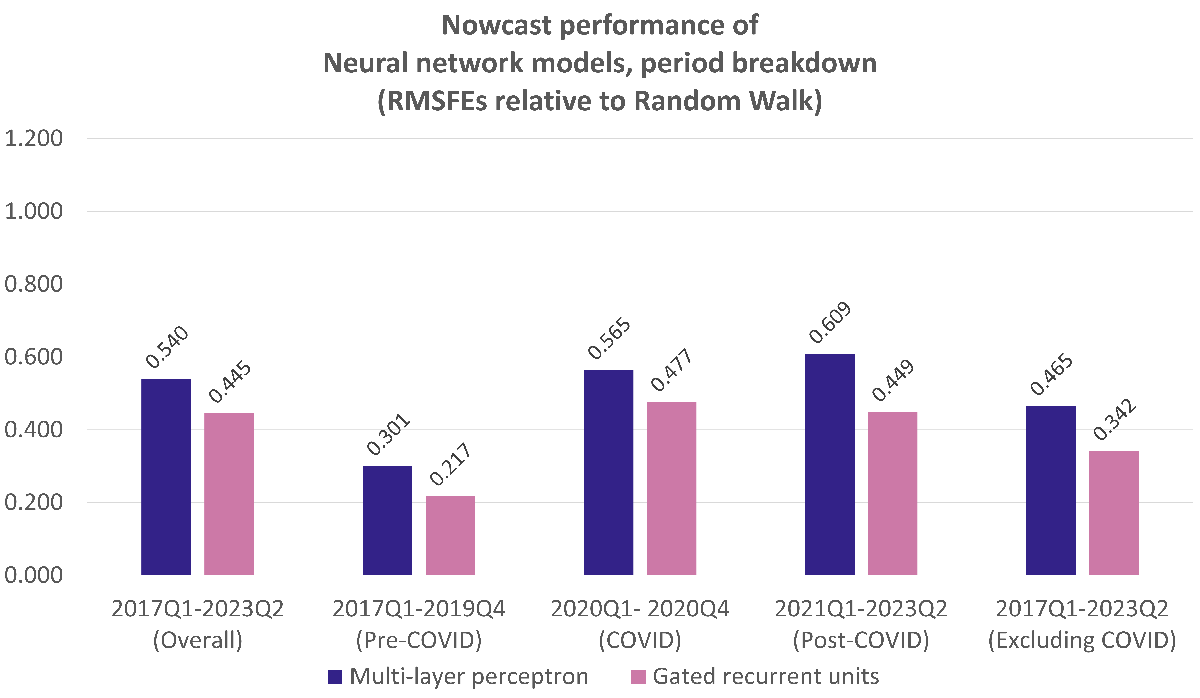}
            \caption{Neural models}
        \end{subfigure}
        \caption{Nowcast accuracy (RMSFE ratios relative to Random walk) for Penalized linear, Dimensionality reduction, Ensemble learning, Neural models across different macroeconomic regimes.}
        \label{fig:rw_relrmsfe}
        \end{figure}

    \item \textbf{Compared to AR(3).}
        \begin{itemize}
        \item \textit{Overall.} The penalized ones (especially Ridge and Elastic Net) and the Dimensionality reduction-based models show very significant error reductions compared to AR(3). The ensemble learning and neural solutions remain below 1, although with more moderate gaps.
        \item \textit{Pre-COVID.} Before the crisis, penalized families and factorial models reached ratios below 1, indicating a significant gain in RMSFE compared to AR(3). Neural networks and ensembles also remain below 1, although with minor advantages.
        \item \textit{COVID.} In this phase of strong uncertainty, the performances of Ridge and Elastic Net continue to lead, with very low ratios. Neural networks and ensembles show greater oscillations, although GRU recovers faster than MLP. PCR improves significantly during this crisis phase, while PLSR's ratio rises moderately.
        \item \textit{Post-COVID.} At the end of the crisis, Ridge and Elastic Net underwent significant deteriorations compared to the previous phase. GRU and XGB recover compared to the previous time window while remaining advantageous compared to AR(3). PLSR and PCR significantly worsened compared to the previous sub-period, with PCR notably exceeding the unit threshold, thus underperforming compared to AR(3).
        \item \textit{Excluding COVID.} Excluding the pandemic period, most models are solidly below 1, with penalized models and dimensional reducers almost always in the lead. Neural networks and ensemble learning show gaps that are, on average, smaller but still positive compared to AR(3).
        \end{itemize}

        \begin{figure}[H]
        \centering
        \begin{subfigure}[t]{0.49\textwidth}
            \includegraphics[width=\textwidth]{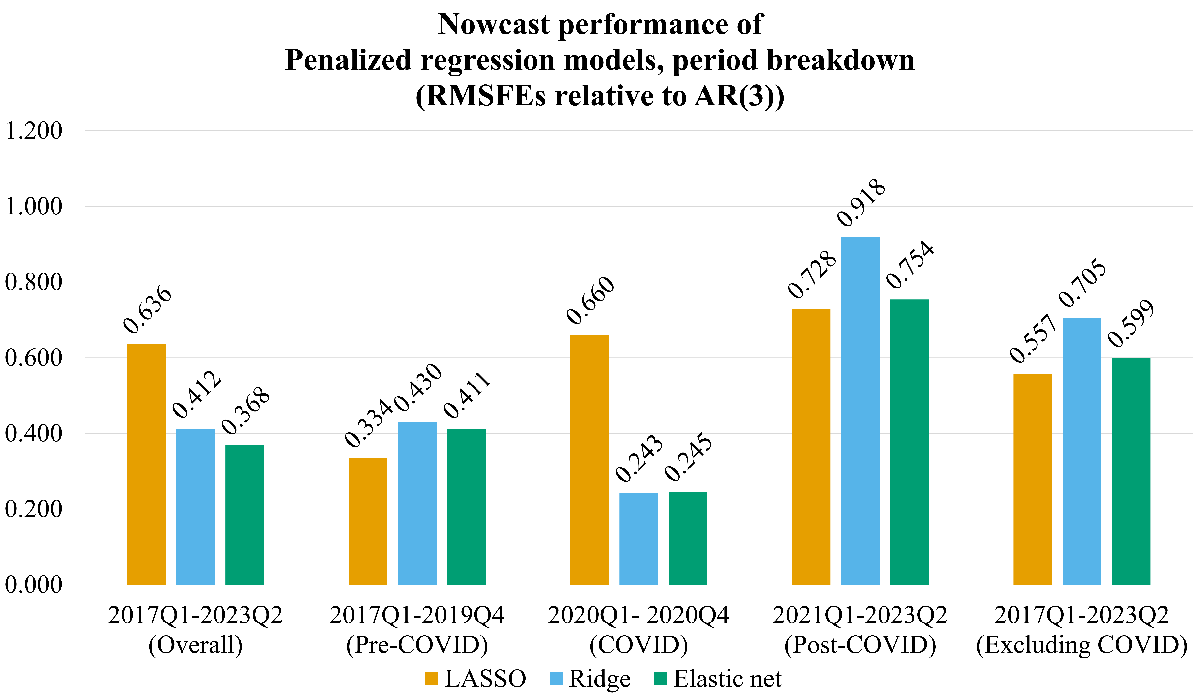}
            \caption{Linear models}
        \end{subfigure}
        \begin{subfigure}[t]{0.49\textwidth}
            \includegraphics[width=\textwidth]{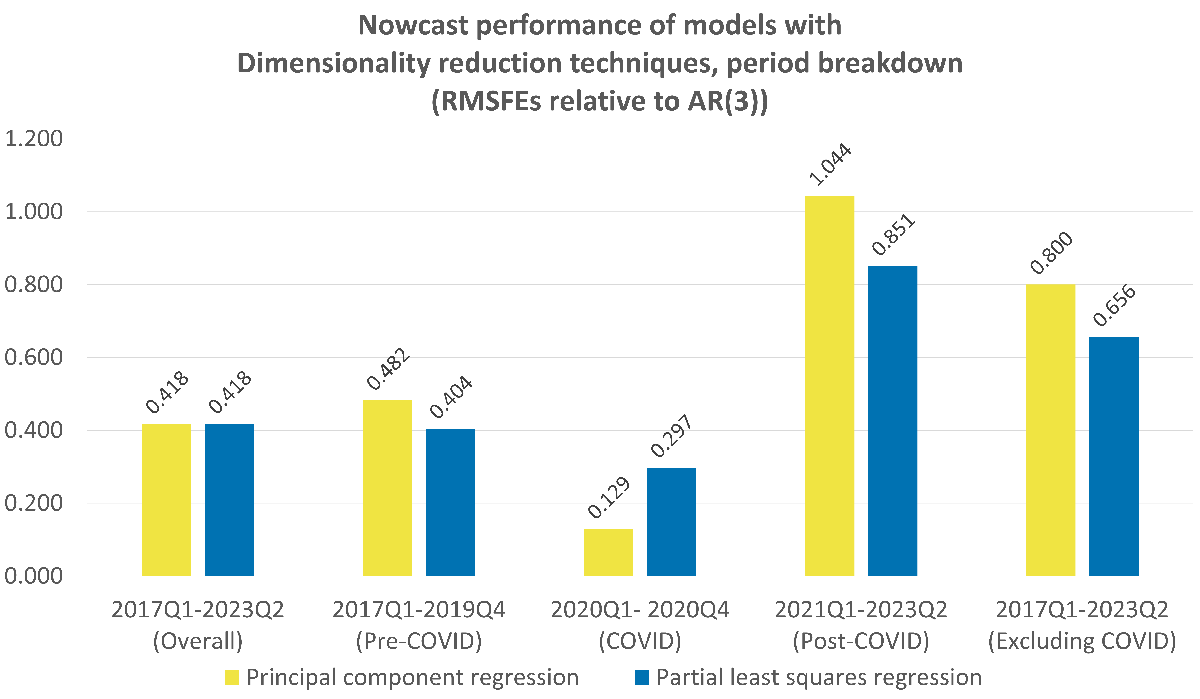}
            \caption{Dimensionality reduction models}
        \end{subfigure}
        
        \vspace{0.2cm}
        
         \begin{subfigure}[t]{0.49\textwidth}
            \includegraphics[width=\textwidth]{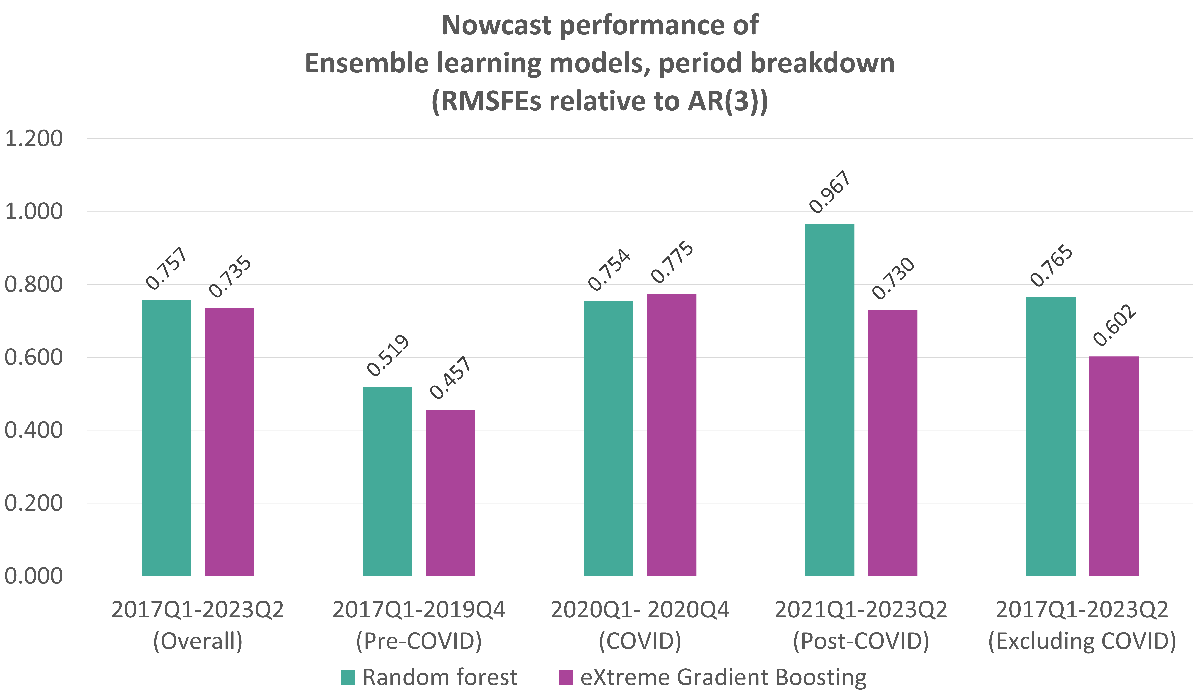}
            \caption{Ensemble learning models}
        \end{subfigure}
        \begin{subfigure}[t]{0.49\textwidth}
            \includegraphics[width=\textwidth]{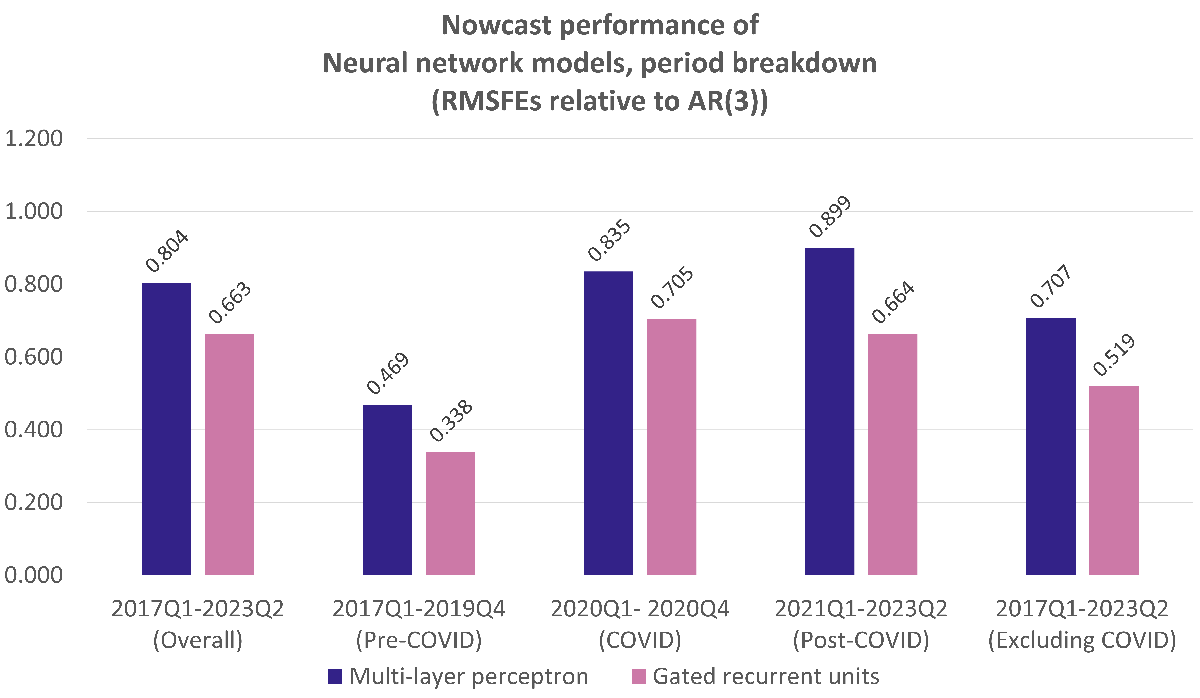}
            \caption{Neural models}
        \end{subfigure}
        \caption{Nowcast accuracy (RMSFE ratios relative to AR(3)) for Penalized linear, Dimensionality reduction, Ensemble learning, Neural models across different macroeconomic regimes.}
        \label{fig:ar3_relrmsfe}
        \end{figure}
        
    \item \textbf{Compared to Dynamic Factor Model.}
        \begin{itemize}
        \item \textit{Overall.} The majority of techniques are competitive or significantly better than DFM: penalized and dimensionally reduced models consistently perform below 1, while ensemble learning and neural models also achieve significant advantages, albeit with generally larger variations across sub-periods.
        \item \textit{Pre-COVID.} In low-volatility periods, nearly all models achieve substantial improvements compared to DFM, although PCR and Random Forest display less pronounced gains, remaining favorable.
        \item \textit{COVID.} During the crisis, penalized models continue to show superior relative robustness. Some methods, such as MLP or Random Forest, experience more significant increases in the ratio but generally remain below 1, then perform better than the benchmark. PCR achieves particularly outstanding relative error reduction during COVID.
        \item \textit{Post-COVID.} As uncertainty is gradually overcome, models such as GRU, MLP, and XGB recover quickly, always maintaining an advantage over DFM. However, Ridge, PCR, and RF undergo considerable relative deteriorations compared to the COVID period, although still retaining performance below the unit threshold.
        \item \textit{Excluding COVID.} By eliminating the most turbulent phase (COVID), penalized linear and factor models consistently are well below 1. However, this does not always improve performance since during COVID, the PCR records a greater reduction in the relative error metric, while ensemble learning and neural show less uniform but still positive improvements compared to the DFM.
        \end{itemize}

        \begin{figure}[H]
        \centering
            \begin{subfigure}[t]{0.49\textwidth}
                \includegraphics[width=\textwidth]{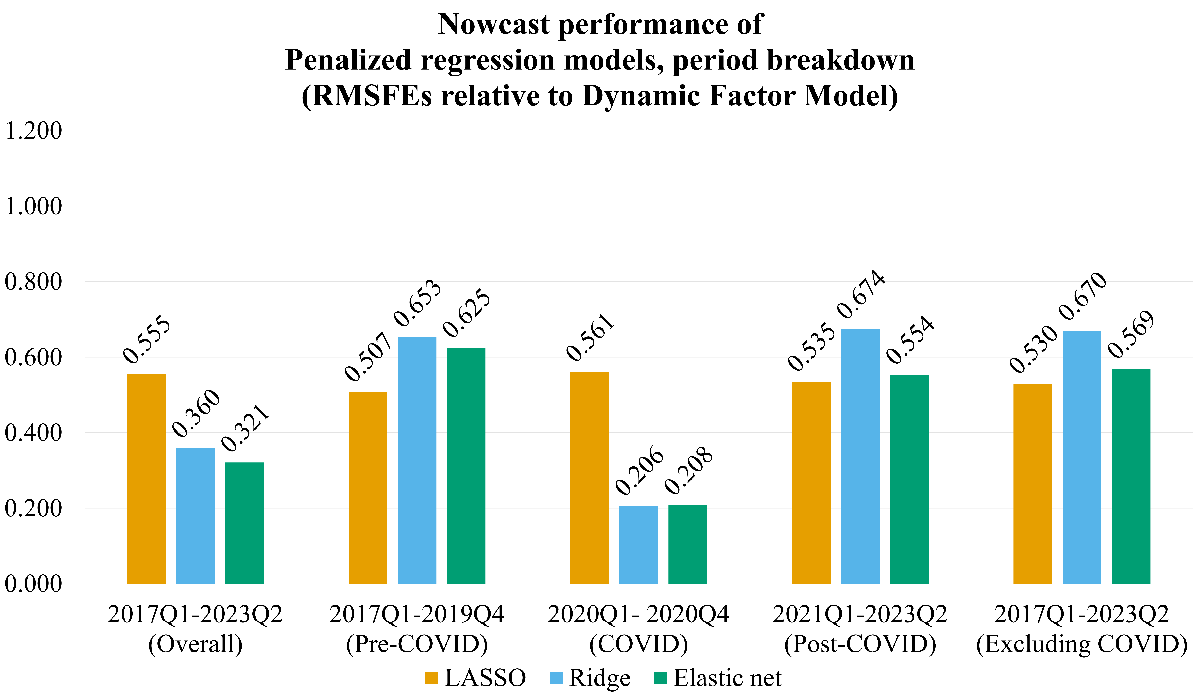}
                \caption{Linear models}
            \end{subfigure}
            \begin{subfigure}[t]{0.49\textwidth}
                \includegraphics[width=\textwidth]{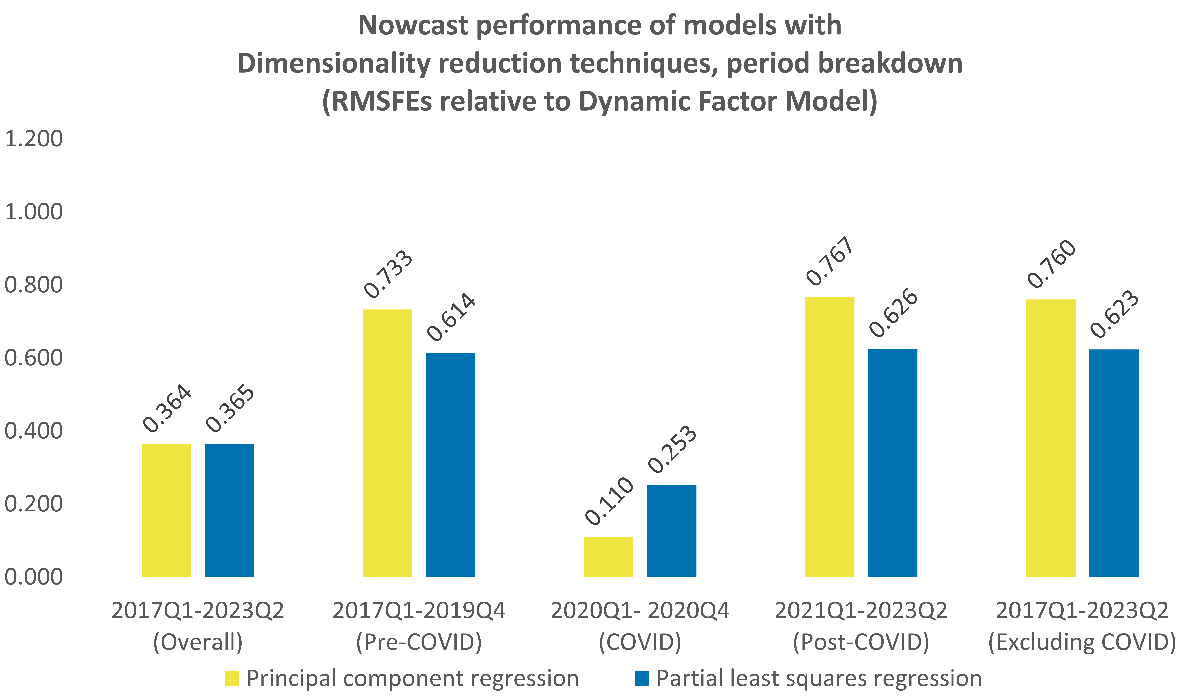}
                \caption{Dimensionality reduction models}
            \end{subfigure}
        
        \vspace{0.2cm}
        
             \begin{subfigure}[t]{0.49\textwidth}
                \includegraphics[width=\textwidth]{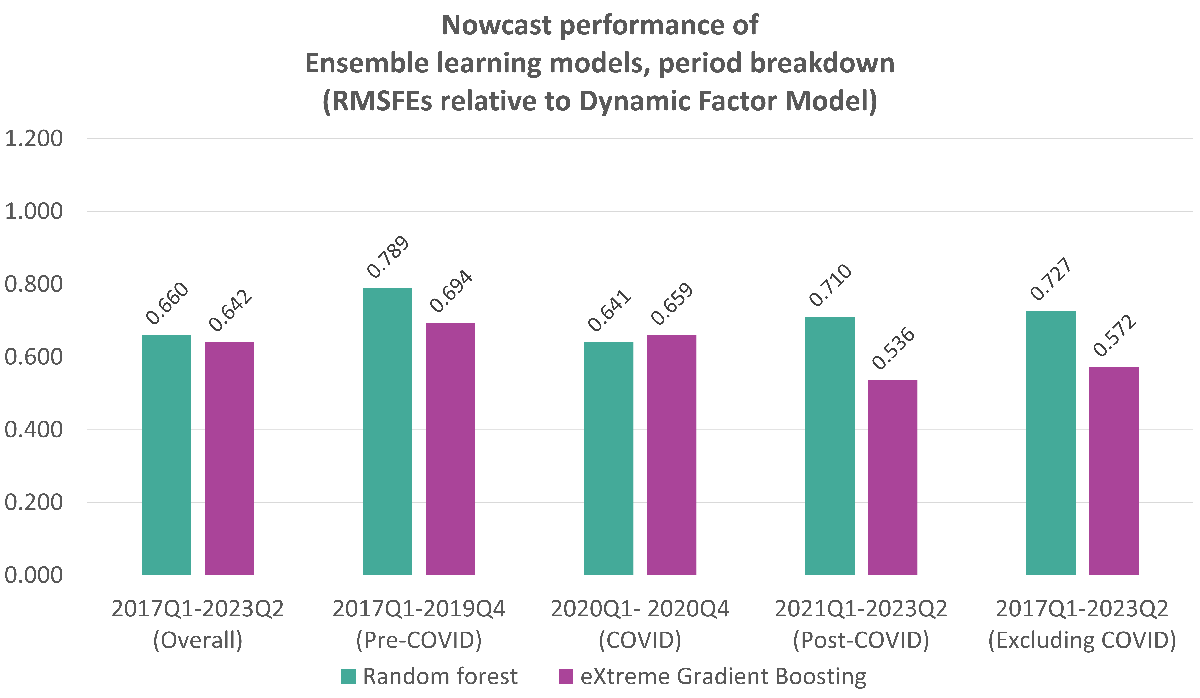}
                \caption{Ensemble learning models}
            \end{subfigure}
            \begin{subfigure}[t]{0.49\textwidth}
                \includegraphics[width=\textwidth]{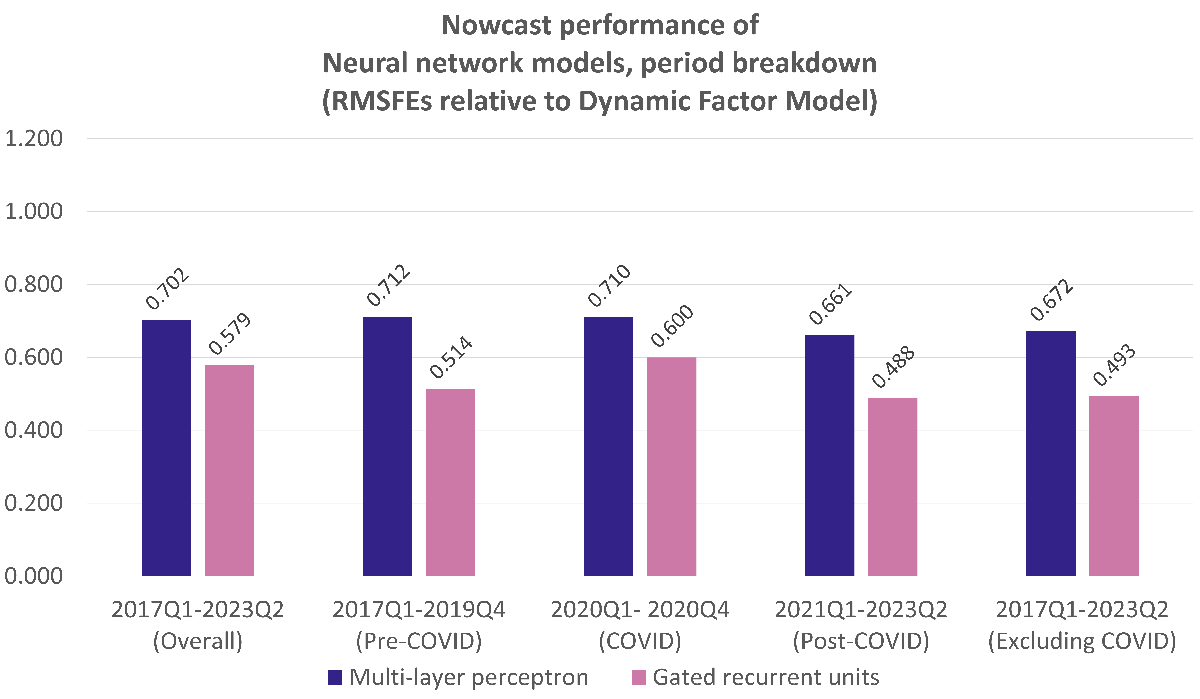}
                \caption{Neural models}
            \end{subfigure}
        \caption{Nowcast performance (RMSFE ratios relative to Dynamic Factor Model) for Penalized linear, Dimensionality reduction, Ensemble learning, Neural models across different macroeconomic regimes.}
        \label{fig:dfm_relrmsfe}
        \end{figure}
        
\end{enumerate}

Penalized linear models (especially Ridge and Elastic Net) and Dimensionality reduction-based models generally achieve RMSFE ratios consistently below benchmarks across most periods. However, PCR shows significant variability: it displays exceptional robustness during COVID but deteriorates substantially in the subsequent Post-COVID phase. In contrast, ensemble learning and neural network models, though frequently outperforming benchmarks, demonstrate higher overall vulnerability to shocks, particularly pronounced during the COVID period. Among neural approaches, GRU systematically outperforms MLP, highlighting better stability and resilience to macroeconomic disturbances.

\vspace{0.35cm}

\subsection{\textbf{\textit{Prediction Uncertainty}}}
\label{subsec:predictionuncertainty}
The reliability of nowcasting methods is inherently linked to their capacity to accurately measure and represent prediction uncertainty, especially in contexts subject to structural changes or unexpected economic shocks. Prediction intervals offer valuable insights, quantifying how model predictions vary under different macroeconomic conditions. Table~\ref{tab:nowcast_uncertainty} presents the average widths of these intervals, computed across the whole prediction period as well as distinct sub-periods (Pre-COVID, COVID, Post-COVID, Excluding COVID), enabling a nuanced assessment of each model family's consistency and resilience.

\begin{table}[htbp]
    \centering
    \caption{Average prediction intervals across models and sub-periods}
    \label{tab:nowcast_uncertainty}
    \resizebox{0.98\textwidth}{!}{%
        \begin{tabular}{lccccc}
        \toprule
              & \textbf{Overall} & \textbf{Pre-COVID} & \textbf{COVID} & \textbf{Post-COVID} & \textbf{Excl. COVID} \\
        \midrule
            \textbf{Penalized linear models} & & & & & \\
            LASSO       & 2.117 & 1.278 & 5.460 & 1.787 & 1.510 \\
            Ridge       & 2.130 & 1.932 & 3.513 & 1.813 & 1.878 \\
            Elastic net & 1.499 & 0.924 & 3.540 & 1.372 & 1.128 \\
            \midrule
            \textbf{Dimensionality reduction-based models} & & & & & \\
            PCR         & 2.157 & 1.443 & 4.819 & 1.950 & 1.674 \\
            PLSR        & 2.125 & 1.124 & 4.918 & 2.209 & 1.617 \\
            \midrule
            \textbf{Ensemble learning models} & & & & & \\
            RF          & 3.059 & 1.190 & 2.230 & 5.632 & 3.209 \\
            XGB         & 1.960 & 1.613 & 2.141 & 2.304 & 1.927 \\
            \midrule
            \textbf{Neural network models} & & & & & \\
            MLP         & 2.785 & 1.584 & 6.296 & 2.822 & 2.147 \\
            GRU         & 2.526 & 1.991 & 3.261 & 2.873 & 2.392 \\
        \bottomrule
        \end{tabular}%
    }
\end{table}

Below, we outline the key patterns characterizing prediction uncertainty, highlighting differences and commonalities within and across model families over the analyzed sub-periods:

\begin{enumerate}
\item \textbf{Penalized linear models.}
    \begin{itemize}
        \item \textit{Overall trends.} Elastic Net consistently provides narrower prediction intervals than LASSO and Ridge across the entire nowcasting horizon. In particular, Elastic Net demonstrates reduced uncertainty during stable economic phases, while Ridge intervals remain relatively wider throughout.
        \item \textit{Period-specific variations.} During the COVID period, prediction intervals significantly expanded for all penalized models, reflecting heightened prediction uncertainty. However, Elastic Net intervals widened comparatively less, indicating better stability under economic volatility. In Post-COVID sub-period, prediction intervals notably contracted, with Elastic Net and LASSO returning closer to pre-crisis levels.
    \end{itemize}

\item \textbf{Dimensionality reduction-based models.}
    \begin{itemize}
        \item \textit{General behavior.} PCR and PLSR show similar average interval widths In the overall prediction period, although exhibiting distinct intra-period dynamics. Specifically, intervals from PCR remained relatively broader than PLSR in most periods, except during the Post-COVID phase, when PLSR intervals notably exceeded PCR.
        \item \textit{Economic shocks impact.} Both models experienced substantial interval widening during COVID, with PCR showing slightly narrower intervals, suggesting marginally better robustness under crisis conditions. Nevertheless, PLSR intervals significantly narrowed after the crisis, highlighting its quicker return to pre-crisis uncertainty levels.
    \end{itemize}

    \item \textbf{Ensemble learning models.}
    \begin{itemize}
        \item \textit{Aggregate results.} Random Forest presented the widest prediction intervals overall, signaling substantial uncertainty across periods. Conversely, XGBoost maintained narrower intervals, demonstrating enhanced stability and precision, especially in stable economic sub-periods.
        \item \textit{Sub-period dynamics.} Random Forest intervals notably expanded during the Post-COVID phase, surpassing considerably the interval widths observed during both Pre-COVID and COVID phases, where intervals were comparatively narrower. XGBoost exhibited less dramatic shifts, maintaining moderate interval widths even during turbulent economic phases.
    \end{itemize}

\item \textbf{Neural models.}
    \begin{itemize}
        \item \textit{Interval widths and volatility.} MLP and GRU displayed elevated uncertainty compared to Penalized and Dimensionality reduction models, especially during COVID. MLP recorded the widest intervals of all models during the crisis, indicating pronounced sensitivity to extreme economic disturbances.
        \item \textit{Post-crisis behavior.} Post-COVID intervals for GRU slightly narrowed compared to the COVID period, while those of MLP narrowed markedly, indicating a faster normalization after the crisis. Considering the whole prediction period but excluding COVID, GRU consistently maintained broader intervals relative to other methods, highlighting ongoing caution in nowcasts even under normal conditions.
    \end{itemize}

\end{enumerate}

\begin{figure}[H]
        \centering
            \begin{subfigure}[t]{0.91\textwidth}
                \includegraphics[width=\textwidth]{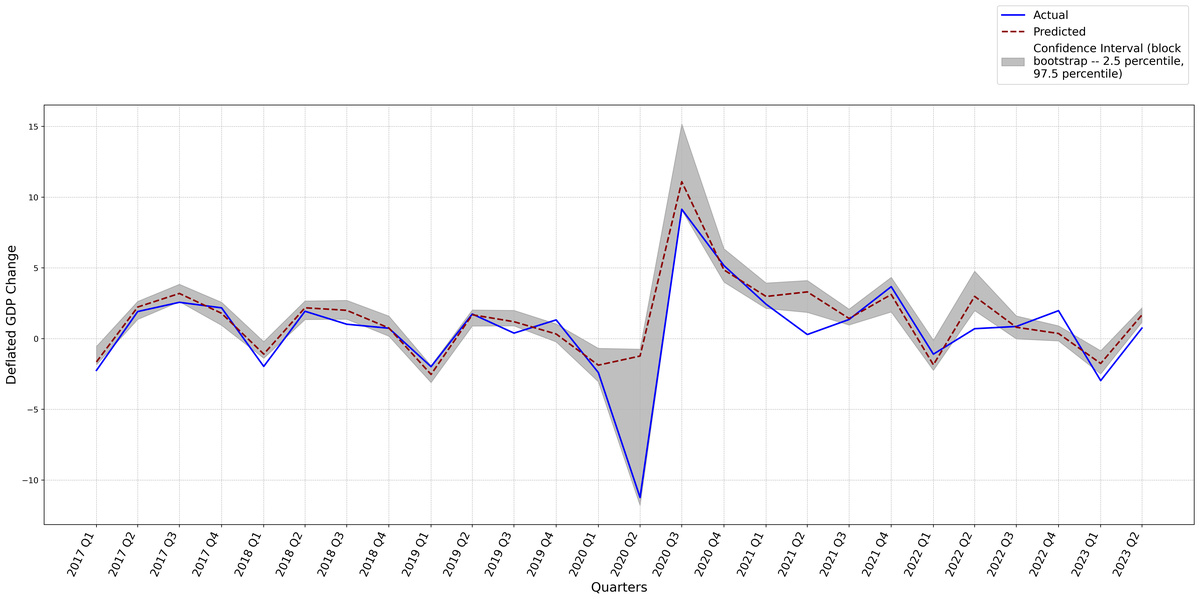}
                \caption{LASSO}
            \end{subfigure}
            
        \vspace{0.2cm}
        
            \begin{subfigure}[t]{0.91\textwidth}
                \includegraphics[width=\textwidth]{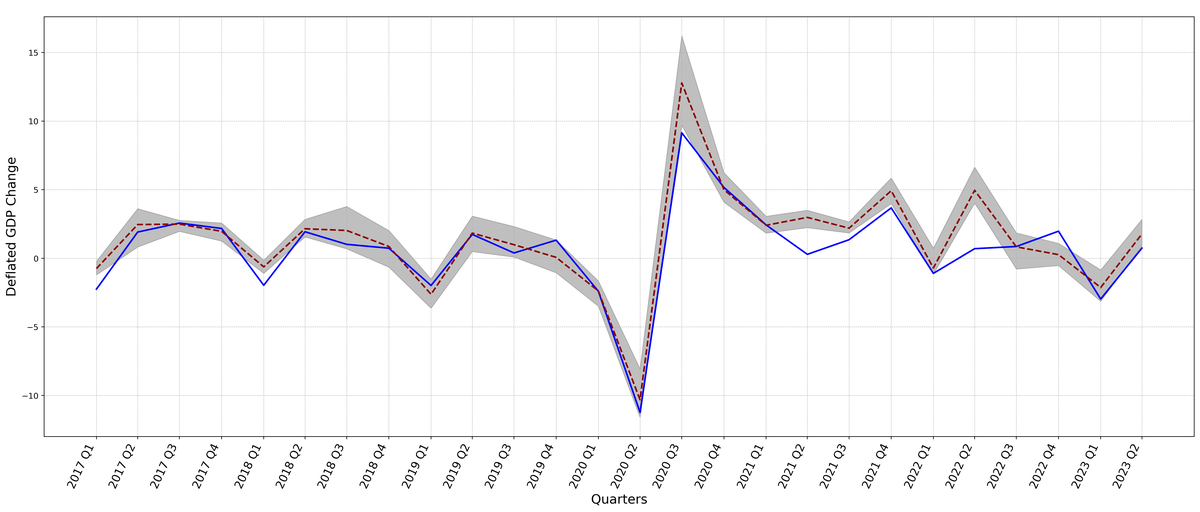}
                \caption{Ridge}
            \end{subfigure}
            
        \vspace{0.2cm}
        
            \begin{subfigure}[t]{0.91\textwidth}
                \includegraphics[width=\textwidth]{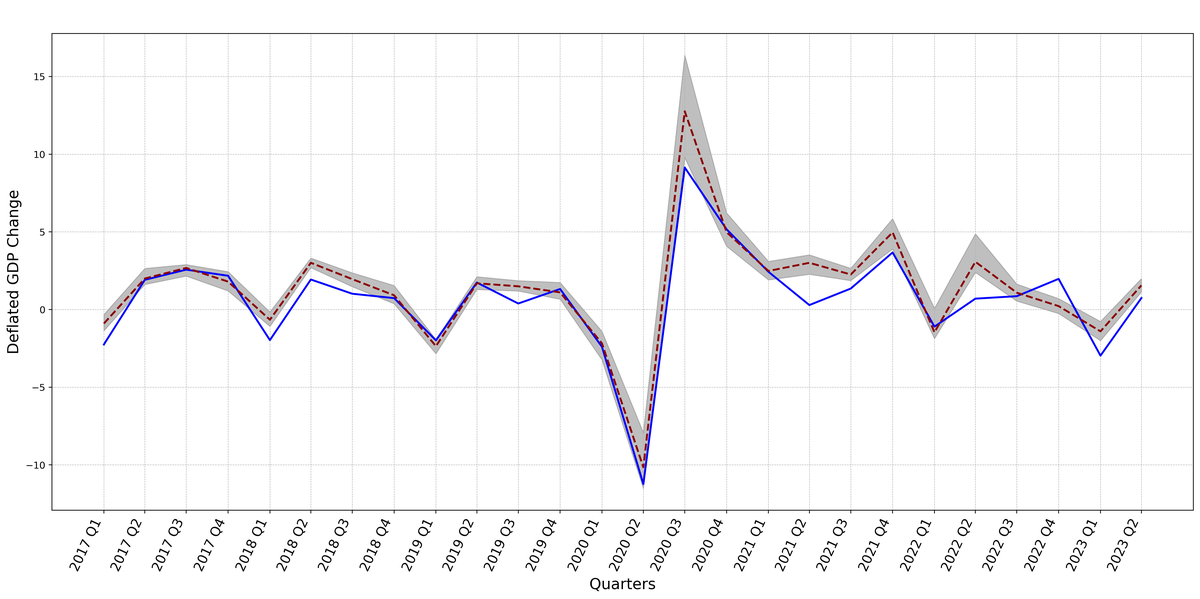}
                \caption{Elastic Net}
            \end{subfigure}
        \caption{Actual and predicted deflated GDP growth rate for Penalized Linear models (LASSO, Ridge, and Elastic Net). Shaded areas represent the 95\% prediction intervals obtained through the block bootstrap method.}
        \label{fig:penalized_predictionintervals}
        \end{figure}

\begin{figure}[H]
        \centering
            \begin{subfigure}[t]{0.91\textwidth}
                \includegraphics[width=\textwidth]{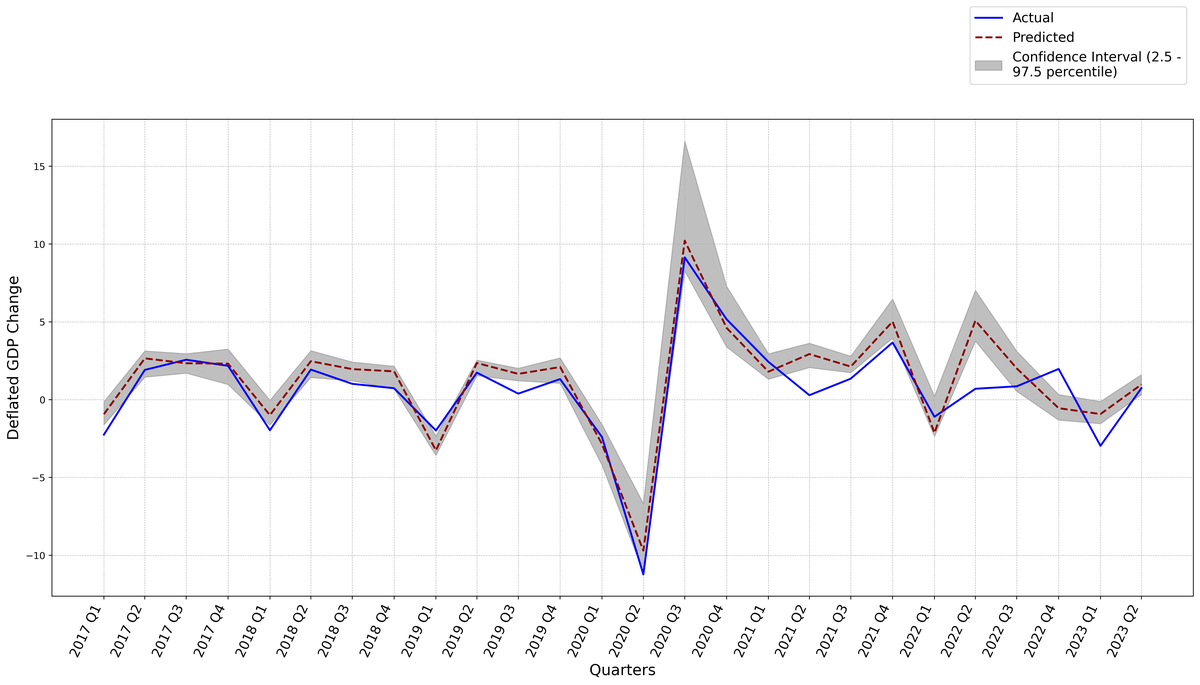}
                \caption{Principal Component Regression}
            \end{subfigure}
            
        \vspace{0.2cm}
        
            \begin{subfigure}[t]{0.91\textwidth}
                \includegraphics[width=\textwidth]{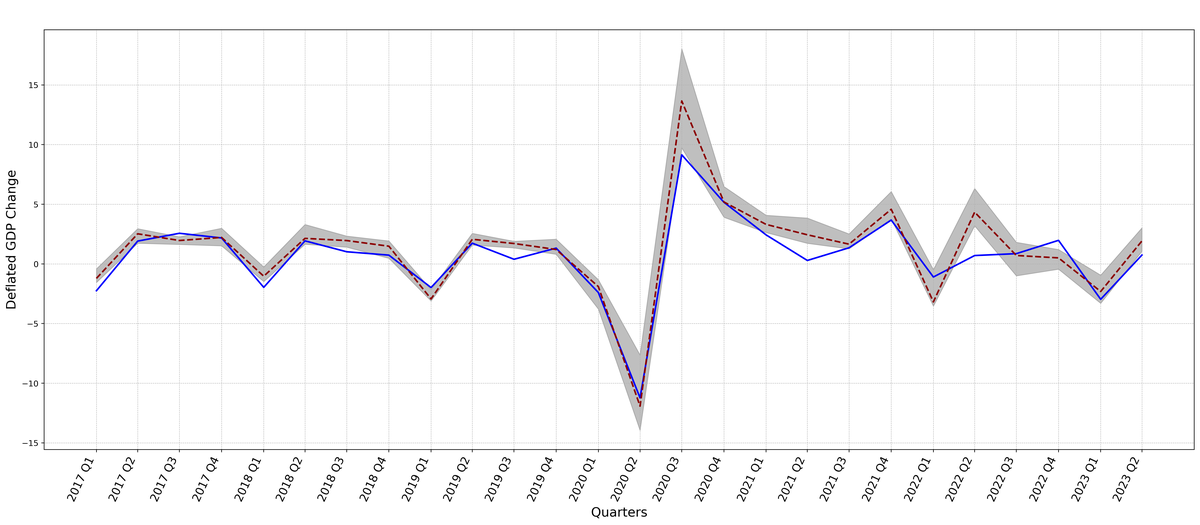}
                \caption{Partial Least Squares Regression}
            \end{subfigure}
        \caption{Actual and predicted deflated GDP growth rate for Dimensionality Reduction-based models (Principal Component Regression and Partial Least Squares Regression). Shaded areas represent the 95\% prediction intervals obtained through the block bootstrap method.}
        \label{fig:dimreduction_predictionintervals}
        \end{figure}

\begin{figure}[H]
        \centering
            \begin{subfigure}[t]{0.91\textwidth}
                \includegraphics[width=\textwidth]{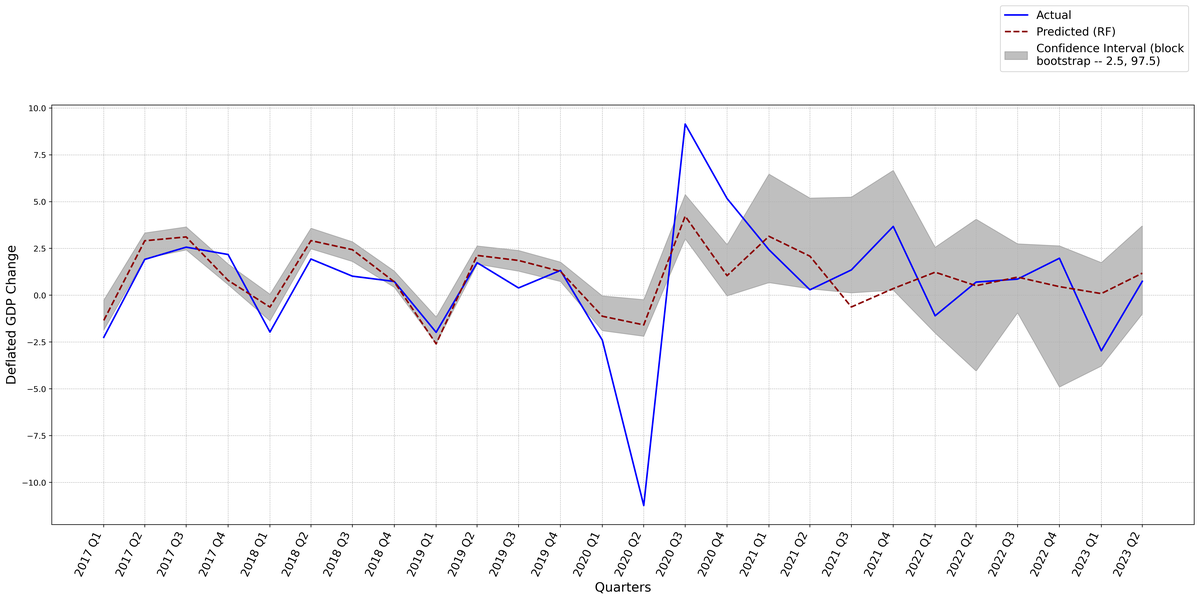}
                \caption{Random Forest}
            \end{subfigure}
            
        \vspace{0.2cm}
        
            \begin{subfigure}[t]{0.91\textwidth}
                \includegraphics[width=\textwidth]{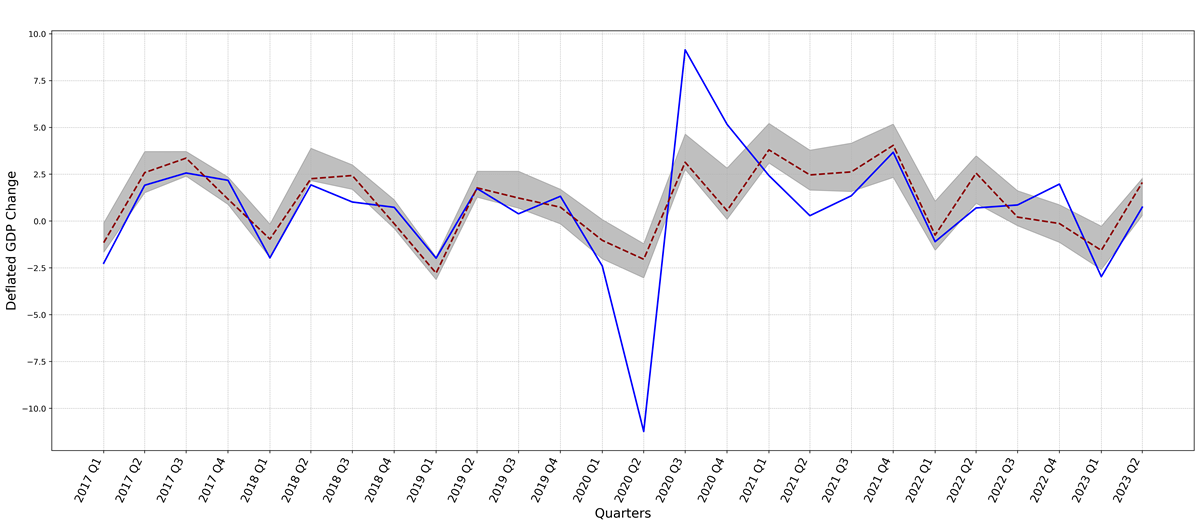}
                \caption{eXtreme Gradient Boosting}
            \end{subfigure}
        \caption{Actual and predicted deflated GDP growth rate for Ensemble Learning models (Random Forest and eXtreme Gradient Boosting). Shaded areas represent the 95\% prediction intervals obtained through the block bootstrap method.}
        \label{fig:ensemblelearn_predictionintervals}
        \end{figure}

\begin{figure}[H]
        \centering
            \begin{subfigure}[t]{0.91\textwidth}
                \includegraphics[width=\textwidth]{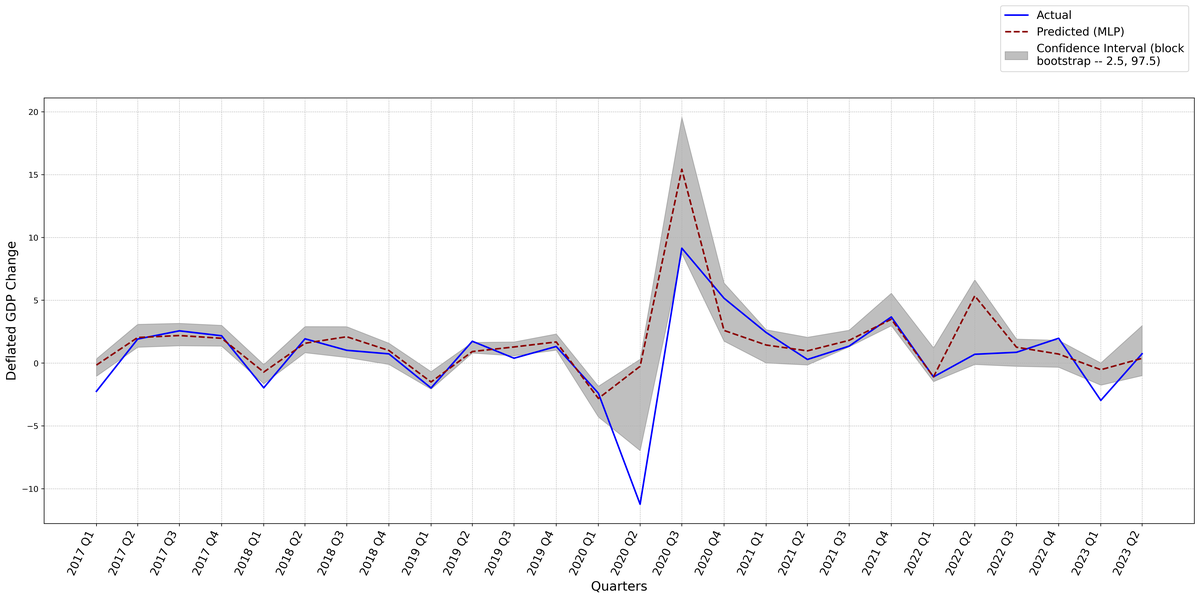}
                \caption{Multilayer Perceptron}
            \end{subfigure}
            
        \vspace{0.2cm}
        
            \begin{subfigure}[t]{0.91\textwidth}
                \includegraphics[width=\textwidth]{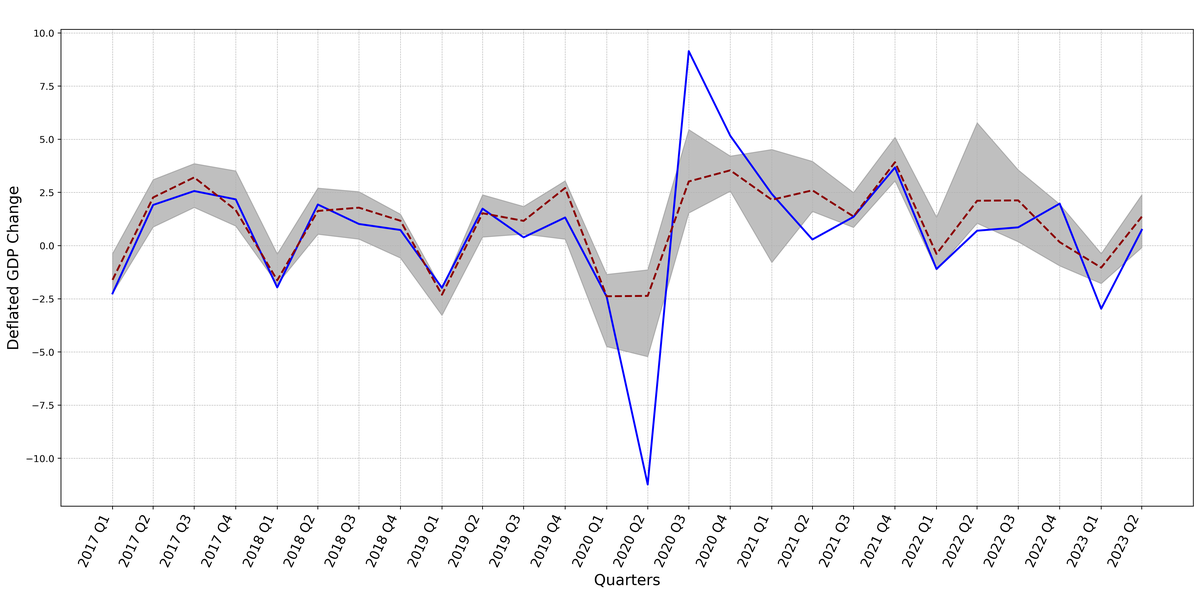}
                \caption{Gated Recurrent Unit}
            \end{subfigure}
        \caption{Actual and predicted deflated GDP growth rate for Neural models (Multilayer Perceptron and Gated Recurrent Unit). Shaded areas represent the 95\% prediction intervals obtained through the block bootstrap method.}
        \label{fig:neural_predictionintervals}
        \end{figure}

A detailed examination of the actual and predicted GDP trajectories across model families, visualized through prediction interval plots (see Figures~\ref{fig:penalized_predictionintervals} to~\ref{fig:neural_predictionintervals}), provides additional insights regarding prediction interval coverage and reliability:

\begin{itemize}
    \item Penalized linear models (Figure~\ref{fig:penalized_predictionintervals}) present narrow prediction intervals closely following the predicted GDP path. Despite their precision, during periods of extreme volatility, actual GDP values frequently fall outside the prediction intervals, signaling potential limitations in capturing exceptional economic fluctuations.
    
    \item Dimensionality reduction-based models (Figure~\ref{fig:dimreduction_predictionintervals}) generate prediction intervals that are slightly broader than penalized linear models, enhancing the coverage of actual GDP values during stable periods. Nonetheless, these intervals still demonstrate challenges in fully encompassing observed GDP dynamics, with actual values occasionally breaching the interval boundaries.
    
    \item Random Forest produces notably wider prediction intervals, especially in the Post-COVID period. Despite this wider uncertainty quantification, intervals sometimes fail to encompass actual GDP data and the model's predictions, suggesting elevated uncertainty together with weaker predictive consistency. Conversely, XGBoost demonstrates moderately wide intervals that consistently envelop the predicted values and, to a greater extent, actual GDP figures across different economic phases (Figure~\ref{fig:ensemblelearn_predictionintervals}).
    
    \item GRU intervals exhibit moderate widths, capturing most of the observed GDP variations, even in volatile contexts. MLP, however, displays the broadest intervals during crisis periods, often encompassing extreme GDP values. This broader uncertainty quantification persists even in less turbulent sub-periods (Figure~\ref{fig:neural_predictionintervals}) .
\end{itemize}

The analysis of prediction intervals reveals distinct uncertainty profiles across model families. Penalized linear models---particularly Elastic Net--- consistently achieve narrower and more stable prediction intervals across most periods, indicating greater reliability in prediction uncertainty quantification, especially during phases of heightened volatility (e.g., COVID). Among Dimensionality reduction-based models, PCR exhibits considerable stability during the COVID crisis but performs less consistently in the subsequent Post-COVID phase. In contrast, PLSR intervals widen notably during the crisis but return to lower levels thereafter. Conversely, ensemble learning methods (particularly Random Forest) and neural network approaches (especially MLP) tend to produce substantially wider prediction intervals, signaling pronounced uncertainty and heightened sensitivity to macroeconomic shocks. Within the neural network family, GRU generally outperforms MLP, maintaining comparatively narrower intervals and exhibiting greater robustness across the evaluated sub-periods.

\vspace{0.35cm}

\subsection{\textbf{\textit{Feature Importance and Explainability}}}
\label{subsec:featureimportance}
Examining feature importance offers a complementary perspective to predictive accuracy and uncertainty, enhancing interpretability and transparency and providing deeper insights into the economic mechanisms underlying model predictions. To this end, we investigate how explanatory variables contribute to nowcasting outcomes across different model families, highlighting general trends and specific variations observed across macroeconomic regimes (Overall, Pre-COVID, COVID, Post-COVID, and Excluding COVID). These results, complementing previously presented analyses on predictive accuracy and uncertainty, offer additional insights on the relative relevance assigned to economic indicators by each modeling framework within the macroeconomic nowcasting context.

\begin{enumerate}
    \item \textbf{Penalized linear models.}
            \begin{itemize}
            \item \textit{General trends.} Persistently ranked among crucial features are industrial production and gross operating surplus. Additionally, labor market indicators such as employee compensation and unit labor costs frequently appear among key predictors, particularly during turbulent economic periods.
            \item \textit{Period-specific variations.} During the COVID sub-period, penalized models significantly increased the importance of labor and production-related variables, effectively capturing volatility. In the Post-COVID interval, the dominance of industrial production notably intensified, clearly reflecting economic stabilization.
            \item \textit{Excluding COVID.} Excluding the COVID period, the hierarchy of relevant variables closely mirrors the overall pattern, confirming stability in predictor selection and importance, particularly regarding industrial production, gross operating surplus, and labor market variables.
        \end{itemize}

        {\fontsize{15pt}{8pt}
        \selectfont
        \begin{table}[htbp]
            \centering
            \caption{Linear models with penalization, top 10 predictors ranked by average feature importance and 95\% confidence intervals, estimated via block bootstrap, across different sub-periods}
            \resizebox{1.00\textwidth}{!}{
            \begin{tabular}{llcc|lcc|lcc}
                  & \multicolumn{3}{c|}{\textbf{LASSO}} & \multicolumn{3}{c|}{\textbf{Ridge}} & \multicolumn{3}{c}{\textbf{Elastic Net}} \\
            \hline
                  & \multicolumn{1}{c}{\textbf{Feature}} & \multicolumn{1}{p{5.78em}}{\textbf{Feature (avg)}} & \multicolumn{1}{p{3.945em}|}{\textbf{CI range}} & \multicolumn{1}{c}{\textbf{Feature}} & \multicolumn{1}{p{5.835em}}{\textbf{Feature (avg)}} & \multicolumn{1}{p{3.61em}|}{\textbf{CI range}} & \multicolumn{1}{c}{\textbf{Feature}} & \multicolumn{1}{p{5.835em}}{\textbf{Feature (avg)}} & \multicolumn{1}{p{3.61em}}{\textbf{CI range}} \\
            \hline
            \multirow{10}[1]{*}{\textbf{Overall}} 
                  & employee compensation change & 1.684 & 3.157 & unit labor cost change & -1.355 & 4.234 & industrial production index change & 0.907 & 0.567 \\
                  & unit labor cost change & -1.657 & 3.139 & employee compensation change & 1.276 & 4.161 & gross operating surplus change & 0.507 & 0.364 \\
                  & industrial production index change & 0.947 & 0.451 & industrial production index change & 0.732 & 0.549 & air cargo loaded change & 0.322 & 0.181 \\
                  & gross operating surplus change & 0.669 & 0.327 & gross operating surplus change & 0.435 & 0.361 & fnb services index change & 0.240 & 0.216 \\
                  & fnb services index change & 0.277 & 0.215 & fnb services index change & 0.237 & 0.276 & taxes less subsidies change & 0.222 & 0.160 \\
                  & air cargo loaded change & 0.272 & 0.253 & air cargo loaded change & 0.232 & 0.254 & pawnshop pledges received change & 0.200 & 0.144 \\
                  & public progress payments change & 0.262 & 0.098 & taxes less subsidies change & 0.229 & 0.083 & public progress payments change & 0.179 & 0.169 \\
                  & taxes less subsidies change & 0.227 & 0.055 & pawnshop pledges received change & 0.210 & 0.275 & private properties available change & -0.160 & 0.130 \\
                  & pawnshop pledges received change & 0.226 & 0.118 & composite leading index change & 0.192 & 0.164 & composite leading index change & 0.158 & 0.102 \\
                  & exchange rate hkd change & -0.173 & 0.163 & exchange rate hkd change & -0.191 & 0.439 & business expectations manufacturing & 0.153 & 0.108 \\
            \hline
            \multirow{10}[0]{*}{\textbf{Pre-COVID}} 
                  & employee compensation change & 3.100 & 0.153 & unit labor cost change & -2.568 & 4.178 & industrial production index change & 0.945 & 0.286 \\
                  & unit labor cost change & -3.075 & 0.153 & employee compensation change & 2.469 & 4.102 & gross operating surplus change & 0.538 & 0.193 \\
                  & industrial production index change & 0.763 & 0.076 & industrial production index change & 0.689 & 0.210 & air cargo loaded change & 0.328 & 0.112 \\
                  & gross operating surplus change & 0.571 & 0.061 & gross operating surplus change & 0.444 & 0.076 & fnb services index change & 0.240 & 0.136 \\
                  & fnb services index change & 0.240 & 0.072 & exchange rate hkd change & -0.266 & 0.474 & taxes less subsidies change & 0.231 & 0.213 \\
                  & public progress payments change & 0.230 & 0.023 & fnb services index change & 0.239 & 0.145 & pawnshop pledges received change & 0.200 & 0.155 \\
                  & taxes less subsidies change & 0.216 & 0.015 & composite leading index change & 0.236 & 0.094 & private properties available change & -0.186 & 0.127 \\
                  & pawnshop pledges received change & 0.204 & 0.011 & taxes less subsidies change & 0.235 & 0.059 & public progress payments change & 0.178 & 0.155 \\
                  & air cargo loaded change & 0.193 & 0.052 & public progress payments change & 0.182 & 0.065 & composite leading index change & 0.169 & 0.097 \\
                  & composite leading index change & 0.192 & 0.030 & pawnshop pledges received change & 0.175 & 0.056 & business expectations manufacturing & 0.153 & 0.144 \\
            \hline
            \multirow{10}[1]{*}{\textbf{COVID}} 
                  & employee compensation change & 1.559 & 3.000 & industrial production index change & 0.755 & 0.208 & industrial production index change & 0.799 & 0.243 \\
                  & unit labor cost change & -1.549 & 3.102 & gross operating surplus change & 0.398 & 0.106 & gross operating surplus change & 0.426 & 0.135 \\
                  & industrial production index change & 0.952 & 0.442 & air cargo loaded change & 0.293 & 0.027 & air cargo loaded change & 0.304 & 0.037 \\
                  & gross operating surplus change & 0.689 & 0.323 & taxes less subsidies change & 0.251 & 0.043 & taxes less subsidies change & 0.243 & 0.050 \\
                  & air cargo loaded change & 0.313 & 0.207 & pawnshop pledges received change & 0.191 & 0.054 & pawnshop pledges received change & 0.189 & 0.052 \\
                  & public progress payments change & 0.275 & 0.078 & unit labor cost change & -0.185 & 0.067 & private properties available change & -0.179 & 0.055 \\
                  & taxes less subsidies change & 0.237 & 0.046 & private properties available change & -0.182 & 0.059 & composite leading index change & 0.175 & 0.018 \\
                  & pawnshop pledges received change & 0.206 & 0.048 & composite leading index change & 0.181 & 0.023 & fnb services index change & 0.174 & 0.052 \\
                  & fnb services index change & 0.205 & 0.064 & business expectations manufacturing & 0.174 & 0.054 & unit labor cost change & -0.164 & 0.055 \\
                  & domestic supply price change & -0.204 & 0.074 & fnb services index change & 0.171 & 0.052 & business expectations manufacturing & 0.162 & 0.046 \\
            \hline
            \multirow{10}[2]{*}{\textbf{Post-COVID}} 
                  & industrial production index change & 1.167 & 0.027 & industrial production index change & 0.775 & 0.550 & industrial production index change & 0.904 & 0.562 \\
                  & gross operating surplus change & 0.780 & 0.108 & gross operating surplus change & 0.439 & 0.362 & gross operating surplus change & 0.503 & 0.361 \\
                  & fnb services index change & 0.351 & 0.208 & unit labor cost change & -0.366 & 0.634 & air cargo loaded change & 0.320 & 0.167 \\
                  & air cargo loaded change & 0.351 & 0.137 & employee compensation change & 0.308 & 0.658 & fnb services index change & 0.266 & 0.216 \\
                  & public progress payments change & 0.295 & 0.046 & air cargo loaded change & 0.277 & 0.084 & pawnshop pledges received change & 0.204 & 0.111 \\
                  & pawnshop pledges received change & 0.261 & 0.112 & fnb services index change & 0.260 & 0.279 & taxes less subsidies change & 0.203 & 0.042 \\
                  & taxes less subsidies change & 0.236 & 0.056 & pawnshop pledges received change & 0.259 & 0.290 & public progress payments change & 0.188 & 0.156 \\
                  & exchange rate hkd change & -0.222 & 0.084 & taxes less subsidies change & 0.211 & 0.050 & business expectations manufacturing & 0.148 & 0.048 \\
                  & business expectations manufacturing & 0.195 & 0.038 & public progress payments change & 0.200 & 0.240 & composite leading index change & 0.139 & 0.063 \\
                  & private properties available change & -0.160 & 0.076 & merchandise exports change & 0.160 & 0.096 & gross fixed capital formation current change & 0.128 & 0.053 \\
            \hline
            \multicolumn{1}{c}{\multirow{10}[2]{*}{\textbf{Excluding COVID}}} 
                  & employee compensation change & 1.707 & 3.161 & unit labor cost change & -1.567 & 4.234 & industrial production index change & 0.926 & 0.576 \\
                  & unit labor cost change & -1.677 & 3.141 & employee compensation change & 1.487 & 4.162 & gross operating surplus change & 0.522 & 0.365 \\
                  & industrial production index change & 0.947 & 0.451 & industrial production index change & 0.728 & 0.550 & air cargo loaded change & 0.325 & 0.184 \\
                  & gross operating surplus change & 0.666 & 0.312 & gross operating surplus change & 0.441 & 0.361 & fnb services index change & 0.252 & 0.216 \\
                  & fnb services index change & 0.290 & 0.200 & fnb services index change & 0.249 & 0.277 & taxes less subsidies change & 0.218 & 0.178 \\
                  & air cargo loaded change & 0.265 & 0.254 & taxes less subsidies change & 0.224 & 0.081 & pawnshop pledges received change & 0.202 & 0.156 \\
                  & public progress payments change & 0.260 & 0.097 & air cargo loaded change & 0.221 & 0.254 & public progress payments change & 0.182 & 0.176 \\
                  & pawnshop pledges received change & 0.230 & 0.110 & pawnshop pledges received change & 0.213 & 0.279 & private properties available change & -0.156 & 0.135 \\
                  & taxes less subsidies change & 0.225 & 0.055 & exchange rate hkd change & -0.209 & 0.452 & composite leading index change & 0.155 & 0.105 \\
                  & exchange rate hkd change & -0.169 & 0.163 & composite leading index change & 0.194 & 0.164 & business expectations manufacturing & 0.151 & 0.119 \\
            \hline
            \end{tabular}
            }
        \end{table}
        }
    
    \item \textbf{Dimensionality reduction-based models.}
            \begin{itemize}
            \item \textit{General trends.} PCR prominently features industrial production and fiscal indicators (taxes less subsidies), consistently maintaining these predictors' importance and relative ranking across all sub-periods. PLSR, on the other hand, notably emphasizes air transport variables such as air passenger arrivals and aircraft movements.
            \item \textit{Period-specific variations.} While PCR maintained relatively stable feature importance throughout all periods, PLSR highlighted air transport variables more intensely during and following the COVID period, capturing critical economic disruptions and subsequent adjustments.
            \item \textit{Excluding COVID.} In the Excluding COVID period, both PCR and PLSR reconfirmed the significance of industrial production and air transport variables. PCR maintained stability in attributing importance to fiscal indicators, while PLSR continued emphasizing air transport and international trade indicators.
        \end{itemize}
    
        {\fontsize{9pt}{10pt}
        \selectfont
        \begin{table}[htbp]
            \centering
            \caption{Dimensionality reduction based models with penalization, top 10 predictors ranked by average feature importance and 95\% confidence intervals, estimated via block bootstrap, across different sub-periods}
            \resizebox{1.01\textwidth}{!}{%
            \begin{tabular}{llcc|lcc}
                  & \multicolumn{3}{c|}{\textbf{Principal Component Regression}} & \multicolumn{3}{c}{\textbf{Partial Least Squares Regression}} \\
            \hline
                  & \multicolumn{1}{c}{\textbf{Feature}} & \multicolumn{1}{p{5.78em}}{\textbf{Feature (avg)}} & \multicolumn{1}{p{3.945em}|}{\textbf{CI range}} & \multicolumn{1}{c}{\textbf{Feature}} & \multicolumn{1}{p{5.835em}}{\textbf{Feature (avg)}} & \multicolumn{1}{p{3.61em}}{\textbf{CI range}} \\
            \hline
            \multirow{10}[1]{*}{\textbf{Overall}} 
                  & industrial production index change & 0.595 & 0.340 & industrial production index change & 2.509 & 1.162 \\
                  & taxes less subsidies change & 0.326 & 0.365 & air passenger arrivals change & 2.081 & 1.778 \\
                  & gross operating surplus change & 0.315 & 0.109 & air cargo loaded change & 2.060 & 0.481 \\
                  & merchandise reexports change & 0.312 & 0.086 & aircraft arrivals departures change & 1.968 & 1.859 \\
                  & air cargo loaded change & 0.309 & 0.118 & merchandise imports change & 1.918 & 0.717 \\
                  & merchandise imports change & 0.268 & 0.096 & gross operating surplus change & 1.829 & 0.926 \\
                  & merchandise exports change & 0.244 & 0.077 & merchandise reexports change & 1.796 & 0.633 \\
                  & composite leading index change & 0.201 & 0.072 & merchandise exports change & 1.728 & 0.704 \\
                  & liquor releases change & 0.175 & 0.110 & fnb services index change & 1.583 & 0.954 \\
                  & cpf due members change & -0.166 & 0.093 & pawnshop pledges received change & 1.327 & 0.632 \\
            \hline
            \multirow{10}[0]{*}{\textbf{Pre-COVID}} 
                  & industrial production index change & 0.657 & 0.313 & industrial production index change & 2.821 & 0.032 \\
                  & gross operating surplus change & 0.349 & 0.042 & air cargo loaded change & 2.211 & 0.102 \\
                  & taxes less subsidies change & 0.330 & 0.387 & merchandise imports change & 2.202 & 0.096 \\
                  & merchandise reexports change & 0.321 & 0.074 & gross operating surplus change & 2.152 & 0.090 \\
                  & air cargo loaded change & 0.303 & 0.047 & merchandise reexports change & 2.060 & 0.091 \\
                  & merchandise imports change & 0.266 & 0.100 & merchandise exports change & 2.018 & 0.121 \\
                  & merchandise exports change & 0.241 & 0.078 & air passenger arrivals change & 1.455 & 0.028 \\
                  & composite leading index change & 0.199 & 0.067 & domestic supply price change & 1.412 & 0.104 \\
                  & air passenger arrivals change & 0.197 & 0.042 & import price index change & 1.395 & 0.135 \\
                  & liquor releases change & 0.187 & 0.104 & aircraft arrivals departures change & 1.320 & 0.057 \\
            \hline
            \multirow{10}[1]{*}{\textbf{COVID}} 
                  & industrial production index change & 0.608 & 0.143 & industrial production index change & 2.545 & 0.609 \\
                  & taxes less subsidies change & 0.353 & 0.152 & aircraft arrivals departures change & 2.131 & 1.648 \\
                  & gross operating surplus change & 0.304 & 0.034 & air cargo loaded change & 2.042 & 0.385 \\
                  & air cargo loaded change & 0.291 & 0.030 & merchandise imports change & 1.903 & 0.542 \\
                  & merchandise reexports change & 0.288 & 0.038 & air passenger arrivals change & 1.872 & 0.984 \\
                  & merchandise imports change & 0.265 & 0.064 & gross operating surplus change & 1.852 & 0.578 \\
                  & merchandise exports change & 0.245 & 0.051 & merchandise reexports change & 1.756 & 0.522 \\
                  & composite leading index change & 0.192 & 0.040 & merchandise exports change & 1.696 & 0.496 \\
                  & air passenger arrivals change & 0.170 & 0.033 & fnb services index change & 1.581 & 0.582 \\
                  & cpf due members change & -0.165 & 0.055 & pawnshop pledges received change & 1.379 & 0.266 \\
            \hline
            \multirow{10}[2]{*}{\textbf{Post-COVID}} 
                  & industrial production index change & 0.516 & 0.163 & air passenger arrivals change & 2.917 & 0.680 \\
                  & air cargo loaded change & 0.324 & 0.125 & aircraft arrivals departures change & 2.680 & 0.800 \\
                  & taxes less subsidies change & 0.311 & 0.107 & new vehicle registration change & 2.451 & 0.691 \\
                  & merchandise reexports change & 0.310 & 0.068 & industrial production index change & 2.121 & 0.577 \\
                  & gross operating surplus change & 0.278 & 0.043 & fnb services index change & 1.920 & 0.465 \\
                  & merchandise imports change & 0.270 & 0.069 & air cargo loaded change & 1.886 & 0.283 \\
                  & merchandise exports change & 0.247 & 0.063 & merchandise imports change & 1.582 & 0.178 \\
                  & composite leading index change & 0.207 & 0.061 & pawnshop pledges received change & 1.507 & 0.346 \\
                  & liquor releases change & 0.174 & 0.107 & merchandise reexports change & 1.495 & 0.115 \\
                  & gross fixed capital formation current change & 0.168 & 0.076 & gross operating surplus change & 1.432 & 0.244 \\
            \hline
            \multicolumn{1}{c}{\multirow{10}[2]{*}{\textbf{Excluding COVID}}} 
                  & industrial production index change & 0.593 & 0.345 & industrial production index change & 2.503 & 1.165 \\
                  & taxes less subsidies change & 0.322 & 0.371 & air passenger arrivals change & 2.120 & 1.779 \\
                  & gross operating surplus change & 0.317 & 0.110 & air cargo loaded change & 2.063 & 0.483 \\
                  & merchandise reexports change & 0.316 & 0.079 & aircraft arrivals departures change & 1.938 & 1.863 \\
                  & air cargo loaded change & 0.313 & 0.117 & merchandise imports change & 1.920 & 0.718 \\
                  & merchandise imports change & 0.268 & 0.098 & gross operating surplus change & 1.825 & 0.927 \\
                  & merchandise exports change & 0.244 & 0.078 & merchandise reexports change & 1.803 & 0.634 \\
                  & composite leading index change & 0.202 & 0.074 & merchandise exports change & 1.734 & 0.706 \\
                  & liquor releases change & 0.181 & 0.111 & fnb services index change & 1.584 & 0.947 \\
                  & cpf due members change & -0.166 & 0.099 & new vehicle registration change & 1.326 & 2.485 \\
            \hline
            \end{tabular}
            }
        \end{table}
        }

    \item \textbf{Ensemble learning models.}
        \begin{itemize}
            \item \textit{General trends.} Consistently appearing as dominant predictors in both RF and XGB models are industrial production index and air cargo loaded. XGB consistently assigns the highest importance to air cargo loaded across all periods, with narrower confidence intervals signaling stability. In contrast, RF significantly emphasizes this indicator only in the Post-COVID period.
            \item \textit{Period-specific variations.} Random Forest notably increased reliance on air cargo loaded in the Post-COVID period, clearly capturing structural shifts in economic dynamics. XGB's predictor importance remains stable across all sub-periods, underscoring its robustness to economic turbulence.
            \item \textit{Excluding COVID.} The Excluding COVID period shows strong convergence in ensemble models on international trade indicators (particularly air cargo) and industrial production, confirming robustness in variable selection outside pandemic conditions.
        \end{itemize}

        {\fontsize{9pt}{10pt}
        \selectfont
        \begin{table}[htbp]
            \centering
            \caption{Ensemble learning models with penalization, top 10 predictors ranked by average feature importance and 95\% confidence intervals, estimated via block bootstrap, across different sub-periods}
            \resizebox{1.0\textwidth}{!}{
            \begin{tabular}{clcc|lcc}
                  & \multicolumn{3}{c|}{\textbf{Random Forest}} & \multicolumn{3}{c}{\textbf{eXtreme Gradient Boosting}} \\
            \hline
                & \multicolumn{1}{c}{\textbf{Feature}} & \multicolumn{1}{c}{\textbf{Feature (avg)}} & \multicolumn{1}{c|}{\textbf{CI range}} & \multicolumn{1}{c}{\textbf{Feature}} & \multicolumn{1}{c}{\textbf{Feature (avg)}} & \multicolumn{1}{c}{\textbf{CI range}} \\
            \hline
            \multirow{10}[1]{*}{\textbf{Overall}} 
                & industrial production index change & 0.224 & 0.355 & air cargo loaded change            & 0.097 & 0.098 \\
                & air cargo loaded change            & 0.217 & 0.431 & merchandise imports change         & 0.066 & 0.116 \\
                & merchandise imports change         & 0.064 & 0.114 & air passenger arrivals change      & 0.062 & 0.083 \\
                & air passenger arrivals change      & 0.046 & 0.091 & merchandise reexports change       & 0.058 & 0.049 \\
                & gross operating surplus change     & 0.042 & 0.104 & industrial production index change & 0.054 & 0.066 \\
                & merchandise reexports change       & 0.036 & 0.071 & gross operating surplus change     & 0.051 & 0.074 \\
                & business expectations manufacturing & 0.030 & 0.057 & aircraft arrivals departures change & 0.050 & 0.080 \\
                & gross fixed capital formation current change & 0.021 & 0.068 & merchandise exports change         & 0.044 & 0.092 \\
                & merchandise exports change         & 0.021 & 0.045 & business expectations manufacturing & 0.027 & 0.024 \\
                & consumer price index change        & 0.020 & 0.047 & fnb services index change          & 0.019 & 0.027 \\
            \hline
            \multirow{10}[1]{*}{\textbf{Pre-COVID}}
                & industrial production index change & 0.262 & 0.383 & air cargo loaded change            & 0.084 & 0.073 \\
                & air cargo loaded change            & 0.093 & 0.070 & gross operating surplus change     & 0.065 & 0.066 \\
                & air passenger arrivals change      & 0.070 & 0.049 & air passenger arrivals change      & 0.062 & 0.082 \\
                & gross operating surplus change     & 0.070 & 0.081 & merchandise reexports change       & 0.061 & 0.048 \\
                & merchandise reexports change       & 0.055 & 0.034 & aircraft arrivals departures change & 0.057 & 0.069 \\
                & merchandise imports change         & 0.043 & 0.033 & industrial production index change & 0.055 & 0.056 \\
                & merchandise exports change         & 0.035 & 0.027 & merchandise imports change         & 0.051 & 0.101 \\
                & business expectations manufacturing & 0.024 & 0.026 & merchandise exports change         & 0.046 & 0.094 \\
                & aircraft arrivals departures change & 0.023 & 0.026 & business expectations manufacturing & 0.028 & 0.022 \\
                & pawnshop pledges received change   & 0.022 & 0.023 & employee compensation change       & 0.021 & 0.035 \\
            \hline
            \multirow{10}[2]{*}{\textbf{COVID}}
                & industrial production index change & 0.154 & 0.064 & air cargo loaded change            & 0.095 & 0.063 \\
                & air cargo loaded change            & 0.117 & 0.098 & merchandise imports change         & 0.066 & 0.052 \\
                & air passenger arrivals change      & 0.065 & 0.017 & air passenger arrivals change      & 0.057 & 0.043 \\
                & merchandise reexports change       & 0.043 & 0.027 & merchandise reexports change       & 0.052 & 0.013 \\
                & gross operating surplus change     & 0.042 & 0.041 & aircraft arrivals departures change & 0.050 & 0.059 \\
                & merchandise imports change         & 0.041 & 0.028 & merchandise exports change         & 0.048 & 0.038 \\
                & business expectations manufacturing & 0.031 & 0.012 & gross operating surplus change     & 0.047 & 0.031 \\
                & aircraft arrivals departures change & 0.027 & 0.023 & industrial production index change & 0.042 & 0.055 \\
                & merchandise exports change         & 0.025 & 0.034 & employee compensation change       & 0.035 & 0.084 \\
                & pawnshop pledges received change   & 0.023 & 0.019 & business expectations manufacturing & 0.026 & 0.011 \\
            \hline
            \multirow{10}[2]{*}{\textbf{Post-COVID}}
                & air cargo loaded change            & 0.405 & 0.254 & air cargo loaded change            & 0.112 & 0.085 \\
                & industrial production index change & 0.208 & 0.109 & merchandise imports change         & 0.085 & 0.106 \\
                & merchandise imports change         & 0.099 & 0.109 & air passenger arrivals change      & 0.066 & 0.068 \\
                & consumer price index change        & 0.038 & 0.042 & industrial production index change & 0.058 & 0.063 \\
                & business expectations manufacturing & 0.037 & 0.063 & merchandise reexports change       & 0.056 & 0.044 \\
                & gross fixed capital formation current change & 0.028 & 0.083 & merchandise exports change         & 0.041 & 0.060 \\
                & pawnshop loans given change        & 0.018 & 0.046 & aircraft arrivals departures change & 0.040 & 0.062 \\
                & pawnshop pledges received change   & 0.017 & 0.034 & gross operating surplus change     & 0.037 & 0.047 \\
                & merchandise reexports change       & 0.011 & 0.060 & business expectations manufacturing & 0.026 & 0.022 \\
                & air passenger arrivals change      & 0.011 & 0.031 & private progress payments change   & 0.024 & 0.046 \\
            \hline
            \multicolumn{1}{c}{\multirow{10}[2]{*}{\textbf{Excluding COVID}}}
                & industrial production index change & 0.237 & 0.348 & air cargo loaded change            & 0.097 & 0.100 \\
                & air cargo loaded change            & 0.235 & 0.438 & merchandise imports change         & 0.066 & 0.119 \\
                & merchandise imports change         & 0.068 & 0.117 & air passenger arrivals change      & 0.063 & 0.084 \\
                & air passenger arrivals change      & 0.043 & 0.092 & merchandise reexports change       & 0.059 & 0.050 \\
                & gross operating surplus change     & 0.042 & 0.109 & industrial production index change & 0.057 & 0.064 \\
                & merchandise reexports change       & 0.035 & 0.072 & gross operating surplus change     & 0.052 & 0.076 \\
                & business expectations manufacturing & 0.030 & 0.059 & aircraft arrivals departures change & 0.050 & 0.075 \\
                & gross fixed capital formation current change & 0.022 & 0.071 & merchandise exports change         & 0.044 & 0.093 \\
                & consumer price index change        & 0.021 & 0.047 & business expectations manufacturing & 0.028 & 0.024 \\
                & merchandise exports change         & 0.020 & 0.045 & fnb services index change          & 0.019 & 0.021 \\
            \hline
            \end{tabular}
            }
        \end{table}
        }
    
    \item \textbf{Neural network models.}
        \begin{itemize}
            \item \textit{General trends.} MLP and GRU prominently feature industrial production, gross operating surplus, and unit labor cost. GRU additionally assigns considerable importance to employee compensation and pawnshop pledges, consistently showing less feature importance variability than MLP across all periods.
            \item \textit{Period-specific variations.} During COVID, neural networks emphasized labor and production indicators, demonstrating high sensitivity to macroeconomic disruptions. GRU exhibited stable predictor importance, whereas MLP's feature relevance displayed notably higher variability, particularly under volatile economic conditions, particularly under volatile conditions.
            \item \textit{Excluding COVID.} Without the COVID period, neural models emphasize the consistent relevance of industrial production, gross operating surplus, and unit labor cost. GRU confirms stability in feature selection, whereas MLP continues to reflect slightly greater variability.
        \end{itemize}
        
        {\fontsize{9pt}{10pt}
        \selectfont
        \begin{table}[htbp]
            \centering
            \caption{Deep learning models with penalization, top 10 predictors ranked by average feature importance and 95\% confidence intervals, estimated via block bootstrap, across different sub-periods}
            \resizebox{1.0\textwidth}{!}{
            \begin{tabular}{clcc|lcc}
        
                  & \multicolumn{3}{c|}{\textbf{Multilayer Perceptron}} & \multicolumn{3}{c}{\textbf{Gated Recurrent Units}} \\
            \hline
                & \multicolumn{1}{c}{\textbf{Feature}} & \multicolumn{1}{c}{\textbf{Feature (avg)}} & \multicolumn{1}{c|}{\textbf{CI range}} & \multicolumn{1}{c}{\textbf{Feature}} & \multicolumn{1}{c}{\textbf{Feature (avg)}} & \multicolumn{1}{c}{\textbf{CI range}} \\
            \hline
            \multirow{10}[1]{*}{\textbf{Overall}} 
                  & industrial production index change & 0.395 & 0.615 & industrial production index change & 0.670 & 0.499 \\
                  & gross operating surplus change & 0.246 & 0.369 & gross operating surplus change & 0.409 & 0.293 \\
                  & air cargo loaded change & 0.180 & 0.231 & unit labor cost change & 0.346 & 0.302 \\
                  & business expectations manufacturing & 0.149 & 0.230 & employee compensation change & 0.285 & 0.303 \\
                  & unit labor cost change & 0.140 & 0.224 & air cargo loaded change & 0.285 & 0.187 \\
                  & composite leading index change & 0.124 & 0.152 & pawnshop pledges received change & 0.265 & 0.238 \\
                  & merchandise imports change & 0.118 & 0.149 & composite leading index change & 0.255 & 0.174 \\
                  & fnb services index change & 0.110 & 0.183 & private properties vacant change & 0.224 & 0.257 \\
                  & merchandise reexports change & 0.109 & 0.122 & fnb services index change & 0.221 & 0.220 \\
                  & merchandise exports change & 0.105 & 0.132 & business expectations manufacturing & 0.218 & 0.161 \\
            \hline
            \multirow{10}[1]{*}{\textbf{Pre-COVID}}
                  & industrial production index change & 0.486 & 0.528 & industrial production index change & 0.653 & 0.473 \\
                  & gross operating surplus change & 0.314 & 0.299 & gross operating surplus change & 0.440 & 0.270 \\
                  & air cargo loaded change & 0.211 & 0.174 & unit labor cost change & 0.370 & 0.209 \\
                  & unit labor cost change & 0.178 & 0.157 & employee compensation change & 0.279 & 0.268 \\
                  & business expectations manufacturing & 0.169 & 0.188 & air cargo loaded change & 0.275 & 0.146 \\
                  & composite leading index change & 0.143 & 0.155 & composite leading index change & 0.269 & 0.147 \\
                  & merchandise imports change & 0.135 & 0.118 & private properties vacant change & 0.248 & 0.217 \\
                  & fnb services index change & 0.115 & 0.164 & pawnshop pledges received change & 0.235 & 0.125 \\
                  & employee compensation change & 0.114 & 0.196 & business expectations manufacturing & 0.213 & 0.172 \\
                  & merchandise reexports change & 0.108 & 0.098 & merchandise imports change & 0.213 & 0.159 \\
            \hline
            \multirow{10}[2]{*}{\textbf{COVID}}
                  & industrial production index change & 0.429 & 0.356 & industrial production index change & 0.590 & 0.434 \\
                  & gross operating surplus change & 0.238 & 0.168 & gross operating surplus change & 0.422 & 0.276 \\
                  & air cargo loaded change & 0.187 & 0.156 & unit labor cost change & 0.370 & 0.273 \\
                  & business expectations manufacturing & 0.172 & 0.123 & employee compensation change & 0.308 & 0.278 \\
                  & composite leading index change & 0.146 & 0.102 & air cargo loaded change & 0.265 & 0.063 \\
                  & unit labor cost change & 0.142 & 0.127 & composite leading index change & 0.194 & 0.054 \\
                  & merchandise imports change & 0.138 & 0.091 & pawnshop pledges received change & 0.183 & 0.112 \\
                  & merchandise exports change & 0.127 & 0.043 & business expectations manufacturing & 0.179 & 0.019 \\
                  & merchandise reexports change & 0.126 & 0.039 & private properties vacant change & 0.163 & 0.030 \\
                  & gross fixed capital formation current change & 0.117 & 0.099 & merchandise imports change & 0.163 & 0.031 \\
            \hline
            \multirow{10}[2]{*}{\textbf{Post-COVID}}
                  & industrial production index change & 0.272 & 0.496 & industrial production index change & 0.722 & 0.200 \\
                  & gross operating surplus change & 0.168 & 0.285 & gross operating surplus change & 0.367 & 0.118 \\
                  & air cargo loaded change & 0.140 & 0.198 & pawnshop pledges received change & 0.334 & 0.131 \\
                  & business expectations manufacturing & 0.116 & 0.208 & unit labor cost change & 0.308 & 0.257 \\
                  & merchandise reexports change & 0.103 & 0.130 & air cargo loaded change & 0.304 & 0.185 \\
                  & fnb services index change & 0.101 & 0.178 & fnb services index change & 0.285 & 0.152 \\
                  & merchandise exports change & 0.095 & 0.133 & employee compensation change & 0.283 & 0.287 \\
                  & unit labor cost change & 0.093 & 0.197 & composite leading index change & 0.264 & 0.142 \\
                  & composite leading index change & 0.093 & 0.123 & business expectations manufacturing & 0.238 & 0.082 \\
                  & merchandise imports change & 0.091 & 0.125 & private properties vacant change & 0.219 & 0.246 \\
            \hline
            \multicolumn{1}{c}{\multirow{10}[2]{*}{\textbf{Excluding COVID}}}
                  & industrial production index change & 0.389 & 0.621 & industrial production index change & 0.684 & 0.445 \\
                  & gross operating surplus change & 0.248 & 0.374 & gross operating surplus change & 0.407 & 0.267 \\
                  & air cargo loaded change & 0.179 & 0.233 & unit labor cost change & 0.342 & 0.269 \\
                  & business expectations manufacturing & 0.145 & 0.233 & air cargo loaded change & 0.288 & 0.195 \\
                  & unit labor cost change & 0.140 & 0.225 & employee compensation change & 0.281 & 0.297 \\
                  & composite leading index change & 0.120 & 0.157 & pawnshop pledges received change & 0.280 & 0.218 \\
                  & merchandise imports change & 0.115 & 0.148 & composite leading index change & 0.266 & 0.160 \\
                  & fnb services index change & 0.109 & 0.186 & private properties vacant change & 0.235 & 0.255 \\
                  & merchandise reexports change & 0.106 & 0.125 & fnb services index change & 0.233 & 0.222 \\
                  & merchandise exports change & 0.101 & 0.133 & business expectations manufacturing & 0.225 & 0.166 \\
            \hline
            \end{tabular}%
            }
        \end{table}
        }

\end{enumerate}

\begin{figure}[H]
\centering
    \begin{subfigure}[t]{0.95\textwidth}
        \includegraphics[width=\textwidth]{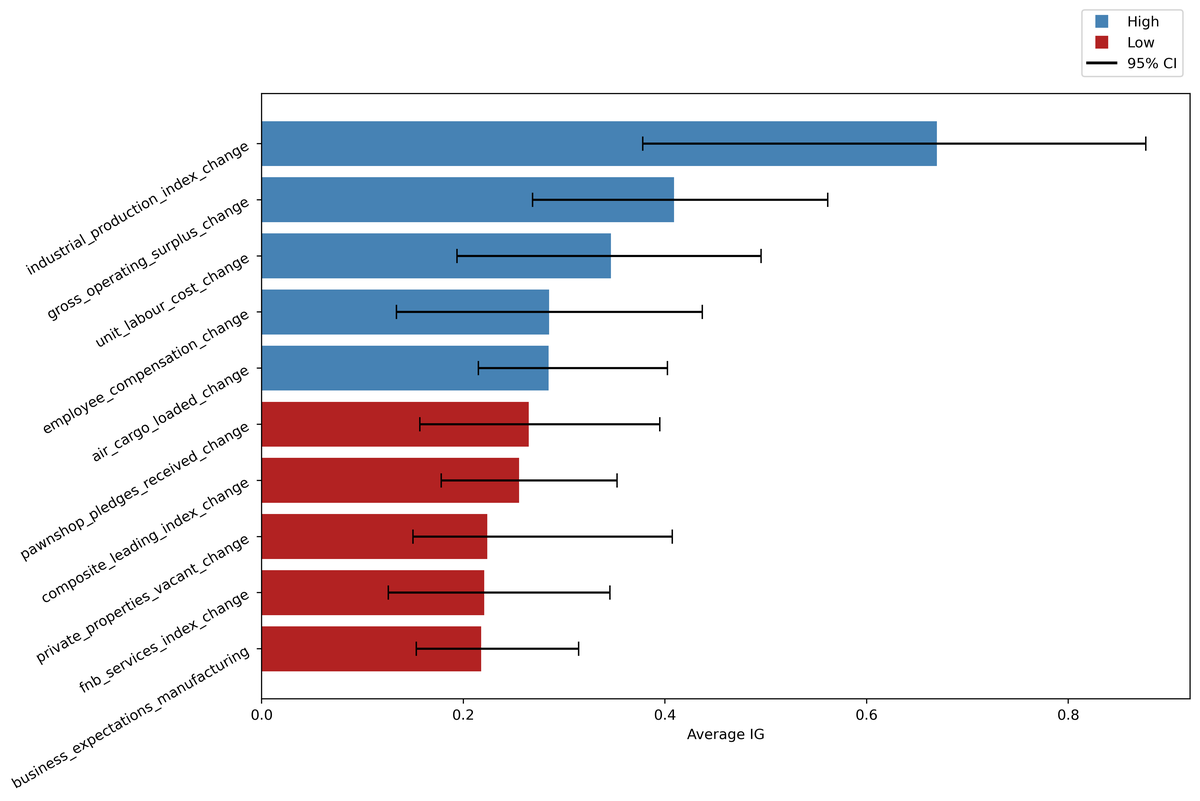}
        \caption{Overall}
    \end{subfigure}

\vspace{0.2cm}
    
    \begin{subfigure}[t]{0.95\textwidth}
        \includegraphics[width=\textwidth]{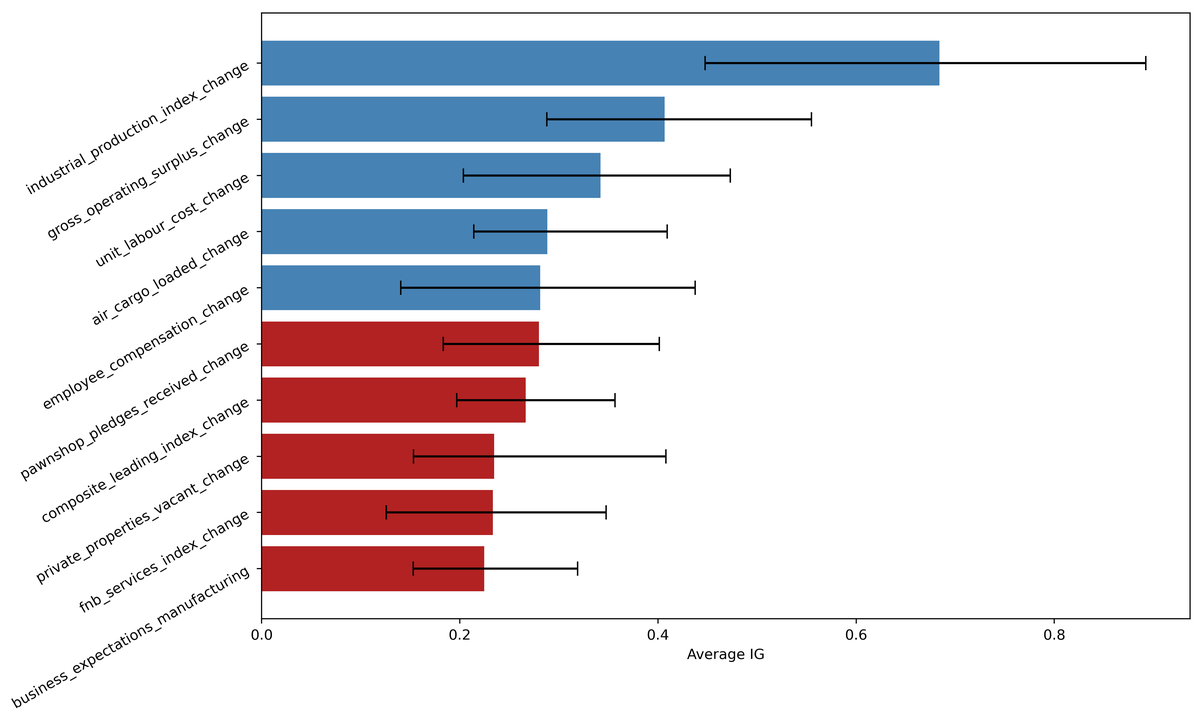}
        \caption{Excluding COVID}
    \end{subfigure}

\caption{GRU model, top 10 explanatory variables ranked by average Integrated Gradients in the Overall and Excluding COVID periods. Variables are color-coded based on importance magnitude (High, Low). Bars represent the average Integrated Gradients values; error bars denote the 95\% confidence intervals obtained via block bootstrap.}
\label{fig:gru_featureimportance_a}
\end{figure}

\begin{figure}[H]
\centering
    \begin{subfigure}[t]{0.95\textwidth}
        \includegraphics[width=\textwidth]{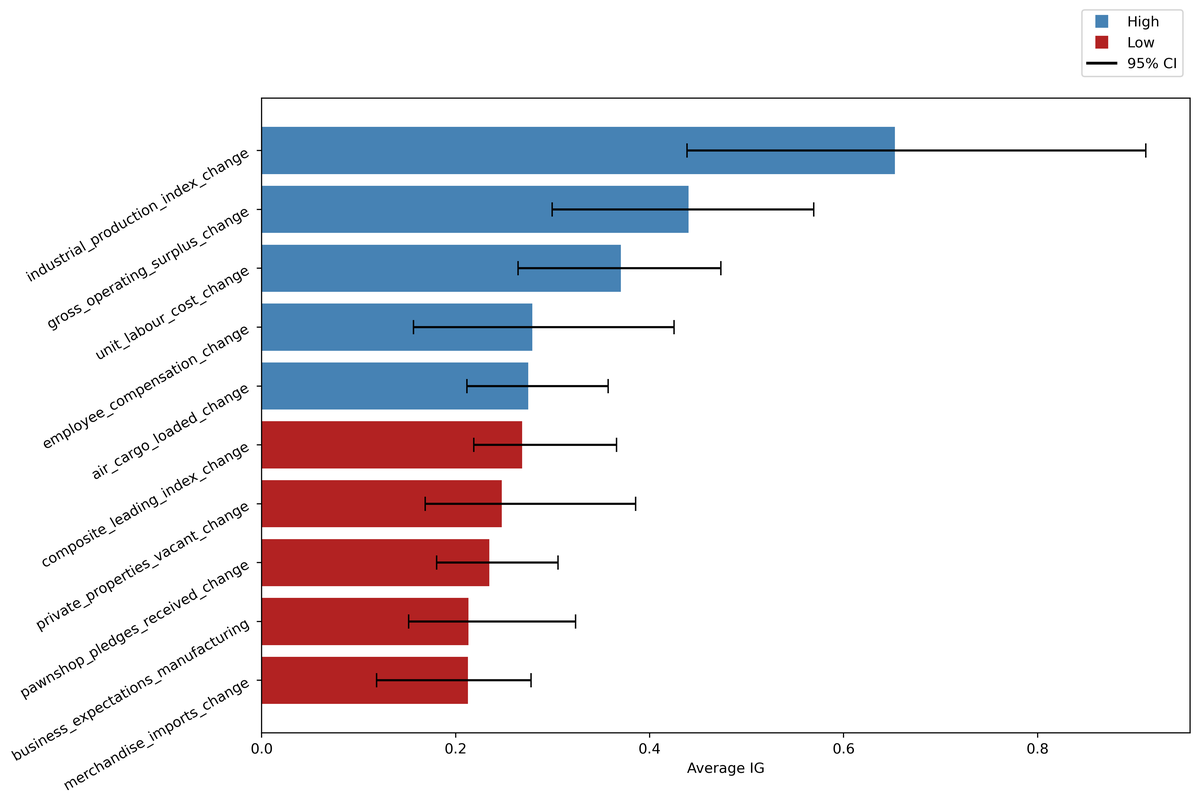}
        \caption{Pre-COVID}
    \end{subfigure}

    \vspace{0.2cm}
    
    \begin{subfigure}[t]{0.95\textwidth}
        \includegraphics[width=\textwidth]{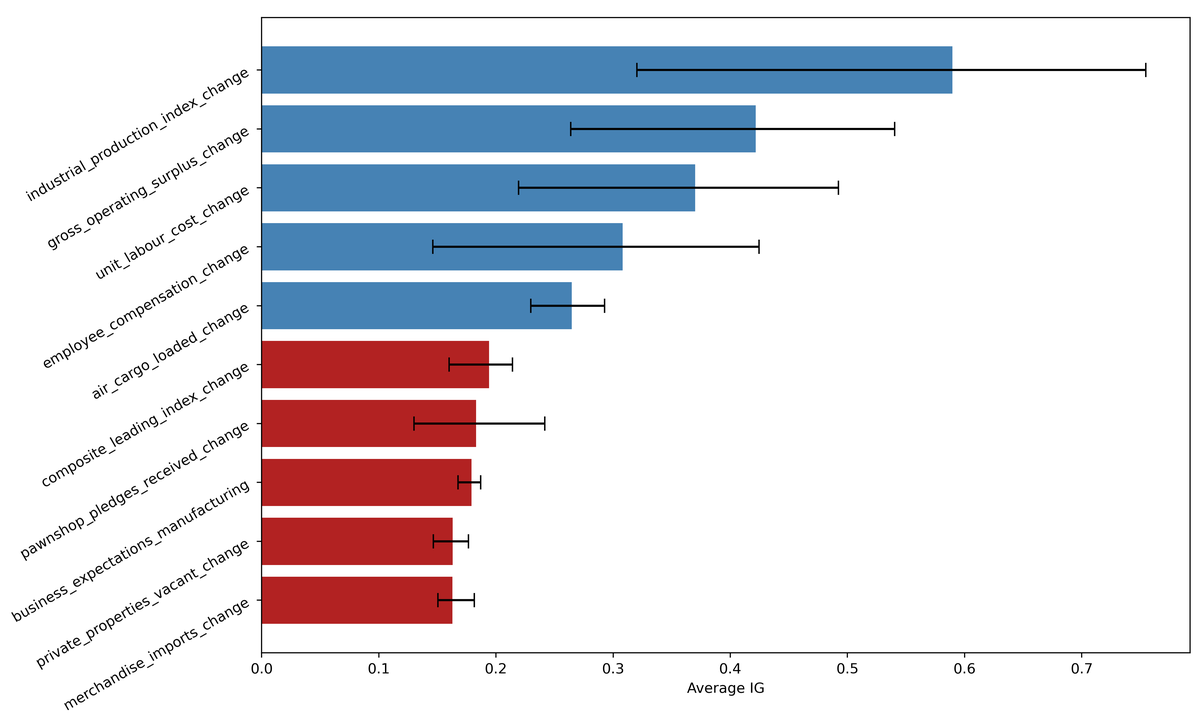}
        \caption{COVID}
    \end{subfigure}
\caption{GRU model, top 10 explanatory variables ranked by average Integrated Gradients in Pre-COVID and COVID periods. Variables are color-coded based on importance magnitude (High, Low). Bars represent the average Integrated Gradients values; error bars denote the 95\% confidence intervals obtained via block bootstrap.}
\label{fig:gru_featureimportance_b}
\end{figure}

\begin{figure}[H]
    \centering
    \begin{subfigure}[t]{0.95\textwidth}
        \includegraphics[width=\textwidth]{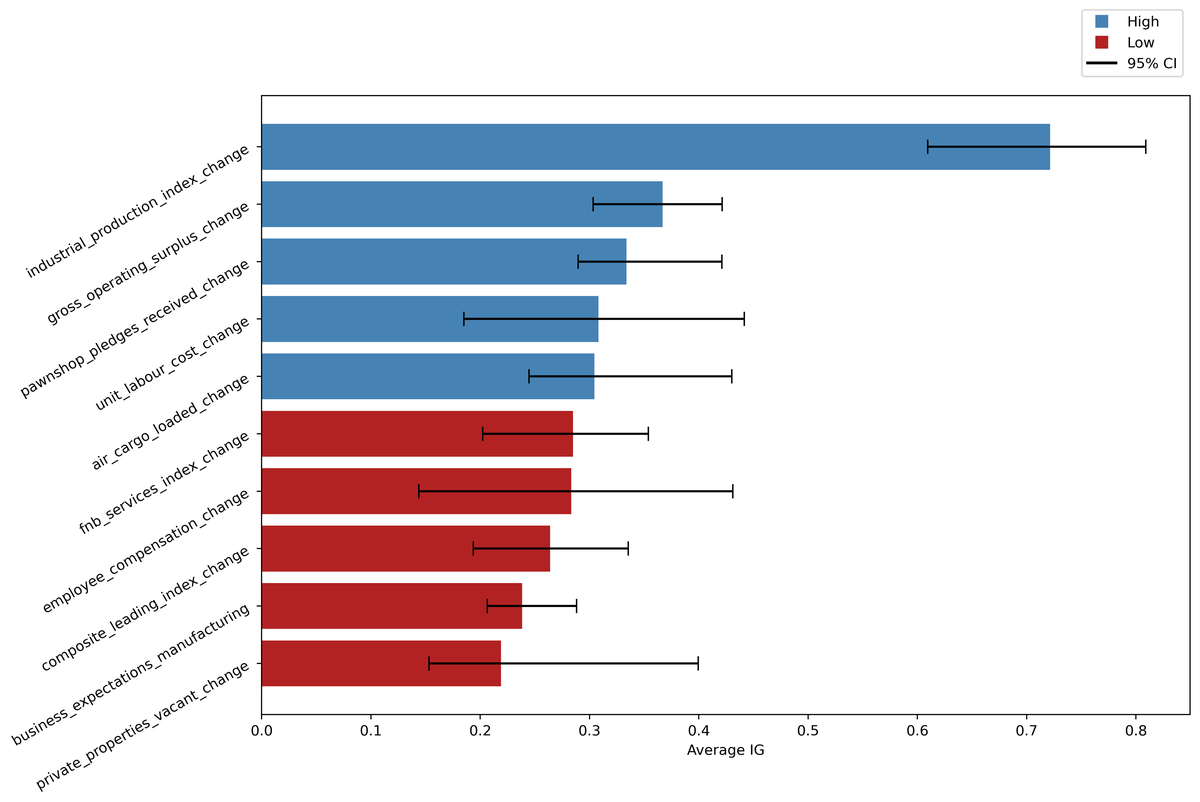}
        \caption{Post-COVID}
    \end{subfigure}
\caption{GRU model, top 10 explanatory variables ranked by average Integrated Gradients in Post-COVID period. Variables are color-coded based on importance magnitude (High, Low). Bars represent the average Integrated Gradients values; error bars denote the 95\% confidence intervals obtained via block bootstrap.}
\label{fig:gru_featureimportance_c}
\end{figure}

Figure~\ref{fig:gru_featureimportance_a} to Figure~\ref{fig:gru_featureimportance_c} illustrate the top 10 explanatory variables identified by GRU for each analyzed period, together with their 95\% confidence intervals.

The graphical representation clearly shows stability and variability in selecting key predictors across macroeconomic regimes, consistently reflecting the descriptive results summarized in Table Y.

The reported patterns highlight differences in feature importance attribution among Penalized linear, Dimensionality reduction-based, Ensemble learning, and Neural network models across varying macroeconomic regimes.

Penalized linear and Dimensionality reduction-based approaches consistently identify industrial production, trade indicators, and labor market variables among their most influential predictors, maintaining stable relevance across various sub-periods.

Ensemble learning and neural network models display variability in feature importance, reflecting differences in how these methodologies select and emphasize predictors across economic regimes. Such variability is particularly evident during significant macroeconomic fluctuations (e.g., COVID).

\begin{figure}[H] 
\centering 
    \includegraphics[width=0.99\textwidth]{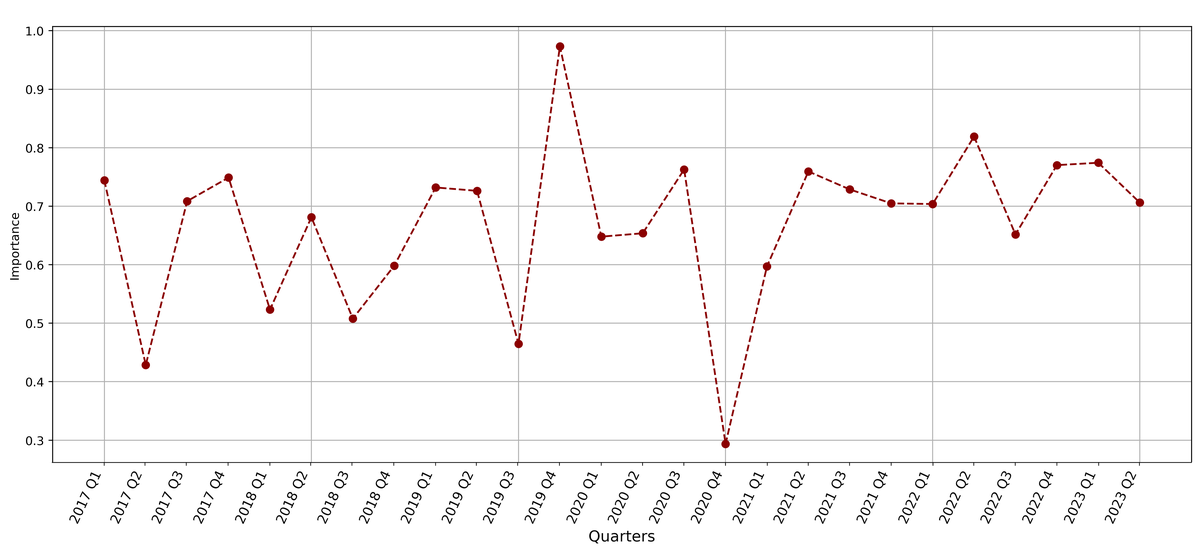} 
    \caption{Quarterly evolution of the feature importance (Integrated Gradients) of the variable industrial production index change in the GRU model. Importance is estimated for each quarter between 2017 Q1 and 2023 Q2.}
    \label{fig:gru_importance_evolution_industrial_production_index_change}
\end{figure}

Figure~\ref{fig:gru_importance_evolution_industrial_production_index_change} provides an additional perspective to the static analyses presented. It demonstrates, for illustrative purposes, the quarterly evolution of the feature importance calculated via Integrated Gradients for the variable industrial production index change in the GRU model. This representation highlights the dynamic reactivity of the model in correspondence with relevant economic events, such as the onset and development of the COVID-19 pandemic.

\begin{figure}[H]
\centering
    \includegraphics[width=0.99\textwidth]{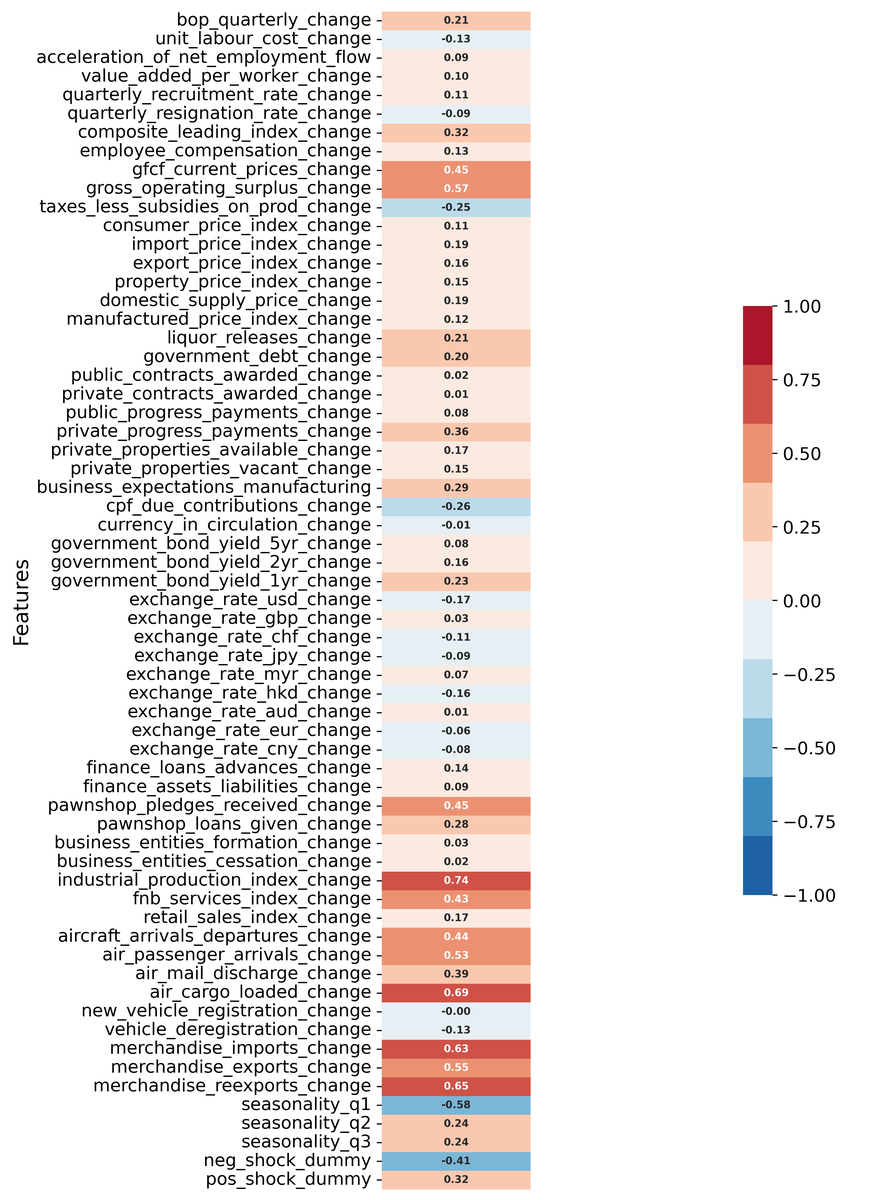}
    \caption{Spearman correlations between the predictors and the target variable. Predictors are sorted in the same order as they appear in the dataset. Positive values (red) indicate positive correlations with GDP, while negative values (blue) indicate inverse correlations.}
    \label{fig:spearman_heatmap}
\end{figure}

To further contextualize the relevance of the explanatory variables that emerged from the previous analyses, Figure~\ref{fig:spearman_heatmap} shows the Spearman correlation between each predictor variable and the target variable (quarterly Singapore's GDP growth). This representation highlights that some of the variables consistently selected by the models, such as industrial production, air cargo loaded and pawnshop pledges received, show statistically significant correlations with GDP, empirically confirming the robustness and coherence of the selections made by the analyzed models.

\vspace{0.35cm}

\subsection{\textbf{\textit{Model Selection and Aggregated Predictions}}}
\label{subsec:modelselection}
Nowcast combination methods enhance prediction accuracy and robustness, mitigating model-specific errors and uncertainties \parencite{BatesGranger1969, Timmermann2006, MakridakisEtAl2020_M4}. In our approach, we combined predictions from models selected through the Model Confidence Set (MCS) procedure \parencite{HansenLundeNason2011} into three aggregation strategies: Simple Average (SA), Weighted Average (WA), and Exponentially Weighted Average (EWA). WA and EWA explicitly track dynamic model weights, substantially enhancing interpretability.

The SA approach assigns equal weights to all selected models, disregarding differences in individual model performance. While simple, this method provides a robust benchmark, effectively mitigating risks associated with over-reliance on any nowcasting approach. In contrast, WA assigns static weights to individual models based on their relative predictive performance over the entire historical period. Differently, EWA dynamically updates model weights, assigning greater importance to recent nowcast performances through exponential smoothing. The dynamic weighting scheme of EWA explicitly captures structural breaks and evolving economic conditions, significantly enhancing explainability by transparently illustrating shifts in each model's importance over time.

Initially, the MCS procedure identified a subset of models demonstrating statistically indistinguishable predictive ability. The MCS selected penalized linear regression models, dimensionality reduction techniques, and GRU among neural networks (excluding ensemble learning models and MLP), leading to the subsequent application of these selected models for aggregation.

\begin{figure}[H]
\centering
    \begin{subfigure}[t]{0.71\textwidth}
        \includegraphics[width=\textwidth]{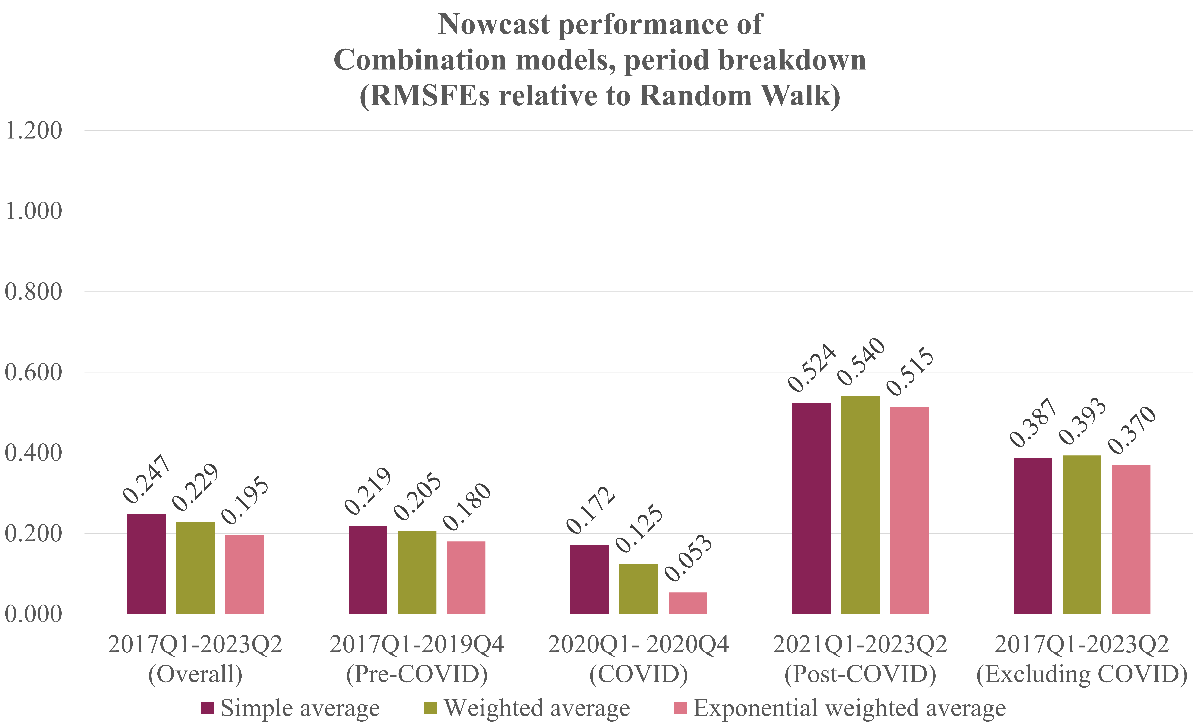}
        \caption{Benchmark: Random Walk}
    \end{subfigure}

    \vspace{0.2cm}
    
    \begin{subfigure}[t]{0.71\textwidth}
        \includegraphics[width=\textwidth]{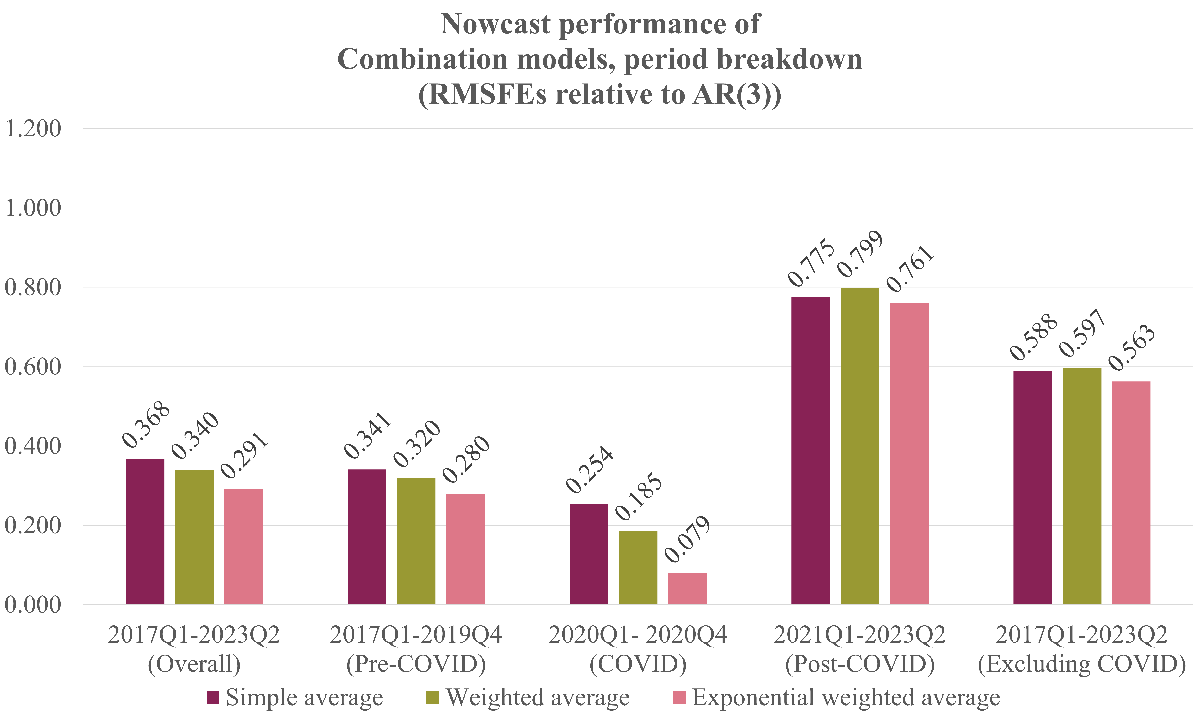}
        \caption{Benchmark: AR(3)}
    \end{subfigure}
    
    \vspace{0.2cm}
    
    \begin{subfigure}[t]{0.71\textwidth}
        \includegraphics[width=\textwidth]{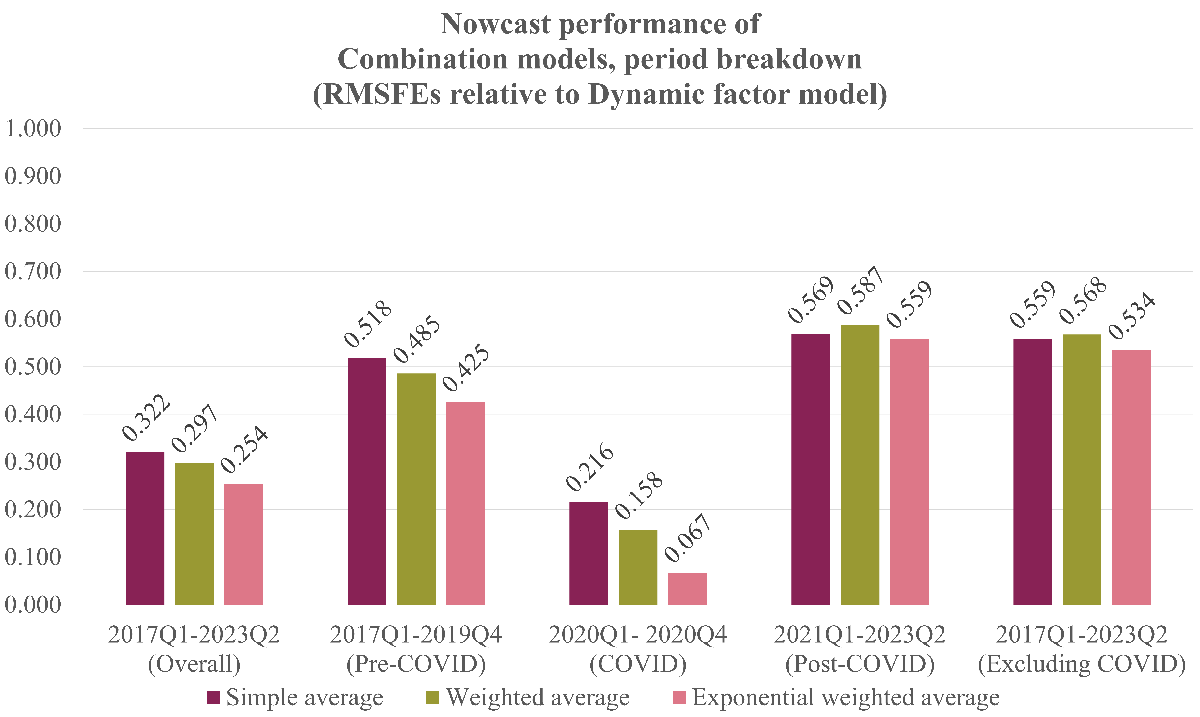}
        \caption{Benchmark: Dynamic Factor Model}
    \end{subfigure}
\caption{Nowcast performance (RMSFE ratios relative to benchmarks) for Simple Average, Weighted Average, and Exponential Weighted Average aggregation strategies across different macroeconomic regimes.}
\label{fig:weighted_relrmsfe}
\end{figure}

Empirical findings, summarized visually in Figure~\ref{fig:weighted_relrmsfe} and detailed quantitatively in Tables~\ref{tab:forecast_combinedmodels} and~\ref{tab:relativePerformance_combinedmodels}, highlighted that:

\begin{itemize}
\item All combined methods (SA, WA, and EWA) improved prediction accuracy relative to individual baseline benchmarks, i.e., the Random Walk, AR(3), and the Dynamic Factor Model, in overall prediction windows and sub-periods.
\item Among the aggregation strategies, EWA consistently provided lower prediction errors compared to WA and SA across various sub-periods, especially during the COVID-19 crisis (2020 Q1-2020 Q4). This improved performance is directly attributable to its dynamically adaptive weighting scheme. It also serves as a valuable interpretative tool by reflecting each model's changing relevance during economic shifts
\item SA offered stable improvements, mitigating misspecification risks through equal weighting. Its static weights, however, led to comparatively higher errors during pronounced instability, underscoring EWA's adaptive advantage. Still, SA occasionally outperformed WA post-COVID, reaffirming its robustness under moderate uncertainty.
\item RMSFE results highlighted notably lower prediction errors for EWA during the COVID-19 crisis than all benchmark models.
\end{itemize}

{
\fontsize{9pt}{10pt}\selectfont
\begin{table}[htbp]
    \centering
    \caption{Forecasting performance metrics (MSFE, RMSFE, and MAFE) of combined models (Simple Average, Weighted Average, and Exponentially Weighted Average) across sub-periods}
    \label{tab:forecast_combinedmodels}

    \resizebox{0.73\textwidth}{!}{
    \begin{tabular}{l c c c c c}
        \multicolumn{6}{l}{\textbf{Combinations models}} \\
    \multicolumn{6}{c}{\textbf{Simple average}} \\
    \midrule
    \textbf{Metric} & \textbf{Overall} & \textbf{Pre-COVID} & \textbf{COVID} & \textbf{Post-COVID} & \textbf{Excl. COVID} \\
    \cmidrule(lr){2-6}
    MSFE  & 1.671 & 0.412 & 3.852 & 2.309 & 1.274 \\
    RMSFE & 1.293 & 0.641 & 1.963 & 1.520 & 1.129 \\
    MAFE  & 0.918 & 0.527 & 1.417 & 1.188 & 0.827 \\
    \addlinespace
    
    \multicolumn{6}{c}{\textbf{Weighted average}} \\
    \midrule
    \textbf{Metric} & \textbf{Overall} & \textbf{Pre-COVID} & \textbf{COVID} & \textbf{Post-COVID} & \textbf{Excl. COVID} \\
    \cmidrule(lr){2-6}
    MSFE  & 1.427 & 0.362 & 2.052 & 2.456 & 1.314 \\
    RMSFE & 1.195 & 0.601 & 1.433 & 1.567 & 1.146 \\
    MAFE  & 0.877 & 0.497 & 1.132 & 1.231 & 0.831 \\
    \addlinespace

    \multicolumn{6}{c}{\textbf{Exponentially weighted average}} \\
    \midrule
    \textbf{Metric} & \textbf{Overall} & \textbf{Pre-COVID} & \textbf{COVID} & \textbf{Post-COVID} & \textbf{Excl. COVID} \\
    \cmidrule(lr){2-6}
    MSFE  & 1.043 & 0.278 & 0.372 & 2.229 & 1.165 \\
    RMSFE & 1.021 & 0.527 & 0.610 & 1.493 & 1.079 \\
    MAFE  & 0.730 & 0.433 & 0.462 & 1.194 & 0.779 \\
    \end{tabular}
    }
\end{table}
}

{
\fontsize{9pt}{10pt}\selectfont

\begin{table}[htbp]
    \centering
    \caption{Forecasting performance (RMSFE ratios) of selected individual models (from MCS) and combination models relative to benchmarks (Random Walk, AR(3), Dynamic Factor Model) across distinct sub-periods.}
    \label{tab:relativePerformance_combinedmodels}

    \resizebox{0.96\textwidth}{!}{
    \begin{tabular}{lc c c c c}

    \multicolumn{6}{l}{\textbf{RMSFE, relative to Random walk}} \\
    & \textbf{Overall} & \textbf{Pre-COVID} & \textbf{COVID} & \textbf{Post-COVID} & \textbf{Excluding COVID} \\
    \cmidrule(lr){2-6}
    \textbf{Best models (MCS)} &       &       &       &       &  \\
    LASSO & 0.427 & 0.215 & 0.446 & 0.493 & 0.367 \\
    Ridge & 0.277 & 0.276 & 0.164 & 0.621 & 0.464 \\
    Elastic net & 0.247 & 0.264 & 0.166 & 0.510 & 0.394 \\
    Principal component regression & 0.280 & 0.310 & 0.088 & 0.706 & 0.526 \\
    Partial least squares regression & 0.281 & 0.260 & 0.201 & 0.576 & 0.432 \\
    Gated recurrent unit & 0.445 & 0.217 & 0.477 & 0.449 & 0.342 \\
    \midrule
    \midrule
    \textbf{Combination models} &       &       &       &       &  \\
    Simple average & 0.247 & 0.219 & 0.172 & 0.524 & 0.387 \\
    Weighted average & 0.229 & 0.205 & 0.125 & 0.540 & 0.393 \\
    Exponential weighted average & 0.195 & 0.180 & 0.053 & 0.515 & 0.370 \\
    \midrule
    \multicolumn{1}{l}{} &       &       &       &       &  \\

    \multicolumn{6}{l}{\textbf{RMSFE, relative to AR(3)}} \\
    & \textbf{Overall} & \textbf{Pre-COVID} & \textbf{COVID} & \textbf{Post-COVID} & \textbf{Excluding COVID} \\
    \cmidrule(lr){2-6}
    \textbf{Best models (MCS)} &       &       &       &       &  \\
    LASSO & 0.636 & 0.334 & 0.660 & 0.728 & 0.557 \\
    Ridge & 0.412 & 0.430 & 0.243 & 0.918 & 0.705 \\
    Elastic net & 0.368 & 0.411 & 0.245 & 0.754 & 0.599 \\
    Principal component regression & 0.418 & 0.482 & 0.129 & 1.044 & 0.800 \\
    Partial least squares regression & 0.418 & 0.404 & 0.297 & 0.851 & 0.656 \\
    Gated recurrent unit & 0.663 & 0.338 & 0.705 & 0.664 & 0.519 \\
    \midrule
    \midrule
    \textbf{Combination models} &       &       &       &       &  \\
    Simple average & 0.368 & 0.341 & 0.254 & 0.775 & 0.588 \\
    Weighted average & 0.340 & 0.320 & 0.185 & 0.799 & 0.597 \\
    Exponential weighted average & 0.291 & 0.280 & 0.079 & 0.761 & 0.563 \\
    \midrule
    \multicolumn{1}{l}{} &       &       &       &       &  \\

    \multicolumn{6}{l}{\textbf{RMSFE, relative to Dynamic factor model}} \\
    & \textbf{Overall} & \textbf{Pre-COVID} & \textbf{COVID} & \textbf{Post-COVID} & \textbf{Excluding COVID} \\
    \cmidrule(lr){2-6}
    \textbf{Best models (MCS)} &       &       &       &       &  \\
    LASSO & 0.555 & 0.507 & 0.561 & 0.535 & 0.530 \\
    Ridge & 0.360 & 0.653 & 0.206 & 0.674 & 0.670 \\
    Elastic net & 0.321 & 0.625 & 0.208 & 0.554 & 0.569 \\
    Principal component regression & 0.364 & 0.733 & 0.110 & 0.767 & 0.760 \\
    Partial least squares regression & 0.365 & 0.614 & 0.253 & 0.626 & 0.623 \\
    Gated recurrent unit & 0.579 & 0.514 & 0.600 & 0.488 & 0.493 \\
    \midrule
    \midrule
    \textbf{Combination models} &       &       &       &       &  \\
    Simple average & 0.322 & 0.518 & 0.216 & 0.569 & 0.559 \\
    Weighted average & 0.297 & 0.485 & 0.158 & 0.587 & 0.568 \\
    Exponential weighted average & 0.254 & 0.425 & 0.067 & 0.559 & 0.534 \\
    \bottomrule
    \end{tabular}
    }
\end{table}
}

\begin{figure}[H]
\centering
    \begin{subfigure}[t]{0.99\textwidth}
        \includegraphics[width=\textwidth]{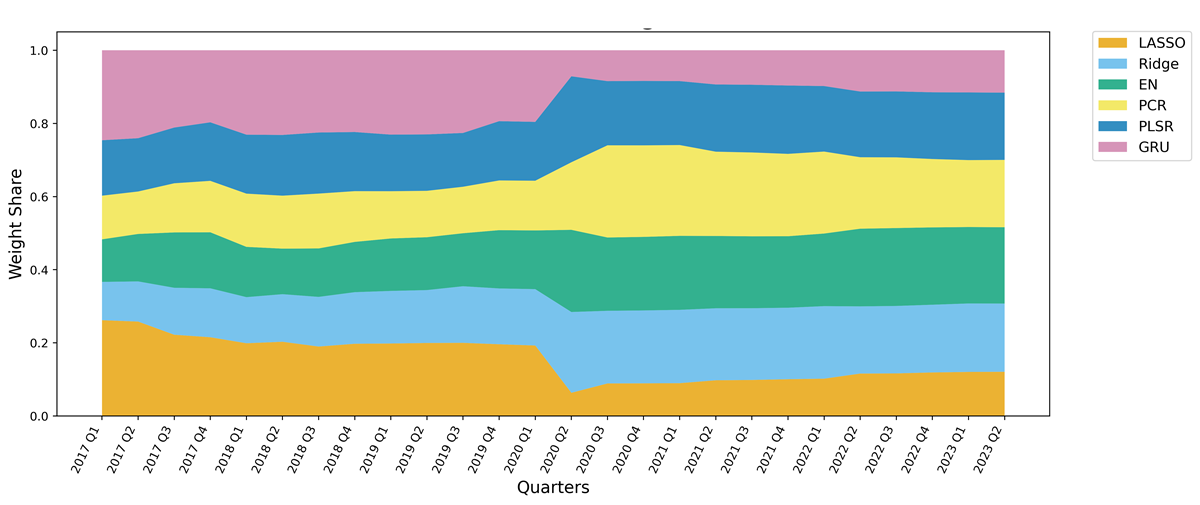}
        \caption{Weighted Average}
    \end{subfigure}
    
    \vspace{0.2cm}
    
    \begin{subfigure}[t]{0.99\textwidth}
        \includegraphics[width=\textwidth]{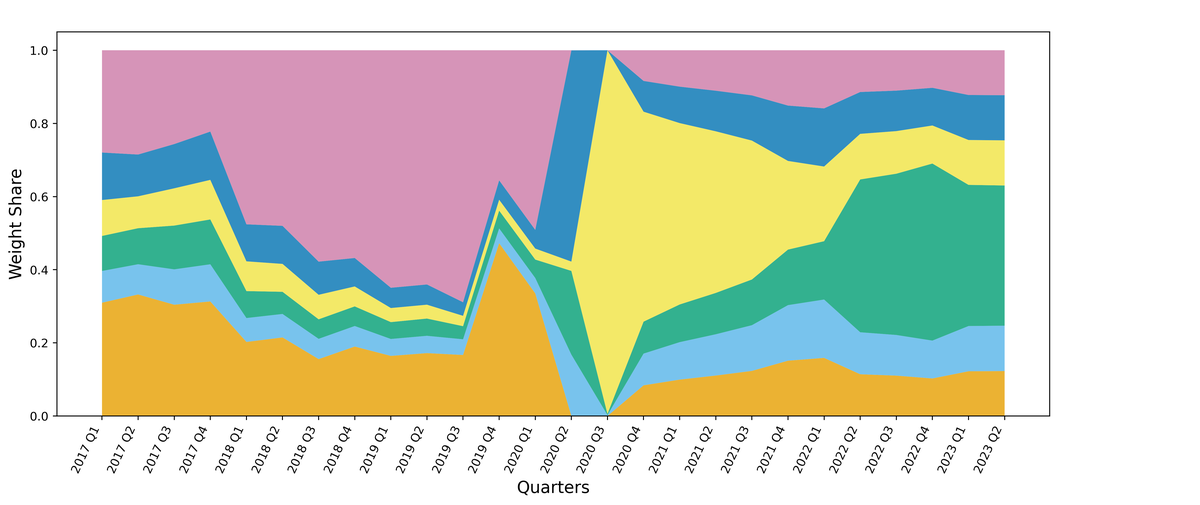}
        \caption{Exponential Weighted Average}
    \end{subfigure}
\caption{Stacked area chart illustrating the evolution of individual model weights within the Weighted Average and Exponential Weighted Average aggregation frameworks, from 2017 Q1 to 2023 Q2.}
\label{fig:weights_stacked}
\end{figure}

Figure~\ref{fig:weights_stacked} visually captures the evolution of model weights in WA and EWA frameworks, providing immediate insight into each model's shifting importance. Complementary detailed quarterly breakdowns in Table~\ref{tab:wa_dominantmodels} (WA) and Table~\ref{tab:ewa_dominantmodels} (EWA) further clarify how individual models predominantly influenced aggregated predictions across different economic regimes, explicitly addressing the explainability objective:

\begin{itemize}
\item Considering the whole prediction period, the GRU model was initially assigned the highest weights in WA and EWA strategies until late 2019. WA maintained essentially stable and constant weights over the entire period, demonstrating minimal fluctuations in model importance. Conversely, EWA dynamically adjusted model weights based on recent nowcasting performance, highlighting substantial shifts in model relevance, particularly during episodes of economic instability such as the COVID-19 period. This adaptive mechanism improved prediction accuracy and interpretability, especially during significant economic disruptions such as COVID-19.
\item During the COVID-19 disruption, dimensionality reduction-based models, particularly PCR, experienced substantial weight increases within the WA and EWA methods. This dynamic adjustment, especially in the EWA plot, reflects PCR's superior predictive reliability under heightened economic volatility. Conversely, SA maintained equal weighting across models without adjustments, implicitly limiting its responsiveness during rapid economic shifts.
\item In the Post-COVID sub-period, Elastic Net consistently received the highest weights within the WA and EWA frameworks. Again, the adaptability of EWA is distinctly evident, showing how quickly and effectively Elastic Net gained prominence as economic conditions stabilized but remained uncertain. WA confirms Elastic Net's high historical average reliability, maintaining constant but elevated weights. With its equal-weight structure, SA continued to allocate identical importance across all models, thus not reflecting these nuanced shifts.
\end{itemize}

\begin{table}[htbp]
\fontsize{9pt}{10pt}
\selectfont
\caption{Evolution of individual model weights within the Weighted Average (WA) aggregation framework, and corresponding dominant model, from 2017 Q1 to 2023 Q2}
\label{tab:wa_dominantmodels}
\resizebox{\textwidth}{!}{
\begin{tabular}{l c c c c c c c}
\toprule
\textbf{Quarters} &
\textbf{LASSO weight} &
\textbf{Ridge weight} &
\textbf{EN weight} &
\textbf{PCR weight} &
\textbf{PLSR weight} &
\textbf{GRU weight} &
\textbf{Dominant model}\\
\midrule
2017 Q1 & 0.262 & 0.105 & 0.116 & 0.119 & 0.152 & 0.246 & LASSO \\
2017 Q2 & 0.258 & 0.110 & 0.130 & 0.116 & 0.146 & 0.240 & LASSO \\
2017 Q3 & 0.222 & 0.128 & 0.151 & 0.134 & 0.152 & 0.211 & LASSO \\
2017 Q4 & 0.215 & 0.134 & 0.153 & 0.141 & 0.160 & 0.197 & LASSO \\
2018 Q1 & 0.199 & 0.126 & 0.137 & 0.146 & 0.161 & 0.231 & GRU \\
2018 Q2 & 0.203 & 0.130 & 0.125 & 0.145 & 0.166 & 0.232 & GRU \\
2018 Q3 & 0.190 & 0.136 & 0.133 & 0.150 & 0.167 & 0.225 & GRU \\
2018 Q4 & 0.198 & 0.141 & 0.137 & 0.139 & 0.162 & 0.224 & GRU \\
2019 Q1 & 0.198 & 0.144 & 0.143 & 0.129 & 0.155 & 0.231 & GRU \\
2019 Q2 & 0.200 & 0.145 & 0.145 & 0.127 & 0.154 & 0.230 & GRU \\
2019 Q3 & 0.200 & 0.155 & 0.145 & 0.127 & 0.147 & 0.226 & GRU \\
2019 Q4 & 0.196 & 0.153 & 0.159 & 0.136 & 0.162 & 0.194 & LASSO \\
2020 Q1 & 0.193 & 0.154 & 0.160 & 0.136 & 0.161 & 0.196 & GRU \\
2020 Q2 & 0.064 & 0.221 & 0.225 & 0.185 & 0.235 & 0.071 & PLSR \\
2020 Q3 & 0.089 & 0.199 & 0.200 & 0.252 & 0.175 & 0.084 & PCR \\
2020 Q4 & 0.089 & 0.199 & 0.201 & 0.250 & 0.176 & 0.084 & PCR \\
2021 Q1 & 0.090 & 0.200 & 0.202 & 0.248 & 0.175 & 0.084 & PCR \\
2021 Q2 & 0.098 & 0.197 & 0.198 & 0.230 & 0.184 & 0.094 & PCR \\
2021 Q3 & 0.099 & 0.196 & 0.197 & 0.229 & 0.185 & 0.094 & PCR \\
2021 Q4 & 0.101 & 0.195 & 0.195 & 0.225 & 0.187 & 0.096 & PCR \\
2022 Q1 & 0.102 & 0.198 & 0.198 & 0.224 & 0.179 & 0.098 & PCR \\
2022 Q2 & 0.116 & 0.184 & 0.212 & 0.195 & 0.180 & 0.113 & EN \\
2022 Q3 & 0.117 & 0.184 & 0.213 & 0.193 & 0.180 & 0.113 & EN \\
2022 Q4 & 0.119 & 0.185 & 0.211 & 0.187 & 0.182 & 0.115 & EN \\
2023 Q1 & 0.121 & 0.187 & 0.209 & 0.183 & 0.185 & 0.115 & EN \\
2023 Q2 & 0.121 & 0.187 & 0.209 & 0.184 & 0.184 & 0.116 & EN \\
\bottomrule
\end{tabular}}
\end{table}

\begin{table}[htbp]
\centering
\fontsize{9pt}{10pt}\selectfont
\caption{Evolution of individual model weights within the Exponential Weighted Average (EWA) aggregation framework, and corresponding dominant model, from 2017 Q1 to 2023 Q2}
\label{tab:ewa_dominantmodels}
\resizebox{\textwidth}{!}{
\begin{tabular}{l c c c c c c c}
\toprule
\textbf{Quarters} & 
\textbf{LASSO weight} & 
\textbf{Ridge weight} & 
\textbf{EN weight} & 
\textbf{PCR weight} & 
\textbf{PLSR weight} & 
\textbf{GRU weight} & 
\textbf{Dominant model}\\
\midrule
2017 Q1 & 0.310 & 0.087 & 0.096 & 0.098 & 0.130 & 0.280 & LASSO \\
2017 Q2 & 0.332 & 0.083 & 0.099 & 0.087 & 0.114 & 0.285 & LASSO \\
2017 Q3 & 0.304 & 0.097 & 0.120 & 0.102 & 0.121 & 0.257 & LASSO \\
2017 Q4 & 0.313 & 0.102 & 0.123 & 0.108 & 0.132 & 0.222 & LASSO \\
2018 Q1 & 0.203 & 0.065 & 0.074 & 0.081 & 0.101 & 0.476 & GRU \\
2018 Q2 & 0.215 & 0.064 & 0.061 & 0.076 & 0.104 & 0.480 & GRU \\
2018 Q3 & 0.156 & 0.055 & 0.053 & 0.067 & 0.091 & 0.578 & GRU \\
2018 Q4 & 0.190 & 0.056 & 0.054 & 0.055 & 0.078 & 0.568 & GRU \\
2019 Q1 & 0.164 & 0.046 & 0.046 & 0.038 & 0.055 & 0.649 & GRU \\
2019 Q2 & 0.172 & 0.047 & 0.047 & 0.038 & 0.055 & 0.640 & GRU \\
2019 Q3 & 0.167 & 0.043 & 0.036 & 0.028 & 0.037 & 0.689 & GRU \\
2019 Q4 & 0.472 & 0.040 & 0.049 & 0.030 & 0.053 & 0.356 & LASSO \\
2020 Q1 & 0.335 & 0.042 & 0.050 & 0.030 & 0.051 & 0.491 & GRU \\
2020 Q2 & 0.000 & 0.168 & 0.229 & 0.025 & 0.578 & 0.000 & PLSR \\
2020 Q3 & 0.000 & 0.003 & 0.003 & 0.994 & 0.000 & 0.000 & PCR \\
2020 Q4 & 0.084 & 0.087 & 0.087 & 0.574 & 0.084 & 0.084 & PCR \\
2021 Q1 & 0.100 & 0.102 & 0.103 & 0.496 & 0.100 & 0.100 & PCR \\
2021 Q2 & 0.111 & 0.113 & 0.113 & 0.442 & 0.111 & 0.111 & PCR \\
2021 Q3 & 0.123 & 0.125 & 0.125 & 0.380 & 0.124 & 0.123 & PCR \\
2021 Q4 & 0.151 & 0.152 & 0.152 & 0.242 & 0.151 & 0.151 & PCR \\
2022 Q1 & 0.159 & 0.159 & 0.159 & 0.204 & 0.159 & 0.159 & PCR \\
2022 Q2 & 0.114 & 0.115 & 0.418 & 0.125 & 0.114 & 0.114 & EN \\
2022 Q3 & 0.110 & 0.111 & 0.441 & 0.116 & 0.111 & 0.110 & EN \\
2022 Q4 & 0.103 & 0.103 & 0.484 & 0.104 & 0.103 & 0.103 & EN \\
2023 Q1 & 0.122 & 0.124 & 0.386 & 0.123 & 0.123 & 0.122 & EN \\
2023 Q2 & 0.123 & 0.124 & 0.383 & 0.123 & 0.123 & 0.123 & EN \\
\bottomrule
\end{tabular}}
\end{table}

A detailed quantitative breakdown of dominant models across each sub-period and aggregation framework (SA, WA, EWA) is provided in Table~\ref{tab:dominantmodels_count_subperiod}. This analysis confirmed the frequent selection of PCR and Elastic Net after the structural break induced by the pandemic, particularly under the WA and EWA aggregation frameworks, while highlighting significant contributions from GRU and PLSR in earlier and transitional periods.

{
\fontsize{9pt}{10pt}\selectfont

\begin{table}[t!]
    \centering
    \caption{Frequency of dominant models selected within each aggregation framework (Simple Average, Weighted Average, Exponentially Weighted Average) by sub-period}
    \label{tab:dominantmodels_count_subperiod}
    \resizebox{0.93\textwidth}{!}{
    \begin{tabular}{lccccc}

    \multicolumn{6}{l}{\textbf{Simple Average}} \\
          & \textbf{Overall} & \textbf{Pre-COVID} & \textbf{COVID} & \textbf{Post-COVID} & \textbf{Excluding COVID} \\
    \cmidrule(lr){2-6}
    \multicolumn{6}{l}{\textbf{Penalized linear models}} \\
    LASSO & 2 & 2 & 0 & 0 & 2 \\
    Ridge & 4 & 2 & 0 & 2 & 4 \\
    Elastic net & 3 & 2 & 0 & 1 & 3 \\
    \midrule
    \multicolumn{6}{l}{\textbf{Dimensionality reduction-based models}} \\
    Principal component regression & 2 & 0 & 1 & 1 & 1 \\
    Partial least squares regression & 8 & 3 & 2 & 3 & 6 \\
    \midrule
    \multicolumn{6}{l}{\textbf{Neural networks}} \\
    Gated recurrent units & 7 & 3 & 1 & 3 & 6 \\
    \midrule

    \multicolumn{1}{l}{} &       &       &       &       &  \\

    \multicolumn{6}{l}{\textbf{Weighted Average}} \\
          & \textbf{Overall} & \textbf{Pre-COVID} & \textbf{COVID} & \textbf{Post-COVID} & \textbf{Excluding COVID} \\
    \cmidrule(lr){2-6}
    \multicolumn{6}{l}{\textbf{Penalized linear models}} \\
    LASSO & 5 & 5 & 0 & 0 & 5 \\
    Ridge & 0 & 0 & 0 & 0 & 0 \\
    Elastic net    & 5 & 0 & 0 & 5 & 5 \\
    \midrule
    \multicolumn{6}{l}{\textbf{Dimensionality reduction-based models}} \\
    Principal component regression   & 7 & 0 & 2 & 5 & 5 \\
    Partial least squares regression  & 1 & 0 & 1 & 0 & 0 \\
    \midrule
    \multicolumn{6}{l}{\textbf{Neural networks}} \\
    Gated recurrent units   & 8 & 7 & 1 & 0 & 7 \\
    \midrule

    \multicolumn{1}{l}{} &       &       &       &       &  \\

    \multicolumn{6}{l}{\textbf{Exponential Weighted Average}} \\
          & \textbf{Overall} & \textbf{Pre-COVID} & \textbf{COVID} & \textbf{Post-COVID} & \textbf{Excluding COVID} \\
    \cmidrule(lr){2-6}
    \multicolumn{6}{l}{\textbf{Penalized linear models}} \\
    LASSO & 5 & 5 & 0 & 0 & 5 \\
    Ridge & 0 & 0 & 0 & 0 & 0 \\
    Elastic net    & 5 & 0 & 0 & 5 & 5 \\
    \midrule
    \multicolumn{6}{l}{\textbf{Dimensionality reduction-based models}} \\
    Principal component regression   & 7 & 0 & 2 & 5 & 5 \\
    Partial least squares regression  & 1 & 0 & 1 & 0 & 0 \\
    \midrule
    \multicolumn{6}{l}{\textbf{Neural networks}} \\
    Gated recurrent units   & 8 & 7 & 1 & 0 & 7 \\
    \bottomrule
    \end{tabular}
    }
\end{table}

}

A final diagnostic check was performed on the residuals from each aggregated nowcasting methodology to ensure robustness and statistical reliability. Table~\ref{tab:combos_residuals_diagnostics} summarizes results from the Shapiro-Wilk test for normality and the Ljung-Box test for residual autocorrelation. The tests suggest that residuals from SA, WA, and EWA approaches largely exhibit acceptable statistical properties and support the statistical validity of the combined predictions across different aggregation strategies. In particular, the Shapiro-Wilk tests do not strongly reject the null hypothesis of residual normality for any of the aggregation methods, indicating that residual distributions from all combinations remain approximately normal. The Ljung-Box test reveals that residual autocorrelation issues are modest, with SA and WA residuals showing marginal evidence of autocorrelation at conventional significance levels, while EWA residuals appear more consistently free of significant autocorrelation.

{
    \fontsize{10pt}{10pt}\selectfont
    \begin{table}[htbp]
    \centering
    \caption{Residual diagnostics (Shapiro-Wilk and Ljung-Box tests) for aggregated models (Simple Average, Weighted Average, Exponentially Weighted Average}
    \label{tab:combos_residuals_diagnostics}
    \resizebox{0.85\textwidth}{!}{%
	\begin{tabular}{l|cc|cc|cc}
            & \multicolumn{2}{c|}{\textbf{SA}}
            & \multicolumn{2}{c|}{\textbf{WA}}
		& \multicolumn{2}{c}{\textbf{EWA}} \\
		\cline{2-7}
		& \textbf{Statistic} & \textbf{p-value}
		& \textbf{Statistic} & \textbf{p-value}
		& \textbf{Statistic} & \textbf{p-value} \\
		\hline
		\textbf{Shapiro-Wilk}      & 0.928  & 0.071 & 0.959 & 0.365 & 0.928 & 0.070 \\
		\textbf{Ljung-Box (lag=4)} & 10.136 & 0.038 & 9.649 & 0.047 & 7.952 & 0.093 \\
		\hline
	\end{tabular}
    }
    \end{table}
}

Aggregating predictions using MCS-selected models systematically enhanced predictive accuracy and robustness compared to baseline benchmarks (Random Walk, AR(3), DFM). The dynamic weighting of EWA proved particularly beneficial during periods of heightened volatility, such as the COVID-19 pandemic, significantly improving both predictive accuracy and interpretability. Meanwhile, the simpler SA approach outperformed WA across various regimes, confirming its reliability as a robust benchmark.

\vspace{0.35cm}

\subsection{\textbf{\textit{Formal Predictive-Ability Tests}}}
\label{subsec:gw_predictive_ability_tests}
Evaluating the predictive performance of nowcasting models necessitates a rigorous statistical approach to determine whether observed improvements in prediction accuracy are statistically meaningful rather than due to random variation. This section presents empirical findings from the Giacomini-White (GW) predictive ability test \parencite{GiacominiWhite2006}. Unlike classic tests, such as the Diebold-Mariano test \parencite{DieboldMariano1995}, the GW test explicitly assesses conditional predictive accuracy, essential when nowcast models adapt dynamically to evolving economic information.\footnote{In this context, predictive ability refers to a model's capacity to consistently deliver more accurate nowcasts compared to alternative models, considering all information available at each prediction step.} We assess whether our models selected by the Model Confidence Set and our nowcast aggregation methods demonstrate statistically significant predictive superiority over our three established benchmarks.

The Giacomini-White test was systematically employed across these comparative scenarios, explicitly accommodating conditional predictive abilities and relying on robust covariance estimation through the Lumley-Heagerty (WEAVE) estimator. This estimator ensures robust covariance estimation, effectively adjusting standard errors for heteroskedasticity and autocorrelation in prediction errors. As summarized in Table~\ref{tab:gw_test_results}, when assessing predictive performance relative to the Random Walk benchmark, all MCS-selected individual models and combined nowcasts consistently demonstrated statistically significant improvements.

{
    \fontsize{9pt}{10pt}\selectfont
    \begin{table}[t!]
        \centering
        \caption{Giacomini-White test results comparing predictive accuracy of MCS-selected individual and aggregated models (SA, WA, EWA) against benchmark models (Random Walk, AR(3), Dynamic Factor Model). Negative intercepts indicate superior nowcast performance relative to benchmarks.}
        \label{tab:gw_test_results}
        \resizebox{0.9\textwidth}{!}{
            \begin{tabular}{l|rr|rr}
                \multicolumn{5}{c}{\textbf{Random Walk}} \\
                \hline
                \textbf{Model} & \textbf{Intercept (Estimate)} & \textbf{p-value (Intercept)} & \textbf{Wald statistic} & \textbf{Wald p-value} \\
                \hline
                LASSO & -22.324630389 & 0.000111374 & 20.954358337 & 0.000111374 \\
                Ridge & -25.207332539 & 0.000019777 & 27.514105600 & 0.000019777 \\
                EN    & -25.628654847 & 0.000015209 & 28.594954554 & 0.000015209 \\
                PCR   & -25.151611279 & 0.000030197 & 25.820526531 & 0.000030197 \\
                PLSR  & -25.146212250 & 0.000015814 & 28.433066364 & 0.000015814 \\
                GRU   & -21.888333365 & 0.000046176 & 24.179216204 & 0.000046176 \\
                \hline\hline
                SA    & -25.627322138 & 0.000020503 & 27.367642584 & 0.000020503 \\
                WA    & -25.245151148 & 0.000021503 & 27.174652336 & 0.000021503 \\
                EWA   & -26.255301547 & 0.000016205 & 28.331950232 & 0.000016205 \\
                \\
                \multicolumn{5}{c}{\textbf{AR(3)}} \\
                \hline
                \textbf{Model} & \textbf{Intercept (Estimate)} & \textbf{p-value (Intercept)} & \textbf{Wald statistic} & \textbf{Wald p-value} \\
                \hline
                LASSO & -7.338380433 & 0.000000494 & 45.035271911 & 0.000000494 \\
                Ridge & -10.221082583 & 0.000041759 & 24.562593668 & 0.000041759 \\
                EN    & -10.642404890 & 0.000021016 & 27.267321898 & 0.000021016 \\
                PCR   & -10.165361322 & 0.000060456 & 23.167166436 & 0.000060456 \\
                PLSR  & -10.159962293 & 0.000036978 & 25.030507111 & 0.000036978 \\
                GRU   & -6.902083408 & 0.000001242 & 40.156413956 & 0.000001242 \\
                \hline\hline
                SA    & -10.641072182 & 0.000012711 & 29.346717758 & 0.000012711 \\
                WA    & -10.258901191 & 0.000013859 & 28.983078148 & 0.000013859 \\
                EWA   & -11.269051591 & 0.000015030 & 28.644198103 & 0.000015030 \\
                \\
                \multicolumn{5}{c}{\textbf{Dynamic Factor Model (DFM)}} \\
                \hline
                \textbf{Model} & \textbf{Intercept (Estimate)} & \textbf{p-value (Intercept)} & \textbf{Wald statistic} & \textbf{Wald p-value} \\
                \hline
                LASSO & -11.186248419 & 0.000008183 & 31.239860475 & 0.000008183 \\
                Ridge & -14.068950569 & 0.000037424 & 24.984211817 & 0.000037424 \\
                EN    & -14.490272876 & 0.000022913 & 26.918502887 & 0.000022913 \\
                PCR   & -14.013229308 & 0.000061349 & 23.112753861 & 0.000061349 \\
                PLSR  & -14.007830280 & 0.000029409 & 25.924740344 & 0.000029409 \\
                GRU   & -10.749951394 & 0.000005111 & 33.339347101 & 0.000005111 \\
                \hline\hline
                SA    & -14.488940168 & 0.000021251 & 27.222283274 & 0.000021251 \\
                WA    & -14.106769177 & 0.000021861 & 27.107988472 & 0.000021861 \\
                EWA   & -15.116919577 & 0.000021619 & 27.152988717 & 0.000021619 \\
            \end{tabular}
        }\\[0.2cm]
        \begin{minipage}{0.85\textwidth}
            \footnotesize\textit{Note}: The intercept estimates represent the difference in prediction errors between the tested models and benchmarks. Negative intercepts indicate that tested models outperform benchmarks.
        \end{minipage}
    \end{table}
}

Specifically:

\begin{itemize}
    \item The Wald test statistics were uniformly high (ranging approximately from 21 to 29), firmly rejecting the null hypothesis of equal predictive accuracy.
    \item All tested models showed statistically significant improvements (all p-values < 0.001), with EN and EWA consistently outperforming across benchmarks.
    \item Among individual models, Elastic Net (EN) yielded the most pronounced reduction in prediction errors (intercept estimate: $-25.63$), closely followed by PLSR and Ridge, underscoring their predictive reliability.
    \item Among combined nowcasts, EWA showed the most significant improvement, indicated by the lowest intercept estimate ($-26.26$) and strongest statistical significance (Wald p-value = $0.000016$), highlighting its capacity to dynamically adapt to periods of higher uncertainty, such as the COVID-19 pandemic.
\end{itemize}

These outcomes indicate that all evaluated models significantly enhance predictive accuracy over the naive Random Walk, with dynamic weighting schemes such as EWA proving especially valuable during volatile economic conditions.

Comparisons against the AR(3) model further reinforced the robustness of our selected nowcasting methodologies. All individual and aggregated models demonstrated statistically significant superiority, according to the GW test:

\begin{itemize}
    \item Wald statistics remained notably large, exceeding 23 across all scenarios, indicative of robust model superiority.
    \item P-values consistently fell far below the standard significance level (all below 0.0001), indicating enhanced nowcasting performance compared to the AR(3) benchmark.
    \item Again, EN emerged as the top-performing individual method, with an intercept estimate of $-10.64$, whereas among aggregations, EWA maintained its leadership in nowcast improvement, reflected by the lowest intercept estimate ($-11.27$).
\end{itemize}

This consistent improvement against an established econometric benchmark underlines the efficacy of penalized regression approaches and adaptive weighting mechanisms in capturing dynamic macroeconomic relationships and structural shifts.

The DFM comparison, frequently adopted in macroeconomic nowcasting literature due to its capacity for dimensionality reduction and common factor extraction, yielded similarly strong support for our modeling strategies. Specifically:

\begin{itemize}
    \item Wald test statistics uniformly exceeded 23, strongly rejecting the null hypothesis of equivalent predictive accuracy between tested models and DFM.
    \item Low p-values (below 0.0001 in all cases) provided definitive evidence of nowcasting improvement.
    \item The best individual models again included EN, Ridge, and PLSR, each showing intercept estimates close to $-14$. Among combination strategies, EWA exhibited the highest predictive enhancement (intercept $-15.12$), reiterating its adaptive strengths across diverse economic regimes.
\end{itemize}

These empirical findings substantiate that both individual MCS-selected models and dynamic aggregations significantly surpass the predictive accuracy of standard factor-based methods, reinforcing their potential for robust macroeconomic nowcasting.

The empirical evidence from the Giacomini-White tests highlights several important implications for macroeconomic nowcasting:

\begin{itemize}
    \item Models selected through rigorous MCS procedures consistently deliver statistically significant nowcasting improvements across benchmarks, validating their empirical relevance.
    \item Adaptive weighting methodologies, particularly EWA, substantially enhance predictive performance in turbulent or regime-switching periods, ensuring robustness and responsiveness.
    \item Simple aggregation methods (SA), despite their straightforward nature, still provide stable and significant improvements over benchmarks and remain valuable as reliable nowcasting baselines.
\end{itemize}

These findings underscore the critical importance of conditional predictive ability testing and adaptive aggregation strategies for effective macroeconomic nowcasting.

These empirical outcomes highlight how conditional predictive ability testing and adaptive aggregation techniques systematically differentiate nowcasting models regarding predictive accuracy. The Giacomini-White test results consistently indicate statistically significant improvements in the predictive accuracy of individual and aggregated models compared to traditional benchmark models typically employed in macroeconomic nowcasting. These findings reinforce the practical value of dynamic aggregation methods (especially EWA) for policymakers, particularly under uncertain economic conditions where adaptability in nowcasting methods is crucial.

\vspace{0.35cm}

\section{\textbf{Discussion and Implications}}
\label{sec:3.2}
This section aims to contextualize the empirical evidence discussed in Section~\ref{sec:3.1} and prepares the ground for the subsequent methodological considerations and policy insights. Specifically, we investigate multiple sub-periods, including Pre-COVID, COVID, and Post-COVID, applying a block bootstrap procedure to enhance interval coverage even under sudden macroeconomic shifts. Our objective here is twofold: first, to distill the central quantitative findings arising from assessments of predictive accuracy, interval-based uncertainty, feature relevance, and nowcast-combination strategies; second, to underscore the significance of these observations for subsequent model evaluations and policy-oriented applications.

From a methodological viewpoint, our empirical results highlight several points of interest. Penalized linear models (particularly Ridge and Elastic Net) exhibit relatively low and stable prediction errors across diverse macroeconomic phases. That suggests regularization can effectively harness key explanatory signals without overfitting, even under heightened volatility. Dimensionality-reduction methods (notably Principal Component Regression and Partial Least Squares Regression) display more pronounced fluctuations around crises such as COVID-19, albeit with phases of commendable robustness. Meanwhile, ensemble learning and neural architectures prove competitive during tranquil periods yet appear more susceptible to abrupt structural shifts.

We further validate these findings through the Giacomini-White test, which confirms the statistical significance of the observed accuracy enhancements over classic benchmarks. The test outcome thus reinforces the reliability of our models under varying macroeconomic scenarios.

Parallel to these performance metrics, the analysis of prediction intervals sheds light on each model's approach to managing uncertainty. While some of our techniques produce tighter intervals in stable phases yet encounter challenges with extreme shocks, others adopt a more conservative stance with broader intervals, achieving more robust coverage in turbulent settings. Moreover, exploring feature importance reveals several consistently influential indicators (e.g., industrial production, trade-related variables, and labor market proxies) whose relative weighting can shape the scale of prediction intervals during volatility spikes.

The investigation of combination strategies, including Simple Average, Weighted Average, and Exponentially Weighted Average (EWA), complements these empirical results. EWA, in particular, demonstrates heightened adaptability in volatile environments by favoring models with strong current performance, thereby achieving either competitive or superior accuracy. Moreover, this dynamic weighting framework enhances transparency by clarifying the individual model's impact on the aggregated prediction.

In the following subsections, we build on these findings to analyze:
\begin{itemize}
    \item how the proposed techniques compare against established benchmarks under distinct macroeconomic regimes,
    \item the reliability and interpretability of the estimated prediction intervals,
    \item the efficacy of static and dynamically weighted nowcasting methods, specifically focusing on how EWA can offer additional insights into model contributions.
\end{itemize}

These perspectives inform broader debates on macroeconomic nowcasting and policymaking, highlighting the potential of flexible model combinations to mitigate the limitations of single-model approaches and reinforce predictive resilience during economic disruptions.

\vspace{0.35cm}

\subsection{\textbf{\textit{Comparative Nowcasting Performance across Macroeconomic Regimes: Evaluating Penalized, Dimensionality Reduction, Ensemble, and Neural Models}}}
\label{subsec:comp_nowcast_performance}

In light of the descriptive evidence presented in Section~\ref{subsec:accuracymodels}, this subsection provides a comparative evaluation of nowcasting model families (penalized linear, dimensionality reduction-based, ensemble learning, and neural network models) across different macroeconomic regimes, offering valuable insights into the relative robustness, stability, and adaptability of various methodological frameworks.

The primary evaluation metric remains the Root Mean Square Forecast Error (RMSFE), which is considered both in absolute terms and relative to traditional benchmarks (Random Walk, AR(3), and Dynamic Factor Model).

Penalized linear approaches, specifically Ridge and Elastic Net \parencite{HoerlKennard1970,ZouHastie2005,HastieTibshiraniFriedman2009,UematsuTanaka2019,SmeekesWijler2018,KimSwanson2018}, consistently demonstrated superior predictive performance relative to traditional benchmarks. The regularization mechanisms inherent to these methods effectively control model complexity, enhancing stability during periods of high volatility, such as the COVID-19 crisis. Conversely, LASSO \parencite{Tibshirani1996} exhibited noticeable vulnerability due to its stringent variable selection criterion, which proved detrimental when structural economic shifts occurred.
Key insights from penalized linear models include:

\begin{itemize}
\item \textit{Stability and robustness}: Ridge and Elastic Net consistently maintained low and stable prediction errors across regimes, highlighting their resilience to macroeconomic volatility.
\item \textit{Relative superiority}: These methods systematically outperformed traditional benchmarks, reinforcing their value for real-time macroeconomic nowcasting applications.
\item \textit{LASSO vulnerability}: Despite widespread use in variable selection, LASSO significantly deteriorated under structural shocks, suggesting limited applicability in volatile economic contexts.
\end{itemize}

Dimensionality reduction approaches (Principal Component Regression and Partial Least Squares Regression) \parencite{Jolliffe1982,Massy1965,JolliffeCadima2016,Wold1985,MehmoodEtAl2012,KraemerSugiyama2011,FuentesPoncelaRodriguez2015,GroenKapetanios2016} provided valuable yet regime-dependent nowcasting capabilities. PCR showed remarkable resilience during extreme volatility periods, effectively capturing critical macroeconomic information while filtering out noise. However, its performance deteriorated significantly during the Post-COVID regime transitions. Conversely, PLSR experienced significant deterioration during the COVID crisis but demonstrated a notable recovery in prediction accuracy during the Post-COVID normalization phase, although it is still not reaching pre-crisis accuracy levels.
Crucial observations include:

\begin{itemize}
\item \textit{Adaptability to economic regimes}: PCR's strong performance in crisis periods contrasted sharply with its reduced adaptability during recovery phases.
\item \textit{Complementary dynamics}: The opposing behavior of PCR and PLSR during and after crises suggests strategic complementarities between these two methodologies.
\item \textit{Overall competitiveness}: Both methods offered substantial improvements over benchmarks, albeit PCR exhibited greater volatility across regimes, suggesting the necessity of regime-specific model selection.
\end{itemize}

Ensemble learning models (Random Forest and eXtreme Gradient Boosting) \parencite{Breiman2001,ChenGuestrin2016,BiauScornet2016,ProbstWrightBoulesteix2019,CoulombeEtAl2022,BorupChristensenMuhlbachNielsen2023,SoybilgenYazgan2021,GouletCoulombe2024} displayed mixed prediction effectiveness. While theoretically well-suited to capturing complex nonlinear relationships, both models showed pronounced vulnerability during abrupt economic disruptions. XGB, however, demonstrated better recovery capabilities post-crisis than RF, reflecting relative strengths in managing transition dynamics.
Notable insights include:

\begin{itemize}
\item \textit{Nonlinearity handling limitations}: Despite inherent flexibility, ensemble methods underperformed in highly volatile periods, highlighting their sensitivity to rapid structural economic changes.
\item \textit{Complexity trade-offs}: High error rates observed during crises underscored the need for dynamic calibration and continuous retraining to maximize their predictive potential. Nevertheless, despite partial recovery, ensemble models remained generally inferior in overall prediction performance compared to penalized linear and dimensionality-reduction-based methods.
\end{itemize}

Neural network models, particularly Gated Recurrent Units (GRU), outperformed traditional Multilayer Perceptrons (MLP), reflecting the advantage of explicitly modeling temporal dependencies \parencite{ChoEtAl2014,ChungGulcehreChoBengio2014,JozefowiczZarembaSutskever2015,Hornik1989,HewamalageBergmeirBandara2021,AlmosovaAndresen2023,CoulombeEtAl2022}. Nonetheless, both neural architectures experienced significant prediction degradation during the COVID crisis, underscoring challenges related to data availability, quality, and model calibration under conditions of economic stress.
Strategic observations include:

\begin{itemize}
\item \textit{Architectural differences}: GRU's systematic superiority emphasizes the necessity of recurrent structures to capture temporal dynamics effectively.
\item \textit{Economic shock sensitivity}: Both neural models demonstrated substantial vulnerability during severe macroeconomic disturbances, highlighting their role primarily as complementary rather than stand-alone prediction tools, especially in volatile environments. Moreover, although neural models consistently outperformed Random Forest, they generally underperformed compared to penalized linear and dimensionality reduction-based models.
\end{itemize}

The collective evidence highlights a nuanced trade-off between model adaptability (the ability to rapidly and flexibly adjust predictions to unexpected and lasting structural changes in the economy) and stability (the capacity to consistently deliver accurate predictions without being overly influenced by short-term volatility or transient noise in economic data) \parencite{Timmermann2006,ElliottTimmermann2016,Rossi2021,PesaranPickPranovich2013,GiraitisKapetaniosPrice2013,ClarkMcCracken2010}.
Highly flexible models (ensemble and neural networks) exhibited sensitivity to extreme shocks, whereas penalized linear approaches offered greater robustness with potential trade-offs in short-term responsiveness to structural changes.
Strategic recommendations derived from this comparative analysis include:

\begin{itemize}
\item \textit{Adaptive stability trade-offs}: Optimal model selection necessitates careful balancing between stability (penalized linear) and adaptability (ensemble, neural networks).
\item \textit{Regime-informed model selection}: Differential performance across economic regimes underlined the importance of dynamically adjusting model choice and calibration based on prevailing macroeconomic conditions.
\item \textit{Benchmarking as strategic guidance}: Consistent outperformance of advanced econometric models relative to traditional benchmarks (RW, AR(3), DFM) underscored the empirical value of these methodological innovations.
\end{itemize}

From this overview, regularization-based models---particularly Ridge and Elastic Net---demonstrated solid and robust performance for quarterly nowcasting of Singapore's GDP growth across different macroeconomic regimes. Dimensionality reduction models (PCR and PLSR) generally achieved good accuracy but exhibited higher variability across sub-periods: PCR notably performed exceptionally well during the COVID crisis yet experienced marked deterioration in the subsequent Post-COVID period, whereas PLSR recovered steadily after initial setbacks. Conversely, ensemble learning and neural network models showed less consistent advantages: while competitive in stable periods, their prediction accuracy significantly deteriorated during turbulent phases (COVID). Among neural networks, GRU consistently outperformed MLP, providing relatively greater stability and resilience during periods of macroeconomic uncertainty.

\vspace{0.35cm}

\subsection{\textbf{\textit{Analysis and Discussion of Prediction Uncertainty}}}
\label{subsec:analysis_and_discussion_prediction_uncertainty}
The prediction intervals presented in Section~\ref{subsec:predictionuncertainty} shed light on each model's ability to quantify uncertainty under evolving economic regimes. In this subsection, we examine how interval width and coverage interact with model structure and macroeconomic volatility, thereby highlighting strengths and limitations of different approaches.

Models producing narrower intervals may occasionally fail to encompass actual values during extreme disruptions, potentially underestimating latent macroeconomic risks. By contrast, methods that generate broader ranges may be more successful in covering tail events, yet they often incur a loss of precision. The pivotal objective is thus to calibrate prediction intervals to reflect prediction uncertainty while avoiding excessive spread.

\begin{enumerate}
    \item \textbf{Penalized linear models.}
    Our results suggest that penalized linear techniques, especially Elastic Net, tend to produce relatively tight intervals, indicative of an efficient regularization strategy that mitigates overfitting. Nevertheless, during acute volatility episodes (e.g., the COVID-19 crisis), Elastic Net may understate uncertainty if the structural break is abrupt and substantial. During COVID-19 specifically, Ridge exhibited notably narrower intervals compared to LASSO, underscoring distinct behaviors within penalized linear methods under heightened volatility conditions. Elastic Net's dual penalty scheme (combining L1 and L2 regularization) appears to support more stable intervals than LASSO or Ridge, balancing variable selection against the risk of overreacting to transient shocks. During abrupt regime changes, Elastic Net intervals expand sufficiently to account for higher uncertainty but not to the point of rendering the nowcast signal uninformative. This feature is valuable for operational nowcasting, where interval interpretability is critical. Penalized linear models show a marked yet relatively moderate expansion of intervals in crisis contexts.
    \item \textbf{Dimensionality reduction-based models.}
    Although PCR and PLSR yield, on average, comparable width measures, their interval patterns diverge under shifting conditions. PCR presents narrower confidence bands during and after the COVID-19 shock, indicating greater resilience in uncertainty recovery compared to PLSR, whose intervals remain comparatively wider in the Post-COVID period. Such contrasting behaviors suggest that the choice between these two techniques should reflect the risk tolerance for crisis versus recovery phases. Dimensionality reduction approaches, such as Principal Component Regression and Partial Least Squares Regression, reveal larger jumps in uncertainty during crises, although they display distinct recovery trajectories.
    \item \textbf{Ensemble learning models.}
    Ensemble methods demonstrate notable interval fluctuations. Random Forest yields the widest intervals, particularly in the Post-COVID phase, without consistently capturing realized values. Random Forest's broad intervals may reflect a heightened reaction to anomalies, occasionally overshooting the true volatility range. Conversely, XGBoost exhibits a more controlled expansion, aided by its intrinsic form of shrinkage, thus preserving steadier prediction bands even under macroeconomic shocks. Random Forest's broader intervals might seem "safer" at first glance, but occasionally fail to align with realized outcomes, hinting at potential oversensitivity to transient or extreme observations. XGBoost, on the other hand, retains tighter intervals through an additional layer of internal shrinkage. This balancing act between flexibility and constraint can be advantageous in volatile settings where moderate expansions of intervals suffice to capture most disturbances without an unwarranted loss of precision.
    \item \textbf{Neural network models.}
    Neural networks exhibit divergent behaviors while capturing nonlinear patterns. MLP and GRU networks expand their prediction intervals markedly during turbulent periods, though the MLP's band growth is often more pronounced. Extensive intervals may hinder practical decision-making, while heightened caution is preferable to missing significant shocks entirely. GRU's less extreme interval widening and its recurrent structure may provide a more controlled uncertainty assessment in real-time contexts. GRU intervals remain marginally broader in the Post-COVID phase than MLP's ones, indicating a continued cautious approach during the recovery period.
    
\end{enumerate}

An important observation is the lack of perfect alignment between point nowcast errors (e.g., Root Mean Squared Forecast Error) and interval widths \parencite{Christoffersen1998}. Some techniques achieve respectable accuracy on average, yet produce intervals that can be disproportionately large. Others exhibit narrower intervals but occasionally exclude realized values, suggesting an underestimation of volatility. As a result, nowcast users should evaluate not only the magnitude of errors but also the alignment of predicted uncertainty with observed economic variations.

In a structurally evolving economy, models may underestimate uncertainty if they rely on historical patterns that fail to anticipate sudden shocks \parencite{Hansen2006}. LASSO, for instance, can be slow to adjust interval scales during the early phase of an unexpected crisis if key covariates are penalized too aggressively. Analogously, MLP networks may show inertia in re-scaling intervals due to stable training scenarios. Detecting early signs of regime shifts is thus pivotal to ensuring accurate uncertainty quantification.

We derived intervals via a block bootstrap procedure that respects temporal dependencies \parencite[see][for the original stationary bootstrap framework]{PolitisRomano1994}, yet each model reacts distinctively to resampled data. More rigid regularization methods, including Ridge, tend to exhibit minor expansions. By contrast, more flexible approaches, such as MLP or Random Forest, can respond sharply to outlying resampled blocks, yielding abrupt shifts in interval width. This heterogeneity underscores why similar resampling techniques may produce dissimilar uncertainty profiles across models.

This analysis underscores distinct uncertainty profiles across model families, revealing their different sensitivity and adaptability to macroeconomic volatility. Penalized linear methods, particularly Elastic Net, and ensemble methods such as XGBoost consistently achieve balanced prediction intervals, effectively responding to regime shifts without excessive uncertainty. Dimensionality reduction models (PCR, PLSR) and neural networks (MLP, GRU) exhibit contrasting behaviors, emphasizing that interval width alone does not always indicate coverage reliability. Ultimately, these insights stress the critical need to carefully calibrate interval estimation methods according to the specific macroeconomic conditions and forecasting objectives.

\vspace{0.35cm}

\subsection{\textbf{\textit{Feature Importance Dynamics across Macroeconomic Regimes}}}
\label{subsec:feature_importance_dynamics}
The interpretability analyses previously described in Section~3.1.3 provide critical insights into the economic dynamics captured by different modeling frameworks. They highlight the robustness and variability of predictor importance across distinct macroeconomic regimes. Below, we systematically discuss key interpretive observations derived from these analyses.

All modeling approaches—penalized linear models, dimensionality-reduction methods, ensemble learning, and neural networks—consistently emphasize a common subset of macroeconomic indicators:

\begin{itemize}
    \item Industrial production index,
    \item Gross operating surplus, and
    \item Key labor market variables, notably employee compensation and unit labor cost.
\end{itemize}

This convergence underscores the structural significance of these variables, suggesting their persistent informational value in tracking Singapore's short-term GDP dynamics, irrespective of economic stability or turmoil.

Although a core set of predictors remains influential across periods, notable shifts in variable importance are observed in response to macroeconomic disturbances, particularly the COVID-19 pandemic. Specifically:

\begin{itemize}
    \item \textit{Penalized linear models} increased emphasis on labor market and production indicators (e.g., employee compensation and unit labor cost), reflecting immediate labor-market disruptions \parencite{CajnerEtAl2020}.
    \item \textit{Dimensionality reduction methods}, especially Partial Least Squares Regression (PLSR), elevated the importance of air transportation metrics \parencite{HakimMerkert2016}, capturing the abrupt contraction and subsequent recovery of global passenger and cargo traffic.
\end{itemize}
 
These findings indicate the capacity of models to internally recalibrate their predictor selection in real-time, effectively capturing structural shifts driven by sudden macroeconomic events.

Modeling methodologies exhibit distinct behaviors concerning feature selection stability and adaptability:

\begin{itemize}
    \item \textit{Penalized linear models} (LASSO, Ridge, Elastic Net) show relatively stable predictor hierarchies, with controlled fluctuations attributable to their regularization mechanisms.
    \item In contrast, \textit{ensemble methods} (Random Forest \parencite{Breiman2001}, XGBoost) and \textit{neural networks} \parencite{Rudin2019} demonstrate substantial variability in their feature importance rankings, significantly altering their internal structures in response to short-term fluctuations. This adaptability enhances responsiveness to regime changes but diminishes the immediate interpretability of economic drivers.
    \item \textit{Principal Component Regression} (PCR) consistently emphasizes production and fiscal indicators, whereas \textit{Partial Least Squares Regression} (PLSR) prioritizes variables related to trade and logistics. Such methodological nuances underline the differential focus inherent in dimensionality-reduction strategies.
\end{itemize}

Our analyses highlight labor-market variables as critical conduits for economic shocks, particularly during crises. Neural network models and penalized regressions notably increased the relevance of employee compensation and unit labor cost during COVID-19 \parencite{CajnerEtAl2020}, reinforcing the labor market's centrality in macroeconomic adjustments.
Concurrently, air transport variables (air cargo loaded and air passenger arrivals) emerged as robust leading indicators. Their consistent prominence, especially in dimensionality-reduction and ensemble methods, underscores Singapore's sensitivity to global trade dynamics and international mobility \parencite{HakimMerkert2016}, which are vital to its economic performance.

The Spearman correlation analysis empirically reinforces the robustness of predictor selection. Variables frequently identified by various models correlate strongly and consistently with GDP. The industrial production index, in particular, consistently displays high and stable correlations, substantiating its role as a primary macroeconomic indicator.

Our results highlight a critical interpretative trade-off:

\begin{itemize}
    \item \textit{Stability in feature selection}, typical of linear and penalized methods, facilitates constructing coherent long-term economic narratives.
    \item \textit{Greater adaptability}, characteristic of ensemble and neural network approaches, captures nonlinearities and sudden economic disruptions, albeit at the cost of interpretative clarity.
\end{itemize}

 Thus, the optimal methodological choice depends on the prevailing macroeconomic context and the specific analytical or operational objectives, whether focused on immediate policy response or longer-term strategic forecasting.

The quarterly evolution analysis of the industrial production index (illustrated through the GRU model's Integrated Gradients \parencite{SundararajanTalyYan2017}) vividly demonstrates its heightened importance during the COVID-19 crisis. This dynamic interpretability underscores industrial production as a reliable barometer of economic resilience and subsequent recovery.

Across all evaluated models, the industrial production index change consistently emerges as the most influential predictor, underscoring its fundamental role in GDP nowcasting. Other consistently important variables include air cargo loaded change, gross operating surplus change, and air passenger arrivals change, reflecting their sensitivity to both cyclical economic fluctuations and structural shifts.

These results emphasizes how different modeling frameworks dynamically recalibrate feature importance in response to evolving macroeconomic conditions. These empirical observations illustrate the inherent methodological trade-off between stability in predictor selection---favoring interpretability---and flexibility in adapting to abrupt economic shifts. A deeper exploration of the econometric and practical implications of these findings, particularly concerning model selection for forecasting and policy purposes, will be provided in the next subsection.

\vspace{0.35cm}

\subsection{\textbf{\textit{Forecast Combination: Robustness, Adaptability, and Implications}}}
\label{subsec:forecast_combination_implications}
The findings presented in Section~\ref{subsec:modelselection} illustrate that combining forecasts from multiple models---via aggregation methods such as Simple Average (SA), Weighted Average (WA), and Exponential Weighted Average (EWA)---can substantially enhance accuracy and resilience in nowcasting tasks \parencite{BatesGranger1969,SmithWallis2009,ClaeskensEtAl2016}. Notably, the initial selection of models via the Model Confidence Set (MCS) procedure ensures that only methods with robust statistical predictive power contribute to these aggregations, strengthening the reliability of combined forecasts. This result holds particular significance in the presence of structural breaks or sudden economic disruptions, corroborating established best practices in the forecasting literature \parencite{HendryClements2004,DieboldLopez1996}.

A key rationale for forecast combination lies in the attenuation of biases and uncertainties exhibited by individual approaches. Notably, the performance of single-model strategies deteriorated sharply during the COVID-19 crisis, characterized by unprecedented volatility \parencite{ForoniMarcellinoStevanovic2020}, whereas aggregated forecasts retained robust predictive power. This result underscores how methodologically diverse ensemble strategies offer greater protection against structural shocks \parencite{MakridakisEtAl2020_M4}. By leveraging each model's unique strengths, combined methods reduce the risk of failure when any single model proves vulnerable to unexpected shifts in macroeconomic dynamics.

EWA emerges as a particularly effective solution among the three examined techniques in volatile contexts. Its dynamic adjustment of weights, based on recent performance (see Figure~\ref{fig:weights_stacked}), enables swift adaptation to regime changes and breaks in the data-generating process \parencite{KoopKorobilis2012}. This adaptability was quantitatively evident during the COVID-19 crisis, with EWA achieving notably lower RMSFE (0.610) compared to SA (1.963) and WA (1.433), as shown in Table~\ref{tab:forecast_combinedmodels}. Additionally, EWA transparently illustrates each model's contribution in real-time, enabling policymakers to identify dominant modeling approaches during rapidly evolving economic conditions \parencite{RafteryEtAl2010}.

The three aggregation methods exhibit distinct trade-offs between simplicity and reactivity:
\begin{itemize}
  \item \textit{Simple Average} assigns equal weights without active reweighting, serving as a robust and stable benchmark. It occasionally surpasses WA only in the post-COVID period (RMSFE: 1.520 vs.\ 1.567, Table~\ref{tab:forecast_combinedmodels}), thus avoiding over-reliance on potentially outdated historical performance metrics \parencite{SmithWallis2009}.
  \item \textit{Weighted Average} employs static weights derived from long-run predictive performance. This approach stabilizes weighting but faces challenges if economic conditions shift abruptly, as historical performance may quickly become obsolete.
  \item \textit{Exponentially Weighted Average} dynamically recalibrates weights, excelling in rapidly changing environments. Nevertheless, it requires vigilant parameter tuning and close monitoring to guard against excessive weight volatility, thus demanding greater implementation effort \parencite{Timmermann2006}.
\end{itemize}

Detailed analyses of WA and EWA weight trajectories (Tables~\ref{tab:wa_dominantmodels},~\ref{tab:ewa_dominantmodels}, Figure~\ref{fig:weights_stacked}) reveal that distinct models temporarily dominate under varying macroeconomic regimes. Principal Component Regression (PCR) notably excelled during peak COVID-19 volatility; Elastic Net (EN) gained prominence during post-crisis stabilization; and the Gated Recurrent Unit (GRU) demonstrated robust performance in pre-crisis and dynamic recovery phases. These complementary strengths suggest that no single approach uniformly outperforms across all scenarios, highlighting the strategic advantage of forecast combinations \parencite{BatchelorDua1995}.

Diagnostic evaluations (Shapiro--Wilk and Ljung--Box tests, Table~\ref{tab:combos_residuals_diagnostics}) confirm that aggregated forecasts do not introduce problematic biases or autocorrelation structures. Indeed, EWA's dynamic approach yields residuals with notably fewer autocorrelation issues (Ljung-Box p-value: 0.093) compared to SA (0.038) and WA (0.047), substantiating its statistical soundness. These characteristics, alongside lower RMSFE values, underscore the empirical robustness of dynamic aggregation methods \parencite{DieboldLopez1996}, particularly under abrupt economic changes.

From an econometric standpoint, results align with the forecasting literature consensus advocating forecast combinations as essential tools in macroeconomic nowcasting \parencite{GenreEtAl2013,KoopKorobilis2012}. Predictive ability tests (e.g., Giacomini-White) reinforce that aggregated methods consistently outperform traditional benchmarks---Random Walk, AR(3), and Dynamic Factor Models---across overall performance and distinct sub-period analyses (Tables~\ref{tab:relativePerformance_combinedmodels}). This evidence further supports their utility in informed policymaking and economic decision-making under heightened uncertainty \parencite{DieboldMariano1995,ClaessensKoseTerrones2011}.

The choice of aggregation method depends on context-specific balances between simplicity and responsiveness:
\begin{itemize}
  \item \textit{Exponentially Weighted Average} is recommended in high-volatility contexts, offering rapid adaptation and interpretability \parencite{KoopKorobilis2012}.
  \item \textit{Simple Average} remains appropriate for stable environments or when operational simplicity and lower computational complexity are prioritized.
  \item \textit{Weighted Average} provides a balanced intermediate solution when historical model performance maintains predictive relevance.
\end{itemize}
Each strategy significantly enhances resilience against shocks while providing interpretative insights into model effectiveness.

These findings substantiate that careful implementation of forecast combination methods---validated through rigorous empirical evaluation and robust statistical diagnostics---significantly enhances predictive accuracy, resilience, and operational transparency. By leveraging complementarities among diverse modeling frameworks, combined forecasting approaches prove highly effective for real-time decision-making, particularly under frequent structural changes or high uncertainty \parencite{HendryClements2004,Timmermann2006}.

\vspace{0.35cm}

\subsection{\textbf{\textit{Methodological and policymaking implications for macroeconomic nowcasting}}}
\label{subsec:method_policy_implications}

The empirical findings in this study offer meaningful insights into methodological practices and policy-oriented applications within macroeconomic nowcasting. This section synthesizes key implications from our analysis, addressing methodological robustness, model selection strategies, interpretability, uncertainty quantification, and strategic policymaking.

\vspace{0.40cm}

\phantomsection
\subsubsection{\textbf{\textit{Methodological Implications}}}
\smallskip
    \begin{itemize}
        \item \textit{Superiority over traditional benchmarks.} Across all macroeconomic regimes analyzed, advanced methods (penalized regression, dimensionality reduction, ensemble, and neural networks) consistently outperformed traditional benchmarks, including Random Walk, AR(3), and Dynamic Factor Models (DFM). Furthermore, formal predictive-ability tests \parencite[e.g.,][]{DieboldMariano1995,GiacominiWhite2006} indicate that these models deliver statistically significant improvements over standard references even under regime shifts, thereby reinforcing the reliability of the recommended modeling frameworks. This superiority validates investments in more sophisticated econometric techniques for enhancing nowcast accuracy \parencite{MakridakisHibon2000,MakridakisEtAl2020_M4,MakridakisEtAl2020_M4}.
        \item \textit{Regularization under volatility.} Penalized regression models, notably Ridge and Elastic Net \parencite{HoerlKennard1970,Tibshirani1996,ZouHastie2005,DeMolGiannoneReichlin2008,HastieTibshiraniFriedman2009,SmeekesWijler2018,UematsuTanaka2019,KimSwanson2018}, demonstrated consistently robust performance during severe economic disruptions, such as the COVID-19 shock. These regularized approaches proved essential in maintaining prediction accuracy and reliability by effectively controlling overfitting and managing extensive macroeconomic datasets. Nevertheless, Elastic Net, characterized by narrower prediction intervals, occasionally risked insufficient coverage of extreme outcomes, highlighting a critical trade-off between precision and comprehensive coverage.
        \item \textit{Dimensionality reduction and structural adaptability.} Principal Component Regression (PCR) revealed significant resilience during crisis periods, despite performance deterioration post-crisis. Conversely, Partial Least Squares Regression (PLSR) experienced immediate accuracy reductions during shocks, though it recovered more swiftly post-crisis. This differential performance underscores the necessity of aligning dimensionality-reduction methods with specific nowcasting objectives, such as initial shock resistance versus accelerated post-crisis normalization \parencite{Jolliffe1982,Wold1985,KraemerSugiyama2011,FuentesPoncelaRodriguez2015}.
        \item \textit{Flexibility--interpretability trade-off.} Neural network models, particularly recurrent architectures such as Gated Recurrent Unit \parencite{ChoEtAl2014,HewamalageBergmeirBandara2021,AlmosovaAndresen2023}, exhibited superior adaptability to nonlinearities and complex temporal structures compared to Multilayer Perceptrons and linear or ensemble methods (XGBoost, Random Forest). Due to its gating mechanism, Gated Recurrent Unit can store and update relevant historical signals more effectively \parencite{ChungGulcehreChoBengio2014}, which proves advantageous when abrupt economic shocks occur. However, this enhanced flexibility comes at the cost of reduced interpretability. Therefore, in such circumstances it is necessary to adopt techniques that allow greater transparency and comprehensibility of the results obtained, for instance Explainable Artificial Intelligence (XAI) techniques (e.g., Integrated gradients, time-series variants of SHapley Additive exPlanations).
        \item \textit{Prediction intervals and uncertainty management.} Penalized regression methods generated comparatively narrower prediction intervals than neural or ensemble approaches, which produced broader and sometimes overly conservative intervals during periods of economic turbulence. Nevertheless, while these methods excel at capturing complex nonlinearities, an overly cautious approach to interval width in neural and ensemble frameworks may occasionally compromise the practical utility of nowcasts, highlighting the importance of carefully balancing coverage and precision in operational settings. In addition, adopting a block bootstrap or related resampling approaches attuned to temporal dependence \parencite{PolitisRomano1994,Li2021} can enhance the reliability of these intervals, particularly in the presence of structural breaks. Future model selection should explicitly consider policymakers' risk tolerance and contextual requirements, balancing interval precision and adequate coverage. Furthermore, incorporating scoring rules that penalize interval under- and over-coverage \parencite{GneitingRaftery2007,GneitingKatzfuss2014} may refine interval evaluation and increase their practical utility.
        \item \textit{Feature importance and interpretability.} Key economic indicators---industrial production, gross operating surplus, labor market variables, and air transport metrics---consistently emerged as central predictors across models, emphasizing their structural relevance. Employing XAI techniques, such as Integrated Gradients \parencite{SundararajanTalyYan2017,Molnar2022}, maintained interpretability even within complex models, facilitating a clearer understanding of underlying economic drivers.
        \item \textit{Forecast combination and model stability.} Our results reinforce established evidence that forecast aggregation---through Simple Average, Weighted Average, and Exponentially Weighted Average---effectively mitigates individual model biases \parencite{BatesGranger1969,SmithWallis2009,Clemen1989,Armstrong2001,Timmermann2006,Atiya2020}. The preliminary use of the Model Confidence Set procedure further improved predictive stability, particularly under volatile macroeconomic conditions \parencite{HansenLundeNason2011,AiolfiTimmermann2006,GenreEtAl2013}.
        \end{itemize}

\vspace{0.40cm}
\phantomsection
\subsubsection{\textbf{\textit{Policy-making Implications}}}
\smallskip
    \begin{itemize}
        \item \textit{Enhanced decision-making in crisis contexts.} Forecast combination approaches, particularly EWA methods, proved crucial in delivering robust and timely nowcasts during high-volatility phases. Policymakers can rely on these forecasts to formulate rapid, targeted counter-cyclical monetary or fiscal interventions \parencite{Clemen1989,HendryClements2004}.
        \item \textit{Prioritizing key economic indicators.} Given Singapore's status as a small, open, and highly industrialized economy \parencite{IMF2024_Singapore}, consistent evidence underscores the predictive centrality of core macroeconomic indicators---industrial production, labor costs, and air cargo volumes. Policymakers should prioritize frequent and granular monitoring of these variables, as timely and accurate data are essential for proactive and effective policy responses \parencite{BanburaRunstler2011,HakimMerkert2016}.
        \item \textit{Modular forecasting and policy dashboards.} Given the variability in model performance across macroeconomic regimes, policymakers should adopt integrated forecasting dashboards that aggregate predictions from diverse methodological frameworks \parencite{BokEtAl2018}. Such dashboards enable clear differentiation between temporary sector-specific disruptions and broader systemic crises, optimizing policy targeting and responsiveness through frequent recalibration \parencite{KoopKorobilis2012,GenreEtAl2013}.
        \item \textit{Transparency and stakeholder communication.} Employing interpretable methodologies and communicating feature importance insights can significantly enhance public and market confidence in policy decisions. Transparent explanation of forecast drivers strengthens institutional credibility, reduces market uncertainty \parencite{BlinderEtAl2008}, and facilitates coordinated economic actions among stakeholders.
        \item \textit{Preparedness for future economic shocks.} The COVID-19 experience highlights the necessity for forecasting frameworks capable of rapid adaptation to structural changes. Strategically employing dynamic weighting mechanisms such as EWA allows rapid recalibration, ensuring forecasting models remain resilient and actionable in future unexpected economic disturbances. Robust and timely forecasts facilitate more efficient allocation of resources and mitigate potential impacts of recessions or financial instability \parencite{BarbagliaEtAl2023}. These findings provide robust methodological guidance and actionable insights for policymakers and practitioners involved in macroeconomic nowcasting, promoting enhanced stability, transparency, and adaptability in economic forecasting and policymaking processes.
    \end{itemize}

\cleardoublepage %

%% file: 03_chapters/chapter4.tex
\chapter{Conclusions}
\label{ch:conclusions}

\section{\textbf{Summary of Contributions and Main Findings}}
\label{sec:4.1}
This dissertation systematically evaluates and advances macroeconomic nowcasting methodologies by integrating state-of-the-art econometric techniques with machine and deep learning approaches. It offers policymakers enhanced predictive tools for navigating economic uncertainty.

Unlike existing studies, which primarily focus on single-model families, our research comparatively assesses penalized regression (LASSO, Ridge, Elastic Net), dimensionality-reduction techniques (PCR, PLSR), ensemble algorithms (Random Forest, XGBoost), and neural network architectures (MLP, GRU).

An essential innovation lies in developing a robust nowcasting pipeline explicitly designed to mitigate look-ahead bias, a frequent source of overly optimistic performance estimates in macroeconomic nowcasting. This methodological framework combines iterative expanding and rolling windows with Bayesian hyperparameter optimization and pruning, effectively balancing dynamic adaptability and predictive stability.

Additionally, we introduce systematic quantification of predictive uncertainty through block bootstrap intervals, responding directly to recent calls for robust prediction methodologies in the presence of structural instabilities and economic shocks. This advancement provides critical insights into the reliability and confidence of macroeconomic nowcasts.

Another significant contribution involves enhancing interpretability in complex macroeconomic ML models using Explainable AI techniques, specifically Integrated Gradients. This innovation fills critical interpretability gaps in recent literature, enabling policymakers to derive actionable insights from traditionally opaque models.

Empirically, our study leverages Singapore's characteristics---small, open, advanced, and highly diversified---as an informative testbed. By analyzing model performances across distinct macroeconomic regimes (Pre-COVID, COVID, Post-COVID), we empirically validate the superior adaptability of advanced econometric and ML models relative to traditional benchmarks, directly addressing recent methodological concerns on robustness and adaptability. Our findings indicate substantial predictive improvements---typically ranging between 40\% and 60\% in terms of RMSFE reductions---over widely established benchmarks such as Random Walk, AR(3), and Dynamic Factor Models. These results underline the empirical validity and practical advantage of our integrated methodological framework.

Furthermore, we introduce a nowcasting combination approach integrating Model Confidence Set procedures with dynamic averaging strategies (Exponentially Weighted Average), significantly enhancing aggregate nowcast accuracy. This method provides critical insights regarding dynamic model performances under shifting economic conditions.

Finally, extensive diagnostic checks (Shapiro-Wilk, Ljung-Box) and formal predictive-ability tests (Diebold-Mariano, Giacomini-White) underscore the adopted approach's methodological rigor and empirical robustness, fully aligning with the rigorous analytical standards required by leading nowcasting literature.

Our work contributes substantially to macroeconomic nowcasting by introducing methodological innovations, robust interpretability frameworks, rigorous uncertainty quantification, extensive empirical validation, and practical nowcasting solutions tailored for effective policy formulation.

\vspace{0.35cm}

\section{\textbf{Methodological and Policy-making Implications}}
\label{sec:4.2}
Leveraging the empirical insights detailed in Chapter 3, we synthesize key implications for methodological practices and policy-oriented applications in macroeconomic nowcasting, explicitly addressing issues of methodological robustness, model selection strategies, interpretability, uncertainty quantification, and strategic policymaking.

\vspace{0.35cm}

\subsection{\textbf{\textit{Methodological Implications}}}
\label{subsec:methodological_implications}

    \begin{itemize}
        \item \textit{Superiority over traditional benchmarks:} Across all macroeconomic regimes analyzed, advanced methods (penalized regression, dimensionality reduction, ensemble, and neural networks) consistently outperformed traditional benchmarks, including Random Walk, AR(3), and Dynamic Factor Models (DFM). Furthermore, formal predictive-ability tests \parencite[e.g.,][]{DieboldMariano1995,GiacominiWhite2006} indicate that these models deliver statistically significant improvements over standard references even under regime shifts, thereby reinforcing the reliability of the recommended modeling frameworks. This superiority validates investments in more sophisticated econometric techniques for enhancing nowcast accuracy \parencite{MakridakisHibon2000,MakridakisEtAl2020_M4}.
        \item \textit{Regularization under volatility:} Penalized regression models, notably Ridge and Elastic Net \parencite{HoerlKennard1970,Tibshirani1996,ZouHastie2005,DeMolGiannoneReichlin2008,HastieTibshiraniFriedman2009,SmeekesWijler2018,UematsuTanaka2019,KimSwanson2018}, demonstrated consistently robust performance during severe economic disruptions, such as the COVID-19 shock. These regularized approaches proved essential in maintaining prediction accuracy and reliability by effectively controlling overfitting and managing extensive macroeconomic datasets. Nevertheless, Elastic Net, characterized by narrower prediction intervals, occasionally risked insufficient coverage of extreme outcomes, highlighting a critical trade-off between precision and comprehensive coverage.
        \item \textit{Dimensionality reduction and structural adaptability:} Principal Component Regression (PCR) revealed significant resilience during crisis periods, despite performance deterioration post-crisis. Conversely, Partial Least Squares Regression (PLSR) experienced immediate accuracy reductions during shocks, though it recovered more swiftly post-crisis. This differential performance underscores the necessity of aligning dimensionality-reduction methods with specific nowcasting objectives, such as initial shock resistance versus accelerated post-crisis normalization \parencite{Jolliffe1982,Wold1985,KraemerSugiyama2011,FuentesPoncelaRodriguez2015}.
        \item \textit{Flexibility--interpretability trade-off:} Neural network models, particularly recurrent architectures such as Gated Recurrent Unit \parencite{ChoEtAl2014,HewamalageBergmeirBandara2021,AlmosovaAndresen2023}, exhibited superior adaptability to nonlinearities and complex temporal structures compared to Multilayer Perceptrons and linear or ensemble methods (XGBoost, Random Forest). Due to its gating mechanism, Gated Recurrent Unit can store and update relevant historical signals more effectively \parencite{ChungGulcehreChoBengio2014}, which proves advantageous when abrupt economic shocks occur. However, this enhanced flexibility comes at the cost of reduced interpretability. Therefore, in such circumstances it is necessary to adopt techniques that allow greater transparency and comprehensibility of the results obtained, for instance Explainable Artificial Intelligence (XAI) techniques (e.g., Integrated gradients, time-series variants of SHapley Additive exPlanations).
        \item \textit{Prediction intervals and uncertainty management:} Penalized regression methods generated comparatively narrower prediction intervals than neural or ensemble approaches, which produced broader and sometimes overly conservative intervals during periods of economic turbulence. Nevertheless, while these methods excel at capturing complex nonlinearities, an overly cautious approach to interval width in neural and ensemble frameworks may occasionally compromise the practical utility of nowcasts, highlighting the importance of carefully balancing coverage and precision in operational settings. In addition, adopting a block bootstrap or related resampling approaches attuned to temporal dependence \parencite{PolitisRomano1994,Li2021} can enhance the reliability of these intervals, particularly in the presence of structural breaks. Future model selection should explicitly consider policymakers' risk tolerance and contextual requirements, balancing interval precision and adequate coverage. Furthermore, incorporating scoring rules that penalize interval under- and over-coverage \parencite{GneitingRaftery2007,GneitingKatzfuss2014} may refine interval evaluation and increase their practical utility.
        \item \textit{Feature importance and interpretability:} Key economic indicators---industrial production, gross operating surplus, labor market variables, and air transport metrics---consistently emerged as central predictors across models, emphasizing their structural relevance. Employing XAI techniques, such as Integrated Gradients \parencite{SundararajanTalyYan2017,Molnar2022}, maintained interpretability even within complex models, facilitating a clearer understanding of underlying economic drivers.
        \item \textit{Forecast combination and model stability:} Our results reinforce established evidence that nowcast aggregation---through Simple Average, Weighted Average, and Exponentially Weighted Average---effectively mitigates individual model biases \parencite{BatesGranger1969,SmithWallis2009,Clemen1989,Armstrong2001,Timmermann2006,Atiya2020}. The preliminary use of the Model Confidence Set procedure further improved predictive stability, particularly under volatile macroeconomic conditions \parencite{HansenLundeNason2011,AiolfiTimmermann2006,GenreEtAl2013}.
        \end{itemize}

\vspace{0.35cm}

\subsection{\textbf{\textit{Policy-making Implications}}}
\label{subsec:policymaking_implications}

    \begin{itemize}
        \item \textit{Enhanced decision-making in crisis contexts:} Nowcast combination approaches, particularly EWA methods, proved crucial in delivering robust and timely nowcasts during high-volatility phases. Policymakers can rely on these nowcasts to formulate rapid, targeted counter-cyclical monetary or fiscal interventions \parencite{Clemen1989,HendryClements2004}.
        \item \textit{Prioritizing key economic indicators:} Given Singapore's status as a small, open, and highly industrialized economy \parencite{IMF2024_Singapore}, consistent evidence underscores the predictive centrality of core macroeconomic indicators---industrial production, labor costs, and air cargo volumes. Policymakers should prioritize frequent and granular monitoring of these variables, as timely and accurate data are essential for proactive and effective policy responses \parencite{BanburaRunstler2011,HakimMerkert2016}.
        \item \textit{Modular nowcasting and policy dashboards:} Given the variability in model performance across macroeconomic regimes, policymakers should adopt integrated nowcasting dashboards that aggregate predictions from diverse methodological frameworks \parencite{BokEtAl2018}. Such dashboards enable clear differentiation between temporary sector-specific disruptions and broader systemic crises, optimizing policy targeting and responsiveness through frequent recalibration \parencite{KoopKorobilis2012,GenreEtAl2013}.
        \item \textit{Transparency and stakeholder communication:} Employing interpretable methodologies and communicating feature importance insights can significantly enhance public and market confidence in policy decisions. Transparent explanation of nowcast drivers strengthens institutional credibility, reduces market uncertainty \parencite{BlinderEtAl2008}, and facilitates coordinated economic actions among stakeholders.
        \item \textit{Preparedness for future economic shocks:} The COVID-19 experience highlights the necessity for nowcasting frameworks capable of rapid adaptation to structural changes. Strategically employing dynamic weighting mechanisms such as EWA allows rapid recalibration, ensuring nowcasting models remain resilient and actionable in future unexpected economic disturbances. Robust and timely nowcasts facilitate more efficient allocation of resources and mitigate potential impacts of recessions or financial instability \parencite{BarbagliaEtAl2023}. These findings provide robust methodological guidance and actionable insights for policymakers and practitioners involved in macroeconomic nowcasting, promoting enhanced stability, transparency, and adaptability in economic nowcasting and policymaking processes.
    \end{itemize}

\vspace{0.35cm}

\section{\textbf{Limitations and Future Developments}}
\label{sec:4.3}
The methodologies presented significantly enhance macroeconomic nowcasting by integrating advanced econometric and machine learning approaches within a comprehensive nowcasting pipeline. Nevertheless, we explicitly acknowledge several critical limitations, clearly defining the scope of our findings and highlighting areas for further methodological refinements and empirical expansion.

\vspace{0.35cm}

\subsection{\textbf{\textit{Main Limitations}}}
\label{subsec:main_limitations}
We identify the following principal limitations:

\begin{itemize}
    \item \textit{Data frequency and granularity constraints:}
    Our empirical analysis primarily relied on monthly and quarterly data, limiting the systematic exploitation of higher-frequency indicators such as daily or weekly data. Furthermore, we did not explicitly incorporate Mixed Data Sampling (MIDAS) methods, potentially constraining predictive timeliness and responsiveness to rapidly evolving economic conditions. Prior research shows mixed-frequency (MIDAS) regressions can yield sizable accuracy gains when high-frequency information is available \parencite{AndreouGhyselsKourtellos2013,GhyselsMarcellino2016}.
    
    \item \textit{Limited temporal coverage:}
    Despite covering significant macroeconomic shocks\linebreak---including the COVID-19 pandemic---our evaluation period (2017 Q1-2023 Q2) remained relatively short. Consequently, this restricted the comprehensive assessment of model robustness and adaptability over extended economic cycles, raising caution concerning broader long-term generalizability. Longer evaluation windows are usually required to detect forecast breakdowns and gauge stability across structural regimes \parencite{GiacominiRossi2009,Rossi2021}.
    
    \item \textit{Geographical generalizability constraints:}
    Our empirical application exclusively focused on Singapore, which is characterized by its small, open, and advanced economic structure. Therefore, generalizing our proposed methodologies to structurally and institutionally diverse economies should be approached cautiously. Cross-country nowcasting evidence indicates that model performance can deteriorate when applied to economies with different data vintages or transmission mechanisms \parencite{GiannoneReichlinSmall2008,BokEtAl2018}.
    
    \item \textit{Operational complexity and computational feasibility:}
    The nowcasting pipeline integrated multiple sophisticated approaches—including penalized regression, dimensionality reduction techniques, ensemble learning methods, neural network models, sequential Bayesian optimization, extensive block-bootstrap procedures, and advanced interpretability techniques. Such methodological comprehensiveness entails significant computational complexity, posing practical challenges for real-time implementation, particularly within resource-constrained policy environments \parencite{BokEtAl2018,ShahriariEtAl2016}.
    
    \item \textit{Sensitivity to initial variable selection:}
    Although we employed rigorous variable selection methods, alternative predictor sets could significantly affect nowcasting stability and predictive performance, highlighting methodological sensitivity as a critical consideration. That mirrors well-documented evidence on the importance of predictor screening in high-dimensional macroeconomic forecasting \parencite{Ng2013,WangEtAl2023_variable_selection}.
    
    \item \textit{Reliability of predictive intervals under extreme events:}
    Despite robust uncertainty quantification via block-bootstrap methods, uncertainties persist regarding predictive interval reliability during unprecedented economic shocks. Additionally, our study did not explicitly account for uncertainties stemming from measurement errors, subsequent data revisions, or the general quality of input data. Recent work shows that conventional predictive intervals often under-cover during crises such as COVID-19 \parencite{Hansen2006,Li2021}.
\end{itemize}

\vspace{0.35cm}

\subsection{\textbf{\textit{Future Methodological and Empirical Developments}}}
\label{subsec:future_developments}
To systematically address these limitations and further advance macroeconomic nowcasting methodologies, we propose several future research avenues:

\begin{itemize}
    \item \textit{Leveraging alternative and granular data sources:} Integrating nontraditional datasets\linebreak---such as social media sentiment analysis, financial news-based uncertainty indices, granular transactional data, satellite imagery, high-frequency mobility indicators (e.g., Google Mobility), and web-scraped economic indicators---could markedly improve nowcast accuracy and responsiveness, thus providing a richer informational basis for timely policy interventions \parencite{BokEtAl2018,CoulombeEtAl2022}.
    \item \textit{Explicit integration of macro-financial risk indicators:}
    Future methodological advancements should explicitly integrate macro-financial stress indicators and systemic risk measures within the nowcasting framework. Such integration would enhance predictive capacity by proactively capturing and managing potential spillovers from financial markets to real economic activities.
    \item \textit{Sensitivity analyses and feature interaction assessments:}
    Detailed sensitivity analyses focusing on the interactive effects among predictor variables could provide valuable insights into underlying macroeconomic dynamics, enhancing predictive robustness and interpretive clarity.
    \item \textit{Extension to multi-step nowcasting horizons:}
    Future research should systematically evaluate model performance over extended nowcasting horizons (e.g., two-step-ahead nowcasts). Such assessments would strengthen operational applicability and test our methodological pipeline's robustness and practical utility for medium-term economic nowcasting scenarios.
    \item \textit{Regional comparative analysis and generalizability assessments:}
    Systematically extending our proposed methodology to Southeast Asian economies lacking sophisticated GDP nowcasting models—such as Brunei Darussalam, Laos, Myanmar, and Timor-Leste—would substantially enhance geographical generalizability. At the same time, several regional economies have recently developed GDP nowcasting frameworks (e.g., Indonesia \parencite{AlifatussaadahEtAl2019}, Malaysia \parencite{JasniEtAl2022,ShabliEtAl2023}, Hong Kong, Thailand, the Philippines, and Japan); however, these studies typically differ in scope, complexity, and forecasting objectives compared to our proposed pipeline. DBS Bank's nowcasting model for Singapore, China, India, and Indonesia focuses primarily on year-over-year (YoY) forecasts using an autoregressive framework with high-frequency proxies, rather than providing quarter-over-quarter (QoQ) nowcasts and comprehensive uncertainty quantification. Explicit comparative analyses between our integrated forecasting pipeline and existing regional models would yield meaningful insights, particularly regarding relative predictive performance, robustness, uncertainty management, and interpretability.
    \item \textit{Explicit integration of high-frequency data via MIDAS models:} Future studies should explicitly incorporate Mixed Data Sampling (MIDAS) methods, enabling systematic integration of daily or weekly indicators with quarterly GDP nowcasts. That would significantly enhance predictive responsiveness, timeliness, and accuracy \parencite[e.g., see][]{GhyselsMarcellino2016,AndreouGhyselsKourtellos2013}.
    \item \textit{Exploratory hybrid econometric-deep learning frameworks:}
    While recent research indicates limited utility of large language models (LLMs) and Transformer architectures for direct time-series forecasting tasks \parencite{TanEtAl2024}, carefully designed hybrid models that combine structural econometric approaches with selected deep-learning components (potentially attention-based architectures) remain a promising—but still exploratory—methodological avenue. Future research should rigorously investigate under which specific conditions, contexts, or model configurations such hybridizations might provide tangible advantages, ensuring that meaningful gains in forecasting performance and interpretability justify any added complexity.
    \item \textit{Comparative assessment between frequentist and Bayesian machine-learning approaches:}
    Explicit comparative analyses of predictive performance and uncertainty quantification between frequentist and Bayesian machine-learning frameworks would yield valuable methodological insights, informing optimal methodological choices for macroeconomic nowcasting.
    \item \textit{Automated real-time nowcasting with dynamic interpretability:}
    Designing fully automated nowcasting dashboards capable of continuous real-time updates and enhanced dynamic interpretability—via advanced Explainable AI (XAI) techniques such as dynamic SHAP values and adaptive Integrated Gradients—would greatly enhance transparency, interpretability, and practical usability for policymakers \parencite{Molnar2022,SundararajanTalyYan2017}.
    \item \textit{Advanced robustness and structural adaptability assessments:}
    Comprehensive Monte Carlo simulations and extensive scenario analyses would further validate nowcasting stability, adaptability, and predictive reliability across diverse macroeconomic conditions, including extreme and unprecedented economic shocks.
    \item \textit{Investigating determinants of predictive-interval characteristics:}
    Further research could systematically examine factors affecting predictive intervals, including interval-width variations, peak behavior, and plateau formations. Such analyses would substantially improve uncertainty-quantification reliability, robustness, and interpretability.
    \item \textit{Expansion of the nowcasting framework to additional macroeconomic indicators as target variables:}
    Future methodological applications could systematically extend our nowcasting pipeline to other critical macroeconomic indicators—including inflation rates and industrial production indices—significantly broadening our research's methodological relevance and practical policy impact.
\end{itemize}

\vspace{0.35cm}

\section{\textbf{Concluding Remarks}}
\label{sec:sec:4.4}
Addressing the explicitly identified limitations and rigorously pursuing the proposed future developments could enhance macroeconomic nowcasting methodologies' robustness, transparency, interpretability, and practical relevance. In particular, integrating high-frequency data, alternative data sources, and advanced hybrid modeling techniques will allow policymakers and practitioners to navigate economic uncertainties more precisely and confidently. Moreover, expanding methodological applications to additional economic indicators and conducting comparative analyses across diverse economic contexts could significantly broaden nowcasting frameworks' generalizability and empirical validity. Ultimately, such comprehensive methodological advancements promise to deliver timely, reliable, and actionable economic insights, thereby improving the quality and effectiveness of macroeconomic policy and decision-making processes in rapidly evolving and increasingly complex global economic environments.

\cleardoublepage %

%% file: 04_appendices/appendices.tex
\chapter*{Appendices}
\addcontentsline{toc}{chapter}{Appendices}
\label{ch:appendices}

\renewcommand{\thetable}{A.\arabic{table}} %
\setcounter{table}{0}

\clearpage
\chapter*{Appendix A. Variables and Descriptive Statistics Used in the GDP Nowcasting Models}
\thispagestyle{plain}
\addcontentsline{toc}{chapter}{A. Variables and Descriptive Statistics Used in the GDP Nowcasting Models}

\input{04_appendices/appendix_A.tex}

\cleardoublepage

%% file: 04_appendices/appendix_A.tex
\begin{landscape} %

\footnotesize %

\begin{longtable}{p{4.5cm} p{10cm} c}
\caption{Table A.1: Variables included in the nowcasting models for Singapore's quarterly GDP growth (categorization by economic indicator type)}
\label{tab:variable_list}\\

\toprule
\multicolumn{1}{c}{\textbf{Variable name}} &
\multicolumn{1}{c}{\textbf{Original name}} &
\multicolumn{1}{c}{\textbf{Category}} \\
\midrule
\endfirsthead

\toprule
\multicolumn{1}{c}{\textbf{Variable name}} &
\multicolumn{1}{c}{\textbf{Original name}} &
\multicolumn{1}{c}{\textbf{Category}} \\
\midrule
\endhead

\midrule
\multicolumn{3}{r}{\textit{Continued on the next page}}\\
\endfoot

\bottomrule
\endlastfoot

business entities formation change 
& Formation Of All Business Entities By Industry, Quartely (Number) 
& \leading \\

business entities cessation change 
& Cessation Of All Business Entities By Industry, Quartely (Number) 
& \leading \\

composite leading index change 
& Composite Leading Index (2015=100), Quarterly (Index) 
& \leading \\

public contracts awarded change 
& Contracts Awarded By Sector And Development Type, Quartely - Total Public Sector (Million Dollars) 
& \leading \\

private contracts awarded change 
& Contracts Awarded By Sector And Development Type, Quartely - Total Private Sector (Million Dollars) 
& \leading \\

landed properties supply change 
& Supply Of Private Residential Properties In The Pipeline By Development Status (End Of Period), Quarter - Total Landed Properties (Number Of Units) 
& \leading \\

non landed properties supply change 
& Supply Of Private Residential Properties In The Pipeline By Development Status (End Of Period), Quarter - Total Non-Landed Properties (Number Of Units) 
& \leading \\

business expectations manufacturing 
& Business Expectations Of The Manufacturing Sector - Forecast For Next 6 Months, Quarterly (In Percentage Terms) 
& \leading \\

business outlook services change 
& Business Expectations For The Services Sector - General Business Outlook For The Next 6 Months, Net Weighted Balance, Quarterly (In Percentage Terms) 
& \leading \\

services employment forecast change 
& Business Expectations For The Services Sector - Employment Forecast For The Next Quarter, Net Weighted Balance, Quarterly (In Percentage Terms) 
& \leading \\

services revenue forecast change 
& Business Expectations For The Services Sector - Operating Revenue Forecast For The Next Quarter, Net Weighted Balance, Quarterly (In Percentage Terms) 
& \leading \\

services outlook weighted change 
& Business Expectations For The Services Sector - General Business Outlook For Next 6 Months, Weighted \% Of Up, Same, Down, Quarterly (In Percentage Terms) 
& \leading \\

services revenue weighted change 
& Business Expectations For The Services Sector - Operating Revenue Forecast Next Quarter, Weighted \% Of Up, Same, Down, Quarterly (In Percentage Terms) 
& \leading \\

services employment weighted change 
& Business Expectations For The Services Sector - Employment Forecast Next Quarter, Weighted Percentages Of Up, Same, Down, Quarterly (In Percentage Terms) 
& \leading \\

\midrule
bop quarterly change 
& Singapore's Balance Of Payments, (BPM6 Format), Quarterly (Million dollars) 
& \coincident \\

household net worth change 
& Household Sector Balance Sheet (End Of Period), Quarterly - Household Net Worth (Million Dollars) 
& \coincident \\

unit labour cost change 
& Unit Labour Cost Index (2015 = 100), Quarterly (Index) 
& \coincident \\

acceleration of net employment flow 
& Acceleration of Net Employment Flow 
& \coincident \\

value added per worker change 
& Changes In Value Added Per Worker In Chained (2015) Dollars, By Industry (SSIC 2020), Quarterly (Per Cent) 
& \coincident \\

quartely recruitment rate change 
& Average Quartely Recruitment Rate, Quarterly (Per Cent) 
& \coincident \\

quartely resignation rate change 
& Average Quartely Resignation Rate, Quarterly (Per Cent) 
& \coincident \\

employee compensation change 
& Compensation Of Employees By Industry At Current Prices, Quarterly (Million Dollars) 
& \coincident \\

gfcf current prices change 
& Gross Fixed Capital Formation At Current Prices, Quarterly (Million Dollars) 
& \coincident \\

gross operating surplus change 
& Gross Operating Surplus By Industry At Current Prices, Quarterly (Million Dollars) 
& \coincident \\

taxes less subsidies on prod change 
& Other Taxes Less Subsidies On Production By Industry At Current Prices, Quarterly (Million Dollars) 
& \coincident \\

personal saving rate change 
& Personal Disposable Income At Current Prices, Quarterly - Personal Saving Rate (Per Cent) 
& \coincident \\

consumer price index change 
& Consumer Price Index (CPI), 2019 As Base Year, Quarterly (Index) 
& \coincident \\

hdb resale price index change 
& Housing And Development Board Resale Price Index (1Q2009 = 100), Quarterly (Index) 
& \coincident \\

property price index change 
& Property Price Index By Type (4th Quarter 1998 = 100), Quarterly (Index) 
& \coincident \\

public progress payments change 
& Progress Payments Certified By Sector And Development Type, Quartely - Total Public (Million Dollars) 
& \coincident \\

private progress payments change 
& Progress Payments Certified By Sector And Development Type, Quartely - Total Private Sector (Million Dollars) 
& \coincident \\

private properties available change 
& Available And Vacant Private Residential Properties (End Of Period), Quarterly - All Types Private Residential Properties Available (Number Of Units) 
& \coincident \\

private properties vacant change 
& Available And Vacant Private Residential Properties (End Of Period), Quarterly - All Types Private Residential Properties Vacant (Number Of Units) 
& \coincident \\

electricity generation change 
& Electricity Generation, Quartely (Gigawatt Hours) 
& \coincident \\

piped gas sales change 
& Piped Gas Sales, Quarterly (Million Kilowatt Hours) 
& \coincident \\

finance loans advances change 
& Loans And Advances Of Finance Companies (End Of Period), Quartely (Million Dollars) 
& \coincident \\

industrial production index change 
& Index Of Industrial Production (2019 = 100), Quarterly (Index) 
& \coincident \\

fnb services index change 
& Food \& Beverage Services Index, (2017 = 100), In Chained Volume Terms, Quarterly (Index) 
& \coincident \\

retail sales index change 
& Retail Sales Index, (2017 = 100), In Chained Volume Terms, Quarterly (Index) 
& \coincident \\

domestic trade index change 
& Domestic Wholesale Trade Index, (2017 = 100), In Chained Volume Terms, Quarterly (Index) 
& \coincident \\

foreign trade index change 
& Foreign Wholesale Trade Index, (2017 = 100), In Chained Volume Terms, Quarterly (Index) 
& \coincident \\

aircraft arrivals departures change 
& Civil Aircraft Arrivals And Departures, Passengers, Air Cargo Tonnage, Direct And Transhipment Tonnage And Mail, Changi Airport, Quartely - Total Aircraft Arrivals And Departures (Number) 
& \coincident \\

air passenger arrivals change 
& Air Passenger Arrivals By Region-Country Of Embarkation, Quartely - Total Passengers (Number) 
& \coincident \\

air mail discharge change 
& Air Cargo Discharged By Region-Country Of Origin, Quartely - Total Mail (Tonne) 
& \coincident \\

air cargo discharge change 
& Air Cargo Discharged By Region-Country Of Origin, Quartely (Tonne) 
& \coincident \\

air cargo loaded change 
& Air Cargo Loaded By Region-Country Of Destination, Quartely (Tonne) 
& \coincident \\

motor vehicle population change 
& Motor Vehicle Population By Type Of Vehicle (End Of Period), Quartely (Number) 
& \coincident \\

new vehicle registration change 
& New Registration Of Motor Vehicles Under Vehicle Quota System, Quartely - Total New Motor Vehicles Registered (Number) 
& \coincident \\

vehicle deregistration change 
& Motor Vehicles De-Registered Under Vehicle Quota System, Quartely (Number) 
& \coincident \\

vehicle population quota change 
& Motor Vehicle Population Under Vehicle Quota System, (As At End Of Period), Quartely (Number) 
& \coincident \\

vessel arrivals change 
& Sea Cargo And Shipping Statistics, Quartely - Vessel Arrivals (Number) 
& \coincident \\

vessel tonnage change 
& Sea Cargo And Shipping Statistics, Quartely - Vessel Arrivals - Shipping Tonnage (Thousand Gross Tonnes) 
& \coincident \\

sea total cargo change 
& Sea Cargo And Shipping Statistics, Quartely - Total Cargo (Thousand Tonnes) 
& \coincident \\

sea container throughput change 
& Sea Cargo And Shipping Statistics, Quartely - Total Container Throughput (Thousand Twenty-Foot Equivalent Units) 
& \coincident \\

bunker sales change 
& Sea Cargo And Shipping Statistics, Quartely - Bunker Sales (Thousand Tonnes) 
& \coincident \\

merchandise imports change 
& Merchandise Trade By Commodity Section, (At Current Prices), Quartely - Total Merchandise Imports, At Current Prices (Thousand Dollars) 
& \coincident \\

merchandise exports change 
& Merchandise Trade By Commodity Section, (At Current Prices), Quartely - Total Domestic Exports At Current Prices (Thousand Dollars) 
& \coincident \\

merchandise reexports change 
& Merchandise Trade By Commodity Section, (At Current Prices), Quartely - Total Re-Exports At Current Prices (Thousand Dollars) 
& \coincident \\

\midrule
import price index change 
& Import Price Index, Base Year 2018 = 100, Quartely - Overall Items (Index) 
& \exogenous \\

export price index change 
& Export Price Index, Base Year 2018 = 100, Quartely - Overall Items (Index) 
& \exogenous \\

domestic supply price change 
& Domestic Supply Price Index, Base Year 2018 = 100, Quartely - Domestic Supply Price Index - Overall Items (Index) 
& \exogenous \\

manufactured price index change 
& Singapore Manufactured Products Price Index, Base Year 2018 = 100, Quartely - Singapore Manufactured Products Price Index - Overall Items (Index) 
& \exogenous \\

government debt change 
& Government Debt, (End Of Period), Quarterly (Million Dollars) 
& \exogenous \\

government bond yield 5yr change 
& Current Banks Interest Rates (End Of Period), Quartely - Government Securities - 5-Year Bond Yield (Per Cent Per Annum) 
& \exogenous \\

government bond yield 2yr change 
& Current Banks Interest Rates (End Of Period), Quartely - Government Securities - 2-Year Bond Yield (Per Cent Per Annum) 
& \exogenous \\

government bond yield 1yr change 
& Current Banks Interest Rates (End Of Period), Quartely -Government Securities - 1-Year Bond Yield (Per Cent Per Annum) 
& \exogenous \\

exchange rate usd change 
& Exchange Rates, (Average For Period), Quartely - US Dollar (Singapore Dollar Per US Dollar) 
& \exogenous \\

exchange rate gbp change 
& Exchange Rates, (Average For Period), Quartely - Sterling Pound (Singapore Dollar Per Pound Sterling) 
& \exogenous \\

exchange rate chf change 
& Exchange Rates, (Average For Period), Quartely - Swiss Franc (Singapore Dollar Per Swiss Franc) 
& \exogenous \\

exchange rate jpy change 
& Exchange Rates, (Average For Period), Quartely - Japanese Yen (Singapore Dollar Per 100 Japanese Yen) 
& \exogenous \\

exchange rate myr change 
& Exchange Rates, (Average For Period), Quartely - Malaysian Ringgit (Singapore Dollar Per Malaysian Ringgit) 
& \exogenous \\

exchange rate hkd change 
& Exchange Rates, (Average For Period), Quartely - Hong Kong Dollar (Singapore Dollar Per Hong Kong Dollar) 
& \exogenous \\

exchange rate aud change 
& Exchange Rates, (Average For Period), Quartely - Australian Dollar (Singapore Dollar Per Australian Dollar) 
& \exogenous \\

exchange rate eur change 
& Exchange Rates, (Average For Period), Quartely - Euro (Singapore Dollar Per Euro) 
& \exogenous \\

exchange rate cny change 
& Exchange Rates, (Average For Period), Quartely - Renminbi (Singapore Dollar Per Renminbi) 
& \exogenous \\

foreign reserves change 
& Official Foreign Reserves (End Of Period), Quartely (Million Dollars) 
& \exogenous \\

temp max change 
& Air Temperature And Sunshine, Relative Humidity And Rainfall - Air Temperature Means Daily Maximum (Degree Celsius) 
& \exogenous \\

temp min change 
& Air Temperature And Sunshine, Relative Humidity And Rainfall - Air Temperature Means Daily Minimum (Degree Celsius) 
& \exogenous \\

total rainfall change 
& Air Temperature And Sunshine, Relative Humidity And Rainfall - Total Rainfall (Millimetre) 
& \exogenous \\

daily sunshine change 
& Air Temperature And Sunshine, Relative Humidity And Rainfall - Bright Sunshine Daily Mean (Hour) 
& \exogenous \\

mean humidity change 
& Air Temperature And Sunshine, Relative Humidity And Rainfall - 24 Hours Mean Relative Humidity (Per Cent) 
& \exogenous \\

\midrule
liquor releases change 
& Duty-Paid Releases Of Liquors, Quartely (Litre) 
& \otherVar \\

tobacco releases change 
& Duty-Paid Releases Of Tobacco, Petroleum, Compressed Natural Gas And Motor Vehicles, Quartely - Tobacco (Kilograms) 
& \otherVar \\

cpf due contributions change 
& Central Provident Fund Contributions, Withdrawals And Amount Due To Members, Quartely - Amount Due To Members (Million Dollars) 
& \otherVar \\

currency in circulation change 
& Currency In Circulation (End Of Period), Quartely - Gross Circulation (Million Dollars) 
& \otherVar \\

finance assets liabilities change 
& Assets And Liabilities Of Finance Companies (End Of Period), Quartely - Assets(=Liabilities) Of Finance Companies (Million Dollars) 
& \otherVar \\

pawnshop pledges received change 
& Pledges At Pawnshops, Quartely - Number Of Pledges Received (Number) 
& \otherVar \\

pawnshop pledges redeemed change 
& Pledges At Pawnshops, Quartely - Number Of Pledges Redeemed (Number) 
& \otherVar \\

pawnshop loans given change 
& Pledges At Pawnshops, Quartely - Amount Of Loans Given Out (Million Dollars) 
& \otherVar \\

pawnshop loans redeemed change 
& Pledges At Pawnshops, Quartely - Amount Of Loans Redeemed Including Interest (Million Dollars) 
& \otherVar \\

registry ships count change 
& Sea Cargo And Shipping Statistics, Quartely - Singapore Registry Of Ships, End Of Period (Number) 
& \otherVar \\

registry ships tonnage change 
& Sea Cargo And Shipping Statistics, Quartely - Singapore Registry Of Ships (End Of Period) ('000 GT - Thousand Gross Tonnes) 
& \otherVar \\

live births change 
& Live-Births By Birth Order, Quarterly 
& \otherVar \\

deaths by ethnicity change 
& Deaths By Ethnic Group And Sex, Quartely 
& \otherVar \\

hospital admissions change 
& Admissions To Public Sector Hospitals, Quartely 
& \otherVar \\

\midrule
lag1 deflated gdp change 
& Lag of Order 1 of Singapore's GDP growth 
& \laggedVar \\

lag2 deflated gdp change 
& Lag of Order 2 of Singapore's GDP growth 
& \laggedVar \\

lag3 deflated gdp change 
& Lag of Order 3 of Singapore's GDP growth 
& \laggedVar \\

lag4 deflated gdp change 
& Lag of Order 4 of Singapore's GDP growth 
& \laggedVar \\

\midrule
seasonality q1 
& Dummy variable tracking of seasonality effect 
& \dummyVar \\

seasonality q2 
& Dummy variable tracking of seasonality effect 
& \dummyVar \\

seasonality q3 
& Dummy variable tracking of seasonality effect 
& \dummyVar \\

neg shock dummy 
& Dummy variable tracking past negative shocks of Singapore's GDP quarter over quarter growth 
& \dummyVar \\

pos shock dummy 
& Dummy variable tracking past positive shocks of Singapore's GDP quarter over quarter growth 
& \dummyVar \\

\specialrule{0.05pt}{1.0pt}{1.0pt} %

\rowcolor[rgb]{ .851,  .851,  .851}
deflated gdp change 
& Deflated Singapore's GDP quarter over quarter growth 
& Target Variable \\

\end{longtable}

\end{landscape}

\begin{landscape}
\fontsize{7.5pt}{9pt}\selectfont
\setlength{\LTcapwidth}{\linewidth} %

\setlength{\tabcolsep}{2pt}
\begin{longtable}{
  p{0.13\linewidth}
  *{10}{S[
    table-format=7.3,
    table-number-alignment=center,
    table-text-alignment=center,
    table-column-width=0.088\linewidth
  ]}
}

\caption{Table A.2: Descriptive statistics of variables employed in GDP growth nowcasting for Singapore (1990 Q1 –2023 Q2)}%
\label{tab:desc_stats}\\

\toprule
\textbf{Variable} &
\multicolumn{1}{c}{\textbf{Mean}} &
\multicolumn{1}{c}{\textbf{SD}} &
\multicolumn{1}{c}{\textbf{Min}} &
\multicolumn{1}{c}{\textbf{Q1}} &
\multicolumn{1}{c}{\textbf{Q2}} &
\multicolumn{1}{c}{\textbf{Q3}} &
\multicolumn{1}{c}{\textbf{Max}} &
\multicolumn{1}{c}{\textbf{Skewness}} &
\multicolumn{1}{c}{\textbf{Kurtosis}} \\
\midrule
\endfirsthead

\toprule
\textbf{Variable} &
\multicolumn{1}{c}{\textbf{Mean}} &
\multicolumn{1}{c}{\textbf{SD}} &
\multicolumn{1}{c}{\textbf{Min}} &
\multicolumn{1}{c}{\textbf{Q1}} &
\multicolumn{1}{c}{\textbf{Q2}} &
\multicolumn{1}{c}{\textbf{Q3}} &
\multicolumn{1}{c}{\textbf{Max}} &
\multicolumn{1}{c}{\textbf{Skewness}} &
\multicolumn{1}{c}{\textbf{Kurtosis}} \\
\midrule
\endhead

\midrule
\multicolumn{11}{r}{\textit{Continued on the next page}}\\
\endfoot

\bottomrule
\endlastfoot

bop\_quarterly\_change & -387.8546752 & 4261.953956 & -49145 & -83.9121993 & -27.6328304 & 70.26038485 & 2361.275755 & -11.42500355 & 131.6617152 \\
unit\_labour\_cost\_change & 0.613493083 & 8.874130893 & -27.72727273 & -6.568417664 & 1.013525082 & 6.631683725 & 17.3553719 & -0.126063224 & -0.381197982 \\
acceleration\_of\_net\_employment\_flow & 9.017399902 & 236.6944683 & -750 & -50.24193549 & -4.613942248 & 35.74973877 & 1700 & 3.488161268 & 25.55836331 \\
value\_added\_per\_worker\_change & 2.270895522 & 4.824433811 & -13 & 0.125 & 1.8 & 4.7 & 19.4 & -0.106163627 & 1.878226406 \\
monthly\_recruitment\_rate\_change & 0.313098592 & 12.5835755 & -36.84210526 & -10 & 0 & 9.303977273 & 33.33333333 & -0.00533343 & -0.170388208 \\
monthly\_resignation\_rate\_change & -0.05525944 & 12.74659899 & -23.52941176 & -11.11111111 & 0 & 10.52631579 & 28.57142857 & 0.109561892 & -0.976900118 \\
composite\_leading\_index\_change & 0.5869634 & 2.350141427 & -9.448818898 & -0.671230441 & 0.475580344 & 1.745476974 & 7.304785894 & -0.070099024 & 2.461584586 \\
employee\_compensation\_change & 1.993777583 & 9.200423498 & -12.23163335 & -6.655604016 & 2.533805821 & 7.527427754 & 22.20082044 & 0.313762723 & -0.808424075 \\
gfcf\_current\_prices\_change & 1.62308286 & 7.016882361 & -27.79906402 & -2.722006996 & 1.893342122 & 5.717518446 & 24.89699483 & -0.112840434 & 2.326867898 \\
gross\_operating\_surplus\_change & 2.117017751 & 7.512638937 & -13.97950709 & -3.68420611 & 1.3612172 & 6.749015413 & 22.60349157 & 0.322347537 & -0.547974335 \\
taxes\_less\_subsidies\_on\_prod\_change & -2.42243572 & 133.6707103 & -677.9702366 & -21.3667892 & 3.883421718 & 14.3928386 & 1148.826816 & 3.137651231 & 48.30100071 \\
consumer\_price\_index\_change & 0.457796233 & 0.639915586 & -1.209926877 & 0.005658907 & 0.438015708 & 0.807985187 & 2.276703058 & 0.326830706 & 0.63728949 \\
import\_price\_index\_change & -0.03331002 & 2.742091273 & -11.59826955 & -1.305820179 & -0.014834849 & 1.216235501 & 10.41239707 & -0.348318822 & 3.656777723 \\
export\_price\_index\_change & -0.350189543 & 2.576767469 & -9.634308084 & -1.633381852 & -0.268828137 & 0.799845362 & 7.824129334 & 0.094362826 & 1.567142525 \\
property\_price\_index\_change & 0.328715501 & 4.696657532 & -12.48999199 & -2.669550926 & 0.380466877 & 2.359447245 & 15.77092511 & 0.250005774 & 1.031299029 \\
domestic\_supply\_price\_change & 0.104788373 & 3.795639587 & -20.16244089 & -1.612827847 & 0.265450935 & 2.180952486 & 11.75079678 & -1.060854289 & 5.948642557 \\
manufactured\_price\_index\_change & -0.249255158 & 2.9737426 & -16.75951291 & -1.938462725 & -0.140074251 & 1.359217122 & 6.377791344 & -1.056888203 & 6.264425634 \\
liquor\_releases\_change & 1.441173092 & 13.19527 & -35.01393356 & -7.245633031 & 2.249524877 & 9.128936175 & 46.16886459 & 0.021213954 & 0.892901959 \\
government\_debt\_change & 2.426707258 & 2.105709554 & -3.449687278 & 1.444290385 & 2.377933072 & 3.099616064 & 13.05155206 & 1.596237894 & 8.090371641 \\
public\_contracts\_awarded\_change & 12.07283249 & 59.29169441 & -66.69839342 & -21.96188835 & 3.498837324 & 29.83639349 & 400.2252796 & 2.810441113 & 13.84828564 \\
private\_contracts\_awarded\_change & 8.312892052 & 41.72095681 & -69.52374239 & -21.49320102 & -1.524659906 & 24.86882985 & 160.0840767 & 1.199261554 & 1.627754459 \\
public\_progress\_payments\_change & 2.396484479 & 13.85959591 & -64.95423954 & -4.331322464 & 3.035643917 & 8.719685296 & 79.64739435 & 0.229991578 & 10.24376976 \\
private\_progress\_payments\_change & 2.164398553 & 11.33698085 & -45.50541923 & -4.468883482 & 1.555718442 & 8.935350625 & 65.32621851 & 0.752289991 & 8.067802379 \\
private\_properties\_available\_change & 1.592232759 & 5.141993402 & -0.072419053 & 0.540828485 & 1.068069247 & 1.55788744 & 59.38317182 & 10.87734017 & 122.522878 \\
private\_properties\_vacant\_change & 1.950766729 & 8.68090058 & -19.92243051 & -3.454811258 & 0.818251722 & 7.6487046 & 33.21793003 & 0.39759581 & 0.570662064 \\
business\_expectations\_manufacturing & 6.417910448 & 17.21177185 & -57 & -2 & 7 & 20 & 39 & -0.851916101 & 1.385690736 \\
cpf\_due\_contributions\_change & 2.069676288 & 1.131382139 & -5.8710217 & 1.794646518 & 2.196864189 & 2.577191059 & 4.17654773 & -2.760962835 & 17.53109951 \\
currency\_in\_circulation\_change & 1.696761705 & 3.201686981 & -12.87625536 & 0 & 1.080103595 & 2.895860174 & 19.68883985 & 1.176338861 & 9.98656581 \\
government\_bond\_yield\_5yr\_change & 2.193732724 & 24.76856116 & -46.46464646 & -9.700887393 & -0.131578948 & 9.216573419 & 116.6666667 & 1.778810036 & 5.552334038 \\
government\_bond\_yield\_2yr\_change & 4.45227576 & 32.7049413 & -67.44186047 & -13.9506932 & -1.475347934 & 17.71209634 & 150 & 1.389850095 & 3.886011299 \\
government\_bond\_yield\_1yr\_change & 3.918579326 & 33.06279268 & -100 & -13.37805966 & 0 & 18.57142857 & 126.4705882 & 0.611371902 & 1.952264035 \\
exchange\_rate\_usd\_change & -0.25282086 & 2.287330847 & -4.923088785 & -1.877587899 & -0.443597568 & 1.1123325 & 7.258227111 & 0.655167564 & 0.609630642 \\
exchange\_rate\_gbp\_change & -0.398657986 & 3.316769985 & -16.02513148 & -1.840728469 & -0.241572683 & 1.273826184 & 9.557800163 & -0.960220479 & 4.263060965 \\
exchange\_rate\_chf\_change & 0.20682673 & 3.266219471 & -9.729270305 & -1.616819844 & 0.080831797 & 2.25014467 & 11.44437779 & 0.160460911 & 1.69708289 \\
exchange\_rate\_jpy\_change & -0.165501635 & 4.098385683 & -10.99710312 & -2.411426896 & -0.363566208 & 1.873748341 & 19.26205322 & 0.81734793 & 3.46849996 \\
exchange\_rate\_myr\_change & -0.634455549 & 2.304754818 & -14.53102721 & -1.588053517 & -0.256233462 & 0.632058242 & 3.618730549 & -2.058295658 & 9.93055277 \\
exchange\_rate\_hkd\_change & -0.255019922 & 2.312941655 & -4.943988051 & -1.959493118 & -0.409841468 & 1.030760044 & 7.337380746 & 0.711587493 & 0.759767785 \\
exchange\_rate\_aud\_change & -0.326687086 & 3.64535156 & -19.36064957 & -2.004432502 & -0.239567885 & 1.257286717 & 11.26190397 & -0.66525355 & 5.301718722 \\
exchange\_rate\_eur\_change & -0.25557265 & 3.12739253 & -10.17295597 & -2.074079591 & -0.160268324 & 1.633082652 & 10.85062799 & 0.069086819 & 1.442327403 \\
exchange\_rate\_cny\_change & -0.624201739 & 4.101480839 & -33.2029719 & -1.764933223 & -0.284254551 & 1.029660184 & 7.682505717 & -4.654516916 & 33.96691724 \\
finance\_loans\_advances\_change & 0.734981564 & 4.361525136 & -25.63117208 & -1.051710244 & 0.980724703 & 3.480693722 & 9.710162545 & -2.669590965 & 13.76609904 \\
finance\_assets\_liabilities\_change & 0.487954517 & 4.356243665 & -28.94661128 & -0.916887409 & 0.535627788 & 2.694332103 & 8.964255397 & -2.972455314 & 17.4506577 \\
pawnshop\_pledges\_received\_change & 0.548312652 & 6.528806046 & -33.53887578 & -2.639146593 & 0.547172897 & 2.452719963 & 39.69346386 & 0.799030402 & 14.41722985 \\
pawnshop\_loans\_given\_change & 2.011362284 & 7.639137422 & -28.95247008 & -0.68725486 & 1.063312109 & 3.181072657 & 40.42152804 & 1.80345055 & 10.73112406 \\
business\_entities\_formation\_change & 1.236486053 & 9.090412555 & -28.47110461 & -4.015830028 & 0.164896204 & 6.257581669 & 39.75637609 & 0.494614477 & 2.957971339 \\
business\_entities\_cessation\_change & 2.409815525 & 18.25649686 & -46.43059367 & -6.866435642 & -0.074536072 & 7.253464435 & 81.27001067 & 1.619108741 & 5.25345683 \\
industrial\_production\_index\_change & 1.511398708 & 6.325005299 & -14.22104632 & -2.149571894 & 0.553769163 & 5.294913637 & 21.42887375 & 0.254491126 & 0.33625133 \\
fnb\_services\_index\_change & 0.305874458 & 7.730516959 & -43.36268297 & -2.479774508 & 0.373705034 & 2.793642595 & 49.38904842 & 0.526878792 & 18.84645464 \\
retail\_sales\_index\_change & 0.981481919 & 8.668689555 & -37.59801805 & -3.785178186 & 0.413191147 & 5.899290757 & 53.78073211 & 0.898765123 & 12.23022401 \\
aircraft\_arrivals\_departures\_change & 1.881354371 & 10.04073711 & -82.71686279 & -0.115661425 & 1.721195246 & 3.388619738 & 39.62816652 & -4.179355805 & 39.80373553 \\
air\_passenger\_arrivals\_change & 6.003984419 & 28.42520266 & -99.1134486 & -1.108301444 & 2.430834018 & 5.118626344 & 185.6741961 & 3.724287545 & 22.60459036 \\
air\_mail\_discharge\_change & 1.568283302 & 15.37647453 & -48.05994662 & -8.02317344 & 0.575225589 & 9.725477178 & 51.76470588 & 0.399675597 & 0.905491696 \\
air\_cargo\_loaded\_change & 1.01930191 & 7.765828576 & -30.75185317 & -4.720149486 & 2.750747164 & 5.941706297 & 20.90275015 & -0.759665672 & 1.160536795 \\
new\_vehicle\_registration\_change & 3.744418608 & 36.37058598 & -79.78728324 & -6.945841404 & -1.215238046 & 6.572880136 & 383.8709677 & 8.641513111 & 91.08260832 \\
vehicle\_deregistration\_change & 2.897811955 & 20.34068456 & -71.26154094 & -5.615022377 & 1.375985576 & 10.40447478 & 130.3612891 & 1.912901227 & 14.84979686 \\
merchandise\_imports\_change & 1.486421583 & 6.716990794 & -20.82507867 & -2.536578328 & 2.057601303 & 6.118227995 & 14.62051005 & -0.57127965 & 0.785019906 \\
merchandise\_exports\_change & 1.438532504 & 7.519382594 & -25.07259868 & -3.611593927 & 1.553458455 & 5.701334767 & 24.19984429 & -0.306972915 & 1.000999499 \\
merchandise\_reexports\_change & 1.928216 & 6.819794891 & -16.58520998 & -2.748901021 & 1.369840661 & 6.601902315 & 25.75602655 & 0.105666303 & 0.771065973 \\
seasonality\_q1 & 0.253731343 & 0.436778486 & 0 & 0 & 0 & 0.75 & 1 & 1.144745027 & -0.700235743 \\
seasonality\_q2 & 0.253731343 & 0.436778486 & 0 & 0 & 0 & 0.75 & 1 & 1.144745027 & -0.700235743 \\
seasonality\_q3 & 0.246268657 & 0.432453521 & 0 & 0 & 0 & 0 & 1 & 1.191229786 & -0.590003067 \\
neg\_shock\_dummy & 0.059701493 & 0.237822011 & 0 & 0 & 0 & 1 & 1 & 3.758858954 & 12.31258675 \\
pos\_shock\_dummy & 0.037313433 & 0.190239913 & 0 & 0 & 0 & 0 & 1 & 4.937943499 & 22.72223746 \\
deflated\_gdp\_change & 1.336099115 & 2.740469915 & -11.23086635 & -0.398795114 & 1.384522168 & 3.060714521 & 9.143570053 & -0.640084835 & 2.781360371 \\

\end{longtable}

\end{landscape}